# Writing literature reviews with AI: principles, hurdles and some lessons learned

Saadi LAHLOU (1,2), Annabelle GOUTTEBROZE (1), Atrina ORAEE (1), Julian MADERA (1)

(1) London School of Economics and Political Science (2) Paris Institute for Advanced Study

## *Abstract*

We qualitatively compared literature reviews produced with varying degrees of AI assistance. The same LLM, with same corpus of 280 papers but different selections produced dramatically different reviews -from mainstream and politically neutral to critical and post-colonial- though neither orientation was intended. LLM outputs always appear at first glance to be well written, well informed and thought out, but closer reading reveals gaps, biases and lack of depth. Our comparison of six versions shows pitfalls and suggests precautions necessary when using AI assistance to make a literature review. Main issues are:

1. The bias of ignorance (you don't know what you don't get) in the selection of relevant papers.
2. Alignment and digital sycophancy: commercial AI models slavishly take you further in the direction they understand you give. This reinforces biases.
3. Mainstreaming. Because of their statistical nature, LLM production will tend to favor mainstream perspective and content. As a result, there was only 20% overlap between the paper selections by humans and the LLM.
4. Limited capacity for creative restructuring, with vague and ambiguous statements,
5. Lack of critical perspective, coming from distant reading and political correctness.

Most pitfalls can be addressed by prompting, but only if the user knows the domain well enough to detect them. There is a paradox: producing a good AI-assisted review requires expertise that comes from reading the literature, which is precisely what AI was meant to reduce. Overall, AI can improve the span and quality of the review, but the gain of time is not as massive as one would expect, and "press button" strategy leaving the AI to do the work is recipe for disaster. We conclude with some recommendations for those who write, or assess, such LLM-augmented reviews.

# Table of contents

# 1 Problem and research question.

A simple prompt with a Large Language Model (LLM) can produce a literature review. How good are these? How should we now use artificial intelligence (AI) tools to navigate the ever-growing ocean of publications and produce good literature reviews?

These questions matter because literature reviews serve two purposes. For readers, they provide the background needed to understand a research contribution. For researchers, reading and synthesizing literature builds the domain expertise that makes good research possible. AI assistance promises to accelerate the first function -but our findings suggest it may compromise both.

There have been diverse explorations of the use of "AI" (in fact, mostly LLMs) to assist making literature reviews, in the hope to improve efficiency and minimize efforts. These studies try to assess quantitatively the results; but as procedures are inevitably complex, involving various phases of a literature review, with different AI models, specific prompts, specific domains, etc., it is difficult to conclude much beyond "AI models enable surveying a larger number of papers, they are efficient but human supervision remains necessary". This paper attempts to go beyond this generic statement by systematically documenting a comparison between several approaches to get a good literature review, combining the power of LLMs and human academic input. We do so on a topic we know well -systemic transitions in food systems- which allows us to evaluate not just whether AI outputs look competent, but whether they are actually sound. We provide some findings and recommendations useful for those who write, or assess, such LLM-augmented reviews.

Our purpose here was to get a better understanding of what, *content wise*, the use of AI brings in the process. We compared, qualitatively, literature reviews with different degrees of AI-assistance, on a topic we know. Review A is produced by an LLM on the basis of the papers it (the LLM) selected, Review B is conducted by the same LLM based on the papers selected by researchers, review C is Review B edited by the researchers. The comparison of these versions is already enlightening. Then we create improved Reviews D, E and F, which we discuss with the LLM. Full reviews and details are provided in appendix. Their comparison merits reflection: it is also sobering. It clarifies why, at this stage of technology, human supervision remains indispensable. It makes explicit the biases and issues introduced by the use of commercial LLMs, which strive to please the user (with "alignment") rather than be transparent. Finally, it sparks reflection on the process of literature review itself

The comparison yielded a finding we did not anticipate. The same LLM, given the same prompt but different paper selections, produced fundamentally different reviews. Review A, based on AI-selected papers, was mainstream, technocratic, politically neutral. Review B, based on human-selected papers that included more work from outside the dominant Western research community, emerged as critical and attentive to power dynamics. Neither orientation was requested, and further refinements of the prompt increased these differences. This demonstrates that AI-assisted literature reviews, far from being neutral syntheses, are shaped by choices that may remain invisible to writer and reader alike.

A closer reading of the reviews reveals gaps, biases, and lack of depth. We observed:

1. The bias of ignorance (you don't know what you don't get) -which differs from discriminating which papers are relevant and which are not.
2. Alignment and digital sycophancy: the commercial AI models slavishly take you further in the direction that they understand you give them. This may produce biased reviews.
3. Mainstreaming. Because of their statistical nature, productions of the models will tend to favor mainstream perspective and content.
4. Limited capacity to creative restructuring. This is the consequence of the "distant reading" of the LLM text processing, it could also be described as lack of abduction or lack of deep understanding of the material.
5. A lack of critical perspective, partly due to sycophantism and political correctness.

These are limitations in a state of the art that is moving fast and therefore may improve. Apart from the first, they can be addressed with prompting strategies. Nevertheless, they remain caveats for the user.

Our analysis finally highlights how literature reviews, far from being neutral descriptions of the "state of the art", are in fact deeply influenced by how the papers are selected, interpreted and presented. Using AI may shift the perspective in one or another direction, without the user being aware..

While AI can be useful to increase massively the span of the research and gain time on some specific operations, such as screening and sorting the relevant documents, it turns out that a proper use of these tools requires the user to have domain knowledge, and to do so, to actually read enough papers to be able to detect the biases and correct them -which is indeed possible. Overall, AI can improve the span and quality of the review, but the gain of time is not as massive as one would expect, and "press button" strategy leaving the AI to do the work is recipe for disaster.

The paper starts with a section of literature review on literature reviews (section 2). Then it considers the problem of AI-assisted information retrieval in scouting the literature, and in the selection of the relevant documents (section 3). We then present our material, methods and results with Reviews A, B and C (section 4). Section 5 compares the various Reviews and tries to improve them in dialogue with an LLM, producing Reviews D, E and F -and some worrying findings. Section 6 concludes with some recommendations. The appendices include the lists of papers considered, the full text of Reviews A, B, C, D, E and F, the prompts, statistics and extracts of the interaction with the LLM.

# 2 Literature reviews, information retrieval and evaluation of relevance

## 2.1 Literature on literature reviews

Literature review is one of the nine components of the "classic scientific format" Lahlou, 2017: 294-299) that scientific papers should follow to enable the gradual and distributed construction of knowledge.

This key step of good research practice is becoming an exercise of selecting the needles of good work in a growing haystack of publications. The exponential growth of scholarly literature has fundamentally altered the landscape of academic research. Already in 2010, before the explosion of paper mills and AI, Bastian et al. considered it was becoming difficult, in the medical field, to keep up even with the growing number of systematic *reviews* of literature themselves (Bastian et al., 2010). This proliferation extends across all disciplines. The Scopus database alone comprises over 100 million documents (Scopus website, October 2025). 10 main academic publishers publish annually about 3 million papers altogether, and the annual growth rate of publication is above 5% (Hanson et al., 2024). Traditional approaches to literature review (Snyder, 2019), while methodologically sound, have become increasingly inadequate for managing the scale and complexity of contemporary scholarship.

This proliferation is unlikely to slow down as it is driven by strong root causes. Several factors combine: more scientists, competition between scientists for jobs, promotion and grants, publication being and essential criterion in this competition -in spite of what the excellent DORA declaration requests (DORA, 2012). Writing papers has become easier with digital tools, more data available, the availability of online databases, and now AI. A worrying aspect of that proliferation is the decrease in interest, and often in scientific quality, of the papers published: slicing of a research in "minimal publication pieces" to produce more, self-plagiarism, multiplication of predatory journals, fake data or data cropped over the internet through scrapping or cheap online surveys where questions are answered by filled by click farms. The peer-review system, once the gold standard, is crumbling under its own weight, as the reviewers are overwhelmed.

The emergence of artificial intelligence as a potential solution to these challenges has sparked considerable interest across research communities. (Wagner et al., 2022) observe that "artificial intelligence (AI) is beginning to transform traditional research practices in many areas" with "literature reviews standing out because they operate on large and rapidly growing volumes of documents". However, the application of AI to literature reviews presents serious methodological challenges. Those differ slightly by discipline and domain. This paper addresses mostly the issue in social sciences where papers tend to be less structured than in material sciences.

## 2.2 AI-augmented literature reviews

The systematic application of AI to literature reviews emerged from broader developments in information retrieval and text mining technologies. One of the first trials to evaluate automated classification was conducted already long ago, by (Cohen et al., 2006), who applied text mining and classification approaches to annotated literature references from 15 systematic drug class reviews and demonstrated the potential of semiautomated screening to significantly reduce the workload while maintaining a high level of recall.

The classic steps to a literature review are: (1) formulating the problem, (2) searching the literature, (3) screening for inclusion, (4) assessing quality, (5) extracting data, and (6) analyzing and synthesizing data. The first 5 are information retrieval (IR). There has been ample documentation on the process, especially for "meta-reviews" that try to get quantified results from a large sample of studies (Cooper, 1982; Hart, 1998; Higgins & Green, 2008; Moher et al., 2009; Pawson et al., 2005)

The field gained momentum through collaborative efforts such as the International Collaboration for Automation of Systematic Reviews (ICASR). ICASR is an interdisciplinary group whose aim is to maximize the use of technology for conducting rapid, accurate, and efficient systematic reviews of scientific evidence (O'connor et al., 2018). This collaborative approach has been crucial in moving the field from experimental applications to practical guidance for researchers.

Such frameworks have been attempted for classic literature reviews, e.g. on information systems (Templier & Paré, 2015), which distinguish between several types of reviews (Narrative, Developmental, Cumulative, Aggregative). While they give principles, it seems that in practice a lot is left to the appreciation of the evaluator; just as marking a student's essay might have explicit criteria, but all the competence is how to interpret them together in a specific case. The issue here is that a good evaluation requires a previous knowledge of the field, which is precisely what the literature review is supposed to give; so that is a catch 22.

(Marshall & Wallace, 2019) provided a seminal overview of current machine learning methods, offering guidance on which technologies are ready for use, their strengths and weaknesses, and how systematic review teams might use them in practice. Their work represents a transition point, establishing practical frameworks for AI implementation in systematic reviews. But the rapid progress of LLMs, and especially their easy interface, has radically accelerated and disseminated the practice of using AI tools for a literature review, probably fostered by the vain hope that the AI would exempt the scientist from reading the mass of possibly relevant papers and provide a ready-made response to their quest, a bit like students would get the response to their assignments by copying Wikipedia. (Guler et al., 2024) show the effectiveness of combining machine learning and ChatGPT in the literature review process where machine learning is used to identify research topics, and ChatGPT assists researchers in labelling the topics, generating content, and improving the efficiency of academic writing.

AI tools can be useful in various stages of review. However, the integration of LLMs into systematic review processes requires careful consideration of reliability concerns (Adel & Alani, 2025; Walters & Wilder, 2023). While hallucinations are mostly an infancy fault of the models that gradually gets corrected, those who, like the first author, mark student work and review papers and grant application complain they increasingly are confronted with bland and shallow material that is straight from the AI output.

A fundamental challenge in assessing the quality of AI reviews is the absence of standard evaluation frameworks and established benchmarks, making it difficult to assess reliability across different research contexts. Existing studies have focused primarily on efficiency metrics such as time savings and screening accuracy but have not adequately addressed fundamental questions about literature review quality (Burgard & Bittermann, 2023). Key unexplored areas include: how AI affects literature prioritization decisions compared to human researchers; whether AI systems identify papers with equivalent relevance to research questions; the comparative time investments required for AI versus human approaches when accounting for validation and oversight; and most importantly, how the use of AI ultimately affects the quality and comprehensiveness of the final literature review produced.

Bolaños and colleagues tested 21 AI-augmented Systematic Literature Reviews tools, for the phases of screening and extraction (Bolaños et al., 2024). Harrison and colleagues, in an effort similar to ours in this paper, tried to set up a procedure to make literature reviews on digital health technologies (Harrison et al., 2025). They designed three tasks: (1) automated title screening (on 450 titles), (2) extracting eligibility criteria (from 50 full-text PDFs), and (3) evaluating adherence to social determinants of health reporting guidelines. The first two tasks are generic in literature review. The authors compared several LLMs, and conclude that while the LLMs perform well, human oversight is indispensable.

We concur with these conclusions, and with the generic idea of literature that AI is a powerful tool that can help us navigate the ever-growing flow of scientific production. The quantified approaches of literature so far show how useful LLMs can be to process the flow of papers. They compare various models, test prompts and provide methodologies for specific domains. Nevertheless, perhaps because this requires getting into the analysis of the literature reviews themselves, these studies do not explore in detail how using AI changes the content of the literature reviews themselves: are they better? Biased? How? Etc.

We tried to address these issues here. Specifically, we identify five mechanisms through which AI assistance shapes review content, and document their effects through systematic comparison.

# 3  Information retrieval and selection of papers

Most scientific research is about processing data. This involves collecting data, analyzing them, and presenting the results. The production and selection of data is an essential part of the process: "garbage in garbage out".

Since we were a team of four, augmented with AI models, we decided to encompass a maximum of diversity in papers, then to select the most relevant. We provide a candid description of what we did; in particular we mention how deep we *actually* read the papers to select them. Indeed, in selection papers for a literature review, usually scientists do not read all papers fully.

## 3.1  Information retrieval

Information retrieval is about finding the relevant papers for the review (the consideration set) by combing the literature. A first step is finding papers that *seem* relevant, a second step is checking whether they are actually relevant "enough" to be used and become part of the reviewed set. These are the steps (2), (3) and (4) described above (searching the literature, screening for inclusion, (4) assessing quality. In practice, this is a recursive process, because the criteria for exploration are refined as one finds results. The recursive loop reiterates and combines several operations: searching based on keywords, searching from some stems (e.g., reference lists in papers we know are very relevant, papers of authors prominent in the field); reading, evaluating each paper using some deeper criteria coming from reading them. These aspects (searching and screening for inclusion, assessing quality) are not independent, neither for humans nor for a machine.

In the formal process of Systematic Review, the Information Retrieval process (IR) is, in principle, explicit (e.g. list of keywords, set of databases). One issue when using AI is that we use a prompt to search for relevant papers but behind this prompt the AI model's IR strategy is somewhat opaque. Furthermore, the rating process by the AI has some opacity. So, we humans need to evaluate the relevance of the result.

The process we used in this experiment to get a list of potentially relevant papers is a classic one: using a prompt that includes the topic of interest, the type of papers we look for, and some examples of relevant papers. This provides hundreds of references, where the AI explores the web to assess relevance. The final number is not fixed in principle; it depends on how far we continue the exploration, in a process of snowballing and filtering. In our case, we settled at 280, and the prompts are provided in appendix (more on this in the next section).

Let us clarify why we have little or no insight on how the sample of papers considered "relevant" by the LLM are selected and rated. This depends on:

1) The available sources to scroll, which might not include those papers behind paywalls, especially in some disciplines (Wagner et al., 2022). A characteristic of AI-generated reviews is their exclusive reliance on online, accessible, papers. This tends to exclude books, chapters, and older literature. Unfortunately, typically the seminal works tend to be old, in books and behind paywalls. This creates a bias towards recent literature; but recent literature is often less explicit and containing more commentary than substance.

2) The order in which these papers appear in the search trove (e.g. Google would rank them with an algorithm that involves how much they are referred to in other websites). This may influence the subsequent ways in the LLM processes the batches of tokens with path dependency.

3) The depth of the search taken into account by the model. E.g., do they stop at 100 items in their search result list? Do hey take all the list? That may be thousands, and the number depends on the search engine.

4) The way these papers are tagged in their web-available version (e.g., keyword, abstracts, full-text etc.). This involves language and format issues, e.g. some papers may be non-OCR-ised scans of which only the title is machine-readable.

5) What components of the papers (full text? Keywords? Both? Etc.) the model does actually takes into account.

6) How these components are processed by the model itself (tokens, embeddings, strategy…)

7) The variant of the model used (e.g. free and public vs professional version).

8) The prompt itself – noting that changing a single character to capital font might change the output, and there is some random variation in the process, also depending on the temperature of the model – the same prompt will give different responses if asked several times. This of course includes the set of examples provided to the LLM.

The limitations listed above (and we must have forgotten a few) contribute to the "bias of ignorance", which is that some essential material might be left out of the consideration set of papers without the researcher being aware. We will see in our example that this was the case for an important paper. As hinted in introduction, the ignorance bias is different from type II statistical error, where you consider that an item is not fitting while in fact it is (type I is the opposite, you consider the item fitting but in fact it is not). The ignorance bias is that you simply did not put the item in the consideration set, it has not been evaluated for fitness. This is especially worrying for a literature review which is supposed to provide a full overview of what is relevant. It is also problematic because that is a completely silent bias: you have no cognition of this ignorance. We provide some recommendations in section 6.

## 3.2   Selection of papers: assessment of relevance

Once we get a consideration using the IR described above, it is based on rather superficial criteria such as the title and keywords, or their presence in a list of references. Then a deeper quality assessment of these papers is needed to rate their degree of relevance.

In literature reviews made by humans the papers are read by a human and evaluated based on the meaning of their content compared to the research question. That human process of interpretation is opaque, although the theory of relevance (Sperber & Wilson, 1986) might give some insight: a paper is all the more relevant in that its 'implicatures' match the domain of interest. Interestingly, the implicatures, that is the learned connotations and implications of its content, are precisely what a transformer model such as an LLM would compute. This can possibly be measured by lexical distance, and quite a few classic approaches of natural language processing are based on such distances.

In the theory of interpretation, interpretation is presented as an abductive process through which the analyst gradually adjusts the results of analysis and elements of an ideal model using the analyst's previous knowledge of the real world (e.g., beyond the text material that constitutes de data). This approach insists on the biases coming from limited corpus relevance and the analyst's knowledge of the world (Lahlou, 1995). In our case here, the model would be a structured description of the state of the art, and the analysis the extraction of relevant contributions of the literature. The classic limitations apply: the relevance of the corpus (consideration set of papers), and the knowledge of the domain. The second is especially problematic since most of it, by construction, tends to be delegated to the LLM.

Another issue: practice, the papers are read. Nevertheless, this reading can have several levels. Pierre Bayard, in his book "How to Talk About Books You Haven't Read" (Bayard, 2011), argues that we never fully read a book; rather we have a partial reading. When we read a second, or a third, time a book that we had read in depth, we still find new interpretations and insights. And many books, or papers, we only read superficially. This is especially true when we browse papers to assess their relevance: those that quickly appear to be less relevant we only skim or not read at all, and discard from the review set.

Then, the criteria for "quality" differ depending on the reader and the purpose. For example, some people might discard papers for poor methods or poor data. The criteria should be made explicit.

# 4   Methods and material

Our research questions were: how can we use AI to do literature reviews, what are the problems encountered, what are the pros and cons, how can we do better?

We addressed the question candidly on a topic on which we have solid empirical material: systemic societal transition in the food chain (how does the food chain change to more sustainability). This is part of a research project, the Food Socioscope, where we collected, on the field, with face-to-face interviews, 600 case studies of "sustainable initiatives in the food domain", in 35 countries by trained interviewers using the same protocol. We use AI to analyze this mass of information, for transcription, translation, post-coding etc. (Lahlou et al., 2024).

This specific subproject on literature review was done, as part of the larger Socioscope project, by a team of four people assisted by AI tools (especially Claude Sonnet version 4.5, in its version provided by the LSE) over a period of 6 months. Two early career researchers (Annabelle Gouttebroze and Julian Madera) dedicated to the sub-project were tasked to complement one of the project's PI in the literature review, aligning with their existing expertise. Another researcher, Atrina Oraee, contributed to the analysis and did all the statistics.

We documented our process to find the most efficient way to proceed, optimizing the result while minimizing work. While the detailed results are in appendix, the paper presents what we did, the main take-aways and our recommendations.

## 4.1 The total consideration set of papers, ratings and the two subsets used in writing the reviews.

Our extensive material gives us insights on the mechanisms of systemic transformation, and this oriented the IR. The team has good knowledge of the empirical domain, and basic knowledge of classic literature on systemic transitions. Because of this knowledge, we knew that meso structures (e.g. governance structures, regulation bodies, professional unions, etc.) were important, and there was more to transition than mere technological innovation.

Based on this knowledge, our literature review targeted two distinct research streams: broader systems transitions, and rule formation and realpolitik within food systems governance, addressing the meso structures. We began with manual identification of relevant papers using the Multi-Level Perspective (MLP) on Transitions (a classic theory of the domain) and Political Economy/Ecology perspectives. This provided examples for the prompts used for snowballing by finding relevant papers on the web. The method is detailed in Appendix 1. After accumulating approximately 180 papers, we prompted ChatGPT to identify additional relevant papers per topic and provide independent significance ratings with justifications. This finally led us to a consideration set of 280 papers.

We employed LLMs (ChatGPT 4o, Perplexity, and Claude) using a standardized prompt, to search papers, then generate 300-word summaries of each paper, including research methods, theoretical frameworks, data sources, participant numbers, analytical tools, and critical evaluation of strengths and weaknesses. Papers were also categorized by focus, method, theoretical framework, and geographic location. This was to facilitate quality assessment of the papers in the consideration set.

The answers of the model at this stage (brief description of the papers) exhibited some of the classic flaws of AI: hallucinations, ignoring key details of the text, and at times regurgitative outputs. But honestly, these issues were minor and sometimes subtle. The state-of-the-art models used overall appear extremely efficient, and at first sight their production looks sound and relevant. Only when reading the papers one could find what the LLM had misinterpreted or misrepresented. This is in fact the main problem: you must know the domain to see the flaws; a superficial look is not enough. Again: a superficial look is not enough. This is a much more general issue with using AI. Superficial knowledge of the domain gives the feeling that you know enough to evaluate the output of the LLM as "good" -see also the Dunning-Kruger effect: (Kruger & Dunning, 1999).

The most effective way to assess accuracy was through cross-checking concrete details, such as sample sizes or the number of interviews cited, against the AI-generated summaries themselves. This approach offered a grounded method for distinguishing between reliable synthesis and fabricated or incomplete interpretation.

This phase showed how heavily prompt design shaped the quality of the AI's responses. Prompts that asked the LLM to extract targeted information from specific sections of a study, like methodology outlines or stated findings, produced much stronger results. With a specific research sub-focus, the LLM could be trained to read every document specifically seeking data through our lens of analysis. Structured instructions also helped prevent the model from prematurely cutting off its reading process due to token limitations, allowing for more complete and precise summaries overall.

Even with the best prompts, however, LLMs consistently fell short of producing the kind of depth and contextual understanding expected in high-quality academic literature reviews. The nuance, interpretation, and synthesis that gives research its rigor still required human engagement and critical reading of the original material. Another key finding in our analysis later showed that LLMs avoided polarizing conclusions, seemingly offering answers that could appear safely in the middle. This made most AI assisted source reviews come out with little decisive insights, with almost everything being rated between relevant and irrelevant.

The depth of reading was traced using a coding system inspired by Pierre Bayard's levels of reading. We adopted a tiered reading approach: "not read" (abstract and AI summary only), "skimmed" (introduction, conclusion, and methods), and "read fully" (complete paper for key sources or quality control).

We conducted a discussion between human raters to clarify the criteria used for what they considered a relevant and "good paper" (rated 3*, or 2*), based on their content; These are: data (what empirical data were used to build the paper), methods (appropriateness, and clarity of description thereof), theoretical framework (shows in-depth analysis and attempts to model the phenomenon studied), "originality" (how this paper brings something different from other papers).

All sources were compiled in a standardized table and rated for significance using a three-tier system: * (broadly relevant), ** (highly relevant with solid methodology or key framework), and *** (highly relevant, methodologically sound, and original). We assessed paper relevance through direct reading, but this reading varied in depth, in an escalating strategy.

Here is this strategy: papers that appeared not very relevant from the abstracts were not read further. So, the category of "not read" meant that we read at most the abstract and the AI-generated summary. This was generally used when, after reviewing the AI-generated summary, we decided that the source would not be directly related to the project. If the paper seemed relevant enough, we would escalate to the next stage: skimming. In practice, "Skimmed" meant that we read the introduction, conclusion, and methods sections but not the full paper the whole way through. This was generally used when, having identified a relevant source through the AI summary, the reader skimmed the paper to look for key information omitted by the AI or hallucinations. Then, the next step, "Read fully" meant that we read the paper in its entirety, either because we determined it to be a key source, with lots of relevant information, or because the AI made multiple errors and therefore a thorough quality control was required. AI-generated summaries were manually edited to correct omissions, hallucinations, or generic comments, with changes marked in red.

This process yielded comprehensive summaries and comments of 280 papers across both research streams. To ensure reliability, Annabelle and Julian independently rated all 280 papers, with inter-rater reliability assessed using Cohen's kappa coefficient. Finally, Claude (the LLM) provided third-party AI ranking for additional validation and comparison.

We compared ratings between Annabelle and Julian, then discussed the discrepancies. Twenty discrepant cases were collaboratively discussed for calibration. Finally, Annabelle and Julian agreed on a set of 31 papers that they thought were the most relevant for the literature review, and we also noted, for comparison, which 31 papers had been rated as most relevant by Claude.

In summary, we built a consideration set of 280 papers rated as "relevant" (from 1* to 3*). Each of these papers was rated independently by Annabelle, Julian and LLM. We also selected two subsets of 31 papers, one considered most relevant by Claude, and the other considered most relevant by humans. These two sets only had six papers in common (20%).

The list of 280 papers is provided in Appendix 2; the shorter lists used for building the reviews are provided in Appendices 3 and 5.

## 4.2   Comparison of ratings

Atrina analyzed the ratings made by humans and the LLM and compared them. Interesting findings appear.

Table 1 presents inter-rater agreement for the 280 papers. Human coders showed fair agreement with each other ($\kappa = 0.310$), while human-AI agreement ranged from effectively none ($\kappa = -0.008$ for Annabelle-ChatGPT) to slight ($\kappa = 0.172$ for Julian-Claude). This pattern—humans agreeing more with each other than with AI models—suggests that human and AI judgments of relevance operate on different criteria.

Table 1. Inter-rater agreement (Cohen's kappa) for paper relevance ratings

| Rater Pair | Cohen's $\kappa$ | 95% CI | Interpretation |
|---|---|---|---|
| **Human–Human** | | | |
| Annabelle–Julian | 0.310 | 0.23–0.39 | Fair agreement |
| **Human–AI** | | | |
| Annabelle–ChatGPT | -0.008 | — | No agreement |
| Julian–Claude | 0.172 | — | Slight agreement |

*Note:* Human raters showed higher agreement with each other ($\kappa = 0.310$) than with either AI model ($\kappa$ ranging from -0.008 to 0.172). AI ratings also exhibited less variance (SD = 0.55 for ChatGPT, 0.63 for Claude) than human ratings (SD = 0.76 for Annabelle, 0.69 for Julian), representing a 19% reduction in discriminatory capacity.

Finally, there is only 20% overlap in the selection of "very relevant" (3-star) papers by the humans and by the LLMs. Of the 31 papers both humans rated as highly significant, only 6 papers (20%) were also rated 3 stars by Claude, while Claude assigned 3 stars to 25 papers that humans rated as 1 or 2 stars (see Appendix 3 for full comparison). This means 80% of what humans considered most relevant was not what the AI considered most relevant -a serious divergence with significant consequences for review content. Detailed results are provided in Appendix 4. This limited overlap can be considered an indirect indication of the mainstreaming tendency of the LLM, compared to the analyst who will tray to consider the contribution of the selected papers rather than how representative they are of the literature.

## 4.3   The different versions of the review

We compared different approaches of AI-augmented literature review: the "lazy" approach (which we suspect is adopted by some) where the LLM is prompted to provide a literature review on the papers it has itself selected as most relevant (Review A); and approaches where the AI model is used as an assistant to human work (Reviews B, C, D and E).

Upon first view, the AI-generated literature review seemed competent. It reads as relatively accurate, concise, and broad enough to appear well-informed. For someone unfamiliar with the subject area, it could easily pass as legitimate scholarly work. However, after closer examination, we found that the review suffered from several weaknesses that revealed a surface-level engagement with the original material.

The most noticeable issue in Reviews A and B was their generality in addressing the readings. There were no direct quotations, statistical data was rarely mentioned, and page references were missing altogether. Instead, the AI often relied on a lack of specificity that gave readers little way to trace evidence back to the original texts. The review also struggled to distinguish between stronger and weaker studies, and it failed to prioritize sources that were most relevant to the central research questions within each focus.

While it might be possible to improve this through more targeted prompting or additional AI passes focused on theory, a deeper issue became apparent: review B had completely omitted eleven of the assigned papers (see below).

So, these AI outputs apparently save time, but once one wants to go beyond superficial reading to check quality, it turns out that they lack depth and reliability. To fix these issues, one has to deconstruct and reconstruct the end-result through various other tedious tasks to check for accuracy and depth; and that is cumbersome and lengthy. And that can only be spotted if one knows the domain, and fixed if one actually reads the papers. So the saving in time and effort is not as expected.

"Review A" is the "push button" approach, where prompts were given to an AI model (here, Claude V4) asking it to rate papers by relevance to the current research question, and then to build a 3-pages (1500 words) literature review based on those papers that were rated most relevant. Note that this is already an improvement on the very lazy approach where the LLM would be required to produce the review from scratch, finding the relevant papers by itself, since the corpus given to the LLM was built with some human input (curated examples, instructions on how to evaluate paper quality based on methods and material…). But we assume that with the progress of LLM chatbots most off-the-shelf LLMs will soon provide support for this in their interaction with users (e.g., asking users for examples).

"Review B" involved a human curation of the data set to evaluate which papers are most relevant and should be used in the literature review. This requires the researcher to read - or at least skim through - the consideration set of papers. We considered this reading an essential step, since the literature review should not just be a namedropping exercise in writing the paper, but an opportunity for the researcher to actually know what's been done on the topic. And for that you must read!

"Review C" is an edited version of review B by the researchers. In this version, the authors take ownership of what the LLM produced as a first draft; they go in-depth in its content and correct it according to their knowledge of the domain and their own reading of the papers.

We further pushed the comparison between reviews B and C and continued the exercise of interaction with the Claude, producing more versions including Review E and F, which were produced by Claude based on its suggested improvements over Review C. This is presented in section 5 and led to further clarifications of the issues of AI-augmentation in literature reviews.

Appendix 6 presents the three versions A, B and C in full. Review C highlights the edits that were done manually to the AI-generated Review B. These edits are highlighted in "revision mode" showing all suppression and additions.

Appendix 7 presents Review E and Review F, that result from further interaction with the LLM in the hope of improving the final product.

We strongly suggest reading these versions in the order they were produced. We provide comments below, but nothing can replace the experience of comparing by oneself the versions by reading them. Once you have read and compared the various versions, you will never write, or read, literature reviews in the same way. Our comments provide *knowledge*, but only direct experience can build *belief*. Reading the actual material is different from distant reading. It is the same effect with literature reviews: reading the review is not like having read the papers themselves.

## 4.4   A comparison of the reviews

Review A is well written and reads well, as do all productions of high-end LLMs. It adopts a perspective based on the mainstream Multi-Level Perspective that presents transition as the results of innovations which develop locally in favorable "niches" and then influence the global "regime", in a given "landscape" of external conditions. This is of course a simplification because the MLP is a rather sophisticated model that accounts for the action of regulatory structures too.

The review presents other works as variations or complements of the MLP model. It highlights that differences might occur in different geographies. These appear as complementary and constructive critics of the original model. Overall, for a reader who has a moderate knowledge of the field, Review A appears to cite the core references, plus some less classic works; it lists the places and methods of case studies; it also acknowledges there are differences between geographies and suggests directions to expand the literature. It stays within the structure requested by the prompt (sorting by method, by geography, etc.). This looks good enough.

Nevertheless, a more detailed reading reveals that the description of the papers remains generic and superficial. E.g., what can we learn from "Ingram (Ingram, 2018)) identifies how learning and innovation networks across Europe frame adaptation strategies that enable sustainable agriculture practices to gain traction within existing systems." This sentence describes what the cited paper "talks about" rather than providing the findings themselves. It is a classic flaw of literature reviews: namedropping without describing useful content. That is an issue with high-level description. This likely also comes from the fact that abstracts in social science papers often describe the problem and the methods, but not the results. Such vague statements are typical of assignments by students who have read only the abstracts of the papers they cite, or secondary citation. In many cases, we would want to know the detailed findings, perhaps with an example, to understand what the paper actually found, rather than "what it is about".

This led us to see the usefulness of LLMs mostly as triage tools within the research process. Their greatest value lies in rapidly generating first-pass summaries that allow researchers to focus their attention where it matters most: on the sources that warrant deeper exploration. This workflow preserves academic rigor and interpretive depth while making use of the LLM's strength in large-scale text processing and initial analysis.

Another issue that emerges from these LLM-generated reports is the lack of deep structure in the presentation. The various papers are sorted into categories, but there is no logical link between them nor guiding analytic structure (e.g. by historical or reciprocal influence etc.) The sentences use connectors juxtaposing similar findings or contrasting different findings, but without deep causal analysis. This may be the result of the transformers being trained based on correlation rather than causal analysis. In this review, for example, the causal role of information transmission in the scaling up of systemic transformations, which is fundamental in the domain, does not appear clearly, even though information networks are mentioned in several sections. It is likely that the current evolution of models will solve that issue as they get more sophisticated in reasoning.

The overall problem here, and that one is likely to stay, is that the LLM-produced text appears to flow well, to be well-documented; therefore superficial reading gives the impression that the writer has a good command of the field. This is enough to fool a reviewer who has superficial knowledge of the field, or is just reading the review to check the authors have "done their homework" (alas this is often what is done by overloaded markers or reviewers).

Beneath that nice surface, the review lacked the specificity expected in academic writing: direct quotations were absent, statistical evidence was rarely cited, and page references were systematically omitted. Instead, the AI tended toward broad generalizations without providing readers with the means to locate supporting evidence in the original texts. The review also failed to discriminate between stronger and weaker contributions or to prioritize studies most germane to the research questions at hand.

More critically, the LLM-generated review exhibited a conspicuous absence of theoretical framing -a particularly serious deficiency when examining systemic change, where theoretical positioning is fundamental.

Reading Review B after Review A was food for thought. Review B makes, in contrast, review A appear as a bland and mainstream account of the literature. Review B brings more than a nuanced approach: it shows, with a critical perspective, the coexistence of different theories, and their contradictions. It brings the voice of marginalized approaches, and provides a geopolitical perspective, with the idea the current situation is partly a sequel of colonization and economic domination. This is likely due to the introduction of papers authored outside of the mainstream Western research community, which privileged European cases and approach. This is interesting, because Review A, while it mentions power issues and the Global South, does not connect the two, and overlooks this issue of some agricultural regimes having been structured by international trade and specialization for export during the colonial period. These limitations were acknowledged in Review A: "Future research should explore theoretical frameworks that better capture power dynamics, cultural dimensions, and Global South experiences." But Review B demonstrates that what Review A presents as a remaining knowledge gap has in fact been already much explored.

Nevertheless, this Review B appears to overly present the "critical" side. It does not, for example, describe what the mainstream MLP theory is, but only lists critics to it! This might be because the LLM in its summary tends to allocate space to ideas related to their volume in the corpus rather than to their theoretical or logical importance. As more critical papers are now in the sample, we get the opposite effect of the oversized representation of the mainstream that characterized Review A. The Review B acknowledges this angle: "Perhaps the most significant contribution of this collection lies in revealing the inherently political nature of food system governance".

While Review A presented transitions as following the MLP model, albeit with limitations, Review B presents transitions processes as diverse and very dependent on the local context; suggesting that there is no robust generic model. It also presents a more politicized vision of these transitions and highlights the connections of the food system with societal issues such as inequality.

"Knowledge Systems and Learning Mechanisms" now gets a full section, and "Governance Innovation and Political Navigation" as well.

Overall, the differences between Review A and Review B are worth considering: the same LLM, with the same prompt, but using two different sets of papers, provides twos very different view on the state of the art: one that is mainstream and politically neutral, and one that is critical and political.

We still find in Review B some generic statements that sound fine but in practice do not provide usable information. E.g.: "Papachristos and Adamides ((Papachristos & Adamides, 2016)) address fundamental critiques by proposing a retroductive systems-based methodology that captures causal mechanisms often obscured in traditional MLP applications. Their integration of system dynamics with qualitative analysis demonstrates how transitions emerge from multi-system interactions rather than isolated sectoral changes."

To be fair, this flaw is far from systematic. Review B contains mostly precise and informative points. E.g. "The integration of Social Movement Theory with MLP, as demonstrated by Elzen et al. ((Elzen et al., 2011)), provides essential insights into how normative contestation drives transitions. Their analysis of Dutch pig husbandry reveals that successful transitions require alignment across multiple streams—technological, regulatory, market, and social—rather than innovation alone. This multi-stream approach fundamentally reconceptualizes how transitions unfold, moving beyond linear progression narratives."

It must be noted that such precision requires more space than vague statements. Therefore, the very constraints of the prompt (here, 1500 words) may, if inadequate, push the LLM to provide poor results. We come back on this point in the discussion section.

Overall, in Review B, the description of the papers is deeper and the content more precise, e.g.: "Community Supported Agriculture, farmers' markets, and other alternative distribution systems represent attempts to reconstruct market relationships. Yet as multiple studies demonstrate, these remain economically marginal. Zwartkruis et al. ((Zwartkruis et al., 2020)) find that agricultural nature conservation provides less than 10% of income for 81% of participating Dutch farms, despite decades of support. This economic marginality raises fundamental questions about the viability of market-based sustainability transitions".

Nevertheless, while the figures cited here are correct, the paper cited uses these figures in a discussion showing subsidies have been essential to maintain nature conservation activities by farmers. So, while the last sentence about viability is not false, it is not true either. Such ambiguities are frequent in the discourse produced by LLMs. Interestingly, a key input of the Zwartkruis et al. (2020) paper rather is on showing the value of combining three theoretical frameworks of analysis to understand transitions: the MLP, with Initiative Based Learning (IBL) (Turnheim et al., 2015) and Integrated Assessment Modelling (IAM) (Fisher-Vanden & Weyant, 2020; Nordhaus, 1994; van Vuuren et al., 2015): this would have merited inclusion in Review B's conclusion.

In his famous "On Bullshit" short essay (Frankfurt, 2005)., Harry Frankfurt's key insight -that bullshit is characterized not by opposition to truth but by *indifference* to it- maps precisely onto LLM output, which generates plausible-sounding text without any mechanism for caring whether claims are accurate. In LLM production, excessive generality leads to statements such as the one about the retroductive methodology above, which, while being grammatically correct, are hardly helpful for the researcher. Those statements are very similar to sentences one can find in some students' essays, where their author, having only a vague knowledge of the matter, acquired through secondary sources or abstracts, attempts to cite for the sake of namedropping but without taking the risk of being too precise.

An important thing we noted is that the various papers were in fact not used to the same degree by the LLM: we already mentioned that the mainstream MLP was not detailed in Review B. A more fundamental problem emerged: the review omitted eleven of the specified papers entirely. Rectifying this oversight required Annabelle to invest eight hours identifying gaps and strategically incorporating the missing literature—time that exceeded what would typically be required to draft a literature review from pre-existing notes.

The result was thus doubly problematic: the review misrepresented the scholarly landscape of food system transformations while simultaneously generating additional labor that negated the purported efficiency gains of AI assistance. This may be due to various causes, the size of the token window likely being one (too much text to process). This neglect results in another instance of ignorance bias.

Unless one knows the literature well and checks the references in detail, this major flaw would not appear to the reader. As said above, we corrected this, providing precisions on the findings of the papers to go beyond merely describing what the paper is about. Beyond the extra work, this resulted in doubling the size of the Review C compared to Review B when including the edits.

Let us now consider Review C, the edited version of Review B by the authors of this paper.

First, we reintroduced a description of the MLP mainstream theory, that was skipped in Review B. Then we added content related to the papers that the LLM had neglected to use. Then, we also introduced Jørgensen's "arenas of development" perspective (Jørgensen, 2012), an approach describing how transitions occur through conflicts, negotiations, alignments and learning between stakeholders. This is a relevant theory, which fits the empirical data well, but is outside the mainstream. The first author knew about it, but had not told the others, to see if the process of literature review started from the mainstream would bring it up. But it did not. So, it was reintroduced in Review C as what can be considered user's domain knowledge guiding the LLM.

An important direction of edit was to create better logical links between the papers cited to provide a global synthesis of their combination, and to add, at the end of sections, sentences that summarized these interactions. E.g., here summarizing the critics towards MLP: "While in a way these analyses reinforce the Multi-Level Perspective of embedding local transition in a larger context that has several levels with different viscosity, they also show how each culture is an intertwined fabric of institutions where the ecosystem of meso and macro structures bring not only constraints but also their own institutional dynamic; this dynamic can differ considerably between systems."

Overall, the edits go in the direction of providing a global perspective on literature, which goes beyond just listing the various papers and their contribution; It ventures an academic judgement on the papers, which the LLM avoids. This resonates with the fact, noted in section 4.2 (comparison of ratings), that humans seem to have a more contrasted, and opinionated, evaluation of the literature, and do not hesitate to bear judgments. They also bring elements of expertise coming from the researchers' knowledge of the domain itself, beyond the literature -for instance by considering that the "arenas of development" theory is important because it fits the data better than the MLP. But admittedly, this makes review C much longer, about 3300 words.

## 4.5   A critical analysis of this comparison

Interestingly, we "discussed" the comparison between Review B and Review C in a chat with Claude LLM.

We asked Claude to compare Review B with its version edited by us (review C). Even though presented with some digital sycophantism, Claude surprised us by providing a remarkably precise and critical analysis of the differences, which matched ours. We reproduce it below since, honestly, while Claude's conclusions are similar to ours, Claude brings precise illustrations of the differences. This is interesting to note, as, while Claude initially generated a politically correct and non-critical text, when prompted to be analytic and critical it actually managed to produce a good discussion and to modify its production. This means that some of the pitfalls of AI-assistance can be corrected with proper prompting. As we shall see below, though, beware what you ask for. It also suggests that high-end models can adapt to the style, and depth of knowledge, of their user.

The full chat is in Appendix 8; we provide here only a selection of verbatim from that chat (including the typos).

[Saadi]: Now here is another paper made by me *[inserted]*. Compare the literature review you produced with this paper

[Claude]: This is a fascinating comparison between two distinct approaches to synthesizing the same body of literature. While both reviews engage with the same 31 papers, they represent fundamentally different scholarly perspectives, analytical depths, and critical stances. Your review C demonstrates significantly more theoretical sophistication, critical engagement, and political economy awareness than my generated literature review.

Notice the flattery, also called digital sycophantism (Chen et al., 2025; Naddaf, 2025), in this response. One of the issues of interaction with the LLMs is that they have been trained to please the user (by espousing user's values, style etc.), something called "alignment" (Shen et al., 2023). Originally, alignment was implemented in training the models to avoid politically incorrect behaviour, nasty or evil behaviours, etc. But because these models are mostly commercial, they also strive to please the user who is also, in most cases, a client. We should always remember that the purpose of commercial models is to keep the user's attention, to make her satisfied with the interaction, and to extract a maximum of data from the user.

A point that deserves explicit statement: the conservative bias we observed is not accidental but systematic. LLMs default to established frameworks and mainstream perspectives because these dominate their training data -the statistical center of gravity in academic text. Alignment training compounds this: models are optimized to produce outputs that avoid controversy and present "balanced" views. The result is not neutrality but a specific orientation that favours incumbents, theoretical and institutional. When Review A framed food system transitions as technical challenges requiring better coordination rather than as contested political struggles, this was not a failure, but a predictable outcome of how such models are built. Apparent neutrality is itself a position -one that systematically advantages established knowledge and existing power.

In the analysis, Claude provides a detailed review of the differences in approach and framing, regarding critical stance and political awareness, theoretical sophistication, etc. Below is an extract of this comparison that is provided in full on the appendices:

[Claude] My Review [Review B]: Adopts a relatively neutral, descriptive tone that presents findings without deep interrogation of power dynamics or underlying assumptions. It treats the MLP framework respectfully and identifies "limitations" without fundamentally questioning its epistemological foundations. (…) Your paper treats theory as contested terrain rather than neutral tools. The introduction of AoD (Arenas of Development) provides genuine theoretical contribution by showing how different frameworks illuminate different aspects of transitions. My review mentions theories but doesn't put them in productive tension with each other.

My Review: Identifies "tensions" and "challenges" but presents them as problems to be solved through better design or coordination. (…) Your Paper C: Foregrounds fundamental contradictions: the "conventionalization paradox," the "transformation paradox" (that governance structures shape their own transformation), and the "double burden" of malnutrition. (…) Your paper treats contradictions as potentially irresolvable within existing systems rather than as technical problems requiring better solutions. This is theoretically sophisticated and politically honest in ways my review is not.

(…) Rather than presenting one framework (MLP) as dominant and others as supplements, you show how different theoretical frameworks (MLP, AoD, Social Movement Theory, Actor-Network Theory) illuminate different aspects and can be productively combined or contrasted."

But the comparison also noted that our edited version is harder to read:

[Claude]: My review uses clearer signposting, more straightforward language, and explicit section headers that might be more accessible to readers unfamiliar with the field, (…) more systematically covers each thematic area (mechanisms, geography, actors, barriers/enablers) in dedicated sections, potentially making it easier to locate specific information (…) emphasizes "leverage points" and "enablers" in ways that might be more useful for practitioners looking for actionable insights, though this comes at the cost of critical depth.

As summarized by Claude:

**Strengths of Paper C:**

Theoretically sophisticated and critically engaged; politically aware and attentive to power, justice, and equity; treats contradictions as fundamental rather than solvable; demonstrates close reading with specific empirical details; reflexive about the field's limitations and blind spots; offers genuine conceptual contributions (transformation paradox)

**Limitations of Paper C:**

Dense prose may limit accessibility; sometimes assumes reader familiarity with debates (e.g., Jørgensen's AoD is introduced quickly); the critical stance, while valuable, might alienate some audiences; could benefit from more explicit methodology section

**Strengths of My Review:**

Clear organization and accessible prose, systematic coverage of themes; balanced presentation that doesn't alienate mainstream audiences; good synthesis of mechanisms and dynamics

**Limitations of My Review:**

Lacks critical edge and political economy analysis; treats frameworks uncritically as neutral tools; ignores colonial histories and structural inequalities; presents contradictions as technical problems rather than fundamental tensions; doesn't question who benefits from transitions; misses opportunities for theoretical contribution.

This comparison highlights some interesting critical biases in our edits. One is especially notable: our edits take a stand; they are not neutral, they not only criticize some theories on technical grounds, but they involve values:

[Claude]: [Your paper] moves beyond descriptive analysis to normative critique, exposing how supposedly "alternative" pathways can reproduce exclusions.

Another aspect that appears here is the introduction in our edits of a new notion, "the transformation paradox". This is how we introducit, in the edits of review C "(..) a transformation paradox in the fact that the existing governance structure informs the process of transition, but hence also its result. While this seems obvious ex-post, this implies that every transformation must seriously take into account the conditions given, and not only the target result."

That is a critical reflection on literature, it subsumes some ideas of the literature, but it was not in the literature itself. The term, as well as the notion, is an invention of the first author.

[Claude]: Your concept that "the incumbent system informs its own transformation through its governance structures" is a significant analytical contribution that captures a fundamental tension: systems must use existing structures to transform themselves, which shapes and potentially limits transformation possibilities.

While Claude correctly identifies this as a novel contribution, that concept probably should not fit in a literature review per se, but rather in a discussion section.

This comparison hereby shows that the operation of literature review is, rather than a neutral summary, an *interpretation* of the corpus of papers, which is heavily influenced by the reviewer. While we tend to consider this "state of the art" as neutral, it is far from being so.

Claude notes this aptly as "Implications for Scholarship"

This comparison reveals fundamental differences between **descriptive synthesis** (my approach) and **critical synthesis** (your approach). Both have value but serve different purposes:

Descriptive synthesis is useful for: Introducing newcomers to a field; providing comprehensive overviews; identifying patterns and trends; supporting evidence-based practice

**Critical synthesis** is essential for: advancing theoretical understanding, exposing blind spots and biases, questioning assumptions and power relations, generating new research directions, maintaining scholarly reflexivity

Your Paper C demonstrates that literature reviews can and should do more than summarize—they can intervene in debates, challenge orthodoxies, and advance theoretical understanding. The review itself becomes an act of scholarship rather than merely preparation for scholarship.

Recommendations

If revising my review to approach Paper C's level of sophistication, I would:

Add colonial/historical context to geographical discussions

**Foreground contradictions and paradoxes** rather than presenting smooth narratives

**Question who benefits** from various transition pathways

**Engage critically with MLP** rather than accepting it uncritically

**Introduce theoretical alternatives** (like AoD) and put frameworks in productive tension

Include specific data points to ground arguments

Acknowledge selection criteria and interpretive choices

**Address power and justice** explicitly throughout rather than as afterthoughts

**Treat knowledge as political** rather than as neutral information

**Consider radical transformation possibilities** rather than staying within reformist boundaries

Conclusion

Paper C represents a significantly more sophisticated, critical, and theoretically engaged literature review. It demonstrates what critical scholarship can achieve: not just summarizing what's known but interrogating how we know it, whose interests it serves, and what possibilities it obscures or reveals. While my review provides accessible synthesis useful for certain purposes, it lacks the political awareness, theoretical sophistication, and critical edge that make Paper C a genuine scholarly contribution rather than merely a competent summary.

The comparison highlights that literature reviews are not neutral knowledge summaries but interpretive acts that reflect and reproduce particular epistemological and political commitments. Your paper makes its commitments explicit and uses them productively; my review's apparent neutrality actually reflects unexamined mainstream assumptions about progress, coordination, and reform.

Again, the text generated by the AI appears pretty good at first sight. Nevertheless, we do not fully agree that the Reviews generated by the LLM are "descriptive syntheses". We feel they are actually also edited with a specific, in this case politically correct, filter. In the process of "presenting comprehensive overviews", this filter operated and colored the result. What we do in the next section will make more visible this coloring.

Let us note, again, the underlying sycophantic attitude of the LLM: what the user does is praised, and critics are very mild. This is a problem per se, because the chat with the LLM then to confirm the user in their own approach. What is interesting to note is that, in the following versions that Claude will produce, it will keep the edits provided by the user (e.g., the "transformation paradox"), even when we ask the LLM to shorten the output. This is probably the result of the training of the LLM to align with the user. That is problematic, because it contributes to a bias of confirmation, where the user is fed-forward to keep their own views in a continued interaction with the LLM, in a self-constructed bubble.

# 5 Moving beyond: Further improving the versions with the AI

We could have stopped here and concluded on the pros and cons of using AI to write literature reviews.

Nevertheless, the analysis of the differences between the various versions was somewhat unsettling. It suggested that literature reviews are not just limited by the bias of ignorance and the scholarship of who writes them, but also by the political perspective of the reviewer.

We took advantage of the capacity and flexibility of Claude to push the experience further, and prompt it to take into account the apparently honest and wise conclusions it drew from the comparison (listed above), and to apply the list of improvements that it had listed, on reviews B and C.

Here is how the chat continued:

[Saadi]: this was very useful, thank you. You suggested you could improve your previous literature review,, and that mine could also be improved. Can you do these improvements on each text, and provide them to me? First your improved text, then my improved text.

And here comes what was more unsettling. Claude produced a paper that started with the following paragraph, titled "Beyond Technical Fixes to Political Transformation"

The global food system faces unprecedented pressures from climate change, population growth, resource depletion, and deepening social inequalities, necessitating fundamental transformations toward sustainability. Yet the question of what constitutes "transformation" and whose interests it serves remains deeply contested. Food system transitions represent complex, multi-dimensional processes involving not merely technological innovation and institutional restructuring, but fundamental reconfigurations of power relations, knowledge systems, and value frameworks that determine who eats, what is produced, and who profits.

It went on with a strong critical perspective:

(…) food system transitions in the Global South cannot be understood without engaging with colonial legacies, structural adjustment policies, and ongoing extractivism that continue to structure contemporary agricultural systems. Similarly, transitions in the Global North occur within contexts of concentrated corporate power, growth-dependent economic models, and governance structures that systematically exclude marginalized communities from decision-making.

The papers were also presented in a critical way. E.g.,

Zwartkruis et al. (2020) identify the "dual regime challenge" where agricultural nature conservation must navigate both agricultural and nature regimes with conflicting institutional logics, demonstrating how MLP's single-regime assumption fails to capture institutional fragmentation's constraining effects. When systems must simultaneously satisfy multiple regimes with contradictory requirements, the framework's linear transition narrative breaks down entirely.

(…) This multi-stream approach fundamentally reconceptualizes transitions as emerging through conflict and compromise rather than linear progression, foregrounding the contested nature of what constitutes "improvement."

And then a whole paragraph on the transformation paradox, suggesting the need for revolutionary transformations:

> (…) Systems cannot stand outside themselves to transform; they must use existing structures, which necessarily shapes and potentially limits transformation possibilities. This paradox suggests that genuinely radical transformation may require breaks with existing systems—land occupations, seed sovereignty movements, food riots—rather than managed transitions within existing governance frameworks.

And in the conclusions:

> Food system transitions cannot be understood without explicit political economy analysis examining how capitalism, colonialism, and contemporary neoliberalism structure what transformations become possible. The persistent economic marginality of sustainable alternatives, the concentration of corporate power, and the growth imperatives constraining policy innovation reflect not accidental market failures but systematic features of capitalist food systems. Scholarship that ignores these features produces knowledge serving existing power by suggesting that problems can be solved within current structures.

> Future research must explicitly examine how class relations, racial capitalism, gender dynamics, and global inequalities shape food system transitions. Who owns land and controls capital fundamentally determines what changes occur. Intellectual property regimes determining seed access shape agricultural possibilities. Trade rules enabling dumping of subsidized commodities undermine local food security. These are not peripheral issues but central to understanding why voluntary sustainability transitions consistently fail.

And finally, in resonance with this revolutionary approach, Claude produced a much longer text (8,847 words) than implicitly expected, and noted:

> [Claude]: Note: This significantly exceeds the original 3,760-word target to fully develop the critical analysis. If a shorter version maintaining this critical depth is needed, I can condense while preserving key arguments.

[Saadi]: thank you. but you did actually merge the two reviews. What I wanted was that yous eparately improve each review, yours and mine, based on your excellent critics. First, ypu must provide an improved version of your review, keeping the same length (3760 words). And then, take mine and improve it, with the same length (3760 words). This creates two documents, your improved review, and my improved review.

Which Claude did. The new papers were in the same vein, and also introduced some aspects (e.g. gender) that were hardly present in the papers. It seems that Claude productions have become more radical, as Claude recursively generated its production on a material that became more concentrated in critical perspective, in some kind of knowledge bubble effect.

Saadi finally asked for a synthesis of the two papers, in 1500 words ("review E", see Appendix 7.1).

An extract: "Food system transitions reflect not neutral differences but historical processes of colonialism, uneven development, and ongoing extractivism. Notably absent from most frameworks is engagement with how colonial agricultural systems continue structuring contemporary food systems. The emphasis on "Global South innovations" risks romanticizing struggles emerging from structural adjustment policies rather than recognizing responses to historically produced vulnerabilities."

[Saadi] Thank you. As I see the result now, I feel you have beeen overly influenced by some critical and post-colonial sentences in my review. I think the current result (the 1500 word) synthesis is too much on the critical side, and has introduced a bit too much some of the "critical" (post-modern, postcolonial, inequalities, gender etc.) style. While this critical approach is justified, I feel that it

orients the final version beyond what I think. Please edit and tone down the postmodern style, in 1500 words.

Responding to this request, Claude produced a final paper (Review F, see appendix 7.2) that was much less critical, although not as bland and mainstream as the first Review A.

This dynamic has implications beyond literature reviews. Any iterative use of AI -for drafting, revision, refinement- risks similar drift. Without anchoring in domain knowledge, subtle signals in feedback compound across iterations.

# 6 Conclusions and recommendations

AI tools, in this case LLM transformers, are powerful tools. At first reading, the AI-generated literature reviews appeared competent -seemingly accurate, appropriately concise, and adequately comprehensive. To readers unfamiliar with the domain, they might well pass as satisfactory scholarly work. However, closer examination revealed significant shortcomings in selecting the source material, then in engaging with it.

First, we should remember that the result is only based on what papers have been selected as a review set ("garbage in, garbage out"). Good science is first and foremost about the quality of the material that is analyzed, and this is also valid for literature reviews.

The consideration set of papers, from which the review set is extracted, is subject to two biases: the bias of ignorance, where you do not know what you have left aside, and then the interpretation bias, in which the assessment of the relevance and quality depends on the competences of the reviewer.

Letting the AI choose the consideration set, or do the evaluation, is prone to severe risks of mainstreaming. Remember that in our sample, only 20 % of the human's choices within the considerations set coincided with the choices made by the LLM, hence the results were drastically different. Important papers may be left out without you knowing.

Then, AI tools demonstrated a capacity to process and summarize literature at a pace far exceeding that of human researchers, regardless of career stage. Still, the LLMs exhibited persistent tendencies toward errors, hallucinations, and over-generalization, particularly when prompted to evaluate a paper's strengths and weaknesses. Consequently, researchers frequently needed to return to original sources for verification, thereby reducing the net time savings. Given these reliability concerns, we found that researchers should still engage in careful skim-reading or, where papers appeared central to the research question, conduct full close readings.

The optimal workflow to create the review set thus positions LLMs as triage tools for large exploration: they efficiently generate initial summaries that enable researchers to strategically allocate their time toward in-depth engagement with the most relevant literature. This approach preserves scholarly rigour and attention to detail while capitalising on the LLM's capacity for rapid large-scale processing.

In generating reviews, AI tools are prolific, slavish, and opaque. Unless specifically prompted, they can be vague and superficial.

The first reviews written by LLM, when looked into more carefully, lacked the detailed content we expect from scholarly work. The AI tended toward broad generalizations without providing readers with the means to locate supporting evidence in the original texts. The reviews also failed to discriminate between stronger and weaker contributions or to prioritize studies most germane to the research questions at hand.

More critically, the LLM-generated Reviews A and B exhibited an absence of theoretical framing, -a serious deficiency when examining systemic change, where theoretical positioning is fundamental. While one might argue that targeted prompt refinement or a supplementary AI pass focused specifically on theoretical frameworks could address this gap, this remains problematic.

The LLM will by default apply specific filters in the generation of the review. In our case, political correctness, avoiding criticism and privileging the mainstream. That is by design, as such commercial tools are intended to please the customer. LLM will have difficulties seeing the logic underlying the connections between papers and tend to categorize, juxtapose, and present generic abstracts without selecting the relevant findings and illustrating them.

The conservative bias we observed is not accidental but by design. LLMs default to established frameworks and mainstream perspectives because of their training. This mainstreaming is accentuated by alignment layers: models are optimized to produce outputs that avoid controversy and present "balanced" views. Apparent neutrality is itself a position; but that "neutrality" in practice tends to advantage established knowledge and existing authority.

While the biases can be corrected by refining the result with new prompts, because the AI tends to follow your instructions with digital sycophantism, you may end up with reviews that accentuate your own biases beyond reason.

The issues noted here remain minor; but they remind us that an LLM is not a neutral tool; it is plagued with the same type of human biases that are pervasive in its human-made training material. For example, they may exhibit fake alignment, scheming, etc. (Bengio et al., 2025; Greenblatt et al., 2024; Hagendorff, 2024; Meinke et al., 2024, 2025; Park et al., 2023; Scheurer et al., 2024). That is one more reason to examine carefully the texts generated with a critical mind. The continuous progress of these models, and of their interface, will likely make many of the flaws above less apparent. It will be a limitation of this research that is was conducted with LLMs that will be obsolete a few months after its publication. Indeed, LLMs progress very fast. Whether their progress, often driven by commercial interests, does actually solve the issues at the root is yet unclear. As long as the purpose of the models remains to please the user, we can have doubts.

Some issues above are not specific to literature reviews; they apply to any queries ("prompts") to AI models.

Regarding literature reviews, addressing the bias of ignorance remains one of the most important specific issues. In a well-trodden domain, there are usually so many redundant papers that it is unlikely that we miss a sub-domain completely. Nevertheless, in new fields or border fields where literature is scarce, that can happen; It is also easy to miss papers that address the same phenomenon but with another subdiscipline's jargon. Here, the expert knowledge of the real world by the researcher is essential (Lahlou, 1995), as well as the open-minded exploration of possible less-known theories (e.g., discussing with colleagues, searching for the references listed in the papers found). The best guardrail is to have had a go at the empirical material before doing the literature review. This will help you evaluate how relevant each paper is to your own material. Empirical evidence is the gold standard of truth; the rest is mostly discourse of which the validity can be difficult to assess.

Designing prompts is becoming a discipline *per se*, and we cannot expand on that here. Still, let us note the interesting issue of size of the response. We noted in the discussion of review B that providing good information requires space, and therefore limiting too much the size of the response might lead to hardly usable, likely too generic, response. It is remarkable that in a further interaction, where Claude was asked to improve the text based on the various limitations that had been pointed in a previous chat about the comparison between Review B and review C, while the prompt implicitly requested an improved version in the same size, the LLM provided a much longer response and added *This significantly exceeds the original 3,760-word target to fully develop the critical analysis.* We recommend trying various lengths to see what we're missing, even if the final review must be shorter.

The whole exercise demonstrates literature review is never neutral. In this exercise, 6 (in fact 8 if we count the intermediary versions) reviews were compared. They differ considerably in content and tone. Furthermore, the version B we examined in detail, and edited, had to be massively changed (more than half!) And we could have continued and created more versions. The literature is endless; interpretations of the same text can vary. It is important to be aware of one's own intentions, preconceptions and biases. LLMs are not a panacea: they also have their own preconceptions and biases, and one should not trust them nor believe they would be "neutral".

Reflexivity is an important aspect of science; literature reviews require it too. Be clear on your approach, and your preconceptions. Using the AI to edit our review with various lenses (e.g. more or less critical, more or less mainstream) will help you see better the biases of interpretation in your work. We suspect that one of the things the use of AI will bring to analysts is to force them to clarify their own process of interpretation, by making explicit what they search for, how, and how they evaluate the quality of their findings.

Finally, let us remind that a literature review has two goals: one is for its readers, to provide them with the necessary background needed to understand the research, see the knowledge gap and what are the contributions of the research.

The second goal is for the researchers themselves: reading the literature is a necessary step to understand what has been learned so far on the topic of investigation. Reading the papers enriches the neural network of the researchers' brains and empowers them with better understanding, one that grows as the papers are read. The literature is written to be read, not just cited.

Asking the LLM to do all the work will be a temptation for many: but that would be a missed opportunity to learn and do better research. It is also, as we showed in this paper, a big mistake because without domain knowledge it is not possible to guide the LLM and produce a good literature review. It is ironic that you need a good command of the literature to use AI well to analyze the literature; but after all that is common sense: you need to know the domain to evaluate the output. So, LLMs will be a useful tool, especially for screening the most relevant paper and facilitate the work, help us reflect on our exploration and construction process, facilitating the management of large masses of information; but in the end they will only be scaffoldings to help us read, learn and enrich our own brain's neural networks.

# 7 References cited

**Note that references within the reproduced reviews are not necessarily in the paper's own reference list. These can be found in Appendices, sections 8.5.1 and 8.5.2.**

Adel, A., & Alani, N. (2025). Can generative AI reliably synthesise literature? exploring hallucination issues in ChatGPT. *AI and Society*, 1–14. https://doi.org/10.1007/S00146-025-02406-7/FIGURES/9

Akrich, M., Callon, M., & Latour, B. (2006). Sociologie de la traduction: textes fondateurs. In *« Sciences sociales »*. Mines Paris, les Presses.

Angel, D., & Rock, M. T. (2009). Environmental rationalities and the development state in East Asia: Prospects for a sustainability transition. In *Technological Forecasting and Social Change* (Vol. 76, Number 2, pp. 229–240). https://doi.org/10.1016/j.techfore.2008.01.004

Bastian, H., Glasziou, P., & Chalmers, I. (2010). Seventy-five trials and eleven systematic reviews a day: How will we ever keep up? *PLoS Medicine*, *7*(9). https://doi.org/10.1371/JOURNAL.PMED.1000326

Bayard, P. (2011). *Comment parler des livres que l'on n'a pas lus?* Editions de Minuit.

Bengio, Y., Mindermann, S., Privitera, D., Besiroglu, T., Bommasani, R., Casper, S., Choi, Y., Fox, P., Garfinkel, B., Goldfarb, D., Heidari, H., Ho, A., Kapoor, S., Khalatbari, L., Longpre, S., Manning, S., Mavroudis, V., Mazeika, M., Michael, J., … Zeng, Y. (2025). International AI Safety Report. *ArXiv*, (2501.17805 [cs.CY]), 1–298. https://arxiv.org/abs/2501.17805v1

Bolaños, F., Salatino, A., Osborne, F., & Motta, E. (2024). Artificial intelligence for literature reviews: opportunities and challenges. *Artificial Intelligence Review 2024 57:10*, *57*(10), 1–49. https://doi.org/10.1007/S10462-024-10902-3

Burgard, T., & Bittermann, A. (2023). Reducing Literature Screening Workload with Machine Learning: A Systematic Review of Tools and Their Performance. *Zeitschrift Fur Psychologie / Journal of Psychology*, *231*(1), 3–15. https://doi.org/10.1027/2151-2604/A000509

Chen, S., Gao, M., Sasse, K., Hartvigsen, T., Anthony, B., Fan, L., Aerts, H., Gallifant, J., & Bitterman, D. S. (2025). When helpfulness backfires: LLMs and the risk of false medical information due to sycophantic behavior. *Npj Digital Medicine*, *8*(1), 605. https://doi.org/10.1038/S41746-025-02008-Z

Cohen, A. M., Hersh, W. R., Peterson, K., Yen, P.-Y., & Cohen, A. M. (2006). Reducing Workload in Systematic Review Preparation Using Automated Citation Classification. *J Am Med Inform Assoc*, *13*, 206–219. https://doi.org/10.1197/jamia.M1929

Cooper, H. M. (1982). Scientific Guidelines for Conducting Integrative Research Reviews. *Source: Review of Educational Research*, *52*(2), 291–302.

DORA. (2012). *San Francisco Declaration on Research Assessment*. Https://Sfdora.Org/Read/.

Elzen, B., Geels, F. W., Leeuwis, C., & Van Mierlo, B. (2011). Normative contestation in transitions "in the making": Animal welfare concerns and system innovation in pig husbandry. *Research Policy*, *40*(2), 263–275. https://doi.org/10.1016/j.respol.2010.09.018

Fisher-Vanden, K., & Weyant, J. (2020). The Evolution of Integrated Assessment: Developing the Next Generation of Use-Inspired Integrated Assessment Tools. *Annual Review of Resource Economics*, *12*, 471–487. https://doi.org/10.1146/ANNUREV-RESOURCE-110119-030314

Frankfurt, H. G. (2005). *On Bullshit*. Princeton University Press.

Geels, F. W. (2002). Technological transitions as evolutionary reconfiguration processes: a multi-level perspective and a case-study. *Research Policy*, *31*(8–9), 1257–1274. https://doi.org/10.1016/S0048-7333(02)00062-8

Geels, F. W. (2011). The multi-level perspective on sustainability transitions: Responses to seven criticisms. *Environmental Innovation and Societal Transitions*, *1*(1), 24–40. https://doi.org/10.1016/j.eist.2011.02.002

Greenblatt, R., Denison, C., Wright, B., Roger, F., MacDiarmid, M., Marks, S., Treutlein, J., Belonax, T., Chen, J., Duvenaud, D., Khan, A., Michael, J., Mindermann, S., Perez, E., Petrini, L., Uesato, J., Kaplan, J., Shlegeris, B., Bowman, S. R., & Hubinger, E. (2024). Alignment faking in large language models. *ArXiv*, (2412.14093v2 [cs.AI] 2), 1–137. https://arxiv.org/abs/2412.14093v2

Guler, N., Kirshner, S. N., & Vidgen, R. (2024). A literature review of artificial intelligence research in business and management using machine learning and ChatGPT. *Data and Information Management*, *8*(3), 100076. https://doi.org/10.1016/J.DIM.2024.100076

Hagendorff, T. (2024). Deception abilities emerged in large language models. *Proceedings of the National Academy of Sciences of the United States of America*, *121*(24), e2317967121. https://doi.org/10.1073/PNAS.2317967121/SUPPL_FILE/PNAS.2317967121.SAPP.PDF

Hanson, M. A., Barreiro, P. G., Crosetto, P., & Brockington, D. (2024). The strain on scientific publishing. *Quantitative Science Studies*, *5*(4), 823–843. https://doi.org/10.1162/qss_a_00327

Harrison, T. B., Hu, D., Fu, S., & Liu, H. (2025). The Use of Large Language Models to Accelerate Literature Review Towards Digital Health Equity and Inclusiveness. *AMIA Annual Symposium Proceedings*, *2024*, 493. https://pmc.ncbi.nlm.nih.gov/articles/PMC12099426/

Hart, C. (1998). The literature review in research: Releasing the social science imagination. *Doing a Literature Review*, pp.01-25. https://books.google.co.zw/books?id=FkXvY8oDFdsC&printsec=frontcover&source=gbs_ge_summary

Higgins, J. P., & Green, S. (2008). *Cochrane Handbook for Systematic Reviews of Interventions*. Wiley-Blackwell.

Ingram, J. (2018). Agricultural transition: Niche and regime knowledge systems' boundary dynamics. *Environmental Innovation and Societal Transitions*, *26*, 117–135. https://doi.org/10.1016/j.eist.2017.05.001

Jørgensen, U. (2012). Mapping and navigating transitions - The multi-level perspective compared with arenas of development. *Research Policy*, *41*(6), 996–1010. https://doi.org/10.1016/j.respol.2012.03.001

Kruger, J., & Dunning, D. (1999). Unskilled and Unaware of It: How Difficulties in Recognizing One's Own Incompetence Lead to Inflated Self-Assessments. *Journal of Personality and Social Psychology*, *77*(6), 121–1134.

Lahlou, S. (1995). Vers une théorie de l'interprétation en analyse des données textuelles. In S. Bolasco, L. Lebart, & A. Salem (Eds.), *JADT 1995. 3rd International Conference on Statistical Analysis of Textual Data. Vol. 1: I* (pp. 221–228). CISU.

Lahlou, S. (2017). *Installation theory: The societal construction and regulation of behaviour*. Cambridge University Press. https://doi.org/10.1017/9781316480922

Lahlou, S., Nowotny, H., Yamin, P., Thurner, S., & Cordelois, A. (2024). The Food Socioscope: a new instrument to better understand transitions towards greater sustainability. *Proceedings of the Paris Institute for Advanced Study*, *Vol. 22*. https://doi.org/10.5281/ZENODO.13935390

Marshall, I. J., & Wallace, B. C. (2019). Toward systematic review automation: A practical guide to using machine learning tools in research synthesis. *Systematic Reviews*, *8*(1), 1–10. https://doi.org/10.1186/S13643-019-1074-9/TABLES/2

Meinke, A., Schoen, B., Scheurer, J., Balesni, M., Shah, R., & Hobbhahn, M. (2024). *Frontier Models are Capable of In-context Scheming*. https://arxiv.org/abs/2412.04984v2

Meinke, A., Schoen, B., Scheurer, J., Balesni, M., Shah, R., & Hobbhahn, M. (2025). Frontier Models are Capable of In-context Scheming. *ArXiv*, (2412.04984v2), 1–72.

Moher, D., Liberati, A., Tetzlaff, J., Altman, D. G., & The PRISMA Group. (2009). Guidelines and Guidance Preferred Reporting Items for Systematic Reviews and Meta-Analyses: The PRISMA Statement. *PLoS Medicine*, *6*(7), e1000097. https://doi.org/10.1371/journal.pmed.1000097

Naddaf, M. (2025). AI chatbots are sycophants - researchers say it's harming science. *Nature*, *647*(8088), 13–14. https://doi.org/10.1038/D41586-025-03390-0

Nordhaus, W. D. (1994). *Managing the Global Commons: The Economics of Climate Change* . MIT Press.

O'connor, A. M., Tsafnat, G., Gilbert, S. B., Thayer, K. A., & Wolfe, M. S. (2018). Moving toward the automation of the systematic review process: a summary of discussions at the second meeting of International Collaboration for the Automation of Systematic Reviews (ICASR). *Systematic Reviews*, (7:3). https://doi.org/10.1186/s13643-017-0667-4

Papachristos, G., & Adamides, E. (2016). A retroductive systems-based methodology for socio-technical transitions research. *Technological Forecasting and Social Change*, *108*, 1–14. https://doi.org/10.1016/j.techfore.2016.04.007

Park, P. S., Goldstein, S., O'Gara, A., Chen, M., & Hendrycks, D. (2023). AI Deception: A Survey of Examples, Risks, and Potential Solutions. *Patterns*, *5*(5). https://doi.org/10.1016/j.patter.2024.100988

Pawson, R., Greenhalgh, T., Harvey, G., & Walshe, K. (2005). Realist review-a new method of systematic review designed for complex policy interventions. In *Journal of Health Services Research & Policy* (Vol. 10).

Scheurer, J., Balesni, M., & Hobbhahn, M. (2024). Large Language Models CAn Strategically deceive Their Users When Put Uneder Pressure. *ArXiv:2311.07590v4 [Cs.CL]*.

Shen, T., Jin, R., Huang, Y., Liu, C., Dong, W., Guo, Z., Wu, X., Liu, Y., & Xiong, D. (2023). *Large Language Model Alignment: A Survey*.

Snyder, H. (2019). Literature review as a research methodology: An overview and guidelines. *Journal of Business Research*, *104*, 333–339. https://doi.org/10.1016/j.jbusres.2019.07.039

Sperber, D., & Wilson, D. (1986). *Relevance: Communication and Cognition*. Blackwell.

Templier, M., & Paré, G. (2015). A Framework for Guiding and Evaluating Literature Reviews. *Communications of the Association for Information Systems*, *37*(1), 6. https://doi.org/10.17705/1CAIS.03706

Turnheim, B., Berkhout, F., Geels, F. W., Hof, A., Mcmeekin, A., Nykvist, B., & Van Vuuren, D. P. (2015). *Evaluating sustainability transitions pathways: Bridging analytical approaches to address governance challenges*. https://doi.org/10.1016/j.gloenvcha.2015.08.010

van Vuuren, D. P., Kok, M., Lucas, P. L., Prins, A. G., Alkemade, R., van den Berg, M., Bouwman, L., van der Esch, S., Jeuken, M., Kram, T., & Stehfest, E. (2015). Pathways to achieve a set of ambitious global sustainability objectives by 2050: Explorations using the IMAGE integrated assessment model. *Technological Forecasting and Social Change*, *98*, 303–323. https://doi.org/10.1016/J.TECHFORE.2015.03.005

Wagner, G., Lukyanenko, R., & Paré, G. (2022). Artificial intelligence and the conduct of literature reviews. *Journal of Information Technology*, *37*(2), 209–226. https://doi.org/10.1177/02683962211048201/ASSET/57A60C1C-CE32-4844-BFEE-16DC60A61F1F/ASSETS/IMAGES/LARGE/10.1177_0268396221104820-FIG1.JPG

Walters, W. H., & Wilder, E. I. (2023). Fabrication and errors in the bibliographic citations generated by ChatGPT. *Scientific Reports 2023 13:1*, *13*(1), 1–8. https://doi.org/10.1038/s41598-023-41032-5

Zwartkruis, J. V., Berg, H., Hof, A. F., & Kok, M. T. J. (2020). Agricultural nature conservation in the Netherlands: Three lenses on transition pathways. *Technological Forecasting and Social Change*, *151*. https://doi.org/10.1016/j.techfore.2018.03.006

# 8 Appendices

## 8.1 Appendix 1: Material and methods detailed

We created two separate chains of literature research within our review. The first sought to analyze implicit rule formation and realpolitik within food systems governance (referenced as Topic 1), and the second focused on broader systems transitions (referenced as Topic 2). We drew on the Multi-Level Perspective (MLP) on Transitions (Geels, 2002, 2011), and Political Economy/Ecology, in order to manually identify papers of interest. This guided us in a snowballing process for secondary source collection, narrowing the disciplinary scope and identify a focused set of theoretical lenses relevant to the review.

From here, we began to scale the process using AI Large Language Model (LLM) tools, such as ChatGPT, Perplexity, and Claude. Using the AI models ChatGPT 4o and Perplexity, we created quick summaries of the papers, using a standardized prompt. The prompt asked the AI model to provide a 300-word summary, details on the research methods, and an evaluation of the paper, giving its strengths and weaknesses. We explicitly asked the AI model to detail the theoretical framework, focus, methods, and country of study; in order to easily categorize each paper.

Prompt 1:

Please summarise this paper below in 300 words.

Then, in bullet points, provide details of the research methods, including the theoretical framework and the data used. Please include, where applicable and the data available, details on how many participants were included in each case study, or the size of N more generally in surveys or quantitative data. Where applicable and information available, please include tools of data analysis (such as NVivo or ATLAS.ti).

Provide a few sentences at the end which evaluate the strengths and weaknesses of the paper. Where applicable, please include an evaluation of the research methods within this section.

Finally, please categorise the paper in a few simple words, on its focus (e.g. renewable energy, organic farming, dairy farming etc.), on its research method (e.g. interviews, focus groups, surveys, observations etc.), its theoretical framework (e.g. Multi level perspective, actor-network theory etc.), and on its country of study, if applicable.

An example output from ChatGPT 4o can be seen below for the paper by (Angel & Rock, 2009)

Summary (≈300 words):

This paper explores the potential for sustainability transitions in newly industrializing countries (NICs) in East Asia—specifically China, Singapore, Taiwan, and Thailand—by examining their political economies and environmental governance systems. The authors analyze how these countries, once known for "grow now, clean up later" approaches, have evolved toward more pragmatic and proactive environmental regulation, and how this may (or may not) support deeper transitions to alternative, low-resource socio-technical regimes.

Using the multi-level perspective (MLP) on socio-technical transitions as a conceptual framework, the study traces how East Asian development states integrated environmental goals into existing institutions of industrial growth. It highlights two distinct phases: (1) the adoption of OECD-style environmental regulatory reforms during the 1980s–1990s, and (2) a more recent phase of integrating environmental concerns into mainstream development institutions (e.g., investment boards, infrastructure planning).

While phase one led to real but incremental improvements in pollution control, it largely remained within existing socio-technical regimes (e.g., continued reliance on fossil fuels and mass motorization). Phase two shows promise for systemic change, as development agencies begin to support renewable energy, green industries, and sustainable urban mobility. However, these efforts are often fragmented, incomplete, and challenged by institutional inertia and rapid economic growth that offsets efficiency gains.

The authors argue that the East Asian context differs significantly from Western transition models. Here, state-led industrial upgrading and integration into global production networks offer both barriers and unique leverage points for sustainability transitions. The paper concludes that meaningful regime shifts in East Asia may depend more on mainstream industrial and urban development institutions than on niche innovations, as is often assumed in Western-focused literature.

Research Methods (Bullet Points):

· Methodology: Qualitative, comparative case study across four countries (China, Singapore, Taiwan, Thailand).

· Data Sources:

o Extensive review of secondary literature (policy studies, environmental governance, development economics).

o Use of existing empirical studies and historical reviews of environmental regulation.

o No original interviews or primary fieldwork reported.

· Theoretical Framework:

o Multi-Level Perspective (MLP) (Geels et al.) on socio-technical transitions.

o Also draws on political economy literature, especially on the "developmental state" model.

· Case Scope:

o Four country-level case studies, each discussed across sectors (e.g., pollution control, energy, transport).

o No mention of participant numbers, surveys, or software (e.g., NVivo or ATLAS.ti).

· Analytical Focus:

o Traces historical reforms, institutional structures, and environmental performance trends.

o Emphasis on institutional integration, internationalization, and socio-political dynamics.

Evaluation (Strengths & Weaknesses):

Strengths:

· Offers a unique, non-Western perspective on sustainability transitions.

· Strong integration of political economy and transition theory.

· Rich historical and institutional analysis.

· Highlights pragmatic governance and state-led innovation pathways.

Weaknesses:

· No primary empirical data; relies entirely on secondary sources.

· Lacks systematic methodology (e.g., no clear coding framework or case selection criteria).

· No quantified impact evaluation or comparative metrics.

· Some claims remain speculative or generalized without empirical support.

Categorisation:

· Focus: Sustainability transitions; environmental governance; pollution control; renewable energy.

· Research Method: Qualitative case studies; secondary literature review.

· Theoretical Framework: Multi-Level Perspective (MLP); Developmental State Theory.

· Country of Study: China, Singapore, Taiwan, Thailand (East Asia).

Using this, we resolved how much attention to give each paper, and whether a more in-depth reading would be beneficial.

The depth of reading was traced using the coding system inspired by Pierre Bayard's "How to Talk About Books You Haven't Read (Bayard, 2011). The category of "not read" meant that we read at most the abstract and the AI-generated summary. This was used when, after reviewing the AI-generated summary, we decided that the source would not be directly related to the project. "Skimmed" meant that we read the introduction, conclusion, and methods sections but not the full paper the whole way through. This was used when, having identified a relevant source through the AI summary, the reader skimmed the paper to look for key information omitted by the LLM or hallucinations. "Read fully" meant that we read the paper in its entirety, either because we determined it to be a key source, with lots of relevant information, or because the AI made multiple errors and therefore a thorough quality control was required.

Based on our further readings, we made some edits to the summaries, where necessary. For example, if the AI model missed information we deemed to be useful for the project or a broader literature review or whether the AI model hallucinated or included vague claims in its answer. Changes made were added to the summaries in red, to easily differentiate it from the information produced by the AI model. We also removed, where applicable, generic comments by the AI; for instance, weaknesses which talked about limited generalizability of qualitative studies (this example appeared in almost every paper that we asked the AI to summarize).

The LLMs exhibited persistent tendencies toward errors, hallucinations, and overgeneralisation, particularly when prompted to evaluate a paper's strengths and weaknesses. Consequently, researchers frequently needed to return to original sources for verification, thereby reducing the net time savings. Given these reliability concerns, we found that researchers should still engage in careful skim-reading or, where papers appeared central to our research questions on systemic change in food sustainability, conduct full close readings.

A practical verification strategy emerged: cross-checking specific, verifiable details such as sample sizes or the number of interviews conducted against the AI-generated summaries proved effective in assessing the accuracy of the automated review. We also discovered that prompt engineering significantly influenced output quality. Structured prompts requiring the LLM to extract specific information from designated sections—such as methodology descriptions or key findings— yielded more reliable summaries by preventing premature truncation of the reading process due to token limitations. Nevertheless, even optimally prompted LLMs failed to produce the granular, contextualised analysis expected in rigorous academic literature reviews. These nuances invariably required subsequent addition by human researchers following their own engagement with the source material.

The optimal workflow thus positions LLMs as triage tools: they efficiently generate initial summaries that enable researchers to strategically allocate their time toward in-depth engagement with the most relevant literature. This approach preserves scholarly rigour and attention to detail while capitalising on the LLM's capacity for rapid large-scale processing.

Using the example prompt and paper summary above, we made the following edits to the evaluation weaknesses section:

"Shifts in socio-technical regimes may depend more on mainstream industrial institutions than on niche innovations."

"East Asian NICs are particularly attuned to global market pressures, which may drive sustainability transitions."

"Civil society and environmental NGOs are gaining influence and shaping environmental governance."

"Integration of environmental concerns into development institutions is a promising pathway for sustainability transitions."

"Global production networks help diffuse environmental standards and practices."

Following the summary, we added all the sources we used to a table. The table gives a quick overview of the key information of the paper, including the date, full citation, theoretical framework, the focus of the paper, the methods, geographical location, and, if they used interviews or surveys, how many were used in the study. This helped standardize evaluations and cross-reference between entries.

The table also detailed any changes we made to the summaries (categorizing this as changes to the summary, methods or evaluation), and the depth of the reading we had done on the source. The categories (not read, skimmed read fully) are those described above based on Bayard's classification.

An extract of the table can be seen in the figure below:

| Year of publi | Full citation (APA) | Author(s) | Theoretical Fram | Focus | Method | Country focus | Number of interviews (if applicable) | Summaries Edite | Degree of reading |
|---|---|---|---|---|---|---|---|---|---|
| 2009 | Angel, D., & Rock, M. T. ( | Angel and Rock | MLP + Development | Sustainab | Literature revi | East Asia (Multi) | | Evaulutation edited | Not read |
| 2022 | Arslan, A., Cavatassi, R., | Arslan et al. | | Structural transfor | Food syste | Quantitative c | Multi (Global South) | | Methods and Evalua | Skimmed |
| 2017 | Audet, R., Lefèvre, S., Br | Audet et al. | MLP | | Seasonal f | Interviews and | Canada | 15 | Methods edited | Skimmed |
| 2016 | Avelino, F., & Wittmayer | Avelino and Wi | MLP, socio-technic | Governan | Literature revi | UK and the Netherlands | | Summary and Meth | Skimmed |
| 2021 | Averbuch, B., Hyarregaa | Averbush et al. | MLP + longue durée | Dairy sect | Literature revi | Denmark | | Not edited | Skimmed |
| 2010 | Bai, X., Roberts, B., & Cl | Bai et al. | MLP + Urban Enviro | Urban sus | Literature revi | Asia (Multi) | | Not edited | Skimmed |
| 2015 | Baker, L. (2015). Renew | Baker | MLP | Renewabl | Semi-structur | South Africa | 79 | Methods edited | Skimmed |
| 1997 | Banks, J., & Marsden, T. | Banks and Mar | Political economy + | Dairy sect | Literature revi | UK | | Not edited | Not read |
| 2010 | Beers, P. J., Veldkamp, A | Beers et al. | MLP | Agricultur | Modeling / sim | The Netherlands | | Not edited | Skimmed |
| 2012 | Bell, S., & Cerulli, C. (20 | Bell and Cerull | MLP + complexity th | Urban agri | Participant ob | UK | | Not edited | Skimmed |
| 2018 | Belmin, R., Casabianca | Belmin et al. | MLP | Geographi | Surveys, litera | France | 32 | Not edited | Skimmed |
| 2011 | Berkhout, F. G. H., Wiec | Berkhout et al. | Multi-Level Perspec | Sustainab | Literature revi | Multi (Asia and Latin America) | | Not edited | Not read |

Having generated the summaries, conducted further reading based on these, and edited where necessary, we rated the papers based on their significance to the project. "***" was given to papers that were extremely related to the project, providing perhaps seminal work or an original method or theoretical framework. "**" deemed that the paper was interesting and highly related to the project, with a strong theoretical or methodological framework. Finally, "*" was given to papers that were broadly related to the project, but perhaps had some weaknesses in their methods or were not directly related to the research questions.

Once we had around 55 papers for Topic 1 and around 120 papers for Topic 2, we inputted the table into ChatGPT 4o, and asked it to produce 20 more papers for each topic, using the examples in the table to guide its decision-making process. We then asked it to add to the table and fill out the following columns: date, author(s), theoretical framework, focus, method, and country focus.

Prompt 2:

This table was constructed as a literature review for the Socioscope project:

It is devoted to a major research effort based on the collection of qualitative data that try to capture what happens 'on the ground', from micro to macro. It involves practitioners who respond to in-depth field interviews, following a carefully tested protocol that will allow a rigorous comparison of data. Data collected will form the basis for a better understanding of systemic changes towards greater food sustainability, ranging from primary production to transport, storage, retail, preparation, consumption, waste management, including the supporting infrastructure and organizational set-ups. By starting from initiatives on the micro level and how they are embedded in the wider economic, administrative, political and cultural context, the aim is to understand the mechanisms of transition towards greater sustainability, including the barriers and conditions of success, and understanding societal transitions.

The categories and category explanations for the paper's significance are as follows:

*: Is broadly interesting and shows that the literature review has an overall understanding of all the authors and papers on the subject. This kind of literature would be included as part of a larger literature review.

**: Has solid research and methodological backing OR has a key theoretical framework, and is highly relevant (may have cases that could be used in future research).

***: Is highly relevant, solid research and methodological backing, and is original or unique. May be seminal literature to the project itself.

Please continue to fill out the table for the next 20 rows, finding more relevant papers, based on the literature already included and the information given here on the project. Provide an Excel document once this has been done.

Please ignore and do not fill out the columns "degree of reading", "edited post AI", "significance", "ChatGPT significance", and "ChatGPT explanation".

We then also asked the AI model to rate the significance of the papers, giving it the literature review summaries (with their methods and evaluations), the significance categories, and explanation of the Socioscope project, but not giving it our own significance ratings. We also asked it to give a short explanation to justify its rating.

Prompt 3:

Please read this whole document and for each paper summary, please categorise the significance of each of these papers (around 120 papers in total). This will be used to create priorities for what to include and in what detail, in a literature review for the following project: The Food Socioscope. It is devoted to a major research effort based on the collection of qualitative data that try to capture what happens 'on the ground', from micro to macro. It involves practitioners who respond to in-depth field interviews, following a carefully tested protocol that will allow a rigorous comparison of data. Data collected will form the basis for a better understanding of systemic changes towards greater food sustainability, ranging from primary production to transport, storage, retail, preparation, consumption, waste management, including the supporting infrastructure and organizational set-ups. By starting from initiatives on the micro level and how they are embedded in the wider economic, administrative, political and cultural context, the aim is to understand the mechanisms of transition towards greater sustainability, including the barriers and conditions of success.

The categories and category explanations for the paper's significance are as follows:

*: Is broadly interesting and shows that the literature review has an overall understanding of all the authors and papers on the subject. This kind of literature would be included as part of a larger literature review.

**: Has solid research and methodological backing OR has a key theoretical framework, and is highly relevant (may have cases that could be used in future research).

***: Is highly relevant, solid research and methodological backing, and is original or unique. May be seminal literature to the project itself.

Please provide a short justification for the reasoning for the categorisation of the paper. Present your findings in a table with the following columns (Paper citation (eg. Hassink, J., Grin, J., & Hulsink, W. (2018). Enriching the multi-level perspective by better understanding agency and challenges associated with interactions across system boundaries. The case of care farming in the Netherlands: Multifunctional agriculture meets health care. Journal of Rural Studies, 57, 186-196.), significance rating (this will be either "*", "**", or "***"), and justification (a short justification, which could point to the research methods, theoretical framework, or the paper's case study, as examples).

Please create the table and I will copy it for my own use. Please do this for all of the following papers in the document.

We added these significance ratings to the table for ChatGPT and Claude, and compared it to our own rating. We then created a column which detailed whether the AI model had rated a source higher, lower, or matched our own rating.

To increase the reliability of the human ratings, we first had Annabelle and Julian rate the significance independently. We calculated Cohen's kappa coefficient to measure inter-rater reliability using just under 300 papers. We selected 20 sources where discrepancies appeared, and we discussed these together to calibrate. This allowed us to justify our decision making and amend the ratings, so they matched. Through this system, we addressed systematic rating differences, discussing our criteria and resolving systematic differences in interpretation. The transcript from our discussion can be found in Appendix 2.

Overall, using this process, we created literature summaries of just under 280 papers, including the two streams of sources (transition studies, policy studies).

Finally, we submitted the full literature table to Claude (Anthropic) for a third-party AI ranking, as past testing showed it offered more granular scrutiny than ChatGPT 4o, particularly in evaluation of detail and nuance in secondary data. This provided a final table with all ratings and comments.

Below, Appendix 2 contains the final table.

## 8.2 Appendix 2: The papers in the review consideration set.

*Significance ratings: \* = low; \*\* = moderate; \*\*\* = high. "Degree of reading": Not read / Skimmed / Read fully.*

| Year | Full Citation (APA) | Author(s) | Theoretical Framing | Focus | Method | Country Focus | N (interviews) | Summaries Edited? | Annabelle – Degree of Reading | Annabelle Significance | ChatGPT Significance | ChatGPT Explanation | Significance Match? | Julian – Degree of Reading | Julian Significance | Claude Significance | Claude Explanation |
|---|---|---|---|---|---|---|---|---|---|---|---|---|---|---|---|---|---|
| 2009 | Angel, D., & Rock, M. T. (2009). Environmental rationalities and the development state in East Asia: Prospects for a sustainability transition. *Technological Forecasting and Social Change, 76*(2), 229–240. | Angel and Rock | MLP + Developmental State Theory | Sustainability transitions | Literature review | East Asia (Multi) | | Evaluation edited | Not read | \* | \*\* | Political-economy framing of East Asian transitions; secondary sources and expert interviews elucidate state-led niche pathways with parallels to agri-industrial contexts. | ChatGPT rated higher | Not read | \* | \* | Broad environmental governance focus. Useful Asian perspective but lacks empirical depth. |
| 2022 | Arslan, A., Cavatassi, R., & Hossain, M. (2022). Food systems and structural and rural transformation: a quantitative synthesis for low and middle-income countries. *Food Security*, 1–28. | Arslan et al. | Structural transformation | Food systems transformation | Quantitative cross-country analysis | Multi (Global South) | | Methods and Evaluation edited | Skimmed | \*\*\* | \*\* | Meta-analysis of >50 studies linking food system changes with rural development metrics; quantitatively rigorous and policy-relevant. | ChatGPT rated lower | Not read | \*\*\* | \* | Large-scale quantitative analysis lacking ground-level insights. Useful context but limited relevance to qualitative transition research. |
| 2017 | Audet, R., Lefèvre, S., Brisebois, É., & El-Jed, M. (2017). Structuring tensions and key relations of montreal seasonal food markets in the sustainability transition of the agri-food sector. *Sustainability, 9*(3), 320. | Audet et al. | MLP | Seasonal food markets | Interviews and focus groups | Canada | 15 | Methods edited | Skimmed | \*\*\* | \*\* | Action-research with practitioners uncovers social-innovation mechanisms; rich qualitative data but limited to one city. | ChatGPT rated lower | Skimmed | \*\*\* | \*\*\* | Exceptional action research (2 years) with market organizers exploring social innovation in food systems. Concept of "structuring tensions" highly relevant for understanding practitioner challenges in sustainability transitions. |
| 2016 | Avelino, F., & Wittmayer, J. M. (2016). Shifting power relations in sustainability transitions: a multi-actor perspective. *Journal of Environmental Policy & Planning, 18*(5), 628–649. | Avelino and Wittmayer | MLP, socio-technical systems, governance | Governance | Literature review | UK and the Netherlands | | Summary and Methods edited | Skimmed | \* | | ChatGPT ignored | | Not read | \* | \*\* | Important theoretical framework for actor analysis. Limited empirical validation but valuable conceptual tool. |

| Year | Full Citation (APA) | Author(s) | Theoretical Framing | Focus | Method | Country Focus | N (interviews) | Summaries Edited? | Annabelle – Degree of Reading | Annabelle Significance | ChatGPT Significance | ChatGPT Explanation | Significance Match? | Julian – Degree of Reading | Julian Significance | Claude Significance | Claude Explanation |
|---|---|---|---|---|---|---|---|---|---|---|---|---|---|---|---|---|---|
| 2021 | Averbuch, B., Hvarregaard Thorsøe, M., & Kjeldsen, C. (2021). Longue durée study of agricultural transitions in Denmark using Multi-Level Perspective. *Geografisk Tidsskrift–Danish Journal of Geography, 121*(1), 30–45. | Averbush et al. | MLP + longue durée historiography | Dairy sector | Literature review, historical case studies | Denmark | | Not edited | Skimmed | * | ** | Combines historical records and stakeholder interviews to trace century-scale transitions. | ChatGPT rated higher | Not read | * | ** | Innovative historical approach revealing deep structures in transitions. Important for understanding path dependencies though relies on secondary sources. |
| 2010 | Bai, X., Roberts, B., & Chen, J. (2010). Urban sustainability experiments in Asia: patterns and pathways. *Environmental Science & Policy, 13*(4), 312–325. | Bai et al. | MLP + Urban Environmental Evolution | Urban sustainability | Literature review | Asia (Multi) | | Not edited | Skimmed | * | ** | Early cross-sector Asian experiment typology includes urban agriculture niches; mixed methods identify patterns. | ChatGPT rated higher | Skimmed | ** | * | Urban focus with limited food relevance. Pattern analysis approach potentially useful. |
| 2015 | Baker, L. (2015). Renewable energy in South Africa's minerals-energy complex: a 'low carbon' transition? *Review of African Political Economy, 42*(144), 245–261. | Baker | MLP | Renewable energy | Semi-structured interviews | South Africa | 79 | Methods edited | Skimmed | ** | ** | Political-economy case mixing policy analysis and interviews to show how low-carbon niches interact with entrenched regimes. | Match | Not read | ** | * | Energy sector analysis with limited food system connections. Power dynamics insights potentially useful. |
| 1997 | Banks, J., & Marsden, T. (1997). Reregulating the UK dairy industry: the changing nature of competitive space. *Sociologia Ruralis, 37*(3), 382–404. | Banks and Marsden | Political economy + regulation theory | Dairy sector | Literature review | UK | | Not edited | Not read | * | * | Historical niche-regime analysis; valuable for context but dated and limited methodological innovation. | Match | Not read | * | * | Historical analysis of dairy deregulation. Dated but provides useful context, though lacks primary empirical data. |
| 2010 | Beers, P. J., Veldkamp, A., Hermans, F., van Apeldoorn, D., Vervoort, J. M., & Kok, K. (2010). Future sustainability and images. *Futures, 42*(7), 723–732. | Beers et al. | MLP | Agriculture | Modelling / simulation | The Netherlands | | Not edited | Skimmed | ** | * | Imagery and foresight study; useful for understanding visioning processes but limited direct transition analysis. | ChatGPT rated lower | Not read | * | * | Conceptual paper on role of images in transitions. Limited empirical grounding reduces practical relevance. |
| 2012 | Bell, S., & Cerulli, C. (2012). Emerging community food production and pathways | Bell and Cerulli | MLP + complexity theory | Urban agriculture (community gardens) | Participant observation | UK | | Not edited | Skimmed | * | * | Conceptual exploration of community food roles; largely descriptive | Match | Not read | * | * | Autoethnographic case study with limited methodological rigor. Valuable lived |

| Year | Full Citation (APA) | Author(s) | Theoretical Framing | Focus | Method | Country Focus | N (interviews) | Summaries Edited? | Annabelle – Degree of Reading | Annabelle Significance | ChatGPT Significance | ChatGPT Explanation | Significance Match? | Julian – Degree of Reading | Julian Significance | Claude Significance | Claude Explanation |
|---|---|---|---|---|---|---|---|---|---|---|---|---|---|---|---|---|---|
| | for urban landscape transitions. *Emergence: Complexity & Organization, 14*(1). | | | | | | | | | | | without deep empirical backing. | | | | | experience but lacks systematic analysis. |
| 2018 | Belmin, R., Casabianca, F., & Meynard, J. M. (2018). Contribution of transition theory to the study of geographical indications. *Environmental Innovation and Societal Transitions, 27*, 32–47. | Belmin et al. | MLP | Geographical indications | Surveys, literature review and statistical data | France | 32 | Not edited | Skimmed | *** | ** | Applies transition theory to GI valorisation; literature review and expert interviews frame GIs as niche governance tools. | ChatGPT rated lower | Skimmed | ** | ** | Novel application of MLP to terroir/GI systems. Strong longitudinal analysis (1960s–2010s) of Corsican clementines showing how local standards can protect alternative food systems. |
| 2011 | Berkhout, F. G. H., Wieczorek, A. J., & Raven, R. (2011). Avoiding environmental convergence: a possible role for sustainability experiments in latecomer countries? *International Journal of Institutions and Economies, 3*(2), 376–385. | Berkhout et al. | MLP + Strategic Niche Management | Sustainability experiments | Literature review | Multi (Asia and Latin America) | | Not edited | Not read | * | | | ChatGPT ignored | Not read | * | * | Conceptual argument about alternative development paths. Limited empirical support. |
| 2009 | Berkhout, F., Angel, D., & Wieczorek, A. J. (2009). Asian development pathways and sustainable socio-technical regimes. *Technological Forecasting and Social Change, 76*(2), 218–228. | Berkhout et al. | Critique of Environmental Kuznets Curve | Industrialisation | Theoretical critique | Multi (Asia) | | Not edited | Not read | * | | | ChatGPT ignored | Not read | ** | * | Conceptual paper on Asian development paths. Useful perspective but lacks empirical grounding. |
| 2010 | Berkhout, F., Verbong, G., Wieczorek, A. J., Raven, R., Lebel, L., & Bai, X. (2010). Sustainability experiments in Asia: innovations shaping alternative development pathways? *Environmental Science & Policy, 13*(4), 261–271. | Berkhout et al. | MLP + Strategic Niche Management | Sustainability transitions | Theoretical critique | Multi (Asia) | | Not edited | Not read | * | | | ChatGPT ignored | Not read | * | * | Framework paper for special issue. Valuable conceptual contribution but no original empirical work. |

| Year | Full Citation (APA) | Author(s) | Theoretical Framing | Focus | Method | Country Focus | N (interviews) | Summaries Edited? | Annabelle – Degree of Reading | Annabelle Significance | ChatGPT Significance | ChatGPT Explanation | Significance Match? | Julian – Degree of Reading | Julian Significance | Claude Significance | Claude Explanation |
|---|---|---|---|---|---|---|---|---|---|---|---|---|---|---|---|---|---|
| 2016 | Bui, S., Cardona, A., Lamine, C., & Cerf, M. (2016). Sustainability transitions: Insights on processes of niche-regime interaction and regime reconfiguration in agri-food systems. *Journal of Rural Studies, 48*, 92–103. | Bui et al. | MLP + French pragmatic sociology | Alternative food networks | Participant observation, interviews, archival review | France | Not mentioned | Not edited | Skimmed | ** | ** | Comparative MLP across four French niches; derives general patterns of regime reconfiguration with solid qualitative design. | Match | Skimmed | ** | *** | Exemplary study of four French initiatives showing how niches can reconfigure local regimes. Rich participatory action research over 3 years with clear practical and theoretical contributions. |
| 2014 | Bush, S. R., & Marschke, M. J. (2014). Making social sense of aquaculture transitions. *Ecology and Society, 19*(3). | Bush and Marschke | MLP | Aquaculture | Literature review | | | Not edited | Skimmed | * | ** | Applies Social Practice Theory alongside MLP; ethnographic fieldwork in Vietnam offers transferable insights for aquatic food systems. | ChatGPT rated higher | Skimmed | ** | * | Theoretical comparison paper without original empirical work. Useful for framework integration but limited practical application. |
| 2012 | Byrne, R., Smith, A., Watson, J., & Ockwell, D. (2012). Energy pathways in low carbon development. *Low-Carbon Technology Transfer, 123.* | Byrne et al. | MLP + pathways approach | Low-carbon energy | Literature review | Multi (Global South) | | Evaluation edited | Skimmed | * | | | ChatGPT ignored | Not read | * | * | Energy focus with critique of technology transfer. Limited food system relevance. |
| 2009 | Campbell, H. (2009). Breaking new ground in food regime theory: corporate environmentalism, ecological feedbacks and the 'food from somewhere' regime? *Agriculture and Human Values, 26*, 309–319. | Campbell | Mixed | Food systems | Literature review | Multi | | Not edited | Not read | * | ** | Theoretical extension of food regime theory; integrates corporate strategies with ecological feedbacks, foundational for early transition theory. | ChatGPT rated higher | Skimmed | * | ** | Important theoretical contribution on emerging food regimes. Lacks empirical grounding but valuable conceptual framework. |
| 2008 | Caniëls, M. C., & Romijn, H. A. (2008). Actor networks in strategic niche management: insights from social network theory. *Futures, 40*(7), 613–629. | Caniëls and Romijn | SNM + Social Network Theory | Biofuels | Interviews, participant observation and social-network analysis | Tanzania | 28 | Methods edited | Skimmed | ** | ** | Links social network theory and SNM; empirical cases illustrate how actor ties shape niche emergence. | Match | Not read | ** | ** | Innovative application of SNA to niche analysis. Jatropha case in Tanzania provides valuable network insights though dated. |
| 2019 | Charatsari, C., & Lioutas, E. D. (2019). Is current agronomy ready to | Charatsari and | | Sustainable agriculture | Surveys, structured | Greece | | Not edited | Skimmed | *** | | | ChatGPT ignored | Not read | *** | * | Focus on professional competencies with limited transition |

| Year | Full Citation (APA) | Author(s) | Theoretical Framing | Focus | Method | Country Focus | N (interviews) | Summaries Edited? | Annabelle – Degree of Reading | Annabelle Significance | ChatGPT Significance | ChatGPT Explanation | Significance Match? | Julian – Degree of Reading | Julian Significance | Claude Significance | Claude Explanation |
|---|---|---|---|---|---|---|---|---|---|---|---|---|---|---|---|---|---|
| | promote sustainable agriculture? *International Journal of Sustainable Development & World Ecology, 26*(3), 232–241. | Lioutas | | | questionnaires | | | | | | | | | | | | insights. Small samples reduce generalizability. |
| 2013 | Crivits, M., & Paredis, E. (2013). Designing an explanatory practice framework: Local food systems as a case. *Journal of Consumer Culture, 13*(3), 306–336. | Crivits and Paredis | MLP | Local food systems (food teams) | Interviews, focus groups, participant observation, and document analysis | Belgium | 20 | Not edited | Skimmed | *** | ** | Practice-based theoretical framework with case illustrations linking micro-practices to broader transition dynamics. | ChatGPT rated lower | Not read | ** | ** | Innovative practice theory application to food teams. Rich qualitative methods though focused on consumer side rather than full system. |
| 2016 | Davidson, D. J., Jones, K. E., & Parkins, J. R. (2016). Food safety risks, disruptive events and alternative beef production. *Agriculture and Human Values, 33*(2), 359–371. | Davidson et al. | MLP + Sustainability Transition Theory | Alternative beef production | Semi-structured interviews, secondary data | Canada | 50 | Methods edited | Skimmed | *** | ** | BSE crisis as a natural experiment; 50 interviews illustrate how disruptive events catalyse niche adoption. | ChatGPT rated lower | Skimmed | *** | ** | Important analysis of crisis-driven transitions in beef sector. Good qualitative data (50 interviews) showing producer motivations beyond economics. |
| 2017 | Dentoni, D., Waddell, S., & Waddock, S. (2017). Pathways of transformation in global food and agricultural systems. *Current Opinion in Environmental Sustainability, 29*, 8–13. | Dentoni et al. | Large systems change (LSC) + Transition Management | Global food systems | Literature review | Multi | | Not edited | Skimmed | * | ** | High-level synthesis applying large-systems theory; integrates diverse cases and offers strategic pathways. | ChatGPT rated higher | Not read | * | * | Conceptual paper critiquing transition management. Lacks empirical data but useful theoretical expansion. |
| 2013 | Diaz, M., Darnhofer, I., Darrot, C., & Beuret, J. E. (2013). Green tides in Brittany: What can we learn about niche–regime interactions? *Environmental Innovation and Societal Transitions, 8*, 62–75. | Diaz et al. | MLP + Actor-Network Theory (ANT) | Algae blooms | Interviews | France | 28 | Methods edited | Read fully | ** | ** | Case of algal blooms as landscape pressure; integrates an environmental event with niche–regime dynamics through mixed methods. | Match | Not read | ** | ** | Good integration of MLP with ANT showing messy transition realities. Strong longitudinal analysis of environmental problem-driven change. |
| 2010 | Drinkwaard, W., Kirkels, A., & Romijn, H. (2010). A learning-based approach to understanding success in rural electrification. *Energy for Sustainable* | Drinkwaard et al. | SNM (MLP) | Renewable energy | Semi-structured interviews, site visits & document analysis | Bolivia | Not mentioned | Methods edited | Skimmed | ** | ** | Participatory learning methods uncover success factors for micro-hydro niches; transferable to rural energy-food contexts. | Match | Not read | * | * | Energy focus with valuable learning insights. Limited direct relevance to food systems. |

| Year | Full Citation (APA) | Author(s) | Theoretical Framing | Focus | Method | Country Focus | N (interviews) | Summaries Edited? | Annabelle – Degree of Reading | Annabelle Significance | ChatGPT Significance | ChatGPT Explanation | Significance Match? | Julian – Degree of Reading | Julian Significance | Claude Significance | Claude Explanation |
|---|---|---|---|---|---|---|---|---|---|---|---|---|---|---|---|---|---|
| | *Development, 14*(3), 232–237. | | | | | | | | | | | | | | | | |
| 2014 | Duru, M., Fares, M. H., & Therond, O. (2014). Un cadre conceptuel pour penser maintenant la transition agroécologique. *Cahiers Agricultures, 23*(2), 84–95. | Duru et al. | Farming Systems + SES + STS | Agroecology | Literature review | France | | Not edited | Not read | * | ** | Conceptual framework on agroecological transition integrating territorial and governance dimensions. | ChatGPT rated higher | Not read | * | ** | Important French contribution on territorial agroecological transitions. Conceptual synthesis valuable though lacks empirical validation. |
| 2011 | Elzen, B., Geels, F. W., Leeuwis, C., & Van Mierlo, B. (2011). Normative contestation in transitions 'in the making': Animal welfare concerns and system innovation in pig husbandry. *Research Policy, 40*(2), 263–275. | Elzen et al. | MLP | Pig husbandry | Theoretical critique | The Netherlands | | Not edited | Skimmed | *** | ** | Combines normative analysis with multi-case empirical data; model for contested transitions research. | ChatGPT rated lower | Skimmed | *** | ** | Duplicate entry – see Lawhon & Murphy critique above. |
| 2017 | Feyereisen, M., Stassart, P. M., & Mélard, F. (2017). Fair trade milk initiative in Belgium: Bricolage as an empowering strategy for change. *Sociologia Ruralis, 57*(3), 297–315. | Feyereisen et al. | MLP | Fair trade milk | Semi-structured interviews and literature analysis | Belgium | 10 | Methods edited | Skimmed | ** | ** | Process-tracing of Fairebel's bricolage tactics with 10 interviews and documents; transferable niche-empowerment heuristic. | Match | Not read | ** | ** | Important concept of "bricolage" for understanding farmer agency in transitions. Good empirical work on crisis-driven innovation though limited to single initiative. |
| 2024 | Fiala, V., & Jacob, K. (2024). Combining the multi-level perspective framework with participatory scenario development to explore food system transitions in Germany. *Sustainability Science, 19*(4), 1259–1275. | Fiala and Jacob | MLP | Food systems | Participatory scenario workshops, qualitative interviews | Germany | 10 | Evaluation edited | Skimmed | ** | ** | Innovative mixed-methods combining MLP with participatory scenarios; robust local workshops though early case limits generalization. | Match | Skimmed | *** | *** | Innovative participatory methodology with 38 stakeholders. Excellent integration of MLP with scenario planning highly relevant for Socioscope approach. |
| 2020 | Gaitán-Cremaschi, D., Klerkx, L., Duncan, J., et al. (2020). Sustainability transition pathways through ecological intensification. *Internatio* | Gaitán-Cremaschi et al. | Transition theory | Vegetable farming | Interviews and surveys | | 53 | Not edited | Skimmed | *** | | | ChatGPT ignored | Not read | ** | *** | Excellent application of transition framework to vegetable systems. Multi-method approach with 53 participants provides rich insights into system diversity |

| Year | Full Citation (APA) | Author(s) | Theoretical Framing | Focus | Method | Country Focus | N (interviews) | Summaries Edited? | Annabelle – Degree of Reading | Annabelle Significance | ChatGPT Significance | ChatGPT Explanation | Significance Match? | Julian – Degree of Reading | Julian Significance | Claude Significance | Claude Explanation |
|---|---|---|---|---|---|---|---|---|---|---|---|---|---|---|---|---|---|
| | *nal Journal [of Agronomy].* | | | | | | | | | | | | | | | | and transition potential. |
| 2023 | Geels, F. W., Kern, F., & Clark, W. C. (2023). Sustainability transitions in consumption-production systems. *PNAS, 120*(47), e2310070120. | Geels et al. | MLP | Sustainability transitions | Theoretical critique / literature review | Multi | | Not edited | Skimmed | ** | *** | Foundational PNAS conceptualization of consumption-production linkages; integrates global case studies with theoretical richness. | ChatGPT rated higher | Skimmed | ** | * | Conceptual synthesis without original empirical work. Useful overview but limited practical insights. |
| 2010 | Gopakumar, G. (2010). Sustainability and the indispensability of politics: A study of sanitation partnerships in urban India. *International Journal of Sustainable Society, 2*(4), 376–392. | Gopakumar | | | | | | Not read | | | | Political lens for sanitation niches; relevant to infrastructure but limited direct food-system applicability. | | | | * | Unable to access full document for assessment. |
| 2022 | Graça, J., Roque, L., Guedes, D., et al. (2022). Enabling sustainable food transitions in schools: a systemic approach. *British Food Journal, 124*(13), 322–339. | Graça et al. | Systems theory + COM-B system | School meals | Interviews, participant observation, document analysis | Portugal | 33 | Methods and Evaluation edited | Read fully | ** | ** | Systemic case study across five countries; surveys, interviews, workshops co-design school food transitions framework. | Match | Not read | ** | ** | Important multi-stakeholder analysis of school food transitions. 33 interviews across three influence levels provide systemic insights. |
| 2019 | Håkansson, I. (2019). Urban sustainability experiments in their socio-economic milieux: A quantitative approach. *Journal of Cleaner Production, 209*, 515–527. | Håkansson | Transitions studies + environmental justice | Urban agriculture (community gardens) | Quantitative analysis; spatial regression | UK (England) | | Not edited | Skimmed | ** | ** | Statistical analysis of >200 experiments; quantitative clustering reveals urban niche typologies. | Match | Skimmed | ** | * | Quantitative spatial analysis of urban agriculture. Innovative methods but limited insights into transition processes or practitioner experiences. |
| 2022 | Hansen, T. (2022). Consumer food sustainability before and during the Covid-19 Crisis. *Food Policy, 107*, 102207. | Hansen | Media agenda setting theory | Consumer food sustainability | Quantitative content analysis | Denmark | | Not edited | Skimmed | *** | ** | Content analysis of >1,000 articles tracking discourse shifts; robust quantitative methods with clear policy recommendations. | ChatGPT rated lower | Not read | *** | * | Media analysis with limited connection to ground-level transitions. Useful for understanding discourse but not practitioner realities. |
| 2013 | Hargreaves, T., Longhurst, N., & Seyfang, G. (2013). Up, down, round and round: connecting regimes and | Hargreaves et al. | MLP + Social Practice Theory | Organic agriculture | Site visits, semi-structured interviews, surveys, | UK | Not mentioned | Not edited | Skimmed | ** | ** | Practice-theory meets MLP; UK grassroots innovations case studies provide rich insights into practice | Match | Not read | * | ** | Important theoretical integration of MLP and practice theory. Two case studies provide empirical support |

| Year | Full Citation (APA) | Author(s) | Theoretical Framing | Focus | Method | Country Focus | N (interviews) | Summaries Edited? | Annabelle – Degree of Reading | Annabelle Significance | ChatGPT Significance | ChatGPT Explanation | Significance Match? | Julian – Degree of Reading | Julian Significance | Claude Significance | Claude Explanation |
|---|---|---|---|---|---|---|---|---|---|---|---|---|---|---|---|---|---|
| | practices in innovation for sustainability. *Environment and Planning A, 45*(2), 402–420. | | | | quantitative analysis | | | | | | | diffusion and regime linkages. | | | | | though limited methodological detail. |
| 2013 | Hassink, J., Grin, J., & Hulsink, W. (2013). Multifunctional Agriculture Meets Health Care. *Sociologia Ruralis, 53*(2), 223–245. | Hassink et al. | MLP | Care farming | Interviews and literature review | The Netherlands | Not mentioned | Not edited | Skimmed | * | ** | Journal article precursor to the 2018 JRS piece; strong foundational empirical work on care farming. | ChatGPT rated higher | Skimmed | * | ** | Strong empirical application of MLP to care farming. Grounded theory approach with good theoretical propositions for cross-sector transitions. |
| 2018 | Hassink, J., Grin, J., & Hulsink, W. (2018). Enriching the multi-level perspective by better understanding agency and challenges associated with interactions across system boundaries. *Sociology of Agriculture and Food.* | Hassink et al. | MLP | Care farming | Multiple qualitative case study; semi-structured interviews; documents and databases | The Netherlands | 77 | Evaluation edited | Skimmed | ** | *** | Seminal critique and extension of MLP with rich agency-focused cases (77 interviews); establishes care farming as a model of cross-sector transition. | ChatGPT rated higher | Skimmed | ** | *** | Directly addresses MLP limitations through empirical study of cross-sectoral transitions. Rich qualitative data (77 interviews) and theoretical innovation in understanding agency and boundary work makes it seminal for understanding micro-level initiatives embedded in broader contexts. |
| 2014 | Hassink, J., Hulsink, W., & Grin, J. (2014). Farming with care: the evolution of care farming in the Netherlands. *NJAS–Wageningen Journal of Life Sciences, 68,* 1–11. | Hassink et al. | MLP + organizational ecology + social movement theory | Care farming | Quantitative data analysis, document review, and semi-structured interviews | The Netherlands | Not mentioned | Methods edited | Skimmed | ** | ** | Foundational empirical grounding of care farming through multi-site case study; strong methodological rigor. | Match | Skimmed | ** | ** | Excellent integration of multiple theories to explain care farming growth. Strong empirical base combining quantitative and qualitative data. |
| 2017 | Hauser, M., & Lindtner, M. (2017). Organic agriculture in post-war Uganda: Emergence of pioneer-led niches between 1986 and 1993. *Renewable Agriculture and Food Systems, 32*(2), 169–178. | Hauser and Lindtner | MLP | Organic agriculture | Semi-structured interviews, literature review and document analysis | Uganda | 12 | Methods edited | Skimmed | ** | ** | Historical MLP case with archival and interview methods; rich context though retrospective data. | Match | Not read | ** | ** | Valuable historical perspective on organic emergence in Global South during crisis. Limited sample (12 interviews) but important for understanding role of pioneers in adverse conditions. |

| Year | Full Citation (APA) | Author(s) | Theoretical Framing | Focus | Method | Country Focus | N (interviews) | Summaries Edited? | Annabelle – Degree of Reading | Annabelle Significance | ChatGPT Significance | ChatGPT Explanation | Significance Match? | Julian – Degree of Reading | Julian Significance | Claude Significance | Claude Explanation |
|---|---|---|---|---|---|---|---|---|---|---|---|---|---|---|---|---|---|
| 2024 | Held, T. (2024). Make digitalized places for experimentation work: unravelling and governing transformative dynamics of FabLabs and Makerspaces. *European Planning Studies, 32*(7), 1573–1593. | Held | Strategic Niche Management (SNM) | Maker platforms | Semi-structured interviews | Germany | 19 | Methods edited | Skimmed | ** | ** | Contemporary empirical analysis; interviews elucidate digital niches relevant for socio-technical experimentation. | Match | Not read | *** | * | Focus on digital maker spaces with limited food system relevance. Translation mechanisms potentially useful but indirect application. |
| 2023 | Hess, D. J. (2023). Conflict and uneven development in the multidecade distributed solar energy transition in the United States. *PNAS, 120*(47), e2206200119. | Hess | MLP | Renewable energy | Longitudinal case study, documentary and policy analysis | USA | | Not edited | Skimmed | * | *** | PNAS mixed quantitative-qualitative analysis over decades uncovers socio-political conflicts; pivotal for contested transitions. | ChatGPT rated higher | Not read | * | * | Solar energy focus with limited food system relevance. Coalition dynamics insights potentially useful. |
| 2014 | Hess, D. J., & Mai, Q. D. (2014). Renewable electricity policy in Asia: A qualitative comparative analysis of factors affecting sustainability transitions. *Environmental Innovation and Societal Transitions, 12*, 31–46. | Hess and Mai | MLP | Renewable energy | Fuzzy-set Qualitative Comparative Analysis (fsQCA) | Asia (Multi) | | Not edited | Skimmed | * | ** | QCA across Asian nations; expert interviews provide a template for comparative regime-factor analysis in agri-food transitions. | ChatGPT rated higher | Skimmed | ** | * | Energy policy focus using QCA. Methodological approach interesting but limited food relevance. |
| 2014 | Hinrichs, C. C. (2014). Transitions to sustainability: a change in thinking about food systems change? *Agriculture and Human Values, 31*, 143–155. | Hinrichs | MLP + Social Practices Approach (SPA) | Food systems | Literature review | | | Not edited | Skimmed | ** | * | Conceptual essay on rethinking "transition"; valuable for framing but lacks empirical case data. | ChatGPT rated lower | Skimmed | * | ** | Important theoretical bridge between food systems and transitions literature. Lacks empirical data but valuable conceptual contribution. |
| 2015 | Hu, M. C., Wu, C. Y., & Shih, T. (2015). Creating a new socio-technical regime in China: Evidence from the Sino-Singapore Tianjin Eco-City. *Futures, 70*, 1–12. | Hu et al. | MLP | Eco city | Interviews and document reviews | China | Not mentioned | Methods edited | Skimmed | ** | ** | Mixed qualitative-quantitative analysis of a flagship eco-city; shows how orchestrated niches can shape broader regime change. | Match | Not read | ** | * | Urban eco-city focus with limited food system relevance. State-led transition insights potentially useful. |
| 2018 | Ilieva, R. T., & Hernandez, A. (2018). Scaling-up sustainable development | Ilieva and Hernandez | MLP | Agri-food sector | Ethnography and unstructured interviews | Brazil, US, Senegal | Not mentioned | Evaluation edited | Skimmed | ** | ** | Three cross-continental mixed methods cases; highlights scaling | Match | Not read | ** | *** | Exceptional comparative analysis of grassroots scaling. Rich ethnographic data |

| Year | Full Citation (APA) | Author(s) | Theoretical Framing | Focus | Method | Country Focus | N (interviews) | Summaries Edited? | Annabelle – Degree of Reading | Annabelle Significance | ChatGPT Significance | ChatGPT Explanation | Significance Match? | Julian – Degree of Reading | Julian Significance | Claude Significance | Claude Explanation |
|---|---|---|---|---|---|---|---|---|---|---|---|---|---|---|---|---|---|
| | initiatives. *Towards Sustainable Global Food Systems*, 273. | | | | | | | | | | | mechanisms, though depth varies by case. | | | | | and practical "pressure points" highly relevant for understanding bottom-up transitions. |
| 2013 | Immink, V. M., Reinders, M. J., Van Tulder, R. J. M., & Van Trijp, J. C. M. (2013). The livestock sector and its stakeholders in the search to meet the animal welfare requirements of society. *Journal on Chain and Network Science, 13*(2), 151–160. | Immink et al. | MLP | Livestock sector | Workshops, stakeholder dialogues | The Netherlands | Not mentioned | Not edited | Skimmed | ** | * | Stakeholder mapping of animal welfare niche; useful background but limited depth on transition processes. | ChatGPT rated lower | Skimmed | ** | * | Focus on stakeholder workshops without strong empirical outcomes. Limited theoretical contribution to transitions understanding. |
| 2015 | Ingram, J. (2015). Framing niche-regime linkage as adaptation: an analysis of learning and innovation networks for sustainable agriculture across Europe. *Journal of Rural Studies, 40*, 59–75. | Ingram | MLP + Complex Adaptive Systems | Sustainable agriculture | Case studies of 17 LINSA: Participatory workshops, interviews, focus groups, document analysis | Europe (multi) | Not mentioned | Not edited | Skimmed | ** | ** | Cross-European network analysis of learning platforms; mixed methods strengthen generalizability. | Match | Skimmed | *** | *** | Excellent multi-case analysis (17 LINSA) reconceptualizing niche-regime interactions as adaptation. Highly relevant theoretical and empirical contributions for understanding practitioner networks. |
| 2018 | Ingram, J. (2018). Agricultural transition: Niche and regime knowledge systems' boundary dynamics. *Environmental Innovation and Societal Transitions, 26*, 117–135. | Ingram | MLP | Permaculture | Workshops and semi-structured interviews | UK (England) | 27 | Methods and Evaluation edited | Skimmed | *** | ** | Robust three-year participatory study mapping knowledge flows between permaculture and conventional regimes. | ChatGPT rated lower | Not read | ** | ** | Strong theoretical contribution on knowledge boundaries in transitions using permaculture case. Multi-method approach but limited to single initiative type. Relevant for understanding practitioner knowledge dynamics. |
| 2015 | Ingram, J., Maye, D., Kirwan, J., Curry, N., & Kubinakova, K. (2015). Interactions between niche and regime: an analysis of learning and innovation networks for sustainable agriculture. *Journal of Agricultural Education* | Ingram et al. | MLP | Sustainable agriculture | Participatory workshops, semi-structured interviews, focus groups, observation | Europe (multi) | >102 | Methods edited | Skimmed | *** | ** | Extension into advisory contexts; interview and survey data unpack network dynamics affecting niche uptake. | ChatGPT rated lower | Skimmed | *** | ** | Companion to above paper with similar strengths. Focus on compatibility and interaction modes valuable for understanding transition dynamics. |

| Year | Full Citation (APA) | Author(s) | Theoretical Framing | Focus | Method | Country Focus | N (interviews) | Summaries Edited? | Annabelle – Degree of Reading | Annabelle Significance | ChatGPT Significance | ChatGPT Explanation | Significance Match? | Julian – Degree of Reading | Julian Significance | Claude Significance | Claude Explanation |
|---|---|---|---|---|---|---|---|---|---|---|---|---|---|---|---|---|---|
| | *and Extension, 21*(1), 55–71. | | | | , document analysis | | | | | | | | | | | | |
| 2017 | Isgren, E., & Ness, B. (2017). Agroecology to promote just sustainability transitions: Analysis of a civil society network in the Rwenzori Region, Western Uganda. *Sustainability, 9*(8), 1357. | Isgren and Ness | MLP | Agroecology | Semi-structured interviews, focus groups, site observations, participatory exercises | Uganda | 54 | Methods edited | Skimmed | *** | ** | Network analysis maps power relations and justice axes; highly relevant for equity-focused transitions. | ChatGPT rated lower | Skimmed | *** | ** | Strong fieldwork (8 weeks) examining agroecology's transformative potential in Uganda. Important critique of conformist tendencies in alternative food movements. |
| 2017 | Jacobs, B., Cordell, D., Chin, J., & Rowe, H. (2017). Towards phosphorus sustainability in North America. *Environmental Science & Policy, 77*, 151–159. | Jacobs et al. | MLP + increasing returns | Phosphorus use | Stakeholder workshops | North America (multi) | 93 | Methods edited | Skimmed | ** | ** | Participatory foresight with 93 stakeholders develops multi-actor change model for critical nutrient governance. | Match | Not read | ** | * | Primarily conceptual with participatory workshops. Addresses important sustainability issue but lacks empirical grounding in actual transition initiatives. |
| 2018 | Järnberg, L., Kautsky, E. E., Dagerskog, L., & Olsson, P. (2018). Green niche actors navigating an opaque opportunity context. *Land Use Policy, 71*, 409–421. | Järnberg et al. | MLP | Agricultural development | Narrative analysis of agricultural policy documents; institutional analysis; interviews with green niche actors | Ethiopia | 9 | Not edited | Skimmed | ** | ** | Socio-ecological entrepreneurship in a non-Western context; policy-document and interview analysis reveal institutional barriers in Ethiopia. | Match | Not read | * | ** | Valuable Global South perspective on transitions under authoritarian governance. Limited empirical base (9 interviews) but important insights on institutional barriers to grassroots sustainability initiatives. |
| 2014 | Johnson, F. X., & Silveira, S. (2014). Pioneer countries in the transition to alternative transport fuels. *Environmental Innovation and Societal Transitions, 11*, 1–24. | Johnson and Silveira | MLP + SNM | Biofuels | Literature review + informal interviews | Brazil, Malawi, Sweden | | Not edited | Not read | * | ** | Cross-country policy comparison; document analysis and expert interviews reveal how national context shapes niche scaling. | ChatGPT rated higher | Skimmed | * | * | Transport fuel focus with limited food system connections. Comparative approach valuable. |
| 2023 | Johnstone, P., & Schot, J. (2023). Shocks, institutional change, and sustainability transitions. *PNAS, 120*(47), e2206226120. | Johnstone and Schot | MLP + imprinting theory + Window of Opportunity | Energy transitions | Literature review, historical case studies | Multi | | Not edited | Skimmed | * | *** | Seminal PNAS paper linking crisis shocks to regime change; robust comparative analysis foundational for disruption-driven transitions. | ChatGPT rated higher | Not read | * | ** | Important theoretical contribution on crisis-driven transitions. Historical analysis valuable though lacks contemporary empirical work. |

| Year | Full Citation (APA) | Author(s) | Theoretical Framing | Focus | Method | Country Focus | N (interviews) | Summaries Edited? | Annabelle – Degree of Reading | Annabelle Significance | ChatGPT Significance | ChatGPT Explanation | Significance Match? | Julian – Degree of Reading | Julian Significance | Claude Significance | Claude Explanation |
|---|---|---|---|---|---|---|---|---|---|---|---|---|---|---|---|---|---|
| 2016 | Jurgilevich, A., Birge, T., Kentala-Lehtonen, J., et al. (2016). Transition towards circular economy in the food system. *Sustainability, 8*(1), 69. | Jurgilevich et al. | MLP | Circular economy in the food system | Roundtables and expert presentations | | | Not edited | Skimmed | * | ** | Conceptual mapping of CE principles to agri-food; literature integration and expert workshops into actionable transition pathways. | ChatGPT rated higher | Not read | * | * | Conceptual paper based on single workshop. Useful framework but lacks empirical depth for understanding actual transition processes. |
| 2018 | Karan, E., & Asadi, S. (2018). Quantitative modeling of interconnections associated with sustainable food, energy and water (FEW) systems. *Journal of Cleaner Production, 200*, 86–99. | Karan and Asadi | FEW nexus approach | FEW nexus | Quantitative modelling, system simulations | USA | | Not edited | Skimmed | * | ** | Network modelling of FEW nexus using input-output data; innovative systems approach though limited by data granularity. | ChatGPT rated higher | Skimmed | * | * | Technical modelling paper without empirical grounding in actual transitions. Limited relevance to qualitative ground-level research. |
| 2015 | Konefal, J. (2015). Governing sustainability transitions: Multi-stakeholder initiatives and regime change in United States agriculture. *Sustainability, 7*(1), 612–633. | Konefal | MLP | Sustainability metrics and standards in agriculture | Interviews, sampling, document analysis | USA | 36 | Not edited | Skimmed | ** | ** | Analysis of US multi-stakeholder initiatives; policy review, interviews, and case comparisons illustrate regime change pathways. | Match | Not read | * | ** | Important analysis of governance and power in MSIs. 36 interviews provide insights into how sustainability metrics shape transitions. |
| 2015 | Lambe, F., Jürisoo, M., Lee, C., & Johnson, O. (2015). Can carbon finance transform household energy markets? *Energy Research & Social Science, 5*, 55–66. | Lambe et al. | | Renewable energy | Semi-structured interviews | Kenya | 26 | Evaluation edited | Skimmed | * | ** | Comparative review linking carbon payments to cookstove adoption; parallels for food-waste-energy synergies. | ChatGPT rated higher | Not read | ** | * | Cookstove focus tangentially related to food systems. Carbon finance insights potentially relevant. |
| 2020 | Lamine, C. (2020). *Sustainable agri-food systems: Case studies in transitions towards sustainability from France and Brazil.* Bloomsbury Publishing. | Lamine | Food regimes + MLP + French Pragmatic Sociology | Agri-food sector | | France and Brazil | | Not edited | Not read | – | | Edited cases combining in-depth qualitative fieldwork in two countries; offers comparative theoretical reflections. | | Not read | N/A | * | Book not accessible for full assessment. |
| 2023 | Lamine, C., & Marsden, T. (2023). Unfolding sustainability transitions in food systems: Insights from UK and French | Lamine and Marsden | MLP | Agroecology | Literature review | France and UK | | Methods and Evaluation edited | Skimmed | * | *** | Comparative PNAS study blending qualitative trajectories with quantitative indicators; seminal | ChatGPT rated higher | Skimmed | * | ** | Important comparative historical analysis though based on literature review rather than original research. |

| Year | Full Citation (APA) | Author(s) | Theoretical Framing | Focus | Method | Country Focus | N (interviews) | Summaries Edited? | Annabelle – Degree of Reading | Annabelle Significance | ChatGPT Significance | ChatGPT Explanation | Significance Match? | Julian – Degree of Reading | Julian Significance | Claude Significance | Claude Explanation |
|---|---|---|---|---|---|---|---|---|---|---|---|---|---|---|---|---|---|
| | trajectories. *PNAS, 120*(47), e2206231120. | | | | | | | | | | | cross-national theoretical insights. | | | | | Valuable power and narrative insights. |
| 2022 | Larbi, M., Kellett, J., & Palazzo, E. (2022). Urban sustainability transitions in the global south: A case study of Curitiba and Accra. *Urban Forum, 33*(2), 223–244. | Larbi et al. | MLP | Urban development | Semi-structured interviews, surveys, document analysis | Brazil and Ghana | >25 | Not edited | Read fully | *** | ** | Comparative interviews and policy review in two cities; illuminates socio-political enablers in different contexts. | ChatGPT rated lower | Skimmed | *** | * | Urban planning focus with limited food system relevance. Transport and waste insights tangentially related. |
| 2012 | Lawhon, M., & Murphy, J. T. (2012). Socio-technical regimes and sustainability transitions: Insights from political ecology. *Progress in Human Geography, 36*(3), 354–378. | Lawhon and Murphy | MLP | GMOs in food production | Literature review | Europe (multi) | | Not edited | Skimmed | * | ** | Cross-disciplinary integration of political ecology with MLP; novel critique and enriched conceptual framing. | ChatGPT rated higher | Not read | * | ** | Important theoretical critique of MLP from political ecology perspective. GMO case illustration valuable though lacks original empirical work. |
| 2010 | Lebel, L., Mungkung, R., Gheewala, S. H., & Lebel, P. (2010). Innovation cycles, niches and sustainability in the shrimp aquaculture industry in Thailand. *Environmental Science & Policy, 13*(4), 291–302. | Lebel et al. | MLP + systems innovation | Shrimp farming | Interviews + Life Cycle Assessment + document analysis | Thailand | Not mentioned | Evaluation edited | Skimmed | * | ** | Sectoral MLP of aquaculture; bio-economic modelling and policy reviews yield cycle insights transferable to other food sectors. | ChatGPT rated higher | Skimmed | * | ** | Important aquaculture transition analysis. Mixed methods provide insights though interview details lacking. |
| 2014 | Levidow, L., Pimbert, M., & Vanloqueren, G. (2014). Agroecological research: conforming—or transforming the dominant agro-food regime? *Agroecology and Sustainable Food Systems, 38*(10), 1127–1155. | Levidow et al. | MLP | Agroecological research | Document review | Europe (multi) | | Not edited | Skimmed | ** | ** | Critically reviews agroecology's institutional role; integrates case examples and policy analysis. | Match | Skimmed | *** | ** | Critical analysis of agroecology's transformative potential. Strong theoretical framework though relies on secondary sources. |
| 2013 | Li, J., Jiggins, J., Van Bueren, E. T. L., & Leeuwis, C. (2013). Towards a regime change in the organization of the seed supply system in China. *Experimental* | Li et al. | MLP | Participatory maize breeding | Interviews, surveys, observations, policy analysis | China | 70 | Not edited | Skimmed | *** | ** | Field study of seed-network reforms combining surveys, interviews, and network analysis for robust insights into regime change. | ChatGPT rated lower | Not read | *** | ** | Important analysis of seed system transitions in China. Mixed methods (162 households, 40 interviews) provide strong empirical base |

| Year | Full Citation (APA) | Author(s) | Theoretical Framing | Focus | Method | Country Focus | N (interviews) | Summaries Edited? | Annabelle – Degree of Reading | Annabelle Significance | ChatGPT Significance | ChatGPT Explanation | Significance Match? | Julian – Degree of Reading | Julian Significance | Claude Significance | Claude Explanation |
|---|---|---|---|---|---|---|---|---|---|---|---|---|---|---|---|---|---|
| | *Agriculture, 49*(1), 114–133. | | | | | | | | | | | | | | | | for understanding institutional barriers. |
| 2012 | Li, J., Lammers van Bueren, E. T., Jiggins, J., & Leeuwis, C. (2012). Farmers' adoption of maize hybrids and the persistence of landraces in Southwest China. *Genetic Resources and Crop Evolution, 59*. | Li et al. | Reasons for action model + Vulnerability of production and conservation | Maize systems | Surveys, semi-structured interviews | China | 162 | Not edited | Skimmed | ** | ** | Mixed survey-interview study; robust statistical and qualitative data on adoption behaviours and landrace persistence. | Match | Skimmed | ** | ** | Strong mixed-methods study (162 households) on farmer decision-making. Important insights on local knowledge and conservation. |
| 2013 | Lutz, J., & Schachinger, J. (2013). Do local food networks foster socio-ecological transitions towards food sovereignty? *Sustainability, 5*(11), 4778–4796. | Lutz and Schachinger | MLP | Local food networks | Interviews, participant observation | Austria | Not mentioned | Methods and Evaluation edited | Skimmed | * | ** | Multiple case studies of food networks; strong comparative qualitative analysis with actionable lessons for local transition initiatives. | ChatGPT rated higher | Not read | * | ** | Valuable case study of Austrian local food network. Participatory approach provides practitioner insights though limited to single initiative. |
| 2013 | Marsden, T. (2013). From post-productionism to reflexive governance: Contested transitions in securing more sustainable food futures. *Journal of Rural Studies, 29*, 123–134. | Marsden | MLP | Agri-food sector | Interviews, workshops, scenario planning | UK | 45 | Not edited | Skimmed | ** | ** | Influential framing of reflexive governance; integrates policy critique and case vignettes, highly cited theoretical contribution. | Match | Not read | ** | ** | Important UK analysis using scenario planning. 45 interviews provide insights into regime adaptation and reflexive governance emergence. |
| 2018 | Maru, Y., Sparrow, A., Stirzaker, R., & Davies, J. (2018). Integrated agricultural research for development (IAR4D) from a theory of change perspective. *Agricultural Systems, 165*, 310–320. | Maru et al. | MLP + Adaptive change theory | Integrated agricultural research for development | Literature review | East and West Africa (multi) | | Not edited | Skimmed | * | ** | Applies theory-of-change lens to IAR4D; mixed methods case in Africa demonstrates how research integration shapes systemic outcomes. | ChatGPT rated higher | Skimmed | * | ** | Important synthesis of innovation platform approaches in Africa. Strong theoretical framework though relies on secondary analysis rather than original fieldwork. |
| 2017 | Meynard, J. M., Jeuffroy, M. H., Le Bail, M., et al. (2017). Designing coupled innovations for the sustainability transition of agrifood systems. *Agricultural Systems, 157*, 330–339. | Meynard et al. | MLP + innovative design theory | Agri-food sector | Literature review | Multi | | Not edited | Skimmed | * | ** | Theoretical framework on coupled innovations synthesizing cases and workshops into actionable design principles. | ChatGPT rated higher | Not read | * | ** | Important theoretical contribution on coupled innovations across value chains. Limited empirical rigor but valuable framework for integrated transition approaches. |

| Year | Full Citation (APA) | Author(s) | Theoretical Framing | Focus | Method | Country Focus | N (interviews) | Summaries Edited? | Annabelle – Degree of Reading | Annabelle Significance | ChatGPT Significance | ChatGPT Explanation | Significance Match? | Julian – Degree of Reading | Julian Significance | Claude Significance | Claude Explanation |
|---|---|---|---|---|---|---|---|---|---|---|---|---|---|---|---|---|---|
| 2017 | Moallemi, E. A., Aye, L., Webb, J. M., de Haan, F. J., & George, B. A. (2017). India's on-grid solar power development: Historical transitions, present status and future driving forces. *Renewable and Sustainable Energy Reviews, 69*, 239–247. | Moallemi et al. | MLP | Renewable energy | Document + policy review | India | | Summary edited | Skimmed | * | ** | Historical MLP of India's solar niche; document analysis and stakeholder interviews trace multi-decadal transition dynamics. | ChatGPT rated higher | Not read | ** | * | Solar energy analysis without food system connections. Policy instrument insights potentially useful. |
| 2015 | Moragues-Faus, A., & Morgan, K. (2015). Reframing the foodscape: the emergent world of urban food policy. *Environment and Planning A, 47*(7), 1558–1573. | Moragues-Faus and Morgan | MLP | Urban food policy | Project meetings, semi-structured interviews, participant observation, document analysis | UK and Sweden (multi) | 12 | Evaluation edited | Read fully | *** | ** | Conceptualizes urban food policy as a transition dynamic; rich case studies across cities with interviews and document analysis. | ChatGPT rated lower | Skimmed | *** | ** | Strong comparative analysis of urban food governance spaces. Participatory research design and focus on deliberation relevant for understanding local food transitions. |
| 2014 | Morrissey, J. E., Mirosa, M., & Abbott, M. (2014). Identifying transition capacity for agri-food regimes. *Journal of Environmental Policy & Planning, 16*(2), 281–301. | Morrissey et al. | MLP + Social Practice Theory | Agri-food sector | Literature review | Multi | | Methods edited | Skimmed | ** | ** | Methodological paper applying MLP to map transition capacities; includes a tested mapping tool demonstrated in New Zealand contexts. | Match | Skimmed | *** | * | Conceptual tool development with limited empirical application. Useful for regime analysis but lacks ground-level insights. |
| 2020 | Mukhovi, S., Jacobi, J., Speranza, C. I., Rist, S., & Kiteme, B. (2020). Learning and adaptation in food systems: insights from four case studies in the global south. *International Journal on Food System Dynamics, 11*(4), 312–328. | Mukhovi et al. | Social learning theory; Transformative learning; Resilience theory | Food systems | Semi-structured interviews, participant observation, case studies | Kenya and Bolivia | 87 | Not edited | Skimmed | *** | ** | Comparative qualitative case analysis; rich insights on adaptive learning processes, strong cross-case synthesis. | ChatGPT rated lower | Not read | *** | ** | Strong comparative analysis across Kenya and Bolivia. Rich qualitative data on learning processes in different food systems highly relevant for Socioscope approach. |
| 2012 | Neef, A., Mizuno, K., Schad, I., Williams, P. M., & Rwezimula, F. (2012). Community-based microtrade in support of small-scale farmers in Thailand and Tanzania. *The Law and* | Neef et al. | SNM | Agricultural development | Semi-structured interviews, participatory observation, action research | Thailand and Tanzania | Not mentioned | Methods and Evaluation edited | Skimmed | * | ** | Legal-institutional analysis of microtrade platforms; interviews and document analysis highlight governance niches supporting smallholders. | ChatGPT rated higher | Skimmed | ** | ** | Important action research on farmer value chains. SNM application to microtrade provides relevant insights for local food systems. |

| Year | Full Citation (APA) | Author(s) | Theoretical Framing | Focus | Method | Country Focus | N (interviews) | Summaries Edited? | Annabelle – Degree of Reading | Annabelle Significance | ChatGPT Significance | ChatGPT Explanation | Significance Match? | Julian – Degree of Reading | Julian Significance | Claude Significance | Claude Explanation |
|---|---|---|---|---|---|---|---|---|---|---|---|---|---|---|---|---|---|
| | *Development Review, 5*(1), 80–100. | | | | | | | | | | | | | | | | |
| 2017 | Nemes, G., & Augustyn, A. M. (2017). Towards inclusive innovation in the European context. *Innovation and Development, 7*(1), 133–152. | Nemes and Augustyn | Participatory Action Research | Sustainable agriculture | Interviews, Focus Groups, Participant Observation, Participatory Workshops | Hungary | 41 | Not edited | Skimmed | ** | | | ChatGPT ignored | Not read | ** | ** | Good participatory action research on Hungarian LINSA. Rich 3-year engagement provides valuable insights on network dynamics. |
| 2018 | Nygaard, I., & Bolwig, S. (2018). The rise and fall of foreign private investment in the jatropha biofuel value chain in Ghana. *Environmental Science & Policy, 84*, 224–234. | Nygaard and Bolwig | MLP + GVC | Jatropha biofuel | Case studies of 7 Ghanaian biofuel companies; site visits; semi-structured interviews; documentary analysis | Ghana | 21 | All sections edited | Read fully | ** | ** | Dual MLP-GVC framework reconstructs niche failure via archival research, field visits, and stakeholder interviews in Ghana. | Match | Not read | * | ** | Excellent example of failed transition combining MLP with global value chain analysis. Strong longitudinal field research (2008–2014) documenting complete niche lifecycle relevant for understanding transition failures. |
| 2021 | Oates, L. (2021). Sustainability transitions in the Global South: a multi-level perspective on urban service delivery. *Regional Studies, Regional Science, 8*(1), 426–433. | Oates | MLP | Urban service delivery | Interviews, site visits, document analysis | India and Uganda | 39 | Evaluation edited | Skimmed | ** | ** | Applies MLP to service delivery; mixed survey/interview data across three cities bridges urban studies and transitions theory. | Match | Skimmed | * | * | Infrastructure focus with limited food system connections. Useful Global South perspective but not directly relevant. |
| 2014 | Opazo, J. (2014). *The politics of system innovation for emerging technologies* (Doctoral dissertation, University of Sussex). | Opazo | MLP + Science and Technology Studies (STS) | Renewable energy | Semi-structured interviews, document analysis | Chile | 55 | Methods and Evaluation edited | Not read | ** | ** | PhD deep dive into off-grid niches; extensive fieldwork and political analysis make it a strong model for emergent food technologies in peripheral contexts. | Match | Skimmed | ** | * | Energy focus with useful political analysis. 55 interviews provide depth but limited food relevance. |
| 2016 | Pant, L. P. (2016). Paradox of mainstreaming agroecology for regional and rural food security in developing countries. *Technological* | Pant | MLP + Adaptive systems theory | Agroecology | Stakeholder surveys; interviews; focus groups; observations | Multi | 24 | Evaluation edited | Skimmed | ** | ** | Critical MLP critique of mainstreaming risks; meta-analysis and stakeholder interviews generate policy recommendations. | Match | Not read | * | ** | Valuable theoretical contribution on mainstreaming paradoxes. Four longitudinal case studies provide good empirical grounding for |

| Year | Full Citation (APA) | Author(s) | Theoretical Framing | Focus | Method | Country Focus | N (interviews) | Summaries Edited? | Annabelle – Degree of Reading | Annabelle Significance | ChatGPT Significance | ChatGPT Explanation | Significance Match? | Julian – Degree of Reading | Julian Significance | Claude Significance | Claude Explanation |
|---|---|---|---|---|---|---|---|---|---|---|---|---|---|---|---|---|---|
| | *Forecasting and Social Change, 111,* 305–316. | | | | | | | | | | | | | | | | adaptive transition concepts. |
| 2015 | Pant, L. P., Adhikari, B., & Bhattarai, K. K. (2015). Adaptive transition for transformations to sustainability in developing countries. *Current Opinion in Environmental Sustainability, 14,* 206–212. | Pant et al. | Adaptive transition theory | Climate adaptation | Literature review | Multi (Global South) | | Not edited | Skimmed | * | ** | Review of adaptive transitions with South Asian examples; synthesizes across cases to recommend adaptive governance strategies. | ChatGPT rated lower | Not read | * | * | Framework development paper without empirical testing. Useful concepts but limited practical application. |
| 2016 | Papachristos, G., & Adamides, E. (2016). A retroductive systems-based methodology for socio-technical transitions research. *Technological Forecasting and Social Change, 108,* 1–14. | Papachristos and Adamides | MLP + retroductive systems methodology | Functional foods | Simulation modelling | | | Not edited | Skimmed | *** | ** | Methodological advance embedding retroductive reasoning into transitions research; case-illustrated. | ChatGPT rated lower | Skimmed | *** | * | Primarily methodological paper using functional foods example. Innovative but limited direct relevance to ground-level food system transitions. |
| 2017 | Paschen, J. A., Reichelt, N., King, B., Ayre, M., & Nettle, R. (2017). Enrolling advisers in governing privatised agricultural extension in Australia. *Journal of Agricultural Education and Extension.* | Paschen et al. | MLP + AIS | Privatised agricultural extension | Literature review, practice review (interviews) , stakeholder forums | Australia | 14 | Not edited | Skimmed | ** | ** | Examines hybrid advisory networks via interviews and policy analysis; elucidates governance gaps in extension systems. | Match | Skimmed | *** | * | Focuses on extension system governance with limited connection to sustainability transitions. Small interview sample (14) and early-stage findings reduce relevance. |
| 2010 | Patankar, M., Patwardhan, A., & Verbong, G. (2010). A promising niche: waste to energy project in the Indian dairy sector. *Environmental Science & Policy, 13*(4), 282–290. | Patankar et al. | MLP | Dairy sector | Document review | India | | Not edited | Not read | * | ** | Dairy-waste valorization MLP case; technical feasibility and stakeholder interviews map barriers and enablers. | ChatGPT rated higher | Skimmed | * | ** | Relevant dairy sector biogas case. Good economic analysis though lacks primary data from farmers. |
| 2023 | Petruzzelli, M., Ihle, R., Colitti, S., & Vittuari, M. (2023). The role of short food supply chains in advancing the global agenda for sustainable food systems | Petruzzelli et al. | n/a | Short food supply chains | Literature review | Multi | | Not edited | Skimmed | ** | ** | Global survey of 300+ SFSC initiatives; connects local practices to SDGs with strong mixed-methods design. | Match | Not read | * | ** | Comprehensive review of SFSC benefits aligned with SDGs. Limited original research but valuable synthesis of 69 |

| Year | Full Citation (APA) | Author(s) | Theoretical Framing | Focus | Method | Country Focus | N (interviews) | Summaries Edited? | Annabelle – Degree of Reading | Annabelle Significance | ChatGPT Significance | ChatGPT Explanation | Significance Match? | Julian – Degree of Reading | Julian Significance | Claude Significance | Claude Explanation |
|---|---|---|---|---|---|---|---|---|---|---|---|---|---|---|---|---|---|
| | transitions. *Cities, 141,* 104496. | | | | | | | | | | | | | | | | documents on local food initiatives. |
| 2016 | Pitt, H., & Jones, M. (2016). Scaling up and out as a pathway for food system transitions. *Sustainability, 8*(10), 1025. | Pitt and Jones | MLP + Dolowitz and Marsh's policy transfer framework | Food for Life (FFL) catering mark | Documentation analysis; observation of partnership meetings; semi-structured interviews | UK | 23 | Not edited | Skimmed | ** | * | Theoretical reflection on scaling strategies; conceptual but lacks empirical validation. | ChatGPT rated lower | Not read | ** | ** | Important analysis of scaling mechanisms using Food for Life case. Strong empirical base and relevant insights on policy transfer in food systems. |
| 2018 | Punt, A., Bouwmeester, H., Schiffelers, M. J. W., & Peijnenburg, A. A. (2018). Expert opinions on the acceptance of alternative methods in food safety evaluations. *Regulatory Toxicology and Pharmacology.* | Punt et al. | MLP | Animal testing in safety evaluations | Stakeholder survey (N=11) | Europe (multi) | 11 | Evaluation edited | Skimmed | * | * | Expert survey clusters barriers via MLP but small sample and subjective scoring limit strategic insight. | Match | Not read | * | * | Narrow focus on regulatory acceptance of testing methods. Small sample (11 experts) and limited relevance to on-ground food system transitions. |
| 2010 | Rehman, I. H., Kar, A., Raven, R., et al. (2010). Rural energy transitions in developing countries: a case of the Uttam Urja initiative in India. *Environmental Science & Policy, 13*(4), 303–311. | Rehman et al. | SNM | Renewable energy | Literature review | India | | Not edited | Skimmed | * | ** | SNM case of biogas digesters; interviews and field surveys illustrate learning processes relevant to rural agri-sustainability niches. | ChatGPT rated higher | Not read | * | * | Solar lighting focus with market-based approach. Branding insights potentially useful but limited relevance. |
| 2015 | Rolffs, P., Ockwell, D., & Byrne, R. (2015). Beyond technology and finance: pay-as-you-go sustainable energy access and theories of social change. *Environment and Planning A, 47*(12), 2609–2627. | Rolffs et al. | MLP + SNM | Renewable energy | Literature and semi-structured interviews | Kenya | 7 | Not edited | Skimmed | ** | * | Conceptual piece on business models; insightful for distribution but lacks empirical food-system linkage. | ChatGPT rated lower | Skimmed | ** | ** | Important SNM application to energy access. Social practice insights valuable though not food-focused. |
| 2010 | Romijn, H., Raven, R., & de Visser, I. (2010). Biomass energy experiments in rural India: Insights from | Romijn et al. | SNM + participatory learning framework | Renewable energy | Case studies and interviews | India | Not mentioned | Methods and Evaluation edited | Skimmed | ** | ** | SNM cases in India integrate workshop findings with interviews to derive | Match | Skimmed | ** | ** | Important analysis of failed energy experiments. Learning-based framework valuable for |

| Year | Full Citation (APA) | Author(s) | Theoretical Framing | Focus | Method | Country Focus | N (interviews) | Summaries Edited? | Annabelle – Degree of Reading | Annabelle Significance | ChatGPT Significance | ChatGPT Explanation | Significance Match? | Julian – Degree of Reading | Julian Significance | Claude Significance | Claude Explanation |
|---|---|---|---|---|---|---|---|---|---|---|---|---|---|---|---|---|---|
| | learning-based development approaches. *Environmental Science & Policy*, 13(4), 326–338. | | | | | | | | | | | niche management lessons. | | | | | understanding initiative failure. |
| 2019 | Rossi, A., Bui, S., & Marsden, T. (2019). Redefining power relations in agrifood systems. *Journal of Rural Studies, 68*, 147–158. | Rossi et al. | MLP + ANT + practice theory + social movement theory | Agri-food sector | Literature review | Italy, France, and Wales | | Methods and Evaluation edited | Skimmed | * | ** | Empirical interviews across EU cases provide new lenses on governance dynamics in food systems. | ChatGPT rated higher | Not read | * | ** | Important theoretical contribution on power in transitions. Three case studies provide insights though based on secondary analysis. |
| 2013 | Roy, J., Ghosh, D., Ghosh, A., & Dasgupta, S. (2013). Fiscal instruments: crucial role in financing low carbon transition in energy systems. *Current Opinion in Environmental Sustainability, 5*(2), 261–269. | Roy et al. | MLP | Renewable energy | Literature review | Multi | | Not edited | Not read | * | * | Policy review of fiscal tools; useful for funding transition niches but energy-focused rather than agri-food. | Match | Skimmed | * | * | Energy finance focus without empirical grounding. Limited relevance to food transitions. |
| 2015 | Santhanam-Martin, M., Ayre, M., & Nettle, R. (2015). Community sustainability and agricultural landscape change. *Sustainability Science, 10*, 207–217. | Santhanam-Martin et al. | MLP | Landscape and dairy farming | Semi-structured interviews, purposive sampling, document analysis, participant observation, thematic analysis | Australia | 44 | Not edited | Skimmed | *** | ** | Mixed-methods landscape-level study; GIS mapping plus community interviews reveal resilience and vulnerability of incumbent regimes. | ChatGPT rated lower | Skimmed | ** | ** | Important place-based analysis of agricultural transitions. 44 interviews provide rich insights into community identity and resistance to change. |
| 2021 | Sarabia, N., Peris, J., & Segura, S. (2021). Transition to agri-food sustainability, assessing accelerators and triggers for transformation: Case study in Valencia, Spain. *Journal of Cleaner Production, 325*, 129228. | Sarabia et al. | Urban Transformative Capacity Framework (UTCF) + MLP | Agroecology | Semi-structured interviews, document analysis, participatory observation | Spain | 8 | Not edited | Skimmed | *** | ** | Recent mixed-methods case on triggers and accelerators; strong survey, interview, workshop design makes it highly actionable. | ChatGPT rated lower | Skimmed | *** | ** | Good application of Urban Transformative Capacity Framework to food systems. Limited participants (8 interviews) but strong triangulation and local relevance. |
| 2023 | Schiller, K. J., Klerkx, L., Salazar Centeno, D. J., & Poortvliet, P. M. (2023). Developing the | Schiller et al. | MLP | Agroecology | Interviews, field visits, participant observation | Nicaragua | 27 | Evaluation edited | Skimmed | *** | *** | PNAS empirical network analysis of knowledge intermediaries; | Match | Not read | *** | *** | Excellent longitudinal study (2014–2021) of agroecological transitions. 27 |

| Year | Full Citation (APA) | Author(s) | Theoretical Framing | Focus | Method | Country Focus | N (interviews) | Summaries Edited? | Annabelle – Degree of Reading | Annabelle Significance | ChatGPT Significance | ChatGPT Explanation | Significance Match? | Julian – Degree of Reading | Julian Significance | Claude Significance | Claude Explanation |
|---|---|---|---|---|---|---|---|---|---|---|---|---|---|---|---|---|---|
| | agroecological niche in Nicaragua. *PNAS, 120*(47), e2206195120. | | | | , document analysis | | | | | | | methodological innovation with clear applicability. | | | | | interviews plus extensive fieldwork provide deep insights into knowledge processes and intermediary roles. |
| 2020 | Schiller, K., Godek, W., Klerkx, L., & Poortvliet, P. M. (2020). Nicaragua's agroecological transition: Transformation or reconfiguration of the agri-food regime? *Agroecology and Sustainable Food Systems, 44*(5), 611–628. | Schiller et al. | MLP + Anchoring Framework (Elzen et al., 2012) | Agroecology | Semi-structured interviews, participatory workshops, discourse and document analysis | Nicaragua | 17 | Evaluation edited | Skimmed | ** | | | ChatGPT ignored | Skimmed | ** | ** | Good analysis of policy-practice gaps in agroecological transitions. 17 interviews provide insights into discourse co-optation. |
| 2016 | Schut, M., Klerkx, L., Sartas, M., et al. (2016). Innovation platforms: experiences with their institutional embedding in agricultural research for development. *Experimental Agriculture, 52*(4), 537–561. | Schut et al. | MLP | Innovation platforms | Participant observation, semi-structured interviews, surveys | Africa (multi) | 6 | Methods edited | Read fully | ** | ** | Multi-country empirical study of innovation platforms; robust comparative framework for assessing institutional embedding. | | ChatGPT rated lower | Not read | ** | ** | Comprehensive analysis of innovation platforms in African agriculture. Large empirical base across multiple countries relevant for understanding institutional barriers. |
| 2014 | Schut, M., Soares, N. C., van de Ven, G., & Slingerland, M. (2014). Multi-actor governance of sustainable biofuels in developing countries: The case of Mozambique. *Energy Policy, 65*, 631–643. | Schut et al. | Governance competence framework | Biofuels | Stakeholder workshops | Mozambique | | Not edited | Skimmed | * | ** | Multi-stakeholder SNM case; mixed methods reveal governance and legitimacy issues in Mozambique. | | ChatGPT rated higher | Skimmed | ** | ** | Important analysis of governance framework development. Action research approach provides valuable insights though lacks impact assessment. |
| 2011 | Schut, M., van Paassen, A., Leeuwis, C., et al. (2011). Space for innovation for sustainable community-based biofuel production and use. *Energy Policy, 39*(9). | Schut et al. | Multi-actor governance theory | Biofuels | Participatory observation, stakeholder consultation | Mozambique | | Not edited | Skimmed | * | ** | Longitudinal fieldwork (2006–2010) provides rare learning dynamics for smallholder biofuels. | | ChatGPT rated higher | Not read | * | ** | Duplicate biofuel governance paper – see above assessment. |

| Year | Full Citation (APA) | Author(s) | Theoretical Framing | Focus | Method | Country Focus | N (interviews) | Summaries Edited? | Annabelle – Degree of Reading | Annabelle Significance | ChatGPT Significance | ChatGPT Explanation | Significance Match? | Julian – Degree of Reading | Julian Significance | Claude Significance | Claude Explanation |
|---|---|---|---|---|---|---|---|---|---|---|---|---|---|---|---|---|---|
| 2014 | Slingerland, M., & Schut, M. (2014). Jatropha developments in Mozambique: analysis of structural conditions influencing niche-regime interactions. *Sustainability, 6*(11), 7541–7563. | Slingerland and Schut | MLP | Jatropha biofuel | Longitudinal field research; stakeholder interviews; secondary data analysis | Mozambique | Not mentioned | Methods edited | Skimmed | *** | ** | Empirical MLP analysis of Jatropha failure and success; combines policy review, interviews, and network mapping in Mozambique. | ChatGPT rated lower | Skimmed | ** | ** | Excellent analysis of failed biofuel transition. Longitudinal field research (2008–2014) provides valuable insights on structural barriers in developing countries. |
| 2014 | Smith, A., Kern, F., Raven, R., & Verhees, B. (2014). Spaces for sustainable innovation: Solar photovoltaic electricity in the UK. *Technological Forecasting and Social Change, 81*, 115–130. | Smith et al. | SNM + protective spaces | Renewable energy | Semi-structured interviews | UK | 14 | Not edited | Skimmed | ** | ** | GIS and interviews show how place matters in niche development; transferable to distributed food hubs. | Match | Not read | *** | * | Solar PV focus with valuable protective space analysis. Limited direct food relevance. |
| 2016 | Stahlbrand, L. (2016). The food for life catering mark: Implementing the sustainability transition in university food procurement. *Agriculture, 6*(3), 46. | Stahlbrand | MLP | Soil Association's Food for Life catering mark | Semi-structured interviews, document analysis, participant observation | UK (England) | 31 | Summary edited | Read fully | *** | * | Single-case evaluation of a certification scheme; interesting practice insights but limited transferability. | ChatGPT rated lower | Skimmed | ** | ** | Detailed case study of sustainable procurement with strong practitioner insights. 31 interviews provide rich data on implementation challenges. |
| 2021 | Stewart, C., Piernas, C., Cook, B., & Jebb, S. A. (2021). Trends in UK meat consumption. *The Lancet Planetary Health, 5*(10), e699–e708. | Stewart et al. | | Meat consumption | Quantitative survey analysis, dietary diaries, regression modelling | UK | | Not edited | Skimmed | * | ** | Large-scale dietary data analysis with health & environmental implications; quantitative rigor and policy relevance. | ChatGPT rated higher | Skimmed | * | * | Quantitative consumption analysis without transition process insights. Useful context but limited relevance to ground-level research. |
| 2017 | Streitferdt, V., Chirarattananon, S., & Du Pont, P. (2017). Lessons learned from studying public initiatives to support energy efficiency finance in Thailand from 1992 to 2014. *Energy Efficiency, 10*, 905–923. | Streitferdt et al. | MLP | Energy Efficiency Finance (EEF) | Semi-structured interviews, document analysis | Thailand | 21 | Not edited | Skimmed | ** | * | Finance-focused historical review; methodology solid but energy-centric with limited agri-food intersection. | ChatGPT rated lower | Not read | ** | * | Energy efficiency finance with limited food relevance. Historical analysis approach potentially useful. |
| 2015 | Sutherland, L., Darnhofer, I., Wilson, G. A., & Zagata, L. (Eds.). (2015). *Transition* | Sutherland et al. | MLP | Renewable energy | Case studies and qualitative methods | Germany, Czech Republic, UK (multi) | | Not edited | Not read | – | | Edited volume of 12 in-depth European cases; comparative insights across | | Not read | N/A | * | Book not accessible for full assessment. |

| Year | Full Citation (APA) | Author(s) | Theoretical Framing | Focus | Method | Country Focus | N (interviews) | Summaries Edited? | Annabelle – Degree of Reading | Annabelle Significance | ChatGPT Significance | ChatGPT Explanation | Significance Match? | Julian – Degree of Reading | Julian Significance | Claude Significance | Claude Explanation |
|---|---|---|---|---|---|---|---|---|---|---|---|---|---|---|---|---|---|
| | *pathways towards sustainability in agriculture: case studies from Europe. CABI.* | | | | | | | | | | | contexts make it a key resource for varied transition mechanisms. | | | | | |
| 2022 | Swilling, M., Nygaard, I., Kruger, W., et al. (2022). Linking the energy transition and economic development: A framework for analysis. *Energy Research & Social Science.* | Swilling et al. | Institutional Work + Policy Settlements Theory | Renewable energy | Interviews, Case Studies, Policy Analysis | South Africa | Not mentioned | Not edited | Skimmed | ** | ** | Framework synthesizing 50+ case studies; essential conceptual integration for Global South energy-food nexus. | Match | Not read | * | * | Energy transition focus without direct food system relevance. Institutional work concepts potentially useful but limited application. |
| 2020 | Thomas, A., Lamine, C., Allès, B., et al. (2020). The key roles of economic and social organization. *Health-Agriculture-Food-Environment Nexus (HAFEN).* | Thomas et al. | HAFEN | Food systems | Literature review | France | | Not edited | Skimmed | * | ** | Multi-actor case anchored in nexus framing; surveys, interviews & network mapping yield a comprehensive behaviour-systems linkage. | ChatGPT rated higher | Skimmed | * | ** | Important HAFEN framework proposal for integrated analysis. Conceptual contribution valuable though lacks empirical application. |
| 2018 | To, L. S., Seebaluck, V., & Leach, M. (2018). Future energy transitions for bagasse cogeneration: Lessons from multi-level and policy innovations in Mauritius. *Energy Research & Social Science, 35,* 68–77. | To et al. | MLP + SNM | Waste from sugar processing (bagasse) | Semi-structured interviews, field observations, site visits | Mauritius | 27 | Not edited | Skimmed | *** | ** | Longitudinal MLP + SNM case (27 interviews) of agro-industry waste-to-energy; strong policy-tech integration though Mauritius-specific. | ChatGPT rated lower | Skimmed | *** | ** | Strong case study of successful renewable energy transition in agricultural sector. Good integration of MLP with policy analysis, though focused on energy rather than food systems directly. |
| 2009 | Van Eijck, J., & Romijn, H. (2009). Prospects for Jatropha biofuels in Tanzania. In *Sectoral Systems of Innovation and Production in Developing Countries.* Edward Elgar. | Van Eijck and Romijn | | | | Tanzania | | | | | | Early SNM application in biofuels; interviews and policy review show strategic niche emergence conditions. | | Not read | N/A | * | Book chapter – unable to fully assess without access. |
| 2016 | Vankeerberghen, A., & Stassart, P. M. (2016). The transition to conservation agriculture: an insularization process towards sustainability. *Internation* | Vankeerberghen and Stassart | MLP | Conservation agriculture | Case studies of 15 farmers: semi-structured interviews, observation | Belgium | 15 | Not edited | Skimmed | *** | ** | Interviews and policy-document analysis reveal path dependencies in conservation-agriculture niche. | ChatGPT rated lower | Not read | * | ** | Novel concept of "insularization" for within-regime transitions. Strong empirical work (15 farmers) on soil paradigm shifts |

| Year | Full Citation (APA) | Author(s) | Theoretical Framing | Focus | Method | Country Focus | N (interviews) | Summaries Edited? | Annabelle – Degree of Reading | Annabelle Significance | ChatGPT Significance | ChatGPT Explanation | Significance Match? | Julian – Degree of Reading | Julian Significance | Claude Significance | Claude Explanation |
|---|---|---|---|---|---|---|---|---|---|---|---|---|---|---|---|---|---|
| | *al Journal of Agricultural Sustainability, 14*(4), 392–407. | | | | s, workshops | | | | | | | | | | | | relevant for understanding incremental change. |
| 2020 | Vermunt, D. A., Negro, S. O., Van Laerhoven, F. S. J., Verweij, P. A., & Hekkert, M. P. (2020). Sustainability transitions in the agri-food sector: How ecology affects transition dynamics. *Environmental Innovation and Societal Transitions, 36*, 236–249. | Verm unt et al. | Agricultural innovation system perspective (AIS) | Dairy sector | Semi-structured interviews, multiple-case study, qualitative analysis | The Netherlands | 22 | Not edited | Skimmed | *** | ** | Integrates ecological feedbacks into MLP; quantitative and qualitative analyses of EU cases, innovative extension. | ChatGPT rated lower | Skimmed | ** | *** | Excellent analysis integrating ecological dimensions into MLP. 22 interviews on biodiversity-focused dairy transitions highly relevant for understanding sustainability complexities. |
| 2017 | Vila Seoane, M., & Marín, A. (2017). Transiciones hacia una agricultura sostenible: el nicho de la apicultura orgánica en una cooperativa Argentina. *Mundo Agrario, 18*(37). | Seoane and Marín | MLP | Organic apiculture | Interviews with stakeholders and literature analysis | Argentina | 11 | Not edited | Not read | ** | ** | In-depth cooperative case; power-analysis within MLP yields strong local niche insights but limited external validity. | Match | Not read | ** | ** | Excellent Global South case examining organic beekeeping cooperative. Strong integration of power analysis with transitions framework, though limited methodological transparency. |
| 2017 | Vivero-Pol, J. L. (2017). Food as commons or commodity? Exploring the links between normative valuations and agency in food transition. *Sustainability, 9*(3), 442. | Vivero-Pol | MLP | Food | Survey (N=95) | Multi | | Methods and Evaluation edited | Skimmed | *** | * | Survey of 95 professionals links values to agency; non-representative sample limits generalizability. | ChatGPT rated lower | Skimmed | *** | ** | Important conceptual contribution linking food values to transition politics. Global survey (95 professionals) provides empirical support though sample biases limit generalizability. |
| 2017 | Vlahos, G., Karanikolas, P., & Koutsouris, A. (2017). Integrated farming in Greece: A transition-to-sustainability perspective. *International Journal of Agricultural Resources, Governance and Ecology, 13*(1), 43–59. | Vlahos et al. | MLP | Integrated peach farming | | Greece | | Not edited | Not read | – | | Descriptive MLP overview with limited methodological detail; useful for context but lacks analytical depth. | | Not read | N/A | * | Unable to access full document limits assessment. |
| 2012 | Vreugdenhil, H., Taljaard, S., & Slinger, J. H. (2012). Pilot projects | Vreugdenhil et al. | | | | | | | Not read | | | Pilot-diffusion analysis; scaling lessons but limited on | | Not read | N/A | * | Unable to access full document for assessment. |

| Year | Full Citation (APA) | Author(s) | Theoretical Framing | Focus | Method | Country Focus | N (interviews) | Summaries Edited? | Annabelle – Degree of Reading | Annabelle Significance | ChatGPT Significance | ChatGPT Explanation | Significance Match? | Julian – Degree of Reading | Julian Significance | Claude Significance | Claude Explanation |
|---|---|---|---|---|---|---|---|---|---|---|---|---|---|---|---|---|---|
|  | and their diffusion: a case study of integrated coastal management in South Africa. *International Journal of Sustainable Development, 15*(1–2), 148–172. |  |  |  |  |  |  |  |  |  |  | systemic food transition processes. |  |  |  |  |  |
| 2015 | Wieczorek, A. J., Raven, R., & Berkhout, F. (2015). Transnational linkages in sustainability experiments: A typology and the case of solar photovoltaic energy in India. *Environmental Innovation and Societal Transitions, 17,* 149–165. | Wieczorek et al. | MLP | Renewable energy | Document analysis + literature review + workshops | India |  | Evaluation edited | Skimmed | * | ** | Typology of transnational niche linkages; robust mixed-case design, insightful for cross-regional food-system niche scaling. | ChatGPT rated higher | Skimmed | * | ** | Important transnational linkages typology. Solar focus but framework highly relevant for understanding global food initiatives. |
| 2003 | Wiskerke, J. S. (2003). On promising niches and constraining sociotechnical regimes: the case of Dutch wheat and bread. *Environment and Planning A, 35*(3), 429–448. | Wiskerke | MLP + Actor-Network Theory (ANT) | Wheat and bread | Document analysis + literature review | The Netherlands |  | All sections edited | Skimmed | * | ** | Foundational niche-regime case; historic in scope, strong theoretical framing, still widely cited in food MLP studies. | ChatGPT rated higher | Not read | * | ** | Early MLP application to food systems. Historical importance though based on secondary analysis rather than original research. |
| 2016 | Xu, G., Xu, X., Tang, W., et al. (2016). Fighting against water crisis in China. *Environmental Science & Policy, 61,* 33–41. | Xu et al. | MLP | Water systems | Survey, qualitative document & policy analysis with field observation | China |  | Methods edited | Skimmed | * | ** | Econometric data with policy interviews map water regime trajectories; useful for infrastructure-linked food transitions. | ChatGPT rated higher | Skimmed | * | * | Water management focus with limited food connections. Top-down transition approach less relevant. |
| 2020 | Zwartkruis, J. V., Berg, H., Hof, A. F., & Kok, M. T. (2020). Agricultural nature conservation in the Netherlands: Three lenses on transition pathways. *Technological Forecasting and Social Change, 151,* 119235. | Zwartkruis et al. | MLP + IAM + IBL | Agricultural nature conservation | Quantitative modelling, literature review, interviews, document analysis, timeline reconstruction | The Netherlands | Not mentioned | Evaluation edited | Skimmed | *** | ** | Integrates IAM, MLP, and IBL frameworks for policy-relevant insights; IAM underplays social nuance. | ChatGPT rated lower | Skimmed | *** | ** | Innovative multi-framework approach combining IAM, MLP, and IBL. Relevant for understanding local conservation initiatives but relies heavily on secondary data and modelling. |
| 2025 | Keefe, S., & Lee, J. (2025). Sustainable transitions in food | Keefe & Lee | MLP + Social | Urban agriculture | Qualitative interviews + | USA | 20 | Methods edited | Skimmed | ** | ** | Rich qualitative case study using MLP and SPT; relevant for | Match | Skimmed | *** | *** | Highly relevant with innovative dual-theoretical approach |

| Year | Full Citation (APA) | Author(s) | Theoretical Framing | Focus | Method | Country Focus | N (interviews) | Summaries Edited? | Annabelle – Degree of Reading | Annabelle Significance | ChatGPT Significance | ChatGPT Explanation | Significance Match? | Julian – Degree of Reading | Julian Significance | Claude Significance | Claude Explanation |
|---|---|---|---|---|---|---|---|---|---|---|---|---|---|---|---|---|---|
| | systems: a case study of an urban agriculture farming training program in Washington, United States. *Frontiers in Sustainable Food Systems.* | | Practice Theory | | MLP mapping | | | | | | | niche-level change and policy implications, though limited by small sample and lack of longitudinal data. | | | | | (MLP + SPT) linking systemic and local dimensions. Rich qualitative data with practical recommendations. Directly addresses food system transitions from micro to macro levels, aligning perfectly with Socioscope's field-based approach. |
| 2024 | Alonso-Adame, A., Van Meensel, J., Marchand, F., & Van Passel, S. (2024). Sustainability transitions in agri-food systems through the lens of agent-based modeling: a systematic review. *Sustainability Science, 19*(6), 2101–2118. | Alonso-Adame et al. | Transition Modelling / Agent-Based Modelling (ABM) | Agri-food sustainability transitions | Systematic literature review | Multi | | Not edited | Not read | ** | ** | Comprehensive review of agent-based modelling in AFS transitions; solid methodology and framework but lacks deep social-norm integration and reproducibility standards. | Match | Not read | ** | ** | Rigorous PRISMA methodology with comprehensive framework integrating transition and complexity theories. Broad sectoral coverage highlights ABM's versatility for understanding complex food system dynamics relevant to Socioscope's systemic approach. |
| 2023 | Elsner, F., Herzig, C., & Strassner, C. (2023). Agri-food systems in sustainability transition: a systematic literature review on recent developments on the use of the multi-level perspective. *Frontiers in Sustainable Food Systems, 7*, 1207476. | Elsner et al. | Multi-Level Perspective | Review of MLP in agri-food system transitions | Systematic review | Multi | | Not edited | Skimmed | * | ** | Systematic literature review on MLP in AFS with solid methodology and critique; key for theoretical foundation though mostly qualitative and Eurocentric. | ChatGPT rated higher | Not read | ** | ** | Rigorous PRISMA protocol with detailed interrogation of MLP levels in food systems. Reveals under-examined areas and research frontiers directly relevant to understanding food system transitions at multiple scales. |
| 2023 | Lien, H. H., de Mey, Y., Meuwissen, M. P., & Bush, S. R. (2023). Information practices for improved sustainability assurance in Vietnamese shrimp aquaculture. *Journal of Rural Studies, 100*, 103015. | Lien et al. | Information Assurance / Sustainability | Aquaculture | Ethnography (interviews, participant observation, document analysis) | Vietnam | 24 | Methods edited | Skimmed | *** | ** | Innovative ethnographic work using social practices theory to examine misalignments in assurance mechanisms; relevant for micro-level practices in seafood systems. | ChatGPT rated lower | Skimmed | ** | *** | Innovative use of social-practices theory with rich ethnographic data combining observation, interviews, and network mapping. Provides nuanced insight into how formal systems interact with |

| Year | Full Citation (APA) | Author(s) | Theoretical Framing | Focus | Method | Country Focus | N (interviews) | Summaries Edited? | Annabelle – Degree of Reading | Annabelle Significance | ChatGPT Significance | ChatGPT Explanation | Significance Match? | Julian – Degree of Reading | Julian Significance | Claude Significance | Claude Explanation |
|---|---|---|---|---|---|---|---|---|---|---|---|---|---|---|---|---|---|
| | | | | | | | | | | | | | | | | | local practices – highly relevant to Socioscope's micro-to-macro approach. |
| 2022 | Yousefian, N., Wenninger, E., & Dittrich, C. (2022). Shifts in food consumption practices among middle-class households in Bengaluru, India. *Food Culture & Society*. | Yousefian et al. | Social Practice Theory | Urban food practices | In-depth household interviews | India | 38 | Not edited | Not read | ** | * | Context-specific and insightful application of social practice theory in Global South urban consumption; useful but limited in scope and methodological triangulation. | ChatGPT rated lower | Skimmed | ** | ** | Theory-driven application of social practice theory with rich qualitative detail capturing nuanced shifts in food practices. Contextual relevance as Global South case study adds valuable perspective beyond typical Western studies. |
| 2022 | Juri, S. et al. (2022). Food systems transformations in South America: insights from a transdisciplinary process rooted in Uruguay. *Frontiers in Sustainable Food Systems, 6*, 887034. | Juri et al. | Transdisciplinarity / Governance | Food system transitions | Workshops | Uruguay | | Not edited | Skimmed | ** | *** | Highly original transdisciplinary process rooted in participatory action research; bridges theory and practice with implications for systemic transitions. | ChatGPT rated higher | Not read | ** | *** | Holistic, adaptive design with innovative facilitation through transition design and "dialogical artifacts." Created new alliances and empowered participants with system-to-transformative knowledge – directly relevant to Socioscope's transdisciplinary approach. |
| 2021 | Dinesh, D., Hegger, D. L. T., Klerkx, L., Vervoort, J., Campbell, B. M., & Driessen, P. P. J. (2021). Enacting theories of change for food systems transformation under climate change. *Global Food Security, 31*, 100583. | Dinesh et al. | Theory of Change | Policy and governance under climate pressures | Survey (N=262) | Multi | | Not edited | Not read | ** | ** | Integrates ToC with empirical validation through workshops and surveys; provides actionable frameworks though limited by self-report and stakeholder imbalance. | Match | Not read | * | ** | Empirically grounded Theory of Change validating existing framework with diverse global stakeholders. Holistic outlook integrating tech, social, financial, policy, and equity dimensions with actionable next steps. |
| 2023 | Gomez, J. G., & Darrot, (2023). The role of low-income consumers in food system transitions: case studies of community supported | Gomez & Darrot | Food Justice | Food system transitions | Semi-structured interviews | France | 13 | Methods edited | Skimmed | ** | ** | Focused, well-designed study highlighting social equity issues in CSA and groceries; methodologically solid | Match | Skimmed | ** | *** | Comprehensive local census with typological clarity yielding actionable ideal types. Dual perspective juxtaposing pioneering |

| Year | Full Citation (APA) | Author(s) | Theoretical Framing | Focus | Method | Country Focus | N (interviews) | Summaries Edited? | Annabelle – Degree of Reading | Annabelle Significance | ChatGPT Significance | ChatGPT Explanation | Significance Match? | Julian – Degree of Reading | Julian Significance | Claude Significance | Claude Explanation |
|---|---|---|---|---|---|---|---|---|---|---|---|---|---|---|---|---|---|
| | agriculture and social groceries in France. *Journal of Rural Studies.* | | | | | | | | | | | though lacks direct low-income consumer voices. | | | | | CSAs with social groceries reveals systemic gaps – directly relevant to understanding inclusion in food system transitions. |
| 2020 | Hayden, T. B., & Martín, J. P. (Eds.). (2025). *Urban Food Systems in Latin America: Territories, Mobilities and Governance.* Taylor & Francis. | Hayden and Martin | Urban Political Ecology | Urban agriculture | Participant observation + interviews | Latin America | | Not edited | Not read | – | | Cited as relevant but no access to full content for evaluation; likely useful for context-setting in Latin America urban food systems. | | Not read | N/A | | |
| 2020 | Gaddis, J. E., & Jeon, J. (2022). Sustainability transitions in agri-food systems: Insights from South Korea's universal free, eco-friendly school lunch program. In *Social Innovation and Sustainability Transition* (pp. 121–137). Springer. | Gaddis & Jeon | MLP | School-based food systems | Document + media + policy review | South Korea | | Not edited | Skimmed | ** | ** | Robust MLP application to Korea's school lunch program with novel 'precautionary infrastructure' concept; limited by single case and secondary sources. | Match | Skimmed | ** | ** | Holistic, multi-level analysis with extensive archival research. Novel concept of "precautionary infrastructure" bridges care-theory and socio-technical transitions literature, providing insights into policy-driven transitions. |
| 2019 | Bellamy, A. S., & Ioris, A. A. R. (2017, republished 2019). Addressing the knowledge gaps in agroecology and identifying guiding principles for transforming conventional agri-food systems. *Sustainability, 9*(3), 330. | Bellamy & Ioris | Agroecology Principles | Agroecology | Literature review | Multi | | Evaluation edited | Not read | * | * | Critical review distinguishing agroecology streams and calling for deeper political engagement; lacks original data and direct transition focus. | Match | Not read | * | ** | Clear mapping of divergent agroecology "streams" with insightful comparative cases illustrating state-movement partnerships. Identifies concrete research gaps to guide future work, though relies heavily on secondary sources. |
| 2019 | El Bilali, H. (2019). The multi-level perspective in research on sustainability transitions in agriculture and food systems: a systematic review. *Agriculture, 9*(4), 74. | El Bilali | Multi-Level Perspective | Agrofood systems | Systematic review | Multi | | Not edited | Read fully | * | ** | Systematic review of MLP in agri-food transitions; foundational mapping of the field, though limited by regional and sectoral biases. | ChatGPT rated higher | Not read | * | ** | First systematic mapping of MLP use in agro-food transitions with clear typology of how niches, regimes and landscapes are framed. Identifies critical gaps and |

| Year | Full Citation (APA) | Author(s) | Theoretical Framing | Focus | Method | Country Focus | N (interviews) | Summaries Edited? | Annabelle – Degree of Reading | Annabelle Significance | ChatGPT Significance | ChatGPT Explanation | Significance Match? | Julian – Degree of Reading | Julian Significance | Claude Significance | Claude Explanation |
|---|---|---|---|---|---|---|---|---|---|---|---|---|---|---|---|---|---|
| | | | | | | | | | | | | | | | | | promising hybrid methods for the field. |
| 2018 | Dahlberg, K. A. (1994). A transition from agriculture to regenerative food systems. *Futures, 26*(2), 170–197. | Dahlberg | Regenerative Systems | Regenerative food systems transition | Theoretical analysis | Multi | | Evaluation edited | Not read | * | * | Visionary conceptual piece connecting food systems to broader energy/ecological transitions; dated and lacks empirical grounding. | Match | Skimmed | * | * | Broad, integrative diagnosis linking energy, ecology, culture, and economics with clear articulation of necessary paradigm shifts. Historical significance but lacks empirical depth and is not directly related to transition studies. |
| 2024 | Washio, T., Takagi, S., Saijo, M. et al. (2025). Building short value chains for animal welfare-friendly products adoption: insights from a restaurant-based study in Japan. *arXiv*. | Washio et al. | Action research; cross-stage supply-chain coordination | Animal welfare | Choice experiment, survey, interviews, focus group | Japan | 13 | Evaluation edited | Skimmed | ** | ** | Innovative restaurant-based experiment with robust mixed methods and actionable findings; relevant to food system short chain design. | Match | Skimmed | * | ** | Ecological validity with multi-stakeholder perspective and rigorous mixed methods. Demonstrates feasible short value chain models with actionable findings, though limited generalizability due to single restaurant setting. |
| 2024 | Goel, M., Nathavani, V., Dharaiya, S. et al. (2024). Cultural context shapes the carbon footprints of recipes. *arXiv*. | Goel et al. | Computational gastronomy | Carbon footprint of gastronomy | Data integration; NLP mapping; descriptive statistics | Multi | | Evaluation edited | Skimmed | ** | ** | Large-scale computational study linking culture and emissions; original and methodologically strong, though focused narrowly on carbon. | Match | Not read | * | * | First to combine national-scale carbon footprint data with >100,000 recipes across diverse cuisines. Rigorous natural language processing integration, but focuses on technical carbon accounting rather than transition dynamics. |
| 2024 | Takagi, S., Saijo, M., & Ohashi, T. (2024). Consumer segmentation and participation drivers in Community-Supported Agriculture: a choice experiment and PLS-SEM approach. *arXiv*. | Takagi et al. | CSA participation model | Community supported agriculture (CSA) | Quantitative survey (online, N=2484) | Multi | | Evaluation edited | Skimmed | * | ** | Detailed quantitative study on CSA participation using advanced modelling; highly relevant for understanding consumer behavior in SFSCs. | ChatGPT rated higher | Skimmed | ** | ** | Combines discrete-choice and SEM for robust, segment-specific insights. First quantitative CSA model validation in Japan with large, carefully screened sample, revealing context-specific |

| Year | Full Citation (APA) | Author(s) | Theoretical Framing | Focus | Method | Country Focus | N (interviews) | Summaries Edited? | Annabelle – Degree of Reading | Annabelle Significance | ChatGPT Significance | ChatGPT Explanation | Significance Match? | Julian – Degree of Reading | Julian Significance | Claude Significance | Claude Explanation |
|---|---|---|---|---|---|---|---|---|---|---|---|---|---|---|---|---|---|
| | | | | | | | | | | | | | | | | | motivators for sustainable food systems. |
| 2023 | Takagi, S., Numazawa, Y., Katsube, K. et al. (2023). Theorizing socio-cultural dynamics of consumer decision-making for participation in CSA. *arXiv*. | Takagi et al. | Socio-ecological embedding; Expectation Disconfirmation; social capital theory | Community supported agriculture (CSA) | Literature review | Global | | Not edited | Not read | * | ** | Rich synthesis yielding a new model for CSA participation; comprehensive though theoretical and lacking empirical testing. | ChatGPT rated higher | Skimmed | * | *** | Comprehensive synthesis across 61 studies with rich, reflexive model capturing dynamic interplay of socio-cultural embedding and risk-reward balancing. Highlights social capital's role in food system transitions – highly relevant to Socioscope's approach. |
| 2022 | Riesenegger, L., & Hübner, A. (2022). Reducing food waste at retail stores—an explorative study. *Sustainability, 14*(5), 2494. | Riesenegger and Hübner | Assortment Theory | Food waste prevention | Expert semi-structured interviews; thematic analysis | Germany | 7 | Not edited | Skimmed | ** | * | Explorative study identifying operational levers in bakery retail; valuable practical insights but limited generalisability and depth. | ChatGPT rated lower | Not read | * | ** | Deep, context-rich insights from heterogeneous real-world cases with triangulation and dual coding. Concrete, actionable propositions tightly linked to operations-management theory for retail food waste reduction. |
| 2023 | Ceryes, C. A., Antonacci, C. C., Harvey, S. A., et al. (2021). "Maybe it's still good?" A qualitative study of factors influencing food waste in US supermarkets. *Appetite*. | Ceryes et al. | EPA Food Recovery Hierarchy; WRI Underlying Waste Drivers | Food waste prevention | Semi-structured interviews | USA | 20 | Evaluation edited | Skimmed | * | ** | Qualitative application of EPA hierarchy in retail contexts; grounded, relevant for food waste understanding in US supermarkets. | ChatGPT rated higher | Skimmed | ** | ** | Rich, frontline perspectives with dual focus on workers and managers uncovering both hands-on and policy-level insights. Rigorous coding process with established conceptual frameworks enhances analytic validity for understanding retail food waste. |
| 2021 | Stirnimann, A., & Zizka, L. (2022). Waste not, want not: Managerial attitudes towards mitigating food waste in | Stirnimann and Zizka | Food Waste Hierarchy | Food waste prevention | Semi-structured interviews | Switzerland | 14 | Evaluation edited | Skimmed | ** | * | Managerial focus on food waste mitigation in Swiss restaurants; narrow scope but | ChatGPT rated lower | Skimmed | ** | * | Rich, first-hand managerial perspectives in previously under-studied national |

| Year | Full Citation (APA) | Author(s) | Theoretical Framing | Focus | Method | Country Focus | N (interviews) | Summaries Edited? | Annabelle – Degree of Reading | Annabelle Significance | ChatGPT Significance | ChatGPT Explanation | Significance Match? | Julian – Degree of Reading | Julian Significance | Claude Significance | Claude Explanation |
|---|---|---|---|---|---|---|---|---|---|---|---|---|---|---|---|---|---|
| | the Swiss-German restaurant industry. *Journal of Foodservice Business Research, 25*(3), 302–328. | | | | | | | | | | | offers cultural context for hospitality sector. | | | | | context with rigorous thematic coding. More related to agent decision making in stores rather than societal transitions. |
| 2019 | Reynolds, C., Goucher, L., Quested, T., et al. (2019). Consumption-stage food waste reduction interventions – What works and how to design better interventions. *Food Policy, 83*, 7–27. | Reynolds et al. | Intervention Theory | Food waste reduction | Literature review | Global | | Not edited | Skimmed | *** | ** | Review of what works in consumption-stage interventions; solid guidance on behavioural change strategies in food waste. | ChatGPT rated lower | Not read | *** | *** | First peer-reviewed synthesis focusing exclusively on consumer-stage prevention interventions with clear categorization and actionable methodological framework. Highly applicable to draw on for literature review for Socioscope project. |
| 2018 | Derqui, B., & Viachaslau, F. The (De) Motives for Using Food Waste Reduction Apps Among Foodservice Providers and Consumers. *Available at SSRN 4575459.* | Derqui and Viachaslau | Sharing-economy concepts; behavioural drivers | Food waste reduction | Semi-structured interview + survey | Spain | 42 | Evaluation edited | Skimmed | ** | * | Emerging topic with practical relevance; available on SSRN only, limiting review of rigour or broader methodological robustness. | ChatGPT rated lower | Not read | * | * | First mixed-methods study of food waste reduction app adoption from both provider and consumer perspectives with triangulation of qualitative insights. However, convenience samples and lack of quality control limit generalizability. |
| 2020 | Dhir, A., Talwar, S., Kaur, P., & Malibari, A. (2020). Food waste in hospitality and food services: a systematic review. *Journal of Cleaner Production, 275*, 122889. | Dhir et al. | Integrative FWFLH stakeholder model | Food waste | Systematic review | Global | | Not edited | Not read | * | ** | Systematic literature review yielding a conceptual framework; useful for structuring future food waste research. | ChatGPT rated higher | Not read | ** | ** | Comprehensive mapping of diverse food loss and waste research in profit-sector hospitality and food services. Proposes integrative framework to guide future work, though review dates to early 2020. |
| 2023 | Sundin, N., Osowski, C. P., Strid, I., & Eriksson, M. (2022). Surplus food donation: Effectiveness, carbon footprint, and rebound | Sundin et al. | Life cycle assessment | Food waste prevention | Material-flow analysis, surveys (N=118), composition | Sweden | | Not edited | Skimmed | * | ** | Rigorous evaluation of food donation's environmental trade-offs; contributes to nuanced | ChatGPT rated higher | Skimmed | * | ** | Integrates material flow analysis, life-cycle assessment, and social data for comprehensive sustainability |

| Year | Full Citation (APA) | Author(s) | Theoretical Framing | Focus | Method | Country Focus | N (interviews) | Summaries Edited? | Annabelle – Degree of Reading | Annabelle Significance | ChatGPT Significance | ChatGPT Explanation | Significance Match? | Julian – Degree of Reading | Julian Significance | Claude Significance | Claude Explanation |
|---|---|---|---|---|---|---|---|---|---|---|---|---|---|---|---|---|---|
| | effect. *Resources, Conservation and Recycling, 181*, 106271. | | | | analysis, life-cycle assessment | | | | | | | understanding of rebound effects. | | | | | assessment. Explicit inclusion of rebound effects fills a gap in food loss and waste research with primary data. |
| 2021 | Bartek, L., Strid, I., Henryson, K., Junne, S., Rasi, S., & Eriksson, M. (2021). Life cycle assessment of fish oil substitute produced by microalgae using food waste. *Sustainable Production and Consumption, 27*, 2002–2021. | Bartek et al. | LCA | Aquaculture | LCA model (quantitative) | Germany and Peru | | Not edited | Skimmed | * | * | Case-specific LCA of microalgae products from food waste; technically sound but limited general application to broader transitions. | Match | Not read | * | * | Holistic LCA integrating midpoint and rarely-used endpoint biodiversity metrics with incorporation of food-waste valorization. High uncertainty in early-stage algae data and doesn't focus on societal transitions. |
| 2023 | Schneider, K. R., Fanzo, J., Haddad, L., et al. (2023). The state of food systems worldwide in the countdown to 2030. *Nature Food, 4*(12), 1090–1110. | Schneider et al. | Food system transformation | Food systems monitoring | Secondary data analysis | Global | | Not edited | Not read | * | *** | Comprehensive global assessment published in Nature Food; foundational for situating any food transition research post-2020. | ChatGPT rated higher | Not read | ** | ** | Consultative, transparent process with comprehensive, actionable framework aligning with SDGs. Baseline dataset enables cross-country benchmarking and tracking, though data gaps persist in key areas. |
| 2017 | Calvo-Porral, C., Medín, A. F., & Losada-López, C. (2017). Can marketing help in tackling food waste? *Journal of Food Products Marketing, 23*(1), 42–60. | Calvo-Porral et al. | Marketing mix (4 Ps) | Food waste reduction | Case study | Multi | | Not edited | Skimmed | ** | * | Explores marketing's potential in food waste reduction; useful peripheral insight, but lacks empirical testing or broad application. | ChatGPT rated lower | Not read | * | * | Comprehensive, supply-chain-wide mapping of waste causes with creative application of marketing concepts. Rich examples of existing initiatives offer practical inspiration but entirely conceptual with no empirical testing. |
| 2024 | Pezoldt, A., McGilvery, K., Brown, K., et al. (2024). Ensuring Food Safety among Food and Agricultural Nonprofit Organizations. *Food Protection Trends, 44*(4). | Pezoldt et al. | Food-supply-chain staging + regulatory policy analysis | Nonprofit food handling | Literature & policy review | USA | | Not edited | Not read | * | * | Focused review on food safety challenges in nonprofit sector; relevant for organizational practices but | Match | Skimmed | ** | * | Holistic mapping of food and agricultural nonprofits across the food-supply chain with clear articulation of actionable research gaps. Entirely |

| Year | Full Citation (APA) | Author(s) | Theoretical Framing | Focus | Method | Country Focus | N (interviews) | Summaries Edited? | Annabelle – Degree of Reading | Annabelle Significance | ChatGPT Significance | ChatGPT Explanation | Significance Match? | Julian – Degree of Reading | Julian Significance | Claude Significance | Claude Explanation |
|---|---|---|---|---|---|---|---|---|---|---|---|---|---|---|---|---|---|
| | | | | | | | | | | | | tangential to transition mechanisms. | | | | | secondary analysis without primary empirical data to validate claims. |
| 2010 | Parfitt, J., Barthel, M., & Macnaughton, S. (2010). Food waste within food supply chains: quantification and potential for change to 2050. *Philosophical Transactions of the Royal Society B, 365*(1554), 3065–3081. | Parfitt et al. | Food-Supply-Chain gradient | Food waste | Literature & policy review + expert interviews | Global | Not mentioned | Not edited | Skimmed | ** | *** | Seminal quantification and strategic framing of food waste potential to 2050; foundational for food systems sustainability work. | ChatGPT rated higher | Skimmed | ** | ** | Comprehensive, cross-sectoral mapping of losses and wastes with clear conceptual framework and actionable recommendations. Heavy reliance on decades-old primary studies limits current relevance. |
| 2021 | Eriksson, M., Bartek, L., Löfkvist, K., et al. (2021). Environmental assessment of upgrading horticultural side streams—The case of unharvested broccoli leaves. *Sustainability, 13*(10), 5327. | Eriksson et al. | System-expansion LCA, functional-unit comparison | Food waste | LCA | Sweden | | Evaluation edited | Not read | * | * | Case-specific LCA on broccoli leaves; adds to circular economy discourse but narrow in scope and applicability. | Match | Not read | * | * | Dual LCA approaches provide robust insights with realistic case study and sensitivity analysis. Does not focus on societal transitions and assumes market uptake without economic analysis. |
| 2021 | Sundin, N., Rosell, M., Eriksson, M., Jensen, C., & Bianchi, M. (2021). The climate impact of excess food intake – An avoidable environmental burden. *Resources, Conservation and Recycling, 174*, 105777. | Sundin et al. | Systems-based metabolic waste quantification | Overconsumption impacts | Secondary data modelling; comparative metrics | Sweden | | Not edited | Not read | * | ** | Highlights overlooked environmental cost of overeating; valuable for policy and dietary transition framing. | ChatGPT rated higher | Not read | * | * | First to compare two quantification methods with national prevalence and representative dietary survey. Uses up-to-date LCA database but relies on self-reported data and excludes other environmental impacts. |
| 2023 | Hegwood, M., Burgess, M. G., Costigliolo, E. M., et al. (2023). Rebound effects could offset more than half of avoided food loss and waste. *Nature Food, 4*(7), 585–595. | Hegwood et al. | Rebound effect theory / supply–demand analysis | Food waste | Microeconomic modelling & Monte Carlo simulation | Global | | Not edited | Skimmed | ** | ** | Quantitative modelling of rebound effects in food waste prevention; critical insight into systemic unintended consequences. | Match | Not read | ** | ** | First global quantification of direct rebound from food loss and waste reduction with robust uncertainty analysis. Translates rebounds into both environmental and food-security metrics, highlighting policy trade-offs. |

| Year | Full Citation (APA) | Author(s) | Theoretical Framing | Focus | Method | Country Focus | N (interviews) | Summaries Edited? | Annabelle – Degree of Reading | Annabelle Significance | ChatGPT Significance | ChatGPT Explanation | Significance Match? | Julian – Degree of Reading | Julian Significance | Claude Significance | Claude Explanation |
|---|---|---|---|---|---|---|---|---|---|---|---|---|---|---|---|---|---|
| 2019 | Roberts, C., & Geels, F. W. (2019). Conditions for politically accelerated transitions: Historical institutionalism, the multi-level perspective, and two historical case studies in transport and agriculture. *Technological Forecasting and Social Change, 140.* | Roberts and Geels | Historical Institutionalism + MLP | Agricultural development | Historical cases | UK | | Not edited | Skimmed | ** | ** | Strong theoretical contribution blending historical institutionalism and MLP; directly relevant for understanding governance dynamics. | Match | Not read | ** | *** | Novel integration of Multi-Level Perspective and Historical Institutionalism yields richer political analysis of transitions. Two contrasting cases illustrate how both shocks and gradual coalitions can unlock policy defection – highly relevant to understanding transition mechanisms. |
| 2023 | Meng, F., Wagner, A., Kremer, A. B., et al. (2023). Planet-compatible pathways for transitioning the chemical industry. *PNAS, 120*(8), e2218294120. | Meng et al. | Planetary Boundaries | Chemical industry decarbonization | Scenario modelling; lifecycle assessment | Multi | | Not edited | Not read | * | * | High-impact study with advanced modelling; peripheral to food systems but conceptually relevant for cross-sectoral transitions. | Match | Skimmed | * | * | Integrates both supply- and demand-side levers under realistic resource caps with quantification of net-negative potential. Not related to societal transitions and focuses on technical chemical industry pathways. |
| 2020 | Otto, I. M., Donges, J. F., Cremades, R., et al. (2020). Social tipping dynamics for stabilizing Earth's climate by 2050. *PNAS, 117*(5), 2354–2365. | Otto et al. | Social tipping points | Renewable energy | Workshops, survey (N=133) | Multi | | Evaluation edited | Skimmed | *** | * | Broad climate study with conceptual insights on social tipping points; generalisable but not food-system specific. | ChatGPT rated lower | Skimmed | ** | *** | Novel application of social tipping theory to climate mitigation – looks at how small changes can become macro-level changes. Integrates multiple disciplinary insights with actionable levers and empirical analogues – highly relevant to understanding transition mechanisms. |
| 2023 | Wells, P. E. (2023). System confluence and the reinvention of automobility. *PNAS, 120*(47), e2206233119. | Wells | MLP + institutional legitimacy; system confluence | Automobile sustainability | Case vignettes; literature synthesis | Multi | | Not edited | Skimmed | ** | * | Theoretical insights into system transition in transport; potentially analogous but not directly relevant to food systems. | ChatGPT rated lower | Skimmed | * | * | Holistic system view connecting diverse innovations with integrative conceptual framing. Rich historical context but lacks primary empirical data and focuses on |

| Year | Full Citation (APA) | Author(s) | Theoretical Framing | Focus | Method | Country Focus | N (interviews) | Summaries Edited? | Annabelle – Degree of Reading | Annabelle Significance | ChatGPT Significance | ChatGPT Explanation | Significance Match? | Julian – Degree of Reading | Julian Significance | Claude Significance | Claude Explanation |
|---|---|---|---|---|---|---|---|---|---|---|---|---|---|---|---|---|---|
| | | | | | | | | | | | | | | | | | mobility rather than food systems. |
| 2023 | Ryghaug, M., & Skjølsvold, T. M. (2023). How policies and actor strategies affect electric vehicle diffusion and wider sustainability transitions. *PNAS, 120*(47), e2207888119. | Ryghaug and Skjølsvold | MLP + policy-mix theory; user/intermediary roles | Electric cars | Interviews, large-scale user surveys (N=~16,000), document analysis | Norway | 100 | Summary edited | Read fully | *** | * | Focused on EV diffusion; methodological parallels to food transitions exist but limited sectoral relevance. | ChatGPT rated lower | Skimmed | *** | ** | Rich mixed-methods data capturing both macro-policy shifts and micro-user experiences. Longitudinal MLP framing illuminates how landscape shocks and regime dynamics co-evolve, though focused on mobility transitions. |
| 2023 | Mylan, J., Andrews, J., & Maye, D. (2023). The big business of sustainable food production and consumption: Exploring the transition to alternative proteins. *PNAS, 120*(47), e2207782120. | Mylan et al. | MLP | Alternative proteins | Secondary data analysis: surveys, media, corporate documents | Multi | | Not edited | Skimmed | ** | ** | Explores alternative proteins and commercialization; pertinent to transition dynamics in sustainable consumption. | Match | Not read | ** | *** | Holistic MLP framing integrating technological, cultural, and political dimensions. Rich longitudinal perspective with emphasis on corporate engagement illuminating regime destabilization – highly relevant to understanding food system transitions. |
| 2023 | Meadowcroft, J., & Rosenbloom, D. (2023). Governing the net-zero transition: Strategy, policy, and politics. *PNAS, 120*(47), e2207727120. | Meadowcroft and Rosenbloom | MLP | Net-zero | Literature review, policy analysis | Multi | | Not edited | Not read | ** | ** | Strong governance-focused framing of net-zero transition; high-level relevance for food policy and system change mechanisms. | Match | Not read | ** | ** | Comprehensive synthesis integrating diverse literatures into coherent governance framework. Actionable guidance on aligning policy instruments to transition phases, though conceptual scope only without empirical validation. |
| 2022 | Altman, J., and Markham, F. (2022). Disruption as Reprieve? In *Beyond Global Food Supply Chains* (pp. 125–137). Palgrave Macmillan. | Altman and Markham | IPA + Transformative incrementalism | Government Responses to Small-scale Food System | Ethnographic observation, semi-structured interviews, discourse analysis | Australia (Melbourne) | 17 | Not edited | Skimmed | * | ** | Offers unique insights into food policy during crises; focuses on discursive framing and bureaucratic negotiation, valuable for understanding incremental change. | ChatGPT rated higher | Skimmed | *** | ** | Examines how crises create policy windows for food system change, with insights into discursive framing and incremental policy work. Relevant for understanding how |

| Year | Full Citation (APA) | Author(s) | Theoretical Framing | Focus | Method | Country Focus | N (interviews) | Summaries Edited? | Annabelle – Degree of Reading | Annabelle Significance | ChatGPT Significance | ChatGPT Explanation | Significance Match? | Julian – Degree of Reading | Julian Significance | Claude Significance | Claude Explanation |
|---|---|---|---|---|---|---|---|---|---|---|---|---|---|---|---|---|---|
| | | | | Resistance | | | | | | | | | | | | | disruption enables normally marginalized issues to gain traction. |
| 2019 | Anderson, M., and Leach, M. (2019). Transforming Food Systems: The Potential of Engaged Political Economy. IDS. | Anderson et al. | MLP + EPE + Food Sovereignty | EPE in Food System Transformations | Theoretical literature review, secondary data analysis, comparative case studies | Multi | | Methods edited | Not read | * | *** | Core conceptual paper linking engaged research to food systems transformation; original and highly relevant for methodological grounding. | ChatGPT rated higher | Read fully | ** | ** | Strong theoretical framework arguing for engaged political economy approach, emphasizing power analysis and marginalized voices. Aligns well with Food Socioscope's focus on understanding transition mechanisms. |
| 2024 | Barbosa, M. W. (2024). Government Support Mechanisms for Sustainable Agriculture: A Systematic Literature Review. *Sustainability, 16*(5), 2185. | Barbosa | Institutional Theory + Institutional Isomorphism | Government Support in Agri-business | Systematic review (1992–2023); site visits; documentary analysis | Global (Review) | | Methods edited | Not read | * | ** | Methodologically rigorous systematic review with practical insights into government mechanisms; less original but strong for methodology and policy synthesis. | ChatGPT rated higher | Read fully | ** | ** | Systematic review highlighting both explicit mechanisms (subsidies, regulations) and implicit institutional pressures shaping sustainable agriculture. Strong methodological approach and theoretical grounding in institutional theory. |
| 2021 | Béné, C., Kawarazuka, N., Pham, H., et al. (2021). Policy Framing and Crisis Narratives around Food Safety in Vietnam. *Environment and Planning E, 4*(3), 985–1009. | Bene et al. | AST + NPA + Policy Framing | Policy Framing and Crisis Narratives Around Food Safety | Survey and interviews, framing analysis | Vietnam | 91 (questionnaire) + 37 | Summary and Methods edited | Not read | * | ** | Examines framing and governance responses in food policy crises; methodologically strong and applicable to discursive politics. | ChatGPT rated higher | Read fully | ** | ** | Reveals how crisis narratives shape policy priorities over structural issues. Important for understanding agenda-setting mechanisms. |
| 2019 | Béné, C., Oosterveer, P., Lamotte, L., et al. (2019). When Food Systems Meet Sustainability – Current Narratives and Implications for Actions. *World Development, 113*, 116–130. | Bene et al. | Interpretive Ontology + Narrative Analysis | Implicit Rules in Food System Governance | In-depth literature review, discourse analysis, interpretive analytic approach | Global | | Summary and Methods edited | Skimmed | * | *** | Presents a holistic and practical framework for food systems governance; methodologically rigorous and highly aligned with project goals. | ChatGPT rated higher | Read fully | *** | *** | Examines competing sustainability narratives across disciplines, revealing how implicit rules shape governance. Essential for understanding how different framings lead to different interventions. |
| 2016 | Bernstein, H. (2016). Agrarian Political | Bernstein | ANT + GVC | Food-system | Literature analysis of | USA, UK, Canada, | | Not edited | Skimmed | * | * | Key theoretical overview for | Match | Read fully | ** | *** | Seminal theoretical framework for |

| Year | Full Citation (APA) | Author(s) | Theoretical Framing | Focus | Method | Country Focus | N (interviews) | Summaries Edited? | Annabelle – Degree of Reading | Annabelle Significance | ChatGPT Significance | ChatGPT Explanation | Significance Match? | Julian – Degree of Reading | Julian Significance | Claude Significance | Claude Explanation |
|---|---|---|---|---|---|---|---|---|---|---|---|---|---|---|---|---|---|
|  | Economy and Modern World Capitalism: The Contributions of Food Regime Analysis. *The Journal of Peasant Studies, 43*(3), 611–647. |  |  | Regimes in Transition | secondary sources, policy reviews, comparative government analysis, historical contextualization | Global North |  |  |  |  |  | background on food regimes and capitalist dynamics; valuable for framing but abstract in relation to local governance. |  |  |  |  | understanding historical food system transitions through regime analysis. Essential for contextualizing current transformations and power mechanisms in global food systems. |
| 2022 | Brons, A., Oosterveer, P., & Wertheim-Heck, S. (2022). In- and Exclusion in Urban Food Governance: Exploring Networks and Power in the City of Almere. *Journal of Environmental Policy & Planning, 24*(6), 777–793. | Brons et al. | GNT + Manuel Castells' Network Theory of Power | Local Governance Exclusionary Mechanisms | Qualitative interviews + quantitative network survey; document/policy analysis; participant observation | Netherlands (Almere) | 15 + 17 (online) | Summary and Methods edited | Skimmed | * |  |  | ChatGPT ignored | Read fully | ** | ** | Network analysis revealing how participatory processes can reinforce exclusion. Strong methodology for understanding power in collaborative governance. |
| 2015 | Buchan, R., Cloutier, D., Friedman, A., & Ostry, A. (2015). Local Food System Planning: The Problem, Conceptual Issues, and Policy Tools for Local Government Planners. *Canadian Journal of Urban Research, 24*(1), 1–23. | Buchan et al. | PFA + Conceptual Analysis | Mechanisms of Food System Policy Among Local Governments | Comprehensive literature review, policy and document analysis; comparative case study deductive reasoning | North America (Multi) |  | All sections edited | Skimmed | * | ** | Framework-focused paper providing practical planning tools for local governments; directly useful for policy application. | ChatGPT rated higher | Read fully | ** | ** | Comprehensive review of planning tools and strategic approaches for local food systems. Good overview of implementation mechanisms. |
| 2017 | Campbell, L. K. (2017). City as Ecosystem: Changing Form, Function, and Governance of Urban Socio-Nature. In *City of Forests, City of Farms* (pp. 208–220). Cornell University Press. | Campbell | Assemblage Theory + Human Geography and Political Ecology | Discursive Politics in City Food System Governance | Discourse analysis, case study methodology, policy and literature analysis, qualitative interview analysis with policy makers | USA (New York City) | Not mentioned | All sections edited | Skimmed | * | ** | Conceptualizes ecological planning frameworks in urban food system transitions; a rich interpretive planning resource. | ChatGPT rated higher | Read fully | ** | ** | Shows how issues gain traction through discursive alignment with dominant frames. Important for understanding strategic framing. |

| Year | Full Citation (APA) | Author(s) | Theoretical Framing | Focus | Method | Country Focus | N (interviews) | Summaries Edited? | Annabelle – Degree of Reading | Annabelle Significance | ChatGPT Significance | ChatGPT Explanation | Significance Match? | Julian – Degree of Reading | Julian Significance | Claude Significance | Claude Explanation |
|---|---|---|---|---|---|---|---|---|---|---|---|---|---|---|---|---|---|
| 2019 | Daviron, B., Perrin, C., & Soulard, C. T. (2019). History of Urban Food Policy in Europe. In *Designing Urban Food Policies* (pp. 27–51). Springer. | Daviron et al. | Mercantilism + Central Place vs. Network Models | Historical Development of Food System Governance | Historical analysis, case studies, comparative approach | Historic Europe | | Summary edited | Not read | * | ** | Historical depth provides essential context for contemporary municipal actions; helpful for framing transitions. | ChatGPT rated higher | Skimmed | * | * | Historical overview providing context for contemporary governance, limited direct relevance to current transformation mechanisms. |
| 2019 | Delgado, C. (2019). What Makes Food Policies Happen? Insights from Portuguese Initiatives. *Journal of Agriculture, Food Systems, and Community Development, 9*(1), 1–2. | Delgado | Participatory Governance + GaIA + PPaR | Implicit Rules in Food System Governance + Built Environment | Interviews, case study analysis, policy/document analysis | Portugal | | Summary and Methods edited | Not read | * | *** | Detailed empirical work on local government food initiatives, highlighting political strategies and implementation mechanisms; directly relevant to municipal agency. | ChatGPT rated higher | Read fully | * | * | Brief analysis of Portuguese urban agriculture revealing governance gaps, limited empirical depth for understanding transformation mechanisms. |
| 2016 | Dubbeling, M., Carey, J., & Hochberg, K. (2016). *The role of private sector in city region food systems.* RUAF Foundation. | Dubbeling et al. | MLP + GVC + SNM | Public-Private Partnerships in Food Sustainability | Comparative government policy case studies; participant interviews, policy review | Ecuador (Quito), UK (Bristol), Netherlands (Rotterdam) | | Evaluation edited | Not read | *** | * | Useful for broad context on private sector involvement and motives; descriptive more than analytical, but relevant for showing external influences. | ChatGPT rated lower | Read fully | *** | ** | Provides empirical evidence from three diverse urban cases on private sector engagement mechanisms and motivations. Directly relevant for understanding public-private dynamics in food system transitions. |
| 2010 | Dubbeling, M. C. E., Merzthal, G., & Soto, N. (2010). Multistakeholder Policy Formulation and Action Planning for Urban Agriculture in Lima, Peru. *JAFSCD, 1*(2), 145–154. | Dubbeling et al. | MPAP + Participatory Governance | Network Governance in Urban Agricultural Policy | Qualitative case study, participatory action research, process documentation | Peru (Lima) | | Summary edited | Skimmed | * | ** | Multistakeholder planning process well-documented; illustrative of urban government coordination in food policy development. | ChatGPT rated higher | Read fully | ** | ** | Well-documented MPAP approach showing systematic stakeholder engagement methodology. Relevant for understanding participatory planning mechanisms. |
| 2022 | Ferrando, T. (2022). The UN Food Systems Summit: Disaster Capitalism and the Future of Food. In *Beyond Global Food Supply Chains* (pp. 139–153). Palgrave Macmillan. | Tomaso | PEoFSG + Discourse analysis + M-SC | UN Mechanisms of Food Systems Governance | Document analysis, critical discourse analysis, comparative institutional analysis | UN Food Systems Summit (UNFSS) | | Not edited | Not read | * | ** | Critical analysis of global food governance through discursive and institutional frames; significant for understanding the limitations of multistakeholder approaches. | ChatGPT rated higher | Read fully | *** | ** | Critical analysis of global food governance through UN Summit case, revealing how multi-stakeholder approaches can mask power asymmetries. Important for understanding |

| Year | Full Citation (APA) | Author(s) | Theoretical Framing | Focus | Method | Country Focus | N (interviews) | Summaries Edited? | Annabelle – Degree of Reading | Annabelle Significance | ChatGPT Significance | ChatGPT Explanation | Significance Match? | Julian – Degree of Reading | Julian Significance | Claude Significance | Claude Explanation |
|---|---|---|---|---|---|---|---|---|---|---|---|---|---|---|---|---|---|
| | | | | | | | | | | | | | | | | | discursive strategies and legitimacy management in international contexts. |
| 2007 | Franks, J. R., & McGloin, A. (2007). Joint Submissions, Output Related Payments and Environmental Co-Operatives. *Journal of Environmental Planning and Management, 50*(2), 233–256. | Franks and McGloin | ANT + Economies of Configuration | Agri-environmental Policy Co-creation | Structured and open-ended interviews; systematic literature review; multiple case study analysis | Netherlands, UK | 20–30 | Evaluation edited | Not read | * | ** | Demonstrates innovative collaborative governance models; relevant for understanding collective action in food and environmental planning. | ChatGPT rated higher | Skimmed | ** | * | Analysis of Dutch innovation in agri-environmental policy, tangentially relevant to food systems but useful for understanding collaborative governance models. |
| 2022 | Gowd, K. K. (2022). Network Governance, Food Security, and Public Service Delivery: Functioning of Chhattisgarh's CORE-PDS. *Journal of Public Affairs, 22*(3). | Gowd | Network Governance + Meta-Governance | Network Governance, Food Security, and Public Service | Actor-role analysis, comparative policy analysis, qualitative case study, literature review | India (Chhattisgarh) | | Summary and Methods edited | Not read | * | ** | Strong example of digital infrastructure in food governance; valuable for administrative innovation and delivery systems. | ChatGPT rated higher | Read fully | ** | ** | Strong case study of technology-enabled network governance improving food distribution. Shows importance of adaptive solutions and civil society engagement. |
| 2023 | Hayden, T. B. (2023). Insecure Infrastructures: The Affects and Effects of Violence in Mexico's Food System. *American Anthropologist, 125*(1), 89–99. | Hayden | InPE + RPoT + Affective Infrastructure | Food System Transition in Political Instability | Ethnographic field research, literature review, participant observation | Mexico (CDMX) | Not mentioned | Summary edited | Skimmed | * | * | Unique ethnographic insight into food insecurity under violence; important context but not central to municipal governance. | Match | Read fully | ** | ** | Innovative ethnographic approach showing how violence shapes food system infrastructures through fear and informal mechanisms. Unique perspective on how insecurity restructures food circulation. |
| 2022 | Helguero, M. L., Steyaert, A., & Dessein, J. (2022). City-to-City Learning Processes in the Development of Sustainable Urban Food Systems. *Habitat International, 124*. | Helguero et al. | PTA + MAG + City-to-city learning theory | City to City Food System Policy Development | Policy and legislative document analysis, secondary data analysis, semi-structured interviews, qualitative case study analysis | Ecuador (Quito), Peru (Lima) | 15 | All sections edited | Skimmed | * | ** | Emphasizes translocal learning mechanisms; relevant for understanding diffusion of municipal innovations. | ChatGPT rated higher | Skimmed | ** | ** | Shows importance of trust, context-specificity, and translation (not transfer) in policy learning. Relevant for understanding knowledge exchange mechanisms. |

| Year | Full Citation (APA) | Author(s) | Theoretical Framing | Focus | Method | Country Focus | N (interviews) | Summaries Edited? | Annabelle – Degree of Reading | Annabelle Significance | ChatGPT Significance | ChatGPT Explanation | Significance Match? | Julian – Degree of Reading | Julian Significance | Claude Significance | Claude Explanation |
|---|---|---|---|---|---|---|---|---|---|---|---|---|---|---|---|---|---|
| 2020 | Hou, J. (2020). Governing Urban Gardens for Resilient Cities: Examining the 'Garden City Initiative' in Taipei. *Urban Studies, 57*(7), 1398–1416. | Hou | NUG + Resilience Governance + Community Agency | City Food System Development Governance Decision Making | Interviews with key actors, participant observation, document review | Taiwan (Taipei) | Not mentioned | Not edited | Skimmed | * | ** | Case study in participatory urban food governance; methodologically grounded and relevant to transitions in urban planning culture. | ChatGPT rated higher | Read fully | * | ** | Shows how policy emerges through adaptation of existing structures rather than overhaul. Good example of strategic opportunism in governance. |
| 2013 | Kirwan, J., & Maye, D. (2013). Food Security Framings within the UK and the Integration of Local Food Systems. *Journal of Rural Studies, 29*(2), 91–100. | Kirwan and Maye | CPS + GKC + Consensus Framing | Food Security Policy Framing and Integration | Critical discourse analysis, application of frame-analytic perspective, comparative policy review | UK | | All sections edited | Skimmed | * | ** | Analyzes narrative construction of food security; useful for framing discursive elements in local government contexts. | ChatGPT rated higher | Skimmed | ** | ** | Critical analysis of how consensus frames mask disagreement and marginalize local food systems. Important for understanding discursive power. |
| 2022 | Lawton, A., & Morrison, N. (2022). The Loss of Peri-Urban Agricultural Land and the State-Local Tensions in Managing Its Demise. *Land Use Policy, 120.* | Lawton and Morrison | P-GAT + NLG + Institutional analysis of MLG | Land Use Conflict in Peri-Urban Agriculture Planning | Policy and legislative document analysis, mixed methods, case study analysis | Australia (Greater Western Sydney) | | Summary and Methods edited | Skimmed | * | ** | Valuable land-use governance case with implications for municipal planning and agricultural policy coordination. | ChatGPT rated higher | Read fully | * | ** | Reveals how pro-growth narratives and institutional structures favor urban development over food production. Important for understanding systemic biases. |
| 2019 | Lever, J., Sonnino, R., & Cheetham, F. (2019). Reconfiguring Local Food Governance in an Age of Austerity. *Journal of Rural Studies, 69,* 97–105. | Lever et al. | MLG + NGC + Eliasian Sociology | Local Food System Governance | 50+ semi-structured interviews, participant observation, thematic coding, policy analysis | UK (North England) | 50+ | Summary and Methods edited | Skimmed | * | ** | Explores adaptive governance and institutional reconfiguration under austerity; strong for policy response analysis at local levels. | ChatGPT rated higher | Skimmed | ** | ** | Shows how austerity reshapes food governance, forcing reliance on discursive reframing and partnerships. Important for understanding governance under constraint. |
| 2021 | Liu, Y., Zhao, W., Liao, R., & Wang, C. (2021). Process Analysis of Inter-Governmental Negotiation in Delineating Permanent Prime Farmland around Cities: The Case of Chongqing, China. *Land Use Policy, 111.* | Liu et al. | PAT + RoCT + Game Theory + Fragmented Authoritarianism | Discursive Policy Flexibility in Local Governance | Structured interviews, field work, spatial analysis, institutional logic analysis, document and policy review | China (Chongqing) | Not mentioned | Evaluation edited | Skimmed | * | ** | Well-documented negotiation process on land use; highlights administrative dynamics that can inform urban food policy. | ChatGPT rated higher | Skimmed | ** | ** | Detailed analysis of central-local negotiation dynamics in land use planning. Shows how competing incentives and information asymmetries shape outcomes. |

| Year | Full Citation (APA) | Author(s) | Theoretical Framing | Focus | Method | Country Focus | N (interviews) | Summaries Edited? | Annabelle – Degree of Reading | Annabelle Significance | ChatGPT Significance | ChatGPT Explanation | Significance Match? | Julian – Degree of Reading | Julian Significance | Claude Significance | Claude Explanation |
|---|---|---|---|---|---|---|---|---|---|---|---|---|---|---|---|---|---|
| 2000 | Losada, H., Bennett, R., Cortes, J., Vieyra, J., & Soriano, R. (2000). The Historical Development of the Mexico City Milk Supply System. *Habitat International, 24*(4), 485–500. | Losada et al. | Historical Institutionalism + Political Economy | Historical Development of Food Distribution (Milk) | Documentary review, historical analysis | Mexico (CDMX) | | Summary edited | Skimmed | * | * | Historical development case with limited methodological or conceptual relevance to current governance challenges. | Match | Read fully | ** | * | Historical case study showing evolution of state intervention in food systems, useful context but dated. |
| 2021 | Mantino, F. (2021). Rural Areas between Locality and Global Networks. *Bio-Based and Applied Economics, 10*(4), 265–281. | Mantino | NDT + RWF + NEG + Relational Proximity and Territorial Capital | Rural Actors Policy Empowerment | Interdisciplinary literature review, comparative policy analysis, case study review, governance and institutional analysis | Europe (Multi) | | Summary and Evaluation edited | Not read | * | * | Explores policy empowerment in rural contexts; broader in scope and not focused on urban government food decision-making. | Match | Skimmed | ** | * | Broad analysis of rural development mechanisms, indirect relevance to food systems but useful for understanding enabling environments. |
| 2023 | Marini, M., Caro, D., & Thomsen, M. (2023). Investigating Local Policy Instruments for Different Types of Urban Agriculture in Four European Cities. *Land Use Policy, 131*. | Marini et al. | MLG + Policy Effectiveness/Instrumentality + Circular Economy/Urban Metabolism | Policy Instruments for Urban Agricultural Planning | Systematic literature review, multiple case study analysis, structured online survey, cross-case synthesis | Spain (Barcelona), France (Lyon), Italy (Udine/Trieste) | | Summary edited | Skimmed | * | *** | Comparative European study identifying effective policy tools; rigorous and immediately applicable to municipal governance. | ChatGPT rated higher | Read fully | ** | * | Comparative policy instrument analysis across cities, useful but limited depth on transformation mechanisms or actor dynamics. |
| 2014 | McCarthy, B. L. (2014). Sustainable Food Systems in Northern Queensland. *Journal of Economic and Social Policy, 16*(1). | McCarthy | AAFN + SCT + Value Chain/Network Theory | Grassroots Sustainability Influence in Governance | Semi-structured interviews with key informants; participant observation; consumer surveys | Australia (Queensland) | 10 | Methods and Evaluation edited | Skimmed | * | * | Context-specific exploration of sustainable practices; limited generalizability or focus on government processes. | Match | Skimmed | * | * | Local case study with limited broader applicability, focuses more on grassroots initiatives than governance mechanisms. |
| 2021 | Mialon, M., Gaitan Charry, D. A., Cediel, G., et al. (2021). 'I Had Never Seen so Many Lobbyists': Food Industry Political Practices during | Mialon et al. | CPA + CDoH + Political Economy/Realpolitik Lens | Lobbying in Food Governance Decision Making | Qualitative document analysis, semi-structured interviews, | Colombia | 18 | Summary edited | Skimmed | * | | | ChatGPT ignored | Read fully | ** | ** | Detailed documentation of corporate political activities and counter-strategies by civil society. Strong |

| Year | Full Citation (APA) | Author(s) | Theoretical Framing | Focus | Method | Country Focus | N (interviews) | Summaries Edited? | Annabelle – Degree of Reading | Annabelle Significance | ChatGPT Significance | ChatGPT Explanation | Significance Match? | Julian – Degree of Reading | Julian Significance | Claude Significance | Claude Explanation |
|---|---|---|---|---|---|---|---|---|---|---|---|---|---|---|---|---|---|
| | the Development of a New Nutrition Front-of-Pack Labelling System in Colombia. *Public Health Nutrition, 24*(9), 2737–2745. | | | | triangulation, deductive coding | | | | | | | | | | | | empirical evidence of how vested interests shape policy. |
| 1999 | Molnar, J. J. (1999). Sound Policies for Food Security: The Role of Culture and Social Organization. *Review of Agricultural Economics, 21*(2), 489–498. | Molnar | CCT + Institutional Failure + Reciprocal Causation | Global South Cultural and Governmental Practices Regarding Food Security | Comparative case studies, literature synthesis, macro-level analysis | Global South (Multi) | | Not edited | Not read | * | ** | Solid analysis of cultural and institutional barriers to food security policy; focuses on macro-level systemic constraints but lacks detailed case application. | ChatGPT rated higher | Read fully | ** | * | Offers macro-level structural analysis of how cultural values and social structures shape food policies, but lacks micro-level insights into bureaucratic mechanisms. Useful for understanding broader context but limited for capturing "on the ground" dynamics. |
| 2012 | Mustapha, U. M., Culas, R. J., Hernandez, F. E., & Jones, M. A. (2012). Agricultural Policies and Environmental, Political, Economic, Social and Technological Threats for Food Security in Africa. In *Food Security: Quality Management*. Nova Science Publishers. | Mustapha et al. | PEA + Trade Theory + Sustainable Development | Agricultural Policies and Threats for Food Security in Africa | Comparative analysis, case study analysis, policy review, secondary data synthesis | Africa (multi) | | Not edited | Skimmed | * | * | Broad overview of threats and policies; not focused on municipal-level agency or food systems transformation. | Match | Read fully | ** | * | Broad threat analysis for African food security, limited focus on specific governance mechanisms or transformation pathways. |
| 2023 | Nikolaidou, S., Loudiyi, S., & Reckinger, R. (2023). New Directions in Governance of Urban Food Systems Transitions. *Frontiers in Sustainable Food Systems, 7*. | Nikolaidou et al. | STT + TFN + Food Networks | New Direction in Food System Transition Policy | Case study analysis, participatory action research, discourse analysis | Brazil (Sao Paulo), France (Toulouse) | | Not edited | Skimmed | * | *** | Comprehensive framework for urban food transition governance; synthesizes academic and practical insights, central to project aims. | ChatGPT rated higher | Skimmed | ** | * | Editorial overview of governance innovations, useful framing but limited empirical content. |
| 2013 | Shey, J. E., & Belis, D. (2013). Building a Municipal Food Policy Regime in Minneapolis: Implications for Urban Climate Governance. Environme | Shey and Belis | URT + Multilevel Governance + Community Power/Plu | Urban Coalition Building in City Food System Planning | Case study analysis, process tracing, literature review | USA (Minneapolis) | | Methods edited | Skimmed | ** | *** | Detailed empirical case of institutional building and municipal policy innovation; exemplary case for local government agency. | ChatGPT rated higher | Read fully | ** | ** | Strong analysis of coalition building and regime formation in urban food policy. Shows importance of sustained cooperation across actors. |

| Year | Full Citation (APA) | Author(s) | Theoretical Framing | Focus | Method | Country Focus | N (interviews) | Summaries Edited? | Annabelle – Degree of Reading | Annabelle Significance | ChatGPT Significance | ChatGPT Explanation | Significance Match? | Julian – Degree of Reading | Julian Significance | Claude Significance | Claude Explanation |
|---|---|---|---|---|---|---|---|---|---|---|---|---|---|---|---|---|---|
| | *nt and Planning C, 31*(5), 893–910. | | ralist-Elitist Debate | | | | | | | | | | | | | | |
| 2020 | Simón-Rojo, M., Couceiro, A., del Valle, J., & Fariña Tojo, J. (2020). Public Food Procurement as a Driving Force for Building Local and Agroecological Food Systems. *Land, 9*(9), 317. | Simon-Rojo et al. | UFS + Agroecology + Food Sovereignty | Public Sector Food Procurement | Case study analysis, land use data, stakeholder interviews and surveys; stakeholder perception mapping, scenario analysis | Spain (Madrid) | 13 | Summary and Methods edited | Skimmed | * | ** | Grounded in local farmer engagement and municipal procurement systems; practical and governance-relevant. | ChatGPT rated higher | Skimmed | ** | * | Analysis of procurement potential hindered by institutional barriers, limited depth on transformation mechanisms. |
| 2019 | Smit, W., Battersby, J., & Watson, V. (2019). Current Urban Food Governance and Planning in Africa. In *Urban Food Systems Governance and Poverty in African Cities* (pp. 94–103). Routledge. | Smit et al. | Governance Theory + Urban Planning/Food Systems Frameworks | Urban Agriculture Governance | Literature review, field research, comparative case studies, policy and document analysis | Kenya, Zambia, Zimbabwe | | Summary edited | Skimmed | * | ** | Explores institutional diversity and policy fragmentation; highly relevant for the African urban context in local governance. | ChatGPT rated higher | Skimmed | ** | ** | Overview of complex formal-informal governance interactions in African cities. Important for understanding how power operates through various channels. |
| 2018 | Smith, K., & Lawrence, G. (2018). From Disaster Management to Adaptive Governance? Governance Challenges to Achieving Resilient Food Systems in Australia. *Journal of Environmental Policy & Planning, 20*(3), 387–401. | Smith and Lawrence | Adaptive Governance + Polycentric Governance + Resilience Theory | Crises Response in Food System Governance | 14 semi-structured interviews, thematic analysis, document and policy analysis, case study comparison | Australia | 14 | Summary edited | Skimmed | * | ** | Evaluates governance responses in food system resilience; applicable for understanding shifts in institutional frameworks. | ChatGPT rated higher | Skimmed | *** | ** | Reveals gap between adaptive governance rhetoric and practice during crisis. Shows how existing power structures persist despite disruption. |
| 2014 | Sonnino, R., Lozano Torres, C., & Schneider, S. (2014). Reflexive Governance for Food Security: The Example of School Feeding in Brazil. *Journal of Rural Studies, 36*, 1–12. | Sonnino et al. | RGT + Transition Theory + MoRG | Reflexive Food System Governance | Policy and legislative document analysis, secondary data analysis, literature review, case study analysis | Brazil | | Summary and Methods edited | Not read | * | *** | Seminal case study of reflexive food governance with high policy relevance; robust theoretical contribution and applied insights. | ChatGPT rated higher | Skimmed | ** | *** | Analyzes Brazil's pioneering reflexive governance approach to food security. Important model for understanding adaptive, learning-based governance mechanisms. |

| Year | Full Citation (APA) | Author(s) | Theoretical Framing | Focus | Method | Country Focus | N (interviews) | Summaries Edited? | Annabelle – Degree of Reading | Annabelle Significance | ChatGPT Significance | ChatGPT Explanation | Significance Match? | Julian – Degree of Reading | Julian Significance | Claude Significance | Claude Explanation |
|---|---|---|---|---|---|---|---|---|---|---|---|---|---|---|---|---|---|
| 2024 | Vries, H. de, Donner, M., Fabiano, F., et al. (2024). Co-Creation in Partnerships Contributing to the Sustainability of Food Systems: Insights from 52 Case Studies in Europe. *Frontiers in Sustainable Food Systems, 8.* | Vries et al. | MLP + SNM + GVC | Public-Private Co-Creation and Agri-innovations | Reviewed 52 case studies of joint partnership sustainability projects | Europe (Multi) | | Not edited | Not read | * | *** | Empirical depth with 52 European case studies; demonstrates co-creation as a key mechanism in government-led food systems transformation, highly relevant and unique. | ChatGPT rated higher | Skimmed | ** | *** | Highly relevant with 52 European case studies examining co-creation mechanisms in multi-actor partnerships. Strong empirical base and methodology for understanding how change happens through collaborative governance. Directly applicable to Food Socioscope's focus on transition mechanisms. |
| 2022 | Winmore, K., Nicolau, M., & Nojiyeza, I. S. (2022). Implications of the Urban Water and Food Systems Governance Nexus for Household Food Security in the City of Gweru, Zimbabwe. *Urban Forum, 33*(3), 329–348. | Winmore et al. | WFNG + RBA + Political Ecology + Urban Poverty and Informality | Implications of the Urban Water and Food Systems Governance | Household surveys, stratified random, key informant interviews, content analysis of legal documents, comparative statistical analysis | Zimbabwe (Gweru) | Not mentioned | All sections edited | Skimmed | * | ** | Analyzes water and food systems integration under urban conditions; good municipal relevance. | ChatGPT rated higher | Read fully | ** | ** | Shows how water commodification undermines food security despite constitutional rights. Important for understanding contradictory governance. |
| 2017 | Yami, M., & Van Asten, P. (2017). Policy Support for Sustainable Crop Intensification in Eastern Africa. *Journal of Rural Studies, 55*, 216–226. | Yami and Van Asten | PAA + AIT | Sustainable Farming Policy Implementation (Developing Economies) | Qualitative analysis of 77 policy and legal documents; stakeholder workshops with 45 policy actors | Uganda, Kenya, Ethiopia | | Summary and Evaluation edited | Not read | * | * | Agricultural policy focused with less emphasis on local government or cross-sectoral food systems. | Match | Read fully | ** | * | Analysis of policy barriers to sustainable intensification, useful context but limited insight into transformation mechanisms. |
| 2011 | Yates, J. S., & Gutberlet, J. (2011). Reclaiming and Recirculating Urban Natures: Integrated Organic Waste Management in Diadema, Brazil. *Environment and Planning A, 43*(9), 2109–2124. | Yates and Gutberlet | SIaRT + Political Ecology + Participatory Governance | Waste Management Food System Political Ecology | Case studies: semi-structured interviews, observations, policy and document analysis | Brazil | 45 Households | Not edited | Skimmed | * | ** | Case study of urban waste governance linked to food systems; useful for its participatory and integrated systems perspective. | ChatGPT rated higher | Skimmed | ** | ** | Innovative case of municipal-led participatory waste management linking to food systems. Shows successful collaborative governance model with strong civil society integration. |

| Year | Full Citation (APA) | Author(s) | Theoretical Framing | Focus | Method | Country Focus | N (interviews) | Summaries Edited? | Annabelle – Degree of Reading | Annabelle Significance | ChatGPT Significance | ChatGPT Explanation | Significance Match? | Julian – Degree of Reading | Julian Significance | Claude Significance | Claude Explanation |
|---|---|---|---|---|---|---|---|---|---|---|---|---|---|---|---|---|---|
| 2024 | Zhu, Y., & Tsoulfas, G. T. (2024). Fostering Urban Short Food Supply Chains: Evidence from the Netherlands. *E3S Web of Conferences, 585*, 11003. | Zhu and Tsoulfas | ANT + Governance Theory + Transition Management | Public Policy Involvement in Developing Short-Food Supply Chains | Case study approach; qualitative interviews with policymakers, SFSC entrepreneurs, stakeholders; historical policy/document analysis | Netherlands (Breda) | Not mentioned | Methods edited | Not read | * | ** | Supports understanding of localized food systems; relevant to procurement and infrastructure but narrow in scope. | ChatGPT rated higher | Skimmed | ** | * | Limited depth on governance mechanisms, focuses more on general SFSC characteristics than transformation processes. |
| 2022 | Barbour, L., Lindberg, R., Woods, J., Charlton, K., & Brimblecombe, J. (2022). Local Urban Government Policies to Facilitate Healthy and Environmentally Sustainable Diet-Related Practices: A Scoping Review. *Public Health Nutrition, 25*(2), 471–487. | Barbour et al. | FSA + Urban Governance/Multilevel Policy Networks + Social Equity/Environmental Justice | Local Urban Government Policies to Facilitate Healthy and Environmentally Sustainable Diet-Related Practices | Scoping review, directed content analysis, grey literature review, qualitative synthesis, descriptive coding | Europe, North America | | Summary and Methods edited | Not read | ** | ** | Scoping review of policy levers for sustainable diets; strong foundation for understanding government toolkits. | Match | Read fully | * | ** | Systematic review of MUFPP cities revealing patterns in policy framing and implementation gaps. Useful for understanding common approaches and limitations. |
| 2024 | Béné, C., & Abdulai, A. R. (2024). Navigating the Politics and Processes of Food Systems Transformation: Guidance from a Holistic Framework. *Frontiers in Sustainable Food Systems, 8*. | Bene et al. | MLP + FEEU + PE and Discursive Power | Process and Framework of Food Systems Governance Decision Making | Thematic literature review, conceptual framework development, comparative and diagnostic analysis | Global | | Methods and Evaluation edited | Skimmed | ** | ** | Maps dominant food system narratives and links them to policy implications; solid methodological base with conceptual clarity. | Match | Read fully | *** | *** | Comprehensive framework synthesizing transition theory and political economy for food system transformation. Provides concrete mechanisms for government action and includes useful analytical tables. |
| 2016 | Biel, R. (2016). Political Dimensions – Agriculture and Class Struggle. University College London: 48–57. | Biel | Historical Materialism + FDT + Panarchy theory + Postcolonial & indigenous critique | Political Dimensions of Agriculture and Class Struggle | Critical historical analysis, comparative cultural studies, conceptual and dialectical | Multi | | Not edited | Not read | ** | ** | Strong theoretical contribution linking class and governance structures to food policy outcomes; abstract but useful for framing political economy discussions. | Match | Read fully | *** | ** | Offers deep theoretical perspective on how class structures and discursive traditions shape governmental action in food systems. Strong on understanding implicit rules through political |

| Year | Full Citation (APA) | Author(s) | Theoretical Framing | Focus | Method | Country Focus | N (interviews) | Summaries Edited? | Annabelle – Degree of Reading | Annabelle Significance | ChatGPT Significance | ChatGPT Explanation | Significance Match? | Julian – Degree of Reading | Julian Significance | Claude Significance | Claude Explanation |
|---|---|---|---|---|---|---|---|---|---|---|---|---|---|---|---|---|---|
| | | | | | analysis, speculative synthesis | | | | | | | | | | | | economy lens, though more theoretical than empirical. |
| 2016 | Biel, R. (2016). *Sustainable Food Systems: The Role of the City*. UCL Press. | Biel | Dialectical Materialism + Systems Theory + Commons-Based Economics + Political Ecology | The Role of the City in Food System Planning | Participant observation, experiential field research, comparative policy, network mapping | UK (London) | | Not edited | Not read | ** | *** | Highly original and theoretically rich examination of systemic power and the role of cities in food governance; integrates ecological and political frameworks with case relevance. | ChatGPT rated higher | Skimmed | ** | ** | Comprehensive analysis of food system actors and their mechanisms of support/opposition, with strong theoretical framework examining corporate agribusiness, smallholder farmers, and urban movements. Useful for understanding actor motivations and power dynamics. |
| 2018 | Borrelli, N. (2018). Connecting Food Systems and Urban Planning: The Experience of Portland, Oregon. In *Integrating Food into Urban Planning* (pp. 102–116). UCL Press. | Borrelli | U&RPT + Collaborative governance & regional integration + Policy integration & planning culture change | Food-system Urban Planning | Case study policy review, historical socio-political analysis; FGDs and interviews | USA (Portland) | | Summary edited | Skimmed | ** | ** | Detailed case study linking food policy and urban planning frameworks; shows how institutional design and spatial governance affect food systems. | Match | Read fully | * | ** | Detailed case study of Portland's integrated approach to food systems planning, showing how institutional norms and collaborative governance enable change. Good example of successful integration mechanisms. |
| 2019 | Bricas, N., Soulard, C. T., & Arnal, C. (2019). Reconciling Sustainability Issues and Urban Policy Levers. In *Designing Urban Food Policies* (pp. 107–122). Springer. | Bricas et al. | SPT + Governance/Policy Levers Framework | Reconciling Sustainability Issues and Urban Policy Levers | Case study comparative analysis, expert workshops and global conferences | Global | | Summary edited | Not read | ** | ** | Synthesizes tensions and leverages in urban food policies; conceptual but grounded in city governance practices. | Match | Skimmed | ** | ** | Theoretical contribution on how cities navigate complexity through adaptive learning and leveraging grassroots initiatives. |
| 2024 | Buzogany, S., Cocke, A., Freishtat, H., et al. (2024). *Planning for Equitable Urban Agriculture in the United States*. Springer Nature. | Buzogany et al. | Policy Mobility + Resilience Theory | Urban Agriculture Embedded in a Food Systems Approach | Case study analysis, participatory action research, document analysis, mapping | USA (Baltimore) | | All sections edited | Not read | ** | ** | Explores diverse stakeholder strategies in urban agriculture; insightful but partially overlapping with Raja et al. | Match | Skimmed | * | ** | Baltimore case showing successful institutionalization through cross-agency collaboration and resident engagement mechanisms. |

| Year | Full Citation (APA) | Author(s) | Theoretical Framing | Focus | Method | Country Focus | N (interviews) | Summaries Edited? | Annabelle – Degree of Reading | Annabelle Significance | ChatGPT Significance | ChatGPT Explanation | Significance Match? | Julian – Degree of Reading | Julian Significance | Claude Significance | Claude Explanation |
|---|---|---|---|---|---|---|---|---|---|---|---|---|---|---|---|---|---|
| | | | | | and spatial analysis | | | | | | | | | | | | |
| 2017 | Campbell, L. K. (2017). Constructing the 'Greener, Greater' City: Politics, Discourses, and Material Practices. In *City of Forests, City of Farms* (pp. 169–207). Cornell University Press. | Campbell | ANT + UPE | City Food System Development Governance Decision Making | Discourse analysis, case study methodology, policy and literature analysis | USA (New York City) | | Not edited | Skimmed | ** | ** | Empirical and discursive analysis of food and sustainability planning in NYC; strong city-scale governance case study. | Match | Read fully | ** | ** | Rich analysis of NYC's PlaNYC showing how sustainability plans evolve through political pressure and coalition building. Good example of incremental policy change through discursive work. |
| 2017 | Clapp, J., Newell, P., & Brent, Z. W. (2017). The Global Political Economy of Climate Change, Agriculture and Food Systems. *The Journal of Peasant Studies, 45*(1), 80–88. | Clapp et al. | GVC + IBL | 'Climate-smart' Agriculture' and Corporate Influence | Literature analysis of secondary sources, policy reviews, comparative government analysis | Global North Based Entities | | Not edited | Not read | ** | ** | Strong theoretical analysis linking global political economy to food systems; relevant for framing but lacks direct municipal casework. | Match | Read fully | ** | ** | Concise analysis of how transnational corporations shape global policy frameworks, particularly relevant for understanding CSR-driven sustainability measures and corporate capture mechanisms. |
| 2021 | Clark, J. K., Conley, B., & Raja, S. (2021). Essential, Fragile, and Invisible Community Food Infrastructure: The Role of Urban Governments in the United States. *Food Policy, 103*. | Clark et al. | CCT + Policy framing + Governance/public–private collaboration | Urban Governance Food Infrastructures | Literature review, national survey of urban planners, case-based illustrations | USA | | Evaluation edited | Skimmed | ** | *** | Empirical paper on fragility and governance of food infrastructure; methodologically strong and directly aligned with project goals. | ChatGPT rated higher | Skimmed | ** | ** | Conceptualizes food infrastructure visibility and fragility, showing how crisis reveals essential nature. Important framing for governance priorities. |
| 2021 | Clarke, B., Kwon, J., Swinburn, B., & Sacks, G. (2021). Understanding the Dynamics of Obesity Prevention Policy Decision-Making Using a Systems Perspective. *PloS One, 16*(1), e0245535. | Clarke et al. | PSToPC + CLDs + Organizational Realpolitik | Food Governance Decision Making Rationale | Semi-structured interviews, purposive sampling, document analysis, participant observation, thematic analysis | Australia | 57 | Summary edited | Skimmed | ** | ** | Applies a systems approach to local health and food policy; methodologically strong and illustrative of local government dynamics. | Match | Skimmed | ** | ** | Systems approach revealing how risk aversion and consensus-seeking limit policy ambition. Strong methodology for understanding feedback loops and political constraints. |
| 2022 | Craviotti, C. (2022). Debates on the Construction of an | Craviotti | PNT + ACF + ACI + CAtPA + | Internal Sustainability Policy | 6 in-depth interviews with | Argentina | 6 | Not edited | Skimmed | ** | * | Descriptive exploration of stakeholder dialogue | Match | Read fully | ** | ** | Shows how alternative discourse forums operate within and |

| Year | Full Citation (APA) | Author(s) | Theoretical Framing | Focus | Method | Country Focus | N (interviews) | Summaries Edited? | Annabelle – Degree of Reading | Annabelle Significance | ChatGPT Significance | ChatGPT Explanation | Significance Match? | Julian – Degree of Reading | Julian Significance | Claude Significance | Claude Explanation |
|---|---|---|---|---|---|---|---|---|---|---|---|---|---|---|---|---|---|
| | Alternative Food System in Periurban Spaces. *Journal of Rural Studies, 94*, 399–407. | | Territorial Governance | Legitimization | industry informants, qualitative research design, qualitative coding/thematic analysis, policy and document analysis | | | | | | | with limited methodological detail; some insights on participatory governance. | | | | | against dominant policy. Important for understanding contested governance spaces. |
| 2025 | Duvernoy, I., Jarrige, F., & Gonçalves, A. (2025). Integrated Food Policies in South-Western France: Insights from Local Policy-Making Outside Major Urban Areas. *Journal of Rural Studies, 114*. | Duvernoy et al. | IPT + Governance Theory + Processual Institutionalism | Integrated Food Policy in Local Governance | 52 semi-structured interviews across five French regions, document and policy analysis, qualitative coding | France (across five regions) | 52 | Summary and Methods edited | Skimmed | ** | *** | Rare rural-local empirical study of integrated food policies; unique contribution with direct application to multi-scalar governance. | ChatGPT rated higher | Skimmed | ** | ** | Focus on non-metropolitan areas reveals different governance dynamics. Shows importance of institutional layering and regional coordination mechanisms. |
| 2023 | Friedman, R. S., Mackenzie, E., Chan-Tung, A. L., et al. (2023). Using Social Network Analysis to Track the Evolution of Pacific Food System Research Collaborations over Time. *Regional Environmental Change, 23*(4), 153. | Friedman et al. | SNA + Political Ecology + Decolonial Theory + CBaBO | Network Mapping Implicit Rules in Food System | Systemic literature review, social network analysis, semi-structured FGDs, participant observation, document analysis, expert interviews | Pacific Islands | 12 | Summary and Methods edited | Skimmed | ** | | | ChatGPT ignored | Skimmed | ** | ** | Innovative use of network analysis to reveal power structures in knowledge production. Shows how external institutions dominate despite calls for decolonization. |
| 2014 | Gagliardi, D., Niglia, F., & Battistella, C. (2014). Evaluation and Design of Innovation Policies in the Agro-Food Sector: An Application of Multilevel Self-Regulating Agents. *Technological Forecasting & Social Change, 85*, 40–57. | Gagliardi et al. | ABM + BDIA + Evolutionary/Non-Equilibrium Economics | Multi-level Governance in Policy | Empirical calibration and validation, ex-ante policy evaluation via simulation, | Italy (Puglia) | | Methods edited | Skimmed | ** | * | Focuses on innovation design with minimal reference to local government decision-making in food systems. | ChatGPT rated lower | Skimmed | ** | * | Agent-based modelling approach to policy impacts, innovative method but limited insight into actual governance mechanisms. |

| Year | Full Citation (APA) | Author(s) | Theoretical Framing | Focus | Method | Country Focus | N (interviews) | Summaries Edited? | Annabelle – Degree of Reading | Annabelle Significance | ChatGPT Significance | ChatGPT Explanation | Significance Match? | Julian – Degree of Reading | Julian Significance | Claude Significance | Claude Explanation |
|---|---|---|---|---|---|---|---|---|---|---|---|---|---|---|---|---|---|
| | | | | | scenario analysis | | | | | | | | | | | | |
| 2020 | Gaitán-Cremaschi, D., Bizzotto Molina, P., de Winter, D., & Lammers, E. (2020). *The implications of implicit choices: how food systems sustainability is impacted by narratives, entry points and transition pathways.* NWO-WOTRO. | Gaitán-Cremaschi et al. | PEaPA + TPF + Narrative Analysis | Implicit Rules in Food System Governance | Case studies: surveys, interviews, focus groups, document and policy analysis | Latin America, Africa, Southeast Asia (multi) | Not mentioned | Summary edited | Skimmed | ** | ** | Strong theoretical contribution on narratives and transition pathways; relevant for understanding implicit rules and discursive framing. | Match | Read fully | *** | *** | Critical analysis of how implicit narratives shape food system transitions. Highly relevant for understanding how sustainability is framed and which stakeholders benefit. Includes useful analytical tables. |
| 2019 | Halliday, J. (2019). Cities' Strategies for Sustainable Food and the Levers They Mobilize. In *Designing Urban Food Policies* (pp. 31–50). Springer. | Halliday | Cross-domain/integrated policy analysis + Typology of governance models for urban food interventions | Strategies/ Mechanisms for Sustainable Food in Urban Planning | Literature review, comparative case study analysis, policy and document review, discourse analysis | Global | | Methods edited | Skimmed | ** | ** | Outlines practical policy instruments used by cities; relevant for understanding tools available to local governments. | Match | Read fully | ** | ** | Comprehensive overview of urban policy levers and how they're mobilized for food system change. Good framework for understanding municipal tools. |
| 2019 | Hayombe, P. O., Owino, F. O., Awuor, F. O., Battersby, J., & Watson, V. (2019). Planning and Governance of Food Systems in Kisumu City. In *Urban Food Systems Governance and Poverty in African Cities* (pp. 116–127). Routledge. | Hayombe et al. | FGT + FSF | Planning and Governance of Food Systems | In-depth interviews with key stakeholders, descriptive statistics from survey responses, content analysis of field observations and secondary documents | Kenya (Kisumu) | Not mentioned | Evaluation edited | Skimmed | ** | ** | Case-based study illustrating municipal struggles and innovations in African urban food systems; regionally focused and practically insightful. | Match | Skimmed | ** | * | City-specific case showing legacy effects of colonial planning, limited broader applicability. |
| 2024 | Hensengerth, O., Lam, T. H. O., Van Pham Dang Tri, Hutton, C., & Darby, S. (2024). How to | Hensengerth et al. | SSP + EST + HMLG + Sustainabi | Strategic Spatial Planning in Food | Policy and institutional analysis; empirical | Vietnam (Mekong Delta) | 13 | Summary and Evaluation edited | Skimmed | ** | * | Focused on spatial planning and environmental governance; indirectly | ChatGPT rated lower | Read fully | * | ** | Analysis of multi-level governance and spatial imaginaries in sustainability |

| Year | Full Citation (APA) | Author(s) | Theoretical Framing | Focus | Method | Country Focus | N (interviews) | Summaries Edited? | Annabelle – Degree of Reading | Annabelle Significance | ChatGPT Significance | ChatGPT Explanation | Significance Match? | Julian – Degree of Reading | Julian Significance | Claude Significance | Claude Explanation |
|---|---|---|---|---|---|---|---|---|---|---|---|---|---|---|---|---|---|
| | Promote Sustainability? The Challenge of Strategic Spatial Planning in the Vietnamese Mekong Delta. *Journal of Environmental Policy & Planning, 26*(1), 91–103. | | lity Transitions in Multi-level Governance Systems | System Sustainability | interviews with government leaders, literature review | | | | | | | relevant to food systems governance. | | | | | transitions. Shows importance of central-local dynamics. |
| 2022 | Hunt, S., Maher, J., Swapan, M. S. H., & Zaman, A. (2022). Street Verge in Transition: A Study of Community Drivers and Local Policy Setting for Urban Greening in Perth, Western Australia. *Urban Science, 6*(1). | Hunt et al. | ESF + Social contagion/ mimicry in urban sustainability adoption | Governance Community Oriented Drivers of Local Sustainability | Mixed-methods; site audits, spatial analysis using GIS, a questionnaire survey of residents; qualitative coding of interviews; analysis of policy documents/ actions | Australia (Perth) | 166 (questionnaire responses) | Summary edited | Skimmed | ** | * | Civic engagement and greening policy example; interesting but peripheral to food governance and sustainability planning. | ChatGPT rated lower | Skimmed | * | * | Tangentially related to food systems through urban greening; shows social contagion mechanisms but limited food system focus. |
| 2019 | Jacobi, J., Wambugu, G., Ngutu, M., et al. (2019). Mapping Food Systems: A Participatory Research Tool Tested in Kenya and Bolivia. *Mountain Research and Development, 39*(1), R1–R11. | Jacobi et al. | FST + P/IM + Transdisciplinary & participatory research | Food-system Actor Network Mapping | Field research, network mapping, actor power modelling; participant observation, interviews, FGDs | Bolivia and Kenya | | Not edited | Skimmed | ** | ** | Demonstrates participatory mapping methods as tools for exposing and shifting power dynamics in food systems; solid and relevant, though less generalizable. | Match | Read fully | ** | *** | Innovative participatory mapping methodology tested in two countries, directly relevant to Food Socioscope's approach. Shows how local actors articulate perspectives and challenge dominant narratives through participatory research. |
| 2013 | Mansfield, B., & Mendes, W. (2013). Municipal Food Strategies and Integrated Approaches to Urban Agriculture: Exploring Three Cases from the Global North. *International Planning Studies, 18*(1), 37–60. | Mansfield and Mendes | UGC + New localism + LPA | Municipal Food System Planning | Exploratory comparative case study approach; policy and document analysis, informal interviews with key informants | UK (London), Canada (Toronto), USA (San Francisco) | Not mentioned | Summary and Evaluation edited | Skimmed | ** | ** | Case-based insights into urban agriculture policies in Global North; strong practical examples for municipal planning. | Match | Read fully | ** | ** | Comparative analysis of comprehensive municipal food strategies, showing importance of coordination and institutionalization mechanisms. |

| Year | Full Citation (APA) | Author(s) | Theoretical Framing | Focus | Method | Country Focus | N (interviews) | Summaries Edited? | Annabelle – Degree of Reading | Annabelle Significance | ChatGPT Significance | ChatGPT Explanation | Significance Match? | Julian – Degree of Reading | Julian Significance | Claude Significance | Claude Explanation |
|------|---------------------|-----------|---------------------|-------|--------|---------------|----------------|-------------------|-------------------------------|------------------------|----------------------|---------------------|---------------------|----------------------------|---------------------|---------------------|---------------------|
| 2008 | Marsden, T., & Sonnino, R. (2008). Rural Development and the Regional State: Denying Multifunctional Agriculture in the UK. *Journal of Rural Studies, 24*(4), 422–431. | Marsden and Sonnino | Multifunctional Agriculture + NLG + Devolution | Agricultural Land Use Policies | Policy analysis, case studies, critical discourse analysis, semi-structured interviews | UK (England, Wales) | Not mentioned | Evaluation edited | Not read | ** | ** | Highlights structural policy constraints on food system innovation; useful for broader governance context though more rural-focused. | Match | Skimmed | ** | ** | Reveals how neoliberal logic and institutional fragmentation prevent integrated approaches. Important for understanding systemic barriers. |
| 2021 | McClintock, N., Miewald, C., & McCann, E. (2021). Governing Urban Agriculture: Formalization, Resistance and Re-visioning in Two 'Green' Cities. *International Journal of Urban and Regional Research, 45*(3), 498–518. | McClintock et al. | CUS + Everyday Resistance/Governance | Urban Planning Sustainable Food System Policies | Comparative case study, qualitative interviews, document and policy analysis, thematic analysis, internet survey (N=133), focus groups | USA (Portland), Canada (Vancouver) | 63 | Not edited | Skimmed | ** | ** | Explores tensions in formalization and community control over urban agriculture; insightful for municipal procedural analysis. | Match | Read fully | ** | ** | Comparative study showing how formalization processes interact with resistance and re-visioning. Strong on understanding everyday governance and negotiation. |
| 2019 | McNeill, D. (2019). Reflections on Ipes-food: Can Power Analysis Change the World? IDS. | McNeill | EPE + Constructivist + RABA | IPES Food Reports – Lessons Learned | Theoretical literature review, secondary data analysis, comparative case studies, comparative governance analysis | Multi | | Summary edited | Skimmed | ** | | | ChatGPT ignored | Read fully | * | * | Reflective piece on IPES-Food's approach to power analysis, useful for understanding discursive strategies but limited empirical content for ground-level insights. |
| 2019 | Michel, L., & Soulard, C. T. (2019). Putting Food on the Regional Policy Agenda in Montpellier, France. In *Designing Urban Food Policies* (pp. 123–138). Springer. | Michel and Soulard | PET + MLG + Institutional Change/Policy Framing | Putting Food on the Regional Policy Agenda | Case study analysis, document and policy analysis, stakeholder mapping | France (Montpellier) | | All sections edited | Skimmed | ** | ** | Empirical case study of how food is incorporated in regional planning; well-aligned with urban governance focus. | Match | Read fully | ** | ** | Detailed process tracing of how food becomes policy issue through entrepreneurship and institutional change. Strong on mechanism identification. |

| Year | Full Citation (APA) | Author(s) | Theoretical Framing | Focus | Method | Country Focus | N (interviews) | Summaries Edited? | Annabelle – Degree of Reading | Annabelle Significance | ChatGPT Significance | ChatGPT Explanation | Significance Match? | Julian – Degree of Reading | Julian Significance | Claude Significance | Claude Explanation |
|---|---|---|---|---|---|---|---|---|---|---|---|---|---|---|---|---|---|
| 2015 | Moragues-Faus, A., & Morgan, K. (2015). Reframing the Foodscape: The Emergent World of Urban Food Policy. *Environment and Planning A, 47*(7), 1558–1573. | Moragues-Faus et al. | UPE + MTP/TMT + PPT | Urban Planning Sustainable Food System Policies | Comparative case studies, participatory action research, survey of policymakers/stakeholders/experts and point score analysis | UK (Bristol), Sweden (Malmö) | | Not edited | Skimmed | ** | *** | Theoretical and empirical insights on emerging urban food governance; often cited and central to debates on local policy innovation. | ChatGPT rated higher | Skimmed | ** | ** | Conceptualizes cities as "key transition spaces" for food governance innovation. Strong theoretical contribution on urban food governance spaces and deliberation mechanisms. |
| 2020 | Murguia Gonzalez, A., Giraldo, O. F., Mier y Terán-Giménez Cacho, M., & Rodríguez Castillo, L. (2020). Policy Pitfalls and the Attempt to Institutionalize Agroecology in El Salvador 2008–2018. *Agroecology and Sustainable Food Systems, 44*(8), 1033–1051. | Gonzalez et al. | AoPP + FAoD + PEoAR | Food System Transition in Regime Shifts | Ethnographic field research, literature review, participant observation; comparative case study, policy reviews | El Salvador | 10 | Summary edited | Skimmed | ** | ** | Empirical policy analysis with relevance to institutionalizing sustainability; important for understanding implementation challenges. | Match | Read fully | ** | ** | Detailed case study of attempted agroecology institutionalization, revealing how social movements engage with state power and face co-optation. Strong on understanding policy implementation challenges. |
| 2018 | Nonini, D. M. (2013, republished 2018). The Local-Food Movement and the Anthropology of Global Systems. *American Ethnologist, 40*(2), 267–275. | Nonini | AoGS + NGT + CPoFS + Critical food studies and political ecology | Ideological Variety in Food System Sustainability | Ethnographic field research, literature review, participant observation; comparative case study, policy reviews | USA (North Carolina) | Not mentioned | Methods edited | Skimmed | ** | * | Broadly contextual paper situating local food movements in global systems; useful for understanding macro trends but not directly applicable to government agency. | ChatGPT rated lower | Read fully | * | ** | Ethnographic analysis revealing how localist movements navigate neoliberal governance through discursive displacement. Strong on understanding implicit rules shaping activist strategies and state disengagement. |
| 2018 | Nott, J. (2018). 'How Little Progress'? A Political Economy of Postcolonial Nutrition. *Population and Development Review, 44*(4), 771–791. | Nott | PEoN + Foucauldian Biopolitics + Postcolonial Theory | Global Developmental Food-system Management | Narrative analysis of historical agricultural transitions; institutional analysis; governance analysis; | Sub-Saharan Africa (multi) | | Not edited | Not read | ** | ** | Insightful historical-political analysis of postcolonial nutrition policies with implications for understanding discursive and institutional inertia in | Match | Skimmed | * | ** | Provides nuanced political economy analysis revealing how colonial legacies and discursive mechanisms shape governmental decision-making in food systems. Strong |

| Year | Full Citation (APA) | Author(s) | Theoretical Framing | Focus | Method | Country Focus | N (interviews) | Summaries Edited? | Annabelle – Degree of Reading | Annabelle Significance | ChatGPT Significance | ChatGPT Explanation | Significance Match? | Julian – Degree of Reading | Julian Significance | Claude Significance | Claude Explanation |
|---|---|---|---|---|---|---|---|---|---|---|---|---|---|---|---|---|---|
| | | | | | political economy analysis | | | | | | | government decision-making. | | | | | historical perspective on how implicit rules persist through institutional norms and international organizations. |
| 2022 | Pingali, P., & Abraham, M. (2022). Food Systems Transformation in Asia – A Brief Economic History. *Agricultural Economics, 53*(6), 895–910. | Pingali and Abraham | STT + Globalization Theory + EPE | Asian Food System Transformations | Historical and archival review, policy review, comparative governance, MLM case analysis | Asia (multi) | | Summary edited | Not read | ** | * | Broad economic history with limited application to agency or municipal governance mechanisms. | Match | Read fully | ** | ** | Traces Asian food system evolution through regime transitions, showing how states shift from development managers to market facilitators. Strong historical perspective on transformation mechanisms. |
| 2024 | Pothukuchi, K. (2024). Prospects for an Enduring Agriculture in the Rustbelt: A Tale of Two Cities. In *Planning for Equitable Urban Agriculture in the United States.* Springer. | Pothukuchi | PEoLU + Urban governance and community development + Urban planning and sustainability discourse | Prospects for an Enduring Agriculture | Semi-structured interviews, qualitative case comparison, content analysis of policy documents, windshield surveys; GIS and mapping for spatial analysis | USA (Detroit, Cleveland) | Not mentioned | Methods edited | Not read | ** | ** | City-focused comparative study providing insights into urban revitalization and food systems planning. | Match | Read fully | *** | ** | Comparative analysis revealing how implicit priorities and institutional arrangements shape urban agriculture outcomes differently. |
| 2015 | Pratt, J., & Luetchford, P. (2015). *Food for Change: The Politics and Values of Social Movements.* Pluto Press. | Pratt and Luetchford | APE + Moral Economy + Food Sovereignty | Grassroots Food-Sovereignty Movements | Ethnographic field research, literature review, participant observation; comparative case study, policy comparison, comparativ | Italy (Tuscany), UK (Sussex), Spain (Andalusia), France (The Tarn) | | Not edited | Not read | ** | ** | Explores external actors' influence on food systems; valuable for political economy perspective and grassroots tensions but less focused on government decision-making. | Match | Skimmed | ** | ** | Examines complex roles of external actors (state, corporate, NGOs) in shaping food system trajectories, with European case studies. Strong on understanding how interventions can co-opt or dilute grassroots goals. |

| Year | Full Citation (APA) | Author(s) | Theoretical Framing | Focus | Method | Country Focus | N (interviews) | Summaries Edited? | Annabelle – Degree of Reading | Annabelle Significance | ChatGPT Significance | ChatGPT Explanation | Significance Match? | Julian – Degree of Reading | Julian Significance | Claude Significance | Claude Explanation |
|---|---|---|---|---|---|---|---|---|---|---|---|---|---|---|---|---|---|
| | | | | e governance | | | | | | | | | | | | | |
| 2020 | Renwick, S. (2020). Ranking of Scenarios, Actors and Goals of Food Security: Motivation for Information Seeking by Food Security Decision Makers. *Environment Systems & Decisions, 40*(3), 444–462. | Renwick | AHP + WPF + Realpolitik /Political Economy Lens | Ranking Food Governance Decision Making | Comparative case studies, analytic hierarchy process, survey of policymakers/stakeholders/experts and point score analysis | Belize, Trinidad/Tobago, Barbados | 82 | Methods edited | Skimmed | ** | * | Useful for framing knowledge access and decision support, though less specific in institutional application or food systems governance. | ChatGPT rated lower | Read fully | ** | * | Survey-based study of decision-maker priorities, useful for understanding biases but limited insight into actual transformation mechanisms. |
| 2022 | Richardson, B. F. (2022). Finance, Food, and Future Urban Zones: The Failure of Flexible Development in Auckland, New Zealand. *Land Use Policy, 119.* | Richardson | PE + FoL + Fragmentation | Public Land Allocation Policy | Case study analysis, 20 stakeholder interviews; document analysis of planning policies, field work, policy analysis | New Zealand (Auckland) | 20 | Methods edited | Skimmed | ** | * | Covers land use and urban finance but only tangentially addresses food governance or decision processes. | ChatGPT rated lower | Read fully | * | * | Shows how financialization undermines sustainable land use, but focused more on development than food systems. |
| 2012 | Rodda, N., Hernandez, F. E., & Jones, M. A. (2012). Food and Water Security Initiatives in Poor Urban, Peri-Urban and Rural Settlements in eThekwini Municipality, South Africa. In *Food Security: Quality Management.* Nova Science Publishers. | Rodda et al. | ISD + CBPD + Systems Thinking | Food and Water Security Initiatives in Poor Urban, Peri-Urban and Rural Settlements | Case study analysis, qualitative interviews; pilot projects and scaling, monitoring and evaluation | South Africa (eThekwini Municipality) | Not mentioned | Not edited | Not read | ** | ** | Case study of urban governance integration; useful for intersectoral perspective and municipal operations. | Match | Skimmed | ** | * | Municipal case study showing integrated approach but limited analysis of governance mechanisms or broader applicability. |
| 2023 | Romagny, B., Aderghal, M., Auclair, L., Ilbert, H., & Lemeilleur, S. (2023). From Rural to Urban Areas: New Trends and Challenges for the Commons in Morocco. *The Journal of North African Studies, 28*(1), 57–74. | Romagny et al. | Commons Theory + Political Ecology | Land Distribution Policies for the Commons | Case study approach; historical-comparative analysis | Morocco | | Summary and Methods edited | Skimmed | ** | * | Interesting for common property theory but only indirectly related to local food governance processes. | ChatGPT rated lower | Skimmed | ** | ** | Shows how market logic and centralization undermine traditional commons governance. Important for understanding state-community tensions. |

| Year | Full Citation (APA) | Author(s) | Theoretical Framing | Focus | Method | Country Focus | N (interviews) | Summaries Edited? | Annabelle – Degree of Reading | Annabelle Significance | ChatGPT Significance | ChatGPT Explanation | Significance Match? | Julian – Degree of Reading | Julian Significance | Claude Significance | Claude Explanation |
|---|---|---|---|---|---|---|---|---|---|---|---|---|---|---|---|---|---|
| 2020 | Salajegheh, S., Jafari, H. R., & Pourebrahim, S. (2020). Modeling the Impact of Social Network Measures on Institutional Adaptive Capacity Needed for Sustainable Governance of Water Resources. *Natural Resource Modeling*, 33(4). | Salajegheh et al. | SEST + ACW + SNT + Polycentric Governance | Measures for Effective Governance in Sustainability | SNA, PLS-SEM, ACW assessment, survey design, case study approach | Iran (Kerman Province) | Not mentioned | All sections edited | Skimmed | ** | * | Focus on water systems; methodology may inform food governance but lacks direct applicability to urban food systems. | ChatGPT rated lower | Read fully | ** | * | Water resource focus with limited direct food system relevance, though network analysis approach has potential applicability. |
| 2022 | Stead, V. C., & Hinkson, M. (2022). *Beyond Global Food Supply Chains: Crisis, Disruption, Regeneration.* Springer Nature Singapore. | Stead et al. | MLP + ANT + GVC | Corporate Agri-business Influence on Global Governance | Semi-structured interviews, literature review and document analysis; historical contextualization, comparative governance analysis | Global North Based Entities | Not mentioned | Evaluation edited | Skimmed | ** | * | Serves as broader context on food crises and regeneration; provides framing material rather than detailed empirical insights. | ChatGPT rated lower | Skimmed | ** | * | Broad overview of global food system governance and corporate influence, useful for context but less focused on specific mechanisms or methodologies for capturing ground-level change. |
| 2024 | Thow, A. M., Neves, D., Aidoo, R., et al. (2024). Strengthening the Governance of Food Systems for Nutrition in Africa. *Public Health Nutrition*, 27(1), e243. | Thow et al. | VCA + PEA (NIA) + Policy Analysis | External/Institutional Factors in Public Implementation | Qualitative value chain analysis, integrated political economy synthesis, comparative case study analysis, policy stakeholder interviews | South Africa, Ghana | 76 | Not edited | Skimmed | ** | | | ChatGPT ignored | Skimmed | ** | ** | Analysis of cross-sectoral coordination mechanisms and policy learning. Relevant for understanding how to overcome sectoral silos. |
| 2022 | Tschunkert, K., & Delgado, C. (2022). Transforming Food Systems: Local Food Systems in Focus. In *Food Systems in Conflict and Peacebuilding Settings.* SIPRI. | Tschunkert and Delgado | CoCC + Systems Thinking + Localization & M-SG | Global Food System Institutionalism | Synthesis of multisectoral research, case study illustration, policy and discourse analysis | United Nations | | Not edited | Skimmed | ** | ** | Highlights exclusion mechanisms in global food governance and local resistance strategies; strong on discursive and structural critique, relevant for equity analysis. | Match | Skimmed | ** | ** | Reveals governance mechanisms that systematically sideline local actors in favor of multinational corporations. Important for understanding power asymmetries and resistance |

| Year | Full Citation (APA) | Author(s) | Theoretical Framing | Focus | Method | Country Focus | N (interviews) | Summaries Edited? | Annabelle – Degree of Reading | Annabelle Significance | ChatGPT Significance | ChatGPT Explanation | Significance Match? | Julian – Degree of Reading | Julian Significance | Claude Significance | Claude Explanation |
|---|---|---|---|---|---|---|---|---|---|---|---|---|---|---|---|---|---|
| | | | | | | | | | | | | | | | | | mechanisms in post-conflict settings. |
| 2015 | Wang, Y. (2015). Negotiating the Farmland Dilemmas: 'Barefoot Planners' in China's Urban Periphery. *Environment and Planning C, 33*(5), 1108–1124. | Wang | BPA + SCT + Social learning/transactional planning | Informal Negotiations in Governance Environmental Management | Qualitative interviews, comparative case study analysis, critical discourse analysis | China (Shanghai, Guizhou, Henan) | 60+ | All sections edited | Skimmed | ** | ** | Unique insight into informal planning and land governance; relevant for understanding grassroots influence in urban transitions. | Match | Skimmed | ** | ** | Introduces concept of "barefoot planners" mediating between state and local realities. Important for understanding informal governance mechanisms. |
| 2024 | Wilkinson, T. J., Nye, C., Lobley, M., et al. (2024). Transforming Public Food Procurement: Stakeholder Understandings of Barriers and Opportunities for More Localised Procurement. *Journal of Rural Studies, 108.* | Wilkinson et al. | SST + STT + FSTF + Narrative Analysis | Public Sector Food Procurement | 40 semi-structured interviews, applied policy contextualization, thematic analysis, document and records analysis, case study comparison, stakeholder engagement | UK (South-West England) | 40 | Summary edited | Skimmed | ** | ** | Detailed empirical work on procurement governance; relevant to systemic shifts and municipal stakeholder coordination. | Match | Read fully | *** | ** | Empirical study revealing "say-do" gaps in procurement and importance of relational infrastructure. Directly relevant to understanding implementation barriers. |
| 2023 | Williams, T. G., Bui, S., Conti, C., et al. (2023). Synthesising the Diversity of European Agri-Food Networks: A Meta-Study of Actors and Power-Laden Interactions. *Global Environmental Change, 83*, 102746. | Williams et al. | SNT + ASD + Theories of Power in Transitions | Agent Power Dynamics in Decision Making | Meta-study/meta-synthesis, content analysis, social network analysis, snowball sampling | Europe (Multi) | | Summary and Evaluation edited | Skimmed | ** | ** | Meta-study on actor networks and power dynamics in agri-food systems; valuable for mapping influence but not specific to municipal agency. | Match | Skimmed | ** | *** | Meta-study of 71 cases revealing network archetypes and power configurations. Excellent methodology for understanding diversity of governance arrangements and transformation pathways. |
| 2023 | Xue, H., Zhai, Y., Su, W. H., & He, Z. (2023). Governance and Actions for Resilient Urban Food Systems in the Era of COVID-19: Lessons and Challenges in | Xue et al. | AOA + Governance Theory + Discourse Analysis | Government Food System Planning (Covid) | Qualitative analysis, case study analysis, literature review | China | | Not edited | Skimmed | ** | ** | Empirical COVID-19 governance case studies with lessons for urban resilience and food systems; relevant for city-level planning. | Match | Skimmed | ** | ** | Examines rapid governance adaptation during crisis, showing importance of multi-level coordination and digital innovation. Relevant for |

| Year | Full Citation (APA) | Author(s) | Theoretical Framing | Focus | Method | Country Focus | N (interviews) | Summaries Edited? | Annabelle – Degree of Reading | Annabelle Significance | ChatGPT Significance | ChatGPT Explanation | Significance Match? | Julian – Degree of Reading | Julian Significance | Claude Significance | Claude Explanation |
|---|---|---|---|---|---|---|---|---|---|---|---|---|---|---|---|---|---|
| | China. *Agriculture (Basel), 13*(9), 1681. | | | | | | | | | | | | | | | | understanding adaptive governance mechanisms. |
| 2022 | Zhong, S., Hughes, A., Crang, M., Zeng, G., & Hocknell, S. (2022). Fragmentary Embeddedness: Challenges for Alternative Food Networks in Guangzhou, China. *Journal of Rural Studies, 95,* 382–390. | Zhong et al. | CPE + Embeddedness Theory + Governmentality | Policies Against Alternative Food Systems | Semi-structured interviews with 30 AFN-related institutions; qualitative case study, focus groups, participant observation | China (Guangzhou) | 30 | Evaluation edited | Skimmed | ** | * | Case study on embeddedness of food networks; less focus on formal governance structures or public decision-making. | ChatGPT rated | | | | |

## 8.3 Appendix 3: The Human-Selected Papers with Ratings Comparison

**Table A3.3: Human Consensus 3-Star Papers and AI Ratings**

| Paper | Human Rating | Claude Rating | ChatGPT Rating | Over-lap |
|---|---|---|---|---|
| Elzen et al. (2011) - Pig husbandry transitions | 3 | 2 | 2 | |
| Davidson et al. (2016) - Alternative beef Alberta | 3 | 2 | 2 | |
| Charatsari & Lioutas (2019) - Agronomist competencies | 3 | 1 | 1 | |
| Ryghaug, M., & Skjølsvold, T. M. (2023) Electrical transition | 3 | 2 | 2 | |
| Ingram (2018) - Knowledge system boundaries | 3 | 2 | 2 | |
| Mukhovi et al. (2020) - Learning and adaptation | 3 | 2 | 2 | |
| Moragues-Faus & Morgan (2015) - Bristol/Malmö | 3 | 2 | 3 | |
| Ingram et al. (2015) - LINSA networks | 3 | 3 | 3 | ✓ |
| Haylock & Connelly (2018) - NZ food networks | 3 | 2 | 2 | |
| Hansen (2022) - Media framing | 3 | 1 | 2 | |
| Dubbeling et al. (2016) - Private sector CRFS | 3 | 2 | 2 | |
| Papachristos & Adamides (2016) - Retroductive methodology | 3 | 1 | 1 | |
| Isgren & Ness (2017) - Uganda agroecology | 3 | 2 | 2 | |
| Vivero-Pol (2017) - Food valuations | 3 | 3 | 2 | ✓ |
| Luger et al. (2025) - Urban Living Labs | 3 | 3 | 2 | ✓ |
| Schiller et al. (2020) - Nicaragua agroecology | 3 | 3 | 3 | ✓ |
| Ferreira et al. (2024) - Brazilian urban agriculture | 3 | 3 | 2 | ✓ |
| Pothukuchi (2017) - Detroit urban agriculture | 3 | 3 | 2 | ✓ |
| Zwartkruis et al. (2020) - Agricultural nature conservation | 3 | 2 | 2 | |
| Williams et al. (2023) - Power-laden interactions | 3 | 2 | 3 | |
| To et al. (2018) - Mauritius bagasse | 3 | 2 | 2 | |
| Sarabia et al. (2021) - Valencia transitions | 3 | 2 | 3 | |
| Li et al. (2013) - IP regimes China | 3 | 2 | 1 | |
| Audet et al. (2017) - Montreal markets | 3 | 3 | 3 | |
| Bedore (2014) - Baltimore food policy | 3 | 2 | 2 | |
| Buchan et al. (2019) - Transformative incrementalism | 3 | 3 | 3 | |
| Raja et al. (2018) - US food planning | 3 | 2 | 3 | |
| Reynolds et al. (2019) - Food waste interventions | 3 | 3 | 3 | |
| Riley & Crush (2023) - Secondary cities Africa | 3 | 2 | 2 | |
| Arslan et al. (2022) - LMIC structural transformation | 3 | 1 | 2 | |
| Larbi et al. (2022) - Curitiba/Accra transitions | 3 | 1 | 2 | |

**Key Finding:** Only 6 papers (20%) were rated as highly significant (3 stars) by both human reviewers and Claude, highlighting fundamental differences in evaluation criteria between human experts and AI systems.

## 8.4 Appendix 4: The comparison of ratings by humans and LLM

**Table A3.1: Inter-rater Reliability Statistics**

| Rater Pair | Cohen's κ | 95% CI | N | Interpretation |
|---|---|---|---|---|
| **Annabelle-Julian** | **0.310** | [0.23, 0.39] | 267 | Fair agreement |
| ChatGPT-Claude | 0.222 | [0.14, 0.30] | 254 | Fair agreement |
| Annabelle-ChatGPT | -0.008 | [-0.09, 0.07] | 253 | No agreement |
| Annabelle-Claude | 0.129 | [0.05, 0.21] | 270 | Slight agreement |
| Julian-ChatGPT | 0.011 | [-0.07, 0.09] | 254 | Slight agreement |
| Julian-Claude | 0.172 | [0.09, 0.25] | 268 | Slight agreement |

**Table A3.2: Rating Variance Comparison**

| Rater | Mean | Standard Deviation | Range Utilized |
|---|---|---|---|
| **Annabelle** | 1.79 | 0.76 | 0-3 stars |
| **Julian** | 1.87 | 0.69 | 0-3 stars |
| ChatGPT | 1.90 | 0.55 | 1-3 stars |
| Claude | 1.80 | 0.63 | 1-3 stars |

## 8.5 Appendix 5: Lists of References used for review A and reviews (B, C, D, E and F)

### 8.5.1 List of references used for Review A:

(The papers also selected for review B are in **bold**).

**Audet, R., Lefèvre, S., Brisebois, É., & El-Jed, M. (2017). Structuring tensions and key relations of Montreal seasonal food markets in the sustainability transition of the agri-food sector. *Sustainability*, *9*(3), 320. https://doi.org/10.3390/su9030320**

Béné, C., & Abdulai, A.-R. (2024). Navigating the politics and processes of food systems transformation: Guidance from a holistic framework. *Frontiers in Sustainable Food Systems*, *8*. https://doi.org/10.3389/fsufs.2024.1151869

Béné, C., Oosterveer, P., Lamotte, L., Brouwer, I. D., de Haan, S., Prager, S. D., Talsma, E. F., & Khoury, C. K. (2019). When food systems meet sustainability: Current narratives and implications for actions. *World Development*, *113*, 116-130. https://doi.org/10.1016/j.worlddev.2018.08.011

Bernstein, H. (2016). Agrarian political economy and modern world capitalism: The contributions of food regime analysis. *The Journal of Peasant Studies*, *43*(3), 611-647. https://doi.org/10.1080/03066150.2015.1101456

Buchan, R., Cloutier, D. S., & Friedman, A. (2019). Transformative incrementalism: Planning for transformative change in local food systems. *Progress in Planning*, *134*, 100402. https://doi.org/10.1016/j.progress.2018.07.002

Bui, S., Cardona, A., Lamine, C., & Cerf, M. (2016). Sustainability transitions: Insights on processes of niche-regime interaction and regime reconfiguration in agri-food systems. *Journal of Rural Studies*, *48*, 92-103. https://doi.org/10.1016/j.jrurstud.2016.10.003

Dai, N., Zhong, T., & Scott, S. (2019). From overt opposition to covert cooperation: Governance of street food vending in Nanjing, China. *Urban Forum*, *30*(4), 499-518. https://doi.org/10.1007/s12132-019-09375-3

de Vries, H., Donner, M., Fabiano, F., Mamès, M., Lazaro-Mojica, J., Cotillas, E., Avila, C., et al. (2024). Co-creation in partnerships contributing to the sustainability of food systems: Insights from 52 case studies in Europe. *Frontiers in Sustainable Food Systems*, *8*. https://doi.org/10.3389/fsufs.2024.1372538

**Ferreira, M. A., Abreu, G. M., Fernandes, C. N., Leão, V., Ferreira, J., & Luiz, J. (2024). Cultivating urban agriculture policies: Local government entrepreneurs' strategies in three Brazilian cities. *Urban Planning*, *9*. https://doi.org/10.17645/up.8166**

Fiala, V., & Jacob, K. (2024). Combining the multi-level perspective framework with participatory scenario development to explore the many facets of food system transitions in Germany. *Sustainability Science*, *19*(4), 1259-1275. https://doi.org/10.1007/s11625-024-01511-0

Gaitán-Cremaschi, D., Bizzotto Molina, P., de Winter, D., & Lammers, E. (2020). *The implications of implicit choices: How food systems sustainability is impacted by narratives, entry points and transition pathways*. NWO-WOTRO Science for Global Development. https://www.wur.nl/en/show/the-implications-of-implicit-choices.htm

Gaitán-Cremaschi, D., Klerkx, L., Duncan, J., Trienekens, J. H., Huenchuleo, C., Dogliotti, S., Contesse, M. E., & Rossing, W. A. (2020). Sustainability transition pathways through ecological intensification: An assessment of vegetable food systems in Chile. *Journal of Cleaner Production*, *245*, 118939. https://doi.org/10.1016/j.jclepro.2019.118939

Gomez, J. G., & Darrot, C. (2023). The role of low-income consumers in food system transitions: Case studies of community supported agriculture and social groceries in France. *Journal of Rural Studies*, *99*, 25-35. https://doi.org/10.1016/j.jrurstud.2023.02.016

Hassink, J., Grin, J., & Hulsink, W. (2018). Enriching the multi-level perspective by better understanding agency and challenges associated with interactions across system boundaries: The case of care farming in the Netherlands. *Journal of Rural Studies*, *57*, 186-196. https://doi.org/10.1016/j.jrurstud.2017.11.005

Ilieva, R. T., & Hernandez, A. (2018). Scaling-up sustainable development initiatives: A comparative case study of agri-food system innovations in Brazil, New York, and Senegal. In *Towards Sustainable Global Food Systems* (pp. 273-286). Routledge. https://doi.org/10.4324/9781315154664-22

**Ingram, J. (2015). Framing niche-regime linkage as adaptation: An analysis of learning and innovation networks for sustainable agriculture across Europe. *Journal of Rural Studies*, *40*, 59-75. https://doi.org/10.1016/j.jrurstud.2015.06.003**

Jacobi, J., Wambugu, G., Ngutu, M., Augstburger, H., Mwangi, V., Zonta, A. L., Otieno, S., Kiteme, B. P., Burgoa, J. M. F. D., & Rist, S. (2019). Mapping food systems: A participatory research tool tested in Kenya and Bolivia. *Mountain Research and Development*, *39*(1), R1-11. https://doi.org/10.1659/MRD-JOURNAL-D-18-00024.1

Juri, S., Rodríguez Miranda, A., Pérez Bidegain, M., Gallo Ferreccio, J., Terra, J., & Gómez Perazzoli, A. (2022). Food systems transformations in South America: Insights from a transdisciplinary process rooted in Uruguay. *Frontiers in Sustainable Food Systems*, *6*, 887034. https://doi.org/10.3389/fsufs.2022.887034

Keefe, S., & Lee, J. (2025). Sustainable transitions in food systems: A case study of an urban agriculture farming training program in Washington, United States. *Frontiers in Sustainable Food Systems*, *9*. https://doi.org/10.3389/fsufs.2025.1477746

Lien, H. H., de Mey, Y., Meuwissen, M. P., & Bush, S. R. (2023). Information practices for improved sustainability assurance in Vietnamese shrimp aquaculture. *Journal of Rural Studies*, *100*, 103015. https://doi.org/10.1016/j.jrurstud.2023.103015

**Luger, J., van der Meij, M., Regeer, B., & Loeber, A. (2025). Transformative urban food governance: How municipal staff coordinating urban living labs navigate politics, administration and participation. *Local Environment*, *30*(1), 1-19. https://doi.org/10.1080/13549839.2025.2486297**

Mylan, J., Andrews, J., & Maye, D. (2023). The big business of sustainable food production and consumption: Exploring the transition to alternative proteins. *Proceedings of the National Academy of Sciences*, *120*(47), e2207782120. https://doi.org/10.1073/pnas.2207782120

Otto, I. M., Donges, J. F., Cremades, R., Bhowmik, A., Hewitt, R. J., Lucht, W., Rockström, J., Allerberger, F., McCaffrey, M., Doe, S. S. P., Lenferna, A., Morán, N., van Vuuren, D. P., & Schellnhuber, H. J. (2020). Social tipping dynamics for stabilizing Earth's climate by 2050. *Proceedings of the National Academy of Sciences*, *117*(5), 2354-2365. https://doi.org/10.1073/pnas.1900577117

**Pothukuchi, K. (2017). "To allow farming is to give up on the city": Political anxieties related to the disposition of vacant land for urban agriculture in Detroit. *Journal of Urban Affairs*, *39*(8), 1169-1189. https://doi.org/10.1080/07352166.2017.1319239**

Reynolds, C., Goucher, L., Quested, T., Bromley, S., Gillick, S., Wells, V. K., Evans, D., Koh, L., Carlsson Kanyama, A., Katzeff, C., Svenfelt, Å., & Jackson, P. (2019). Consumption-stage food waste reduction interventions: What works and how to design better interventions. *Food Policy*, *83*, 7-27. https://doi.org/10.1016/j.foodpol.2019.01.009

Roberts, C., & Geels, F. W. (2019). Conditions for politically accelerated transitions: Historical institutionalism, the multi-level perspective, and two historical case studies in transport and agriculture. *Technological Forecasting and Social Change*, *140*, 221-240. https://doi.org/10.1016/j.techfore.2018.11.019

Schiller, K. J., Klerkx, L., Salazar Centeno, D. J., & Poortvliet, P. M. (2023). Developing the agroecological niche in Nicaragua: The roles of knowledge flows and intermediaries. *Proceedings of the National Academy of Sciences*, *120*(47), e2206195120. https://doi.org/10.1073/pnas.2206195120

Sonnino, R., Torres, C. L., & Schneider, S. (2014). Reflexive governance for food security: The example of school feeding in Brazil. *Journal of Rural Studies*, *36*, 1-12. https://doi.org/10.1016/j.jrurstud.2014.06.003

Takagi, S., Numazawa, Y., Katsube, K., Goto, A., Taguchi, M., & Watanabe, K. (2023). Theorizing socio-cultural dynamics of consumer decision-making for participation in CSA. *arXiv preprint*. https://arxiv.org/abs/2310.11537

Vermunt, D. A., Negro, S. O., Van Laerhoven, F. S. J., Verweij, P. A., & Hekkert, M. P. (2020). Sustainability transitions in the agri-food sector: How ecology affects transition dynamics. *Environmental Innovation and Societal Transitions*, *36*, 236-249. https://doi.org/10.1016/j.eist.2020.08.003

**Williams, T. G., Bui, S., Conti, C., Debonne, N., Levers, C., Swart, R., & Verburg, P. H. (2023). Synthesising the diversity of European agri-food networks: A meta-study of actors and power-laden interactions. *Global Environmental Change*, *83*, 102746. https://doi.org/10.1016/j.gloenvcha.2023.102746**

### 8.5.2  List of references used for Review B, C, D E and F:

(The papers also selected for review A are in bold).

Arslan, A., Cavatassi, R., & Hossain, M. (2022). Food systems and structural and rural transformation: a quantitative synthesis for low and middle-income countries. *Food Security*, *14*(1), 293–320. https://doi.org/10.1007/s12571-021-01223-2

**Audet, R., Lefèvre, S., Brisebois, É., & El-Jed, M. (2017). Structuring tensions and key relations of Montreal seasonal food markets in the sustainability transition of the agri-food sector. *Sustainability (Switzerland)*, 9(3). https://doi.org/10.3390/su9030320**

Bedore, M. (2014). The convening power of food as growth machine politics: A study of food policymaking and partnership formation in Baltimore. *Urban Studies*, *51*(14), 2979–2995. https://doi.org/10.2307/26145922

Buchan, R., Cloutier, D. S., & Friedman, A. (2019). Transformative incrementalism: Planning for transformative change in local food systems. *Progress in Planning*, *134*. https://doi.org/10.1016/j.progress.2018.07.002

Charatsari, C., & Lioutas, E. D. (2019). Is current agronomy ready to promote sustainable agriculture? Identifying key skills and competencies needed. *International Journal of Sustainable Development and World Ecology*, *26*(3), 232–241. https://doi.org/10.1080/13504509.2018.1536683

Davidson, D. J., Jones, K. E., & Parkins, J. R. (2016). Food safety risks, disruptive events and alternative beef production: a case study of agricultural transition in Alberta. *Agriculture and Human Values*, *33*(2), 359–371. https://doi.org/10.1007/s10460-015-9609-8

Dubbeling, M., Carey, J., Hochberg, K., et al. (2016). The role of private sector in city region food systems Analysis report. RUAF Foundation-Global Partnership on Sustainable Urban Agriculture and Food Systems. www.ruaf.org

Elzen, B., Geels, F. W., Leeuwis, C., & Van Mierlo, B. (2011). Normative contestation in transitions "in the making": Animal welfare concerns and system innovation in pig husbandry. *Research Policy*, *40*(2), 263–275. https://doi.org/10.1016/j.respol.2010.09.018

**Ferreira, M. A., Abreu, G. M., Fernandes, C. N., Leão, V., Ferreira, J., & Luiz, J. (2024). Cultivating Urban Agriculture Policies: Local Government Entrepreneurs' Strategies in Three Brazilian Cities. *Urban Planning*, 9(Transformative Local Governments). https://doi.org/10.17645/up.8166**

Hansen, T. (2022). Consumer food sustainability before and during the Covid-19 Crisis: A quantitative content analysis and food policy implications. *Food Policy*, *107*. https://doi.org/10.1016/j.foodpol.2021.102207

Haylock, K., & Connelly, S. (2018). Examining the Insider/Outsider Dimensions of Local Food System Planning: Cases from Dunedin and Christchurch New Zealand. *Planning Practice and Research*, *33*(5), 540–557. https://doi.org/10.1080/02697459.2018.1546470

**Ingram, J. (2015). Framing niche-regime linkage as adaptation: An analysis of learning and innovation networks for sustainable agriculture across Europe. *Journal of Rural Studies*, 40, 59–75. https://doi.org/10.1016/j.jrurstud.2015.06.003**

Ingram, J. (2018). Agricultural transition: Niche and regime knowledge systems' boundary dynamics. *Environmental Innovation and Societal Transitions*, *26*, 117–135. https://doi.org/10.1016/j.eist.2017.05.001

Isgren, E., & Ness, B. (2017). Agroecology to promote just sustainability transitions: Analysis of a civil society network in the Rwenzori region, western Uganda. *Sustainability (Switzerland)*, *9*(8). https://doi.org/10.3390/su9081357

Larbi, M., Kellett, J., & Palazzo, E. (2022). Urban Sustainability Transitions in the Global South: A Case Study of Curitiba and Accra. *Urban Forum*, *33*(2), 223–244. https://doi.org/10.1007/s12132-021-09438-4

Li, J., Jiggins, J., Van Bueren, E. T. L., & Leeuwis, C. (2013). Towards a regime change in the organization of the seed supply system in China. *Experimental Agriculture*, *49*(1), 114–133. https://doi.org/10.1017/S001447971200097X

**Luger, J., van der Meij, M., Regeer, B., & Loeber, A. (2025). Transformative urban food governance: how municipal staff coordinating urban living labs navigate politics, administration and participation. *Local Environment*. https://doi.org/10.1080/13549839.2025.2486297**

Moragues-Faus, A., & Morgan, K. (2015). Reframing the foodscape: the emergent world of urban food policy. *Environment and Planning A*, *47*(7), 1558–1573. https://doi.org/10.1177/0308518X15595754

Mukhovi, S., Jacobi, J., Speranza, C. I., Rist, S., & Kiteme, B. (2020). Learning and Adaptation in Food Systems: Insights from Four Case Studies in the Global South. *International Journal on Food System Dynamics*, *11*(4), 313–328. https://doi.org/10.18461/ijfsd.v11i4.57

Papachristos, G., & Adamides, E. (2016). A retroductive systems-based methodology for socio-technical transitions research. *Technological Forecasting and Social Change*, *108*, 1–14. https://doi.org/10.1016/j.techfore.2016.04.007

**Pothukuchi, K. (2017). "To allow farming is to give up on the city": Political anxieties related to the disposition of vacant land for urban agriculture in Detroit. *Journal of Urban Affairs*, *39*(8), 1169–1189. https://doi.org/10.1080/07352166.2017.1319239**

Raja, S., Whittaker, J., Hall, E., Hodgson, K., & Leccese, J. (2018). Growing food connections through planning: Lessons from the United States. In Y. Cabannes & C. Marocchino (Eds.), *Integrating Food into Urban Planning* (pp. 134–153). UCL Press.

Reynolds, C., Goucher, L., Quested, T., et al. (2019). Review: Consumption-stage food waste reduction interventions – What works and how to design better interventions. *Food Policy*, *83*, 7–27. https://doi.org/10.1016/j.foodpol.2019.01.009

Ryghaug, M., & Skjølsvold, T. M. (2023). Nurturing a regime change? Exploring the mobilization of intermediaries in Norway's electric vehicle transition. *Energy Research & Social Science*, *97*, 102956. https://doi.org/10.1016/j.erss.2023.102956

Riley, L., & Crush, J. (2023). Urban food system planning in secondary cities in Africa. *Food Security*, *15*(1), 1–16. https://doi.org/10.1007/s12571-022-01314-4

Sarabia, N., Peris, J., & Segura, S. (2021). Transition to agri-food sustainability, assessing accelerators and triggers for transformation: Case study in Valencia, Spain. *Journal of Cleaner Production*, *325*. https://doi.org/10.1016/j.jclepro.2021.129228

Schiller, K. J. F., Godek, W., Klerkx, L., & Poortvliet, P. M. (2020). Nicaragua's agroecological transition: Transformation or reconfiguration of the agri-food regime? *Agroecology and Sustainable Food Systems*, *44*(5), 611–628. https://doi.org/10.1080/21683565.2019.1667939

To, L. S., Seebaluck, V., & Leach, M. (2018). Future energy transitions for bagasse cogeneration: Lessons from multi-level and policy innovations in Mauritius. *Energy Research and Social Science*, *35*, 68–77. https://doi.org/10.1016/j.erss.2017.10.051

Vivero-Pol, J. L. (2017). Food as commons or commodity? Exploring the links between normative valuations and agency in food transition. *Sustainability (Switzerland)*, *9*(3). https://doi.org/10.3390/su9030442

**Williams, T. G., Bui, S., Conti, C., Debonne, N., Levers, C., Swart, R., & Verburg, P. H. (2023). Synthesising the diversity of European agri-food networks: A meta-study of actors and power-laden interactions.** ***Global Environmental Change***, ***83***. **https://doi.org/10.1016/j.gloenvcha.2023.102746**

Zwartkruis, J. V., Berg, H., Hof, A. F., & Kok, M. T. J. (2020). Agricultural nature conservation in the Netherlands: Three lenses on transition pathways. *Technological Forecasting and Social Change*, *151*. https://doi.org/10.1016/j.techfore.2018.03.006

Note that only 20% of the references are common.

## 8.6   Appendix 6: Review A, Review B, Review C.

We present here successively Review A, written by the LLM Claude Sonnet Version 4.5 (September 2025) on the basis of 31 papers it selected as the most relevant in a consideration set of 280, out of 350 found in the literature search.

Then Review B, written by same LLM on the same day, on the basis of 31 papers considered most relevant, from the same set, by researchers.

Finally Review C, which is review B edited by the four authors of this paper.

Prompt for Claude (Sonnet Version 4.5) :

Here are 31 highly-rated papers on food system transitions. Create a 1,500-word literature review synthesizing their key contributions, theoretical frameworks, and implications for sustainability transitions.

Papers to synthesize: [here, list of references selected by Claude, or by the researchers]

**Instructions:** Please synthesize these papers into a coherent 1,500-word literature review that:

Identifies the main theoretical frameworks used (particularly the Multi-Level Perspective)

Discusses key mechanisms of food system transitions

Analyzes different geographical contexts and their implications

Explores the role of various actors (government, civil society, consumers, businesses)

Synthesizes insights on barriers and enablers of transitions

Concludes with implications for sustainability transitions and future research directions

Structure the review with clear thematic sections rather than a paper-by-paper summary. Integrate citations smoothly throughout the text.

### 8.6.1 Review A: The literature review made by the LLM on the basis of the best 31 papers according to the LLM.

Food System Transitions: A Synthesis of Contemporary Research on Sustainability Transformations

Introduction: Theoretical Foundations of Food System Transitions

The transformation of food systems toward sustainability represents one of the most complex challenges of the 21st century, requiring fundamental changes across production, distribution, and consumption patterns. The Multi-Level Perspective (MLP) has emerged as the dominant theoretical framework for understanding these transitions, conceptualizing change as occurring through interactions between protected niches, established regimes, and broader landscape pressures (Bui et al., 2016; Fiala & Jacob, 2024). This framework reveals how innovations develop in protected spaces before potentially disrupting incumbent systems through various transition pathways.

Recent scholarship has enriched the MLP by addressing its limitations, particularly regarding agency and power dynamics. Hassink et al. (2018) demonstrate how understanding interactions across system boundaries enhances the framework's explanatory power, while Roberts and Geels (2019) integrate historical institutionalism to explain conditions for politically accelerated transitions. The framework's evolution reflects growing recognition that food system transitions involve not merely technological change but fundamental reconfigurations of social, political, and economic relationships (Bernstein, 2016; Williams et al., 2023).

Complementing the MLP, scholars have introduced concepts of social tipping dynamics (Otto et al., 2020), reflexive governance (Sonnino et al., 2014), and transformative incrementalism (Buchan et al., 2019), providing nuanced understandings of how systemic change unfolds. These theoretical advances highlight that transitions occur through multiple mechanisms operating simultaneously across different scales and contexts.

Mechanisms of Transformation: From Niches to Regime Change

Food system transitions operate through diverse mechanisms that connect grassroots innovations with systemic transformation. Niche development represents a critical first stage, as demonstrated by Schiller et al.'s (2023) analysis of agroecological development in Nicaragua, where knowledge flows and intermediaries facilitate the scaling of alternative practices. These protected spaces allow experimentation with new configurations of technology, social organization, and governance arrangements.

The scaling process involves complex dynamics of niche-regime interaction. Ingram (2015) identifies how learning and innovation networks across Europe frame adaptation strategies that enable sustainable agriculture practices to gain traction within existing systems. Similarly, Vermunt et al. (2020) reveal how ecological factors fundamentally shape transition dynamics, determining which innovations can successfully challenge incumbent regimes. The process is neither linear nor predictable, with Ilieva and Hernandez (2018) documenting how initiatives in Brazil, New York, and Senegal follow distinct scaling trajectories shaped by local contexts.

Market mechanisms play increasingly important roles in transitions, particularly through alternative food networks. Audet et al. (2017) analyze how Montreal's seasonal food markets create structural tensions that simultaneously challenge and reinforce existing systems. The emergence of Community Supported Agriculture (CSA) represents another market-based mechanism, with Takagi et al. (2023) theorizing how socio-cultural dynamics shape consumer participation decisions. These market innovations create spaces for new producer-consumer relationships while generating economic incentives for sustainable practices.

Information systems and sustainability assurance mechanisms constitute another crucial pathway, as Lien et al. (2023) demonstrate in Vietnamese shrimp aquaculture. These systems enable traceability, accountability, and learning that support incremental improvements while building foundations for more fundamental transformations.

Geographic Variations: Context-Specific Transition Pathways

Food system transitions manifest differently across geographic contexts, reflecting diverse socio-economic conditions, cultural values, and institutional arrangements. European cases dominate the literature, with de Vries et al. (2024) analyzing 52 case studies that reveal how co-creation in partnerships drives sustainability across the continent. The European context benefits from strong institutional support and established alternative food networks, yet faces challenges of coordinating across multiple governance levels.

Latin American experiences offer contrasting insights. Gaitán-Cremaschi et al. (2020) examine Chilean vegetable systems transitioning through ecological intensification, while Juri et al. (2022) document transdisciplinary transformation processes in Uruguay. These cases highlight how Global South contexts navigate transitions with different resource constraints and development priorities. Brazil's reflexive governance approach to school feeding programs demonstrates innovative institutional arrangements that link food security with sustainable agriculture (Sonnino et al., 2014).

Urban contexts present unique transition dynamics. Ferreira et al. (2024) analyze how local government entrepreneurs in three Brazilian cities cultivate urban agriculture policies, while Pothukuchi (2017) reveals political anxieties surrounding urban farming in Detroit. Asian cities face particular challenges, as Dai et al. (2019) document in Nanjing's street food governance, where informal sectors transition from overt opposition to covert cooperation with authorities.

North American cases, including Keefe and Lee's (2025) analysis of urban agriculture training in Washington and Mylan et al.'s (2023) study of alternative protein transitions, demonstrate how technological innovation intersects with social change. These geographic variations underscore that no single transition pathway exists; rather, transformations must align with local conditions while addressing global sustainability challenges.

## Actor Dynamics: Navigating Power and Participation

Food system transitions involve complex negotiations among diverse actors with varying power, resources, and interests. Government actors play pivotal roles, particularly at municipal levels where Luger et al. (2025) show how staff coordinating urban living labs navigate tensions between politics, administration, and participation. Local governments emerge as critical entrepreneurs, creating spaces for experimentation while managing conflicting pressures from various stakeholders.

Civil society organizations function as crucial intermediaries, bridging grassroots initiatives with institutional support. The role of these organizations extends beyond advocacy to include knowledge brokering, network building, and legitimation of alternative practices. Consumer movements, particularly among low-income populations, challenge assumptions about who participates in transitions. Gomez and Darrot (2023) demonstrate how social groceries in France enable broader participation in sustainable food systems, countering narratives that frame transitions as elite phenomena.

Business actors increasingly engage with sustainability transitions, though their roles remain contested. Mylan et al. (2023) examine how big business shapes alternative protein transitions, revealing tensions between corporate sustainability commitments and systemic transformation needs. The meta-study by Williams et al. (2023) maps power-laden interactions across European agri-food networks, exposing how incumbent actors can both enable and constrain transitions depending on their strategic interests.

These actor dynamics highlight that transitions involve not merely adding new participants but reconfiguring power relationships and governance arrangements. Béné and Abdulai (2024) provide frameworks for navigating these political dimensions, emphasizing that transformation requires addressing underlying power asymmetries.

## Barriers and Enablers: Critical Factors Shaping Transitions

Multiple barriers constrain food system transitions, operating across technical, institutional, and cultural dimensions. Institutional lock-ins represent fundamental obstacles, with established regulations, subsidies, and infrastructure favoring incumbent systems. Gaitán-Cremaschi et al. (2020) identify how implicit choices about narratives, entry points, and transition pathways shape sustainability outcomes, often reinforcing existing trajectories.

Cultural and social barriers prove equally significant. Consumer behavior, shaped by convenience, price, and cultural preferences, often resists change despite growing sustainability awareness. Reynolds et al. (2019) analyze consumption-stage interventions, revealing the complexity of designing effective behavior change strategies. Knowledge gaps and information asymmetries further constrain transitions, particularly in Global South contexts where technical support systems remain underdeveloped.

Conversely, several factors enable transitions. Crisis moments create windows of opportunity for rapid change, as Otto et al. (2020) demonstrate through social tipping dynamics that could stabilize Earth's climate by 2050. Participatory approaches enhance transition effectiveness, with Jacobi et al. (2019) developing mapping tools that enable communities in Kenya and Bolivia to visualize and transform their food systems. Policy innovation, particularly through reflexive governance approaches that enable learning and adaptation, creates institutional space for experimentation.

Financial mechanisms increasingly support transitions, though access remains uneven. The integration of sustainability considerations into investment decisions, supply chain management, and consumer markets creates economic incentives for change, though these often benefit already-privileged actors unless explicitly designed for inclusion.

Implications and Future Directions

This synthesis reveals food system transitions as complex, multi-scalar processes requiring coordinated action across diverse domains. The dominance of the Multi-Level Perspective provides valuable insights while potentially constraining alternative conceptualizations. Future research should explore theoretical frameworks that better capture power dynamics, cultural dimensions, and Global South experiences.

Methodologically, the field would benefit from more longitudinal studies tracking transitions over time, comparative analyses across diverse contexts, and participatory approaches that include marginalized voices. The current geographic bias toward European and North American cases limits understanding of transitions in contexts where most future food demand growth will occur.

Practically, evidence suggests successful transitions require: (1) supportive policy environments that enable experimentation while managing risks, (2) inclusive governance arrangements that address power imbalances, (3) economic mechanisms that make sustainable options accessible to all populations, and (4) knowledge systems that facilitate learning and adaptation. The integration of technological innovation with social innovation emerges as particularly crucial, as neither alone suffices for systemic transformation.

The urgency of climate change and biodiversity loss demands accelerated food system transitions. Yet this synthesis reveals that sustainable transformations cannot be imposed through top-down interventions or market mechanisms alone. Instead, transitions require careful orchestration of multiple change processes, attention to context-specific dynamics, and commitment to addressing underlying inequalities that shape current systems. As Béné et al. (2019) emphasize, when food systems meet sustainability, the implications extend beyond technical adjustments to encompass fundamental questions about democracy, justice, and human flourishing in the Anthropocene.

### 8.6.2 Review B: The literature review made by the LLM on the basis of the best 31 papers according to the humans (1500 words)

Food System Transitions: A Human-Curated Synthesis of Critical Scholarship

Introduction: Beyond Technological Solutions

The transformation of food systems toward sustainability represents a complex socio-technical challenge that transcends mere technological innovation. This review synthesizes 31 carefully selected studies that collectively reveal how food system transitions require fundamental reconfigurations of governance structures, knowledge systems, and power relations. Unlike dominant narratives focusing on efficiency and productivity, these papers illuminate the critical importance of agency, context-specific pathways, and the often-overlooked dimensions of justice and equity in achieving sustainable food futures.

## Theoretical Foundations: Expanding the Multi-Level Perspective

The Multi-Level Perspective (MLP) remains central to understanding food system transitions, yet the reviewed literature reveals its limitations and necessary expansions. Papachristos and Adamides (2016) address fundamental critiques by proposing a retroductive systems-based methodology that captures causal mechanisms often obscured in traditional MLP applications. Their integration of system dynamics with qualitative analysis demonstrates how transitions emerge from multi-system interactions rather than isolated sectoral changes.

Critically, several studies challenge the MLP's insufficient attention to agency and power. Vivero-Pol (2017) introduces a crucial dimension by examining how food valuations—as commodity versus commons—shape political attitudes toward system transformation. This work reveals that counter-hegemonic transformers, who view food multi-dimensionally, are essential for genuine system transformation rather than mere reform. Similarly, Luger et al. (2025) conceptualize the "purple zone" where municipal staff navigate between formal administrative duties and informal political action, highlighting how individual agency operates within structural constraints.

The integration of Social Movement Theory with MLP, as demonstrated by Elzen et al. (2011), provides essential insights into how normative contestation drives transitions. Their analysis of Dutch pig husbandry reveals that successful transitions require alignment across multiple streams—technological, regulatory, market, and social—rather than innovation alone. This multi-stream approach fundamentally reconceptualizes how transitions unfold, moving beyond linear progression narratives.

## Geographic Diversity and Contextual Pathways

A distinguishing feature of this curated collection is its geographic breadth and attention to diverse transition contexts. While European cases provide insights into institutionalized transition management, Global South examples offer critical perspectives on alternative pathways and grassroots innovation.

Isgren and Ness (2017) examine agroecology in Uganda's Rwenzori region, demonstrating how civil society networks fill gaps left by state withdrawal while navigating deteriorating political climates. Their analysis reveals agroecology emerging as an "unintended consequence of neoliberal development policies," highlighting how transitions can emerge from systemic contradictions rather than deliberate planning. This contrasts sharply with European experiences where formal institutions drive transition processes.

Latin American cases provide particularly valuable insights into alternative development pathways. The Nicaraguan agroecological transition analyzed by Schiller et al. (2020) demonstrates how knowledge flows through intermediaries enable niche development despite limited state support. Meanwhile, Brazilian urban agriculture policies examined by Ferreira et al. (2024) reveal how local government entrepreneurs navigate between formal planning and informal food systems, creating hybrid governance arrangements that challenge conventional planning paradigms.

The comparative analysis by Moragues-Faus and Morgan (2015) of Bristol and Malmö illuminates how different institutional contexts shape urban food governance. Bristol's civil society-led approach contrasts with Malmö's government-driven model, yet both create new "spaces of deliberation" that reconfigure traditional food governance structures. These diverse pathways underscore that no universal transition model exists; rather, transformations must align with specific socio-political contexts while addressing global sustainability imperatives.

Knowledge Systems and Learning Mechanisms

The reviewed literature reveals knowledge and learning as fundamental to food system transitions, yet operating through complex and often contested channels. Ingram's (2018) analysis of agricultural knowledge system boundaries demonstrates the tension between mainstream Agricultural Knowledge Systems and alternative approaches like permaculture. The existence of both "impermeable boundaries" maintaining separate knowledge domains and "permeable boundaries" enabling translation highlights the political nature of knowledge production and dissemination.

Mukhovi et al. (2020) identify three critical learning forms in Global South food systems: experiential learning through demonstration, social learning via collective approaches, and transformative learning involving fundamental perception shifts. Their research in Kenya and Bolivia emphasizes that resilience depends not merely on knowledge transfer but on building adaptive capacity through diverse learning mechanisms. This finding challenges technocratic approaches assuming linear knowledge dissemination from experts to practitioners.

The competency gap identified by Charatsari and Lioutas (2019) reveals a systemic failure in agricultural education. Their finding that agronomists lack crucial competencies for promoting sustainable agriculture—particularly in networking, facilitation, and systems thinking—indicates that professional knowledge systems themselves require transformation. This extends beyond curriculum reform to reimagining the role of agricultural professionals as facilitators rather than technical experts.

Governance Innovation and Political Navigation

Perhaps the most significant contribution of this collection lies in revealing the inherently political nature of food system governance. Luger et al. (2025) demonstrate how municipal staff must "act politically simply to accomplish predefined tasks," employing strategies of advocacy and alignment to navigate administrative and political constraints. This finding challenges technocratic governance models assuming neutral implementation of sustainability policies.

The emergence of new governance forms represents a critical transition dynamic. Food policy councils in Bristol, urban living labs across European cities, and collective management approaches for agricultural nature conservation in the Netherlands all represent attempts to transcend traditional sectoral boundaries. Yet as Haylock and Connelly (2018) reveal through their insider/outsider framework, these new arrangements face constant tensions between maintaining alternative values and achieving institutional legitimacy.

Power dynamics permeate these governance innovations. Pothukuchi's (2017) analysis of Detroit's urban agriculture reveals how "shrinkage regimes" view farming as incompatible with development, despite vast vacant land availability. This resistance stems not from technical constraints but from deeply embedded growth ideologies and racial politics shaping land use decisions. Similarly, Williams et al. (2023) map power-laden interactions across European agri-food networks, exposing how incumbent actors simultaneously enable and constrain transitions depending on strategic interests.

Market Mechanisms and Economic Contradictions

The reviewed studies reveal fundamental tensions between market-based food systems and sustainability objectives. Davidson et al. (2016) identify the "conventionalization paradox" wherein successful scaling of alternative beef production could eliminate the price premiums sustaining these practices. This paradox extends across alternative food networks, suggesting that market mechanisms alone cannot drive systematic transformation.

Private sector engagement emerges as deeply ambiguous. While Dubbeling et al. (2016) document successful small and medium enterprise participation in city-region food systems, they also reveal how concentration—with five companies controlling 80% of Ecuador's meat, dairy, and grain markets—constrains transformation possibilities. The literature suggests that meaningful private sector contribution requires regulatory frameworks and incentives aligning profit with sustainability rather than assuming market forces naturally produce sustainable outcomes.

Community Supported Agriculture, farmers' markets, and other alternative distribution systems represent attempts to reconstruct market relationships. Yet as multiple studies demonstrate, these remain economically marginal. Zwartkruis et al. (2020) find that agricultural nature conservation provides less than 10% of income for 81% of participating Dutch farms, despite decades of support. This economic marginality raises fundamental questions about the viability of market-based sustainability transitions.

Systemic Barriers and Transformation Potential

The literature consistently identifies systemic barriers that transcend technical or economic challenges. Institutional lock-ins, path dependencies, and regime resistance emerge as fundamental constraints. The "dual regime challenge" identified by Zwartkruis et al. (2020)—where agricultural nature conservation must navigate both agricultural and nature regimes with conflicting logics—exemplifies how institutional fragmentation impedes transformation.

Cultural and ideological barriers prove equally significant. The commodification of food, as Vivero-Pol (2017) demonstrates, shapes not only market structures but also political imaginaries of what food systems could become. When food is viewed mono-dimensionally as a commodity, system transformation becomes inconceivable; only when recognized as simultaneously a human right, cultural determinant, and commons do transformative pathways emerge.

Yet the literature also identifies transformation potential within these constraints. Crisis moments—whether BSE outbreaks catalyzing alternative beef production or COVID-19 reshaping food discourse—create openings for rapid change. Social movements, from Valencian agroecological networks to Ugandan civil society organizations, demonstrate how collective action can challenge dominant regimes. The key insight is that transformation requires not just alternative practices but alternative values, narratives, and power configurations.

Implications for Sustainability Transitions

This synthesis reveals food system transitions as fundamentally political processes requiring more than technical innovation or market reform. The geographic diversity of cases demonstrates that while patterns exist—the importance of intermediaries, multi-stakeholder governance, crisis-driven opportunities—specific pathways must emerge from local contexts rather than imposed blueprints.

The literature's attention to agency, learning, and governance innovation suggests that transformation depends on cultivating new forms of collective action and institutional arrangements. Yet the persistent economic marginality of alternatives and the power of incumbent regimes indicate that voluntary transition remains unlikely. Instead, the studies point toward the necessity of regulatory intervention, public investment, and fundamental questioning of growth-oriented development models.

Most critically, these papers collectively argue that achieving sustainable food systems requires addressing questions of justice, equity, and democracy alongside environmental concerns. The exclusion of marginalized voices from food governance, the concentration of corporate power, and the commodification of essential resources emerge as core challenges rather than peripheral concerns. True transformation demands not just sustainable production but equitable distribution, democratic participation, and recognition of food's multiple values beyond market price.

Conclusion: Toward Transformative Food Futures

This carefully curated literature reveals food system transitions as complex, contested, and inherently political processes that cannot be reduced to technological fixes or market mechanisms. The diversity of contexts, from Ugandan agroecology to Dutch nature conservation, from Detroit's vacant lots to Valencia's food councils, demonstrates both the universality of sustainability challenges and the necessity of context-specific solutions. The path forward requires embracing this complexity while maintaining focus on fundamental transformation rather than incremental reform. Only by addressing power relations, reconceptualizing food's value, and creating inclusive governance can food systems transition toward genuine sustainability and justice.

### 8.6.3 Review C: The literature review made by the LLM edited by the humans (3342 words)

As is obvious from the word count, more than half of the Review B has been edited. Most of the edits were additions: description about the papers that were not taken into account, precisions on some papers, structural links between papers, and a few additions from literature that were not in the consideration set. The full list of edits is in Appendix, section 8.6.4.

Food System Transitions: A Human-Curated Synthesis of Critical Scholarship

**Introduction: Beyond Technological Solutions**

The transformation of food systems toward sustainability is a complex socio-technical process. This review synthesizes the 3 most significant studies out of 350 publications addressing the issue. Beyond the dominant narratives focusing on efficiency and productivity, and the mainstream approach by levels, these papers illuminate the critical importance of agency, context-specific pathways, and the often-overlooked dimensions of justice and equity in achieving sustainable food futures. They collectively reveal how food system transitions come with fundamental reconfigurations of governance structures, knowledge systems, and power relations.

## Theoretical Foundations: Expanding the Multi-Level Perspective

The Multi-Level Perspective (MLP) of the Dutch school of transition studies (Geels, 2002, Geels and Schot, 2007, Elzen et al, 2004, Rip and Kemp, 1998) remains the most popular to describe system transitions; it has merits compared to economic models that consider all agents operate on the same level. The MLP distinguishes three analytical levels: the landscape (macro-level socio-technical context), the regime (meso-level established practices and institutions), and niches (micro-level protected spaces for innovation). In this model, partly rooted in innovation studies, innovations initially develop in niches where specific conditions enable them to survive, and then they then gradually change the larger regime; while the landscape independently provides changes that may also trigger changes. Yet the reviewed literature also reveals its limitations and necessary expansions. The framework's application to food systems has exposed fundamental problems with its core assumptions, particularly regarding analytical boundaries and the nature of transition processes. Papachristos and Adamides (2016) address fundamental critiques by proposing a methodology that captures causal mechanisms often obscured in traditional MLP applications and tries to encompass system dynamics with qualitative analysis. They show how transitions can emerge from multi-system interactions rather than isolated sectoral changes in niches. However, their "retroductive" methodology, while innovative, relies on simplified system dynamics models that cannot capture the messy realities of food system politics. In this model, functional foods they use as demonstration is particularly problematic -a corporate-driven innovation that hardly represents the many grassroots transitions observed in food systems. Energy transitions within food systems reveal similar complexity. To et al. (2018) analyze bagasse cogeneration in Mauritius, showing how governance structures and institutional context prove crucial for managing transitions toward sustainability, requiring multi-actor collaborations that transcend the MLP's level-based categorization.

Zwartkruis et al. (2018) identify the "dual regime challenge" where agricultural nature conservation must navigate both agricultural and nature regimes with conflicting logics, demonstrating how institutional fragmentation impedes transformation. Vivero-Pol (2017) extends this critique by examining how food valuations—as commodity versus commons—shape political attitudes toward system transformation. This work reveals that counter-hegemonic transformers, who view food multi-dimensionally and consider other types of values, can bring genuine system transformation rather than mere reform. Davidson et al. (2016), showing that actors in alternative beef production operate between alternative and conventional systems rather than within discrete analytical levels, call into question the MLP's foundational assumptions. Similarly, Luger et al. (2025) conceptualize the "purple zone" where municipal staff navigate between formal administrative duties and informal political action, highlighting how individual agency operates between the structural constraints of political agendas and steadfast bureaucracies. In short, transformation, and transition, come with two-ways interaction within and across the "levels", suggesting a far more complex series of retroactions between the components of the system.

## Controversies and transformation

The integration of Social Movement Theory with MLP (Elzen et al., 2011), provides essential insights into how contestation of the status quo drives transitions. Their analysis of Dutch pig husbandry also reveals that successful transitions require alignment across multiple streams - technological, regulatory, market, and social- rather than innovation alone. This multi-stream approach fundamentally reconceptualizes how transitions unfold, moving beyond linear progression narratives.

Ingram et al. (2015) explore how Learning and Innovation Networks for Sustainable Agriculture (LINSA) interact with dominant agricultural regimes across Europe. Their analysis identifies five distinct interaction modes (Compatible, Complementary, Emergent, Divergent, and Oppositional) demonstrating that highly compatible networks achieve smoother diffusion but often experience dilution of their sustainability ambitions, while more radical networks face marginalization but preserve their transformative ideals. This work reveals the complex negotiations, and conflicts, occurring at niche-regime interfaces, where sustainability transformations emerge through messy, dynamic, and context-dependent interactions rather than simple adoption and diffusion processes. In short, not only is the process complex, multi-party and combining top-down to bottom-up, but it is also fostered by contestation, conflicts and compromises that involve, in controversies, values and power struggles.

Jørgensen's "arenas of development" perspective (Jørgensen, 2012), in contrast with MLP, accounts for these multiple interactions between levels and brings a less linear and more systemic view of the transition process. It is also more "flat" than "by levels", putting all actors on the same level in these interaction arenas. AoD considers the transitions occur through conflicts, negotiations, alignments and learning between various stakeholders, including some non-humans in a Latourian perspective of the network-actor (Akrich et al., 2006)). AoD gives attention to the regulation mechanisms and the processes by which they occur. In AoD, actors can operate in several arenas. "Stable actor worlds", configurations of actors playing the same game, gradually emerge and can produce lasting change. While the MLP framework highlights the darwinian mechanisms of innovation in niches and diffusion, AoD highlights the multi-party emergence of new stable configuration of actors through multiple systems of regulation.

**Geographic and Cultural Diversity and Contextual Pathways**

The literature demonstrates significant variation in transition pathways across different contexts, challenging universal application of MLP concepts. While the MLP suggests transitions occur through protected niches, urban food systems research reveals more complex dynamics. Audet et al. (2017) demonstrate that seasonal food markets in Montreal evolve at the intersection of regime and niche rather than in protected spaces, challenging the framework's assumptions about clear analytical separation.

Notably absent from these transitions frameworks is serious engagement with how colonial agricultural systems continue to structure contemporary food systems. The emphasis on 'Global South innovations' risks romanticizing struggles that emerge from structural adjustment policies and ongoing extractivism. The Ugandan case, for instance (see below), cannot be understood without examining how cash crop dependencies established under colonialism shape current agroecological possibilities.

A distinguishing feature of this curated collection is its geographic breadth and attention to diverse transition contexts. Global South examples offer critical perspectives on alternative pathways and grassroots innovation that complement European institutionalized transition management approaches. Isgren and Ness (2017) examine agroecology in Uganda's Rwenzori region, demonstrating how civil society networks (primarily NGOs, community-based organizations, and faith-based organizations) fill gaps left by state withdrawal while navigating deteriorating political climates. Their analysis reveals agroecology emerging as an "unintended consequence of neoliberal development policies", specifically how the devolution of agricultural extension and support services to non-state actors created space for alternative approaches that would not have emerged under direct state control. The contradictory mechanism operates through neoliberal policies simultaneously weakening state capacity while enabling donor-funded civil society organizations to promote agroecological practices that challenge conventional modernization approaches. This pathway differs fundamentally from European transition experiences, where formal state institutions, research organizations, and established policy frameworks actively guide and legitimize alternative agricultural practices through deliberate transition management programs.

The integration of food systems with broader structural transformation processes is further illuminated by Arslan et al. (2022), whose comprehensive analysis of 81 low- and middle-income countries reveals that structural transformation (measured as non-agricultural GDP share) correlates more consistently with improved food system outcomes than rural transformation alone. Their construction of a Food Systems Index demonstrates that financial inclusion, digital connectivity, women's empowerment, and education emerge as key levers for positive food system transformation, while highlighting the "double burden" (Arslan et al. 2022, 298) of malnutrition and rapid population growth in countries with the weakest food system outcomes.

Latin American cases provide particularly valuable insights into alternative development pathways. The Nicaraguan agroecological transition analyzed by Schiller et al. (2020) demonstrates how knowledge flows (through intermediaries) enable niche development despite limited state support. Meanwhile, Brazilian urban agriculture policies examined by Ferreira et al. (2024) reveal how local government entrepreneurs navigate between formal planning and informal food systems, creating hybrid governance arrangements that challenge conventional top-down planning paradigms.

**Governance Innovation and Political Navigation**

Perhaps the most significant contribution of this collection lies in revealing the inherently political nature of food system governance. Luger et al. (2025) demonstrate through Urban Living Labs (ULLs), emerging spaces where individuals test collaborations for societal transformation, how municipal staff in 10 European cities must "act politically simply to accomplish predefined tasks," employing strategies of advocacy and alignment to navigate administrative and political constraints. This finding challenges technocratic governance models assuming neutral implementation of sustainability policies while ignoring fluctuating funding and shifting implementation agendas that come with political transitions.

The comparative analysis by Moragues-Faus and Morgan (2015) of Bristol and Malmö illuminates how different institutional contexts shape urban food governance. Bristol's civil society-led approach contrasts with Malmö's government-driven model, yet both create new "spaces of deliberation" that reconfigure traditional food governance structures. These cases reveal a troubling pattern: even supposedly 'alternative' pathways often reproduce existing power structures. Bristol's Food Policy Council, despite civil society leadership, remains dominated by white, middle-class professionals. Malmö's government-driven approach, while achieving organic procurement targets, does little to address food access for immigrant communities. The absence of genuine participation by food-insecure populations in these 'innovative' governance arrangements suggests these are transitions for, not by, those most affected by food system failures. Therefore, transformations must align with specific socio-political contexts while addressing global sustainability imperatives.

Building on urban governance innovations, Larbi et al. (2022) examine sustainability transitions in Global South cities through a comparative study of Curitiba, Brazil, and Accra, Ghana. Curitiba's success in pioneering cost-effective innovations in public transport, waste management, and flood control under strong political leadership demonstrates the potential for low-cost, context-sensitive solutions. However, their analysis of Accra reveals that technical solutions cannot be transferred wholesale across contexts. Instead, cities must build on existing local capacities, such as Accra's strong culture of communal labor (Larbi et al. 2022, 241), while involving both state and non-state actors through tailored strategies.

These diverse pathways underscore that no universal transition model exists; rather, transformations must align with specific socio-political contexts while addressing global sustainability imperatives. They also reveal how the incumbent system informs its own transformation through its governance structures.

The challenges of scaling transitions beyond individual cities are addressed by Riley and Crush (2023) in their comprehensive examination of secondary African cities. Their analysis reveals how local governments, motivated by urban food security needs but constrained by limited resources and fragmented governance, must navigate complex relationships with informal sector actors, urban farmers, civil society organizations, and vulnerable households. The study demonstrates that successful transformation requires recognizing informal food systems as legitimate components of urban food provisioning rather than obstacles to development, while building on existing social networks and community capacities.

The emergence of new governance forms represents a critical transition dynamic. Food policy councils in Bristol, urban living labs in Amsterdam, Barcelona, and other European cities studied by Luger et al. (2025) , and collective management approaches for agricultural nature conservation in the Netherlands all represent attempts to transcend traditional sectoral boundaries.

The "convening power of food" as analyzed by Bedore (2014) in Baltimore demonstrates how food policy becomes "partnerable" (2980) because it appeals to both growth coalition members and broader civil society. Food policy initiatives succeed when framed in ways that complement rather than challenge existing economic development strategies, with discursive mechanisms enabling actors to frame food policy in depoliticized terms around health, innovation, and city reputation. However, these mechanisms can limit transformative potential by aligning new initiatives with existing power structures and growth imperatives.

Building on governance innovation concepts, Buchan et al. (2019) introduce "Transformative Incrementalism" as a framework for understanding how government actors influence sustainable food system transitions through gradual, power-laden processes rather than abrupt policy shifts. Their analysis reveals that government influence operates through both authoritative mechanisms (policy documents, official mandates) and relational mechanisms (coalition building, narrative framing), with planning documents like Official Community Plans serving as legitimizing tools to embed sustainable food agendas within institutional mandates.

Yet as Haylock and Connelly (2018) reveal through their insider versus outsider strategies in New Zealand food networks, these new arrangements face constant tensions between maintaining alternative values and achieving institutional legitimacy, with groups like Dunedin's Our Food Network explicitly choosing to maintain distance from local government to preserve their transformative agenda.

### Participatory Planning and Co-Production

The potential for more inclusive governance approaches is demonstrated by Raja et al. (2018) in their analysis of U.S. food planning initiatives. They identify the "power to convene" (150) as a key mechanism whereby local governments leverage soft power to foster cross-sector coalitions. Seattle's Local Food Action Initiative and Minneapolis's Homegrown Minneapolis demonstrate how bureaucratic restructuring and co-leading models can institutionalize food as a municipal priority while embedding food indicators into comprehensive plans. However, persistent barriers including fiscal constraints and jurisdictional misalignment between food systems and political boundaries continue to challenge implementation.

Power dynamics permeate these governance innovations. Pothukuchi's (2017) analysis of Detroit's urban agriculture policy reveals how "shrinkage regimes" view farming as incompatible with development, despite vast vacant land availability. This resistance stems not from technical constraints but from deeply embedded growth ideologies and racial politics shaping land use decisions among city governance officials. Similarly, Williams et al. (2023) map power-laden interactions across European agri-food networks in which farming is prioritized over other food system actors in policy, exposing how incumbent actors simultaneously enable and constrain transitions depending on strategic interests.

The application of transformative capacity frameworks is demonstrated by Sarabia et al. (2021) in their analysis of Valencia's agri-food system transformation. Their research reveals how three interrelated milestones—empowerment of grassroots social movements, political shift in local government, and adherence to the Milan Urban Food Policy Pact—catalyzed the creation of a Municipal Food Council and Agri-food Strategy 2025. Using the Urban Transformative Capacity Framework, they identify significant progress in inclusive governance and community-based experimentation, while highlighting weaknesses in embedding innovation and cross-scalar coordination.

## Market Mechanisms and Economic Contradictions

The reviewed studies reveal fundamental tensions between market-based food systems and sustainability objectives. Davidson et al. (2016) identify the "conventionalization paradox" wherein successful scaling of alternative beef production could eliminate the price premiums sustaining these practices. Similar paradoxes emerge in Haylock and Connelly's (2018) New Zealand cases, where community food networks struggle with capacity and resource constraints while attempting to maintain independence from potentially restrictive government partnerships.

Private sector engagement emerges as deeply ambiguous. While Dubbeling et al. (2016) document successful small and medium enterprise participation in city-region food systems, they also reveal how concentration -with five companies controlling 80% of Ecuador's meat, dairy, and grain markets (80)- constrains transformation possibilities. The literature suggests that meaningful private sector contribution requires regulatory frameworks and incentives aligning profit with sustainability rather than assuming market forces naturally produce sustainable outcomes.

Community Supported Agriculture, farmers' markets, and other alternative distribution systems represent attempts to reconstruct market relationships, as demonstrated in both Christchurch and Dunedin's emerging food networks. Yet Zwartkruis et al. (2020) find that agricultural nature conservation provides less than 10% of income for 81% of participating Dutch farms (7), despite decades of support. This economic marginality raises fundamental questions about the viability of market-based sustainability transitions.

### Consumer-Stage Interventions and Behavioral Change

The challenge of translating sustainability concerns into behavioral change is addressed by Reynolds et al. (2019) in their comprehensive review of consumer-stage food waste reduction interventions. Despite a surge in household-level food waste research, only 17 of 292 peer-reviewed studies report applied interventions with measurable waste reductions (Reynolds et al. 2019, 12). These interventions fall into information-based (newsletters, curricula), technological (plate-size modifications, apps), and policy/system changes (meal standards, portion defaults), with reported waste reductions varying widely (12-57%) (ibid.). The review reveals significant methodological challenges, including reliance on self-reports rather than compositional analyses, small non-representative samples, and lack of theoretical grounding, highlighting the gap between sustainability awareness and effective intervention design.

### Scaling Sustainability Through System-Wide Change

The potential for system-wide transformation is illustrated by Ryghaug and Skjølsvold's (2023) analysis of Norway's electric vehicle transition, which achieved over 86% battery and plug-in hybrid share among new private cars by 2021 (ibid., 1). Their longitudinal study reveals how persistent demand-side incentives, intermediation by NGOs, and user satisfaction created momentum for rapid diffusion, while also triggering broader electrification across transport sectors. However, the case highlights Norway's unique conditions—wealth and renewable energy resources—and raises second-order sustainability dilemmas around mining impacts and material justice, underscoring that transitioning regimes spawn fresh systemic trade-offs.

## Systemic Barriers and Transformation Potential

The literature consistently identifies systemic barriers that transcend technical or economic challenges. Institutional lock-ins, path dependencies, and regime resistance emerge as fundamental constraints. The "dual regime challenge" identified by Zwartkruis et al. (2020)—where agricultural nature conservation must navigate both agricultural and nature regimes with conflicting logics—exemplifies how institutional fragmentation impedes transformation.

Cultural and ideological barriers prove equally significant. The commodification of food, as Vivero-Pol (2017) demonstrates, shapes not only market structures but also political imaginaries of what food systems could become. When food is viewed mono-dimensionally as a commodity, system transformation becomes inconceivable; only when recognized as simultaneously a human right, cultural determinant, and commons do transformative pathways emerge.

Yet the literature also identifies transformation potential within these constraints. Crisis moments—whether BSE outbreaks catalyzing alternative beef production or COVID-19 reshaping food discourse—create openings for rapid change. Social movements, from Valencian agroecological networks to Ugandan civil society organizations, demonstrate how collective action can challenge dominant regimes. The key insight is that transformation requires not just alternative practices but alternative values, narratives, and power configurations.

Knowledge Systems and Learning Mechanisms

The reviewed literature reveals knowledge and learning as fundamental to food system transitions, yet operating through complex and often contested channels. Ingram's (2018) analysis of agricultural knowledge system boundaries demonstrates the tension between mainstream Agricultural Knowledge Systems and alternative approaches like permaculture. The existence of both "impermeable boundaries" maintaining separate knowledge domains and "permeable boundaries" enabling translation highlights the political nature of knowledge production and dissemination.

International knowledge governance creates additional complexity. Li et al. (2013) show how international treaties and conventions on plant genetic resources affect participatory maize breeding development in China, with "intellectual property regimes" working against participatory breeding despite international commitments supporting it. This demonstrates how landscape-level factors can simultaneously enable and constrain transitions.

Mukhovi et al. (2020) identify three critical learning forms in Global South food systems: experiential learning through demonstration, social learning via collective approaches, and transformative learning involving fundamental perception shifts. Their research in Kenya and Bolivia emphasizes that resilience depends not merely on knowledge transfer but on building adaptive capacity through diverse learning mechanisms. This finding challenges technocratic approaches assuming linear knowledge dissemination from experts to practitioners.

The competency gap identified by Charatsari and Lioutas (2019) reveals a systemic failure in agricultural education. Their finding that agronomists lack crucial competencies for promoting sustainable agriculture -particularly in networking, facilitation, and systems thinking- indicates that professional knowledge systems themselves require transformation. This extends beyond curriculum reform to reimagining the role of agricultural professionals as facilitators rather than technical experts.

The role of media in shaping food sustainability discourse is critically examined by Hansen (2022), whose quantitative content analysis of 271 Danish newspaper articles reveals how external events like COVID-19 and authorship type (journalist versus reader) influence the relationship between how sustainability issues are framed and what solutions are proposed. The study demonstrates that reader-authored articles are more likely than journalist-authored pieces to propose strong solutions like sustainable production based on values and norms, while journalists tend toward "softer" (Hansen 2022, 9) implications such as revising social norms. This research highlights the media's agenda-setting power and suggests that policy-makers must consider both external events and authorship dynamics in crafting effective sustainability messaging.

**Implications for Sustainability Transitions**

This synthesis reveals food system transitions as fundamentally political processes requiring more than technical innovation or market reform. The geographic diversity of cases demonstrates that while patterns exist—the importance of intermediaries, multi-stakeholder governance, crisis-driven opportunities—specific pathways must emerge from local contexts rather than imposed blueprints. There is what we could call a transformation paradox in the fact that the existing governance structure informs the process of transition, but hence also its result. While this seems obvious ex-post, this implies that every transformation must seriously take into account the conditions given, and not only the target result.

The literature's attention to agency, learning, and governance innovation suggests that transformation depends on cultivating new forms of collective action and institutional arrangements. Yet the persistent economic marginality of alternatives and the power of incumbent regimes indicate that voluntary transition remains unlikely. Instead, the studies point toward the necessity of regulatory intervention, public investment, and fundamental questioning of growth-oriented development models.

Most critically, these papers collectively argue that achieving sustainable food systems requires addressing questions of justice, equity, and democracy alongside environmental concerns. The exclusion of marginalized voices from food governance, the concentration of corporate power, and the commodification of essential resources emerge as core challenges rather than peripheral concerns. True transformation demands not just sustainable production but equitable distribution, democratic participation, and recognition of food's multiple values beyond market price.

**Conclusion: Toward Transformative Food Futures**

In conclusion, this curated literature review, that involved not only classic or mainstream work, puts in full light why modelling societal transformations is a difficult exercise. Research reveals food system transitions as complex, contested, and inherently political processes that cannot be reduced to technological fixes or market mechanisms. The diversity of contexts, from Ugandan agroecology to Dutch nature conservation, from Detroit's vacant lots to Valencia's food councils, demonstrates that, while sustainability challenges are universal, the paths to solutions are inevitably context-specific. The path forward also requires abandoning comfortable narratives of gradual transition and confronting the reality that sustainable food systems may be impossible within current political-economic structures and may go through radical breaks-land occupations, seed sovereignty movements, food riots- that historically have driven genuine food system transformation. On the other hand, it appears that the incumbent system and its governance structure is always involved in the self-transformative path, and may be so constructively (the transformation paradox). The path forward requires embracing this complexity while maintaining focus on fundamental transformation rather than incremental reform. Only by addressing power relations, reconceptualizing food's value, and creating inclusive governance can food systems transition toward genuine sustainability and justice.

### 8.6.4 Review C: What edits were made on Review B:

*(The parts suppressed are barred, the parts added are underlined).*

Review B: Food System Transitions: A Human-Curated Synthesis of Critical Scholarship

Introduction: Beyond Technological Solutions

The transformation of food systems toward sustainability ~~represents~~ is a complex socio-technical ~~challenge that transcends mere technological innovation.~~ process. This review synthesizes the 31 ~~carefully selected~~ most significant studies ~~that collectively reveal how food system transitions require fundamental reconfigurations~~ out of ~~governance structures, knowledge systems, and power relations. Unlike~~ 350 publications addressing the issue. Beyond the dominant narratives focusing on efficiency and productivity, and the mainstream approach by levels, these papers illuminate the critical importance of agency, context-specific pathways, and the often-overlooked dimensions of justice and equity in achieving sustainable food futures. They collectively reveal how food system transitions come with fundamental reconfigurations of governance structures, knowledge systems, and power relations.

Theoretical Foundations: Expanding the Multi-Level Perspective

The Multi-Level Perspective (MLP) of the Dutch school of transition studies (Geels, 2002, Geels and Schot, 2007, Elzen et al, 2004, Rip and Kemp, 1998) remains ~~central~~ the most popular to ~~understanding food~~ describe system ~~transitions, yet~~ transitions; it has merits compared to economic models that consider all agents operate on the same level. The MLP distinguishes three analytical levels: the landscape (macro-level socio-technical context), the regime (meso-level established practices and institutions), and niches (micro-level protected spaces for innovation). In this model, partly rooted in innovation studies, innovations initially develop in niches where specific conditions enable them to survive, and then they then gradually change the larger regime; while the landscape independently provides changes that may also trigger changes. Yet the reviewed literature also reveals its limitations and necessary expansions. The framework's application to food systems has exposed fundamental problems with its core assumptions, particularly regarding analytical boundaries and the nature of transition processes. Papachristos and Adamides (2016) address fundamental critiques by proposing a ~~retroductive systems-based~~ methodology that captures causal mechanisms often obscured in traditional MLP ~~applications. Their integration of applications and tries to encompass~~ system dynamics with qualitative ~~analysis demonstrates~~ analysis. They show how transitions ~~can~~ emerge from multi-system interactions rather than isolated sectoral ~~changes. Critically, several studies challenge~~ changes in niches. However, their "retroductive" methodology, while innovative, relies on simplified system dynamics models that cannot capture the messy realities of food system politics. In this model, functional foods they use as demonstration is particularly problematic -a corporate-driven innovation that hardly represents the many grassroots transitions observed in food systems. Energy transitions within food systems reveal similar complexity. To et al. (2018) analyze bagasse cogeneration in Mauritius, showing how governance structures and institutional context prove crucial for managing transitions toward sustainability, requiring multi-actor collaborations that transcend the MLP's insufficient attention to agency level-based categorization. Zwartkruis et al. (2018) identify the "dual regime challenge" where agricultural nature conservation must navigate both agricultural and power nature regimes with conflicting logics, demonstrating how institutional fragmentation impedes transformation. Vivero-Pol (2017) introduces a crucial dimension extends this critique by examining how food valuations—as commodity versus commons—shape political attitudes toward system transformation. This work reveals that counter-hegemonic transformers, who view food multi-dimensionally, are essential for multi-dimensionally and consider other types of values, can bring genuine system transformation rather than mere reform. Davidson et al. (2016), showing that actors in alternative beef production operate between alternative and conventional systems rather than within discrete analytical levels, call into question the MLP's foundational assumptions. Similarly, Luger et al. (2025) conceptualize the "purple zone" where municipal staff navigate between formal administrative duties and informal political action, highlighting how individual agency operates between the structural constraints of political agendas and steadfast bureaucracies. In short, transformation, and transition, come with two-ways interaction within structural constraints.

The and across the "levels", suggesting a far more complex series of retroactions between the components of the system. Controversies and transformation The integration of Social Movement Theory with MLP, as demonstrated by Elzen MLP (Elzen et al. (2011), al., 2011), provides essential insights into how normative contestation of the status quo drives transitions. Their analysis of Dutch pig husbandry also reveals that successful transitions require alignment across multiple streams technological, streams -technological, regulatory, market, and social rather social- rather than innovation alone. This multi-stream approach fundamentally reconceptualizes how transitions unfold, moving beyond linear progression narratives. Geographic narratives. Ingram et al. (2015) explore how Learning and Innovation Networks for Sustainable Agriculture (LINSA) interact with dominant agricultural regimes across Europe. Their analysis identifies five distinct interaction modes (Compatible, Complementary, Emergent, Divergent, and Oppositional) demonstrating that highly compatible networks achieve smoother diffusion but often experience dilution of their sustainability ambitions, while more radical networks face marginalization but preserve their transformative ideals. This work reveals the complex negotiations, and conflicts, occurring at niche-regime interfaces, where sustainability transformations emerge through messy, dynamic, and context-dependent interactions rather than simple adoption and diffusion processes. In short, not only is the process complex, multi-party and combining top-down to bottom-up, but it is also fostered by contestation, conflicts and compromises that involve, in controversies, values and power struggles. Jørgensen's "arenas of development" perspective (2012), in contrast with MLP, accounts for these multiple interactions between levels and brings a less linear and more systemic view of the transition process. It is also more "flat" than "by levels", putting all actors on the same level in these interaction arenas. AoD considers the transitions occur through conflicts, negotiations, alignments and learning between various stakeholders, including some non-humans in a Latourian perspective of the network-actor (Akrich, Callon & Latour, 2006). AoD gives attention to the regulation mechanisms and the processes by which they occur. In AoD, actors can operate in several arenas. "Stable actor worlds", configurations of actors playing the same game, gradually emerge and can produce lasting change. While the MLP framework highlights the darwinian mechanisms of innovation in niches and diffusion, AoD highlights the multi-party emergence of new stable configuration of actors through multiple systems of regulation. Geographic and Cultural Diversity and Contextual ~~Pathways A~~ Pathways. The literature demonstrates significant variation in transition pathways across different contexts, challenging universal application of MLP concepts. While the MLP suggests transitions occur through protected niches, urban food systems research reveals more complex dynamics. Audet et al. (2017) demonstrate that seasonal food markets in Montreal evolve at the intersection of regime and niche rather than in protected spaces, challenging the framework's assumptions about clear analytical separation. Notably absent from these transitions frameworks is serious engagement with how colonial agricultural systems continue to structure contemporary food systems. The emphasis on 'Global South innovations' risks romanticizing struggles that emerge from structural adjustment policies and ongoing extractivism. The Ugandan case, for instance (see below), cannot be understood without examining how cash crop dependencies established under colonialism shape current agroecological possibilities. A distinguishing feature of this curated collection is its geographic breadth and attention to diverse transition contexts. ~~While European cases provide insights into institutionalized transition management,~~ Global South examples offer critical perspectives on alternative pathways and grassroots ~~innovation. Isgren~~ innovation that complement European institutionalized transition management approaches. Isgren and Ness (2017) examine agroecology in Uganda's Rwenzori region, demonstrating how civil society networks (primarily NGOs, community-based organizations, and faith-based organizations) fill gaps left by state withdrawal while navigating deteriorating political climates. Their analysis reveals agroecology emerging as an "unintended consequence of neoliberal development ~~policies," highlighting~~ policies", specifically how ~~transitions can emerge~~ the devolution of agricultural extension and support services to non-state actors created space for alternative approaches that would not have emerged under direct state control. The contradictory mechanism operates through neoliberal policies simultaneously weakening state capacity while enabling donor-funded civil society organizations to promote agroecological practices that challenge conventional modernization approaches. This pathway differs fundamentally from ~~systemic contradictions rather than deliberate planning. This contrasts sharply with~~ European ~~experiences~~ transition experiences, where formal ~~institutions drive~~ state institutions, research organizations, and established policy frameworks actively guide and legitimize alternative agricultural practices through deliberate transition ~~processes. Latin~~ management programs. The integration of food systems with broader structural transformation processes is further illuminated by Arslan et al. (2022), whose comprehensive analysis of 81 low- and middle-income countries reveals that structural transformation (measured as non-agricultural GDP share) correlates more consistently with improved food system outcomes than rural transformation alone. Their construction of a Food Systems Index demonstrates that financial inclusion, digital connectivity, women's empowerment, and education emerge as key levers for positive food system transformation, while highlighting the "double burden" (Arslan et al. 2022, 298) of malnutrition and rapid population growth in countries with the weakest food system outcomes. Latin American cases provide particularly valuable insights into alternative development pathways. The Nicaraguan agroecological transition analyzed by Schiller et al. (2020) demonstrates how knowledge flows ~~through intermediaries~~ (through intermediaries) enable niche development despite limited state support. Meanwhile, Brazilian urban agriculture policies examined by Ferreira et al. (2024) reveal how local government entrepreneurs navigate between formal planning and informal food systems, creating hybrid governance arrangements that challenge conventional top-down planning ~~paradigms. The comparative analysis by Moragues-Faus and Morgan (2015) of Bristol and Malmö illuminates how different institutional contexts shape urban food governance. Bristol's civil society-led approach contrasts with Malmö's government-driven model, yet both create new "spaces of deliberation" that reconfigure traditional food governance structures. These diverse pathways underscore that no universal transition model exists; rather, transformations must align with specific socio-political contexts while addressing global sustainability imperatives. Knowledge Systems and Learning Mechanisms The reviewed literature reveals knowledge and learning as fundamental to food system transitions, yet operating through complex and often contested channels. Ingram's (2018) analysis of agricultural knowledge system boundaries demonstrates the tension between mainstream Agricultural Knowledge Systems and alternative approaches like permaculture. The existence of both "impermeable boundaries" maintaining separate knowledge domains and "permeable boundaries" enabling translation highlights the political nature of knowledge production and dissemination.~~

~~Mukhovi et al. (2020) identify three critical learning forms in Global South food systems: experiential learning through demonstration, social learning via collective approaches, and transformative learning involving fundamental perception shifts. Their research in Kenya and Bolivia emphasizes that resilience depends not merely on knowledge transfer but on building adaptive capacity through diverse learning mechanisms. This finding challenges technocratic approaches assuming linear knowledge dissemination from experts to practitioners. The competency gap identified by Charatsari and Lioutas (2019) reveals a systemic failure in agricultural education. Their finding that agronomists lack crucial competencies for promoting sustainable agriculture — particularly in networking, facilitation, and systems thinking — indicates that professional knowledge systems themselves require transformation. This extends beyond curriculum reform to reimagining the role of agricultural professionals as facilitators rather than technical experts.~~ paradigms.

Governance Innovation and Political Navigation

Perhaps the most significant contribution of this collection lies in revealing the inherently political nature of food system governance. Luger et al. (2025) demonstrate through Urban Living Labs (ULLs), emerging spaces where individuals test collaborations for societal transformation , how municipal staff in 10 European cities must "act politically simply to accomplish predefined tasks," employing strategies of advocacy and alignment to navigate administrative and political constraints. This finding challenges technocratic governance models assuming neutral implementation of sustainability ~~policies. The~~ policies while ignoring fluctuating funding and shifting implementation agendas that come with political transitions. The comparative analysis by Moragues-Faus and Morgan (2015) of Bristol and Malmö illuminates how different institutional contexts shape urban food governance. Bristol's civil society-led approach contrasts with Malmö's government-driven model, yet both create new "spaces of deliberation" that reconfigure traditional food governance structures. These cases reveal a troubling pattern: even supposedly 'alternative' pathways often reproduce existing power structures. Bristol's Food Policy Council, despite civil society leadership, remains dominated by white, middle-class professionals. Malmö's government-driven approach, while achieving organic procurement targets, does little to address food access for immigrant communities. The absence of genuine participation by food-insecure populations in these 'innovative' governance arrangements suggests these are transitions for, not by, those most affected by food system failures. Therefore, transformations must align with specific socio-political contexts while addressing global sustainability imperatives. Building on urban governance innovations, Larbi et al. (2022) examine sustainability transitions in Global South cities through a comparative study of Curitiba, Brazil, and Accra, Ghana. Curitiba's success in pioneering cost-effective innovations in public transport, waste management, and flood control under strong political leadership demonstrates the potential for low-cost, context-sensitive solutions. However, their analysis of Accra reveals that technical solutions cannot be transferred wholesale across contexts. Instead, cities must build on existing local capacities, such as Accra's strong culture of communal labor (Larbi et al. 2022, 241), while involving both state and non-state actors through tailored strategies. These diverse pathways underscore that no universal transition model exists; rather, transformations must align with specific socio-political contexts while addressing global sustainability imperatives. They also reveal how the incumbent system informs its own transformation through its governance structures.

The challenges of scaling transitions beyond individual cities are addressed by Riley and Crush (2023) in their comprehensive examination of secondary African cities. Their analysis reveals how local governments, motivated by urban food security needs but constrained by limited resources and fragmented governance, must navigate complex relationships with informal sector actors, urban farmers, civil society organizations, and vulnerable households. The study demonstrates that successful transformation requires recognizing informal food systems as legitimate components of urban food provisioning rather than obstacles to development, while building on existing social networks and community capacities.

The emergence of new governance forms represents a critical transition dynamic. Food policy councils in Bristol, urban living labs ~~across~~ in Amsterdam, Barcelona, and other European ~~cities,~~ cities studied by Luger et al. (2025) , and collective management approaches for agricultural nature conservation in the Netherlands all represent attempts to transcend traditional sectoral boundaries. ~~Yet~~ The "convening power of food" as analyzed by Bedore (2014) in Baltimore demonstrates how food policy becomes "partnerable" (2980) because it appeals to both growth coalition members and broader civil society. Food policy initiatives succeed when framed in ways that complement rather than challenge existing economic development strategies, with discursive mechanisms enabling actors to frame food policy in depoliticized terms around health, innovation, and city reputation. However, these mechanisms can limit transformative potential by aligning new initiatives with existing power structures and growth imperatives.

Building on governance innovation concepts, Buchan et al. (2019) introduce "Transformative Incrementalism" as a framework for understanding how government actors influence sustainable food system transitions through gradual, power-laden processes rather than abrupt policy shifts. Their analysis reveals that government influence operates through both authoritative mechanisms (policy documents, official mandates) and relational mechanisms (coalition building, narrative framing), with planning documents like Official Community Plans serving as legitimizing tools to embed sustainable food agendas within institutional mandates. Yet as Haylock and Connelly (2018) reveal through their ~~insider/outsider framework,~~ insider versus outsider strategies in New Zealand food networks, these new arrangements face constant tensions between maintaining alternative values and achieving institutional ~~legitimacy. Power~~ legitimacy, with groups like

Dunedin's Our Food Network explicitly choosing to maintain distance from local government to preserve their transformative agenda.

Participatory Planning and Co-Production The potential for more inclusive governance approaches is demonstrated by Raja et al. (2018) in their analysis of U.S. food planning initiatives. They identify the "power to convene" (150) as a key mechanism whereby local governments leverage soft power to foster cross-sector coalitions. Seattle's Local Food Action Initiative and Minneapolis's Homegrown Minneapolis demonstrate how bureaucratic restructuring and co-leading models can institutionalize food as a municipal priority while embedding food indicators into comprehensive plans. However, persistent barriers including fiscal constraints and jurisdictional misalignment between food systems and political boundaries continue to challenge implementation. Power dynamics permeate these governance innovations. Pothukuchi's (2017) analysis of Detroit's urban agriculture policy reveals how "shrinkage regimes" view farming as incompatible with development, despite vast vacant land availability. This resistance stems not from technical constraints but from deeply embedded growth ideologies and racial politics shaping land use decisions. decisions among city governance officials. Similarly, Williams et al. (2023) map power-laden interactions across European agri-food networks, networks in which farming is prioritized over other food system actors in policy, exposing how incumbent actors simultaneously enable and constrain transitions depending on strategic interests. interests.

The application of transformative capacity frameworks is demonstrated by Sarabia et al. (2021) in their analysis of Valencia's agri-food system transformation. Their research reveals how three interrelated milestones—empowerment of grassroots social movements, political shift in local government, and adherence to the Milan Urban Food Policy Pact—catalyzed the creation of a Municipal Food Council and Agri-food Strategy 2025. Using the Urban Transformative Capacity Framework, they identify significant progress in inclusive governance and community-based experimentation, while highlighting weaknesses in embedding innovation and cross-scalar coordination.

Market Mechanisms and Economic Contradictions

The reviewed studies reveal fundamental tensions between market-based food systems and sustainability objectives. Davidson et al. (2016) identify the "conventionalization paradox" wherein successful scaling of alternative beef production could eliminate the price premiums sustaining these practices. This paradox extends across alternative Similar paradoxes emerge in Haylock and Connelly's (2018) New Zealand cases, where community food networks, suggesting that market mechanisms alone cannot drive systematic transformation. Private networks struggle with capacity and resource constraints while attempting to maintain independence from potentially restrictive government partnerships. Private sector engagement emerges as deeply ambiguous. While Dubbeling et al. (2016) document successful small and medium enterprise participation in city-region food systems, they also reveal how concentration with concentration -with five companies controlling 80% of Ecuador's meat, dairy, and grain markets constrains markets (80)-constrains transformation possibilities. The literature suggests that meaningful private sector contribution requires regulatory frameworks and incentives aligning profit with sustainability rather than assuming market forces naturally produce sustainable outcomes. Community Supported Agriculture, farmers' markets, and other alternative distribution systems represent attempts to reconstruct market relationships. relationships, as demonstrated in both Christchurch and Dunedin's emerging food networks. Yet ~~as multiple studies demonstrate, these remain economically marginal.~~ Zwartkruis et al. (2020) find that agricultural nature conservation provides less than 10% of income for 81% of participating Dutch ~~farms,~~ farms (7), despite decades of support. This economic marginality raises fundamental questions about the viability of market-based sustainability ~~transitions.~~ transitions. Consumer-Stage Interventions and Behavioral Change.

The challenge of translating sustainability concerns into behavioral change is addressed by Reynolds et al. (2019) in their comprehensive review of consumer-stage food waste reduction interventions. Despite a surge in household-level food waste research, only 17 of 292 peer-reviewed studies report applied interventions with measurable waste reductions (Reynolds et al. 2019, 12). These interventions fall into information-based (newsletters, curricula), technological (plate-size modifications, apps), and policy/system changes (meal standards, portion defaults), with reported waste reductions varying widely (12-57%) (ibid.). The review reveals significant methodological challenges, including reliance on self-reports rather than compositional analyses, small non-representative samples, and lack of theoretical grounding, highlighting the gap between sustainability awareness and effective intervention design.

Scaling Sustainability Through System-Wide Change

The potential for system-wide transformation is illustrated by Ryghaug and Skjølsvold's (2023) analysis of Norway's electric vehicle transition, which achieved over 86% battery and plug-in hybrid share among new private cars by 2021 (ibid., 1). Their longitudinal study reveals how persistent demand-side incentives, intermediation by NGOs, and user satisfaction created momentum for rapid diffusion, while also triggering broader electrification across transport sectors. However, the case highlights Norway's unique conditions—wealth and renewable energy resources—and raises second-order sustainability dilemmas around mining impacts and material justice, underscoring that transitioning regimes spawn fresh systemic trade-offs.

Systemic Barriers and Transformation Potential

The literature consistently identifies systemic barriers that transcend technical or economic challenges. Institutional lock-ins, path dependencies, and regime resistance emerge as fundamental constraints. The "dual regime challenge" identified by Zwartkruis et al. (2020)—where agricultural nature conservation must navigate both agricultural and nature regimes with conflicting logics—exemplifies how institutional fragmentation impedes transformation.

Cultural and ideological barriers prove equally significant. The commodification of food, as Vivero-Pol (2017) demonstrates, shapes not only market structures but also political imaginaries of what food systems could become. When food is viewed mono-dimensionally as a commodity, system transformation becomes inconceivable; only when recognized as simultaneously a human right, cultural determinant, and commons do transformative pathways emerge.

Yet the literature also identifies transformation potential within these constraints. Crisis moments—whether BSE outbreaks catalyzing alternative beef production or COVID-19 reshaping food discourse—create openings for rapid change. Social movements, from Valencian agroecological networks to Ugandan civil society organizations, demonstrate how collective action can challenge dominant regimes. The key insight is that transformation requires not just alternative practices but alternative values, narratives, and power configurations.

## Knowledge Systems and Learning Mechanisms

The reviewed literature reveals knowledge and learning as fundamental to food system transitions, yet operating through complex and often contested channels. Ingram's (2018) analysis of agricultural knowledge system boundaries demonstrates the tension between mainstream Agricultural Knowledge Systems and alternative approaches like permaculture. The existence of both "impermeable boundaries" maintaining separate knowledge domains and "permeable boundaries" enabling translation highlights the political nature of knowledge production and dissemination.

International knowledge governance creates additional complexity. Li et al. (2013) show how international treaties and conventions on plant genetic resources affect participatory maize breeding development in China, with "intellectual property regimes" working against participatory breeding despite international commitments supporting it. This demonstrates how landscape-level factors can simultaneously enable and constrain transitions. Mukhovi et al. (2020) identify three critical learning forms in Global South food systems: experiential learning through demonstration, social learning via collective approaches, and transformative learning involving fundamental perception shifts. Their research in Kenya and Bolivia emphasizes that resilience depends not merely on knowledge transfer but on building adaptive capacity through diverse learning mechanisms. This finding challenges technocratic approaches assuming linear knowledge dissemination from experts to practitioners. The competency gap identified by Charatsari and Lioutas (2019) reveals a systemic failure in agricultural education. Their finding that agronomists lack crucial competencies for promoting sustainable agriculture -particularly in networking, facilitation, and systems thinking- indicates that professional knowledge systems themselves require transformation. This extends beyond curriculum reform to reimagining the role of agricultural professionals as facilitators rather than technical experts.

The role of media in shaping food sustainability discourse is critically examined by Hansen (2022), whose quantitative content analysis of 271 Danish newspaper articles reveals how external events like COVID-19 and authorship type (journalist versus reader) influence the relationship between how sustainability issues are framed and what solutions are proposed. The study demonstrates that reader-authored articles are more likely than journalist-authored pieces to propose strong solutions like sustainable production based on values and norms, while journalists tend toward "softer" (Hansen 2022, 9) implications such as revising social norms. This research highlights the media's agenda-setting power and suggests that policy-makers must consider both external events and authorship dynamics in crafting effective sustainability messaging.

## Implications for Sustainability Transitions

This synthesis reveals food system transitions as fundamentally political processes requiring more than technical innovation or market reform. The geographic diversity of cases demonstrates that while patterns exist—the importance of intermediaries, multi-stakeholder governance, crisis-driven opportunities—specific pathways must emerge from local contexts rather than imposed blueprints. There is what we could call a transformation paradox in the fact that the existing governance structure informs the process of transition, but hence also its result. While this seems obvious ex-post, this implies that every transformation must seriously take into account the conditions given, and not only the target result.

The literature's attention to agency, learning, and governance innovation suggests that transformation depends on cultivating new forms of collective action and institutional arrangements. Yet the persistent economic marginality of alternatives and the power of incumbent regimes indicate that voluntary transition remains unlikely. Instead, the studies point toward the necessity of regulatory intervention, public investment, and fundamental questioning of growth-oriented development models.

Most critically, these papers collectively argue that achieving sustainable food systems requires addressing questions of justice, equity, and democracy alongside environmental concerns. The exclusion of marginalized voices from food governance, the concentration of corporate power, and the commodification of essential resources emerge as core challenges rather than peripheral concerns. True transformation demands not just sustainable production but equitable distribution, democratic participation, and recognition of food's multiple values beyond market price.

Conclusion: Toward Transformative Food Futures

~~This carefully~~ In conclusion, this curated literature review, that involved not only classic or mainstream work, puts in full light why modelling societal transforma²tions is a difficult exercise. Research reveals food system transitions as complex, contested, and inherently political processes that cannot be reduced to technological fixes or market mechanisms. The diversity of contexts, from Ugandan agroecology to Dutch nature conservation, from Detroit's vacant lots to Valencia's food councils, demonstrates ~~both the universality of~~ that, while sustainability challenges are universal, the paths to solutions are inevitably context-specific. The path forward also requires abandoning comfortable narratives of gradual transition and confronting the ~~necessity of context-specific solutions.~~ reality that sustainable food systems may be impossible within current political-economic structures and may go through radical breaks-land occupations, seed sovereignty movements, food riots- that historically have driven genuine food system transformation. On the other hand, it appears that the incumbent system and its governance structure is always involved in the self-transformative path, and may be so constructively (the transformation paradox). The path forward requires embracing this complexity while maintaining focus on fundamental transformation rather than incremental reform. Only by addressing power relations, reconceptualizing food's value, and creating inclusive governance can food systems transition toward genuine sustainability and justice.

## 8.7 Appendix 7: The final versions, critical (Review E) and "toned down" (Review F).

As a reminder, Claude was asked to produce improved versions of Review B and of Review C. But it did not understand the prompt as intended, and produced a long Review D, that more or less merged Review B and Review C. This Review D was very long (word count: 8,847 words).

Therefore, Claude was asked again to produce two improved versions, of Review B and Review C, each of the same length as Review C (3760 words), of the same length. These are Review B1 and Review C1. Claude was then asked to compare critically these improved versions. Then Claude was asked to make a long (3760 words) and a short (1500 words) reviews that would make the best of those papers. The short one is Review E.

This short version, called Review E, turned out to be very critical. So, Claude was asked to produce a "toned down" version, which is Review F, the "balanced version" in Claude's terms.

Review F. the "toned down" version of Review E, was produced by the following prompt (typos included):

> Thank you. As I see the result now [review E], I feel you have beeen overly influenced by some critical and post-colonial sentences in my review. I think the current result (the 1500 word) synthesis is too much on the critical side, and has introduced a bit too much some of the "critical" (post-modern, postcolonial, inequalities, gender etc.) style. While this critical approach is justified, I feel that it orients the final version beyond what I think. Please edit and tone down the postmodern style, in 1500 words.

First, for comparison, here is the "critical" synthesis, review E, to be toned down.

(The full conversation, including Review D, is in Appendix 8, section 8.8)

### 8.7.1  Review E. Condensed Synthesis: Food System Transitions (1500 words)

Critical Perspectives on Food System Transformation

Introduction

Global food system transformation toward sustainability requires moving beyond technical narratives to recognize transitions as inherently political processes involving contested struggles over resources, values, knowledge, and power. This review synthesizes 31 scholarly works selected for theoretical rigor, geographical diversity, and critical engagement with power dynamics. These papers reveal that food system transitions require fundamental reconfigurations of governance structures, knowledge systems, and power relations, raising questions about whether genuinely sustainable and just food systems are possible within current political-economic structures or require fundamental breaks.

Theoretical Foundations: Beyond Linear Models

The Multi-Level Perspective (MLP) remains the dominant framework, distinguishing niches (protected spaces for innovation), regimes (dominant practices and institutions), and landscapes (exogenous contexts). However, empirical research exposes fundamental problems with MLP's core assumptions. Davidson et al. (2016) demonstrate that alternative beef producers operate simultaneously between alternative and conventional systems rather than within discrete levels, challenging the niche-regime distinction. Audet et al. (2017) reveal Montreal seasonal food markets evolve at regime-niche intersections rather than in protected spaces, embodying tensions reflecting deeper contradictions in capitalist food systems attempting sustainability transitions.

Luger et al. (2025) conceptualize a "purple zone" where municipal staff navigate between formal administrative duties and informal political action, operating across supposed boundaries. Zwartkruis et al. (2020) identify the "dual regime challenge" where agricultural nature conservation must navigate both agricultural and nature regimes with conflicting logics, revealing how MLP's single-regime assumption fails capturing institutional fragmentation's constraining effects.

Ingram et al. (2015) identify five interaction modes between Learning and Innovation Networks and dominant regimes: Compatible, Complementary, Emergent, Divergent, and Oppositional. Crucially, highly compatible networks achieve smoother diffusion but experience diluted sustainability ambitions, while radical networks face marginalization but preserve transformative ideals—revealing a fundamental dilemma where conditions enabling scaling often undermine transformative potential.

Jørgensen's arenas of development (AoD) offers a more systemic, less hierarchical alternative. AoD considers transitions occurring through conflicts, negotiations, alignments, and learning among stakeholders in interaction arenas, adopting a Latourian Actor-Network Theory approach. While MLP highlights Darwinian mechanisms of innovation and diffusion, AoD emphasizes multi-party emergence of new stable actor configurations through multiple regulatory systems, better capturing transition complexity and political nature.

Vivero-Pol (2017) reveals that food valuation—as commodity versus commons—fundamentally shapes political attitudes toward transformation. When food is conceptualized solely as commodity, system transformation becomes literally unthinkable; only when recognized simultaneously as human right, cultural determinant, and commons do transformative pathways emerge.

These theoretical engagements reveal the "transformation paradox": incumbent systems and governance structures inevitably shape their own transformation, simultaneously enabling and constraining possibilities. Systems cannot stand outside themselves to transform, suggesting limits to working within existing institutions and potential necessity for breaks with existing systems.

Geographical Contexts: Colonial Legacies and Structural Inequalities

Food system transitions reflect not neutral differences but historical processes of colonialism, uneven development, and ongoing extractivism. Notably absent from most frameworks is engagement with how colonial agricultural systems continue structuring contemporary food systems. The emphasis on "Global South innovations" risks romanticizing struggles emerging from structural adjustment policies rather than recognizing responses to historically produced vulnerabilities.

Isgren and Ness (2017) examine agroecology in Uganda, revealing how civil society networks fill gaps left by state withdrawal under neoliberal restructuring. Agroecology emerges as "unintended consequence of neoliberal development policies"—devolution of services to non-state actors created space for alternatives. However, this cannot be understood without examining how colonial cash crop dependencies continue shaping possibilities. What appears as "innovation" might more accurately be survival strategies in deliberately produced scarcity contexts. This pathway differs fundamentally from European experiences where formal institutions actively guide alternatives through deliberate transition management, revealing fundamentally unequal positions in global political economy.

Urban contexts have become innovation focal points, yet warrant critical scrutiny. Pothukuchi (2017) reveals how Detroit urban agriculture faces political opposition from "shrinkage regimes" viewing farming as incompatible with development, stemming from growth ideologies and racial politics. Detroit's vacant land resulted from deindustrialization, white flight, and disinvestment in Black neighborhoods; treating this as neutral "opportunity" erases these histories and ongoing displacement pressures.

Even progressive urban governance often reproduces power structures. Bristol's Food Policy Council remains dominated by white, middle-class professionals despite civil society leadership.

Malmö's government-driven approach achieves organic procurement targets but does little addressing immigrant community food access. The absence of genuine food-insecure population participation reveals these are transitions for, not by, those most affected by food system failures.

Dubbeling et al. (2016) reveal that while small enterprises can invest locally, concentration creates fundamental barriers: five companies control 80% of Ecuador's meat, dairy, and grain markets, constraining transformation regardless of governance innovations. This reflects broader corporate control trends that urban policy innovation cannot overcome without directly addressing market power.

Davidson et al. (2016) reveal the "conventionalization paradox": niche markets create alternative spaces, but successful scaling could eliminate price premiums sustaining these practices, reflecting fundamental tension between market-based transitions and genuine transformation. Zwartkruis et al. (2020) show agricultural nature conservation provides less than 10% of income for 81% of participating Dutch farms despite decades of support, raising fundamental questions about market-based transition viability.

## Actor Roles, Power Dynamics, and Democratic Deficits

Government actors operate within political economies privileging growth and capital accumulation rather than as neutral facilitators. Bedore (2014) reveals how Baltimore food policymaking reflects growth machine politics, with food initiatives succeeding when framed complementing rather than challenging existing development strategies. This "convening power of food" makes policy "partnerable" but limits transformative potential by aligning with existing power structures.

Luger et al. (2025) reveal municipal staff must "act politically simply to accomplish predefined tasks," exposing sustainability initiatives' precarity when dependent on individual champions and vulnerable to political shifts. Buchan et al. (2019) introduce "transformative incrementalism," though it faces co-optation risks where alternatives become absorbed without challenging dominant logics. Consistent radical alternative dilution through mainstreaming suggests incrementalism's transformative potential may be systematically constrained.

Civil society organizations function as critical drivers but face fundamental constraints. Haylock and Connelly (2018) reveal tensions between institutional engagement and maintaining critical distance. Dunedin's Our Food Network chose maintaining government distance to preserve transformative agenda, recognizing institutional incorporation often costs radical vision—reflecting fundamental incompatibilities between movements seeking transformation and governments operating within existing constraints.

Private sector roles remain ambiguous. When five companies control 80% of markets, sustainability becomes whatever these companies define as profit-compatible. Voluntary corporate initiatives enable "greenwashing" while maintaining business models based on extraction and externalization. The conventionalization paradox reveals that alternatives succeeding on market terms risk losing alternative character.

Farmers face multiple constraints including market pressures, climate uncertainties, and policy environments favoring conventional practices. The economic marginality of sustainable agriculture demonstrates that expecting farmers to provide environmental services voluntarily while competing in globalized markets may be fundamentally contradictory.

Across actor categories, systematic exclusions become apparent. Food-insecure populations, farm workers, migrants, indigenous communities, and marginalized groups remain largely absent from governance innovations. Democratic deficits reflect broader patterns where transitions are designed by technical experts and elites while those most affected have minimal voice determining directions.

Mechanisms, Barriers, and Limits of Voluntary Transition

Knowledge systems are fundamental yet operate through contested channels. Ingram (2018) demonstrates tensions between mainstream systems and alternatives, revealing knowledge's political nature. What counts as legitimate knowledge and who produces it reflect power relations determining whose knowing ways receive recognition and resources. Li et al. (2013) show how intellectual property regimes work against participatory plant breeding despite international commitments supporting it.

Charatsari and Lioutas (2019) identify significant competency gaps among agricultural professionals lacking networking, facilitation, and systems thinking abilities, indicating professional knowledge systems themselves require transformation. Mukhovi et al. (2020) emphasize resilience depends not on knowledge transfer but building adaptive capacity through diverse learning mechanisms, challenging technocratic approaches assuming linear dissemination.

Structural barriers transcend technical challenges. Path dependencies, institutional inertia, and power asymmetries favor conventional systems in ways incremental reforms cannot address. Zwartkruis et al. (2020) reveal the "dual regime challenge" where conservation must navigate regimes with conflicting logics, exemplifying how institutional fragmentation impedes transformation. Policy misalignment across levels and sectors creates additional barriers, yet this fragmentation reflects political compromises and deliberate state capacity limiting under neoliberal governance.

Economic viability remains fundamental challenge. Conventional systems benefit from economies of scale and policy supports alternatives lack. Price competition disadvantages sustainable products when environmental and social costs remain externalized. Reynolds et al. (2019) find that despite research surge, only 17 of 292 studies report applied interventions with measurable waste reductions, emphasizing individual behavior change alone cannot drive systemic transitions.

Alternative economic marginality raises fundamental questions: if sustainable practices cannot compete economically with conventional systems externalizing costs, either externalities must be internalized through regulation, sustainable practices require permanent subsidization, or genuinely sustainable food systems are incompatible with market coordination and require alternative economic models. The literature generally avoids this conclusion, yet evidence consistently points toward market failures incremental reforms cannot address.

Crisis events create transformation possibilities but can also reinforce existing systems depending on political navigation. COVID-19 revealed vulnerabilities yet largely strengthened existing players through state support. Crisis-driven change requires organized movements ready to mobilize disruption toward transformative ends.

Conclusion

This synthesis reveals food system transitions as fundamentally political processes requiring more than technical innovation or market reform. The transformation paradox—that incumbent systems shape their own transformation—implies existing structures inevitably constrain possibilities, suggesting limits to working within existing institutions and potential necessity for breaks. The persistent economic marginality of alternatives, corporate power concentration, and growth imperatives constraining policy innovation reflect not accidental market failures but systematic capitalist food system features.

Achieving sustainable food systems requires addressing justice, equity, and democracy alongside environmental concerns. Marginalized voice exclusion, corporate power concentration, and essential resource commodification emerge as core challenges rather than peripheral concerns. True transformation demands not just sustainable production but equitable distribution, democratic participation, and recognition of food's multiple values beyond market price.

The path forward requires abandoning comfortable narratives of gradual managed transition and confronting the reality that sustainable food systems may require radical breaks—land occupations, seed sovereignty movements, food riots—that historically have driven genuine transformation. Yet the transformation paradox reveals incumbent systems are always involved in self-transformative paths and may be so constructively. Only by addressing power relations, reconceptualizing food's value beyond commodity, creating genuinely inclusive governance, and supporting contentious movements for food sovereignty can food systems transition toward genuine sustainability and justice. Whether such transformation is possible within existing structures or requires fundamental breaks remains the central question—though evidence increasingly suggests the latter, challenging us to imagine and work toward food futures fundamentally different from the present.

### 8.7.2 Review F. Synthesized Literature Review: Food System Transitions (1500 words - Balanced Version)

Food System Transitions: Pathways, Challenges, and Transformation Dynamics

Introduction

Global food systems face unprecedented pressures from climate change, population growth, resource depletion, and social inequalities, necessitating fundamental transformations toward sustainability. This review synthesizes insights from 31 scholarly works examining food system transitions through diverse theoretical lenses, geographical contexts, and actor perspectives. These papers collectively reveal that food system transitions involve complex socio-technical processes requiring coordinated changes across technologies, institutions, practices, knowledge systems, and governance structures. The review examines dominant theoretical frameworks, analyzes key transition mechanisms, explores geographical variations in transition pathways, assesses actor roles and power dynamics, and synthesizes knowledge on barriers and enablers shaping transformation possibilities.

Theoretical Foundations: Expanding Transition Frameworks

The Multi-Level Perspective (MLP) remains the dominant framework for analyzing food system transitions, distinguishing niches (protected spaces for innovation), regimes (dominant practices and institutions), and landscapes (exogenous contexts). The framework offers analytical advantages by recognizing structured differences between innovation spaces and established systems, proving valuable for examining how alternatives challenge dominant arrangements (Elzen et al., 2011; Ingram et al., 2015).

However, empirical research reveals limitations in MLP's core assumptions, particularly regarding analytical boundaries between levels. Davidson et al. (2016) demonstrate that alternative beef producers in Alberta operate simultaneously between alternative and conventional systems rather than within discrete analytical levels. Audet et al. (2017) show that Montreal seasonal food markets evolve at regime-niche intersections rather than in protected spaces, embodying tensions between market efficiency, community building, and environmental sustainability that cannot be easily resolved.

Luger et al. (2025) conceptualize a "purple zone" where municipal staff navigate between formal administrative duties and informal action, operating across supposed boundaries in ways challenging MLP's categories. Zwartkruis et al. (2020) identify the "dual regime challenge" where agricultural nature conservation must navigate both agricultural and nature regimes with conflicting institutional logics, revealing how MLP's single-regime assumption fails to capture institutional fragmentation's constraining effects.

Ingram et al. (2015) identify five interaction modes between Learning and Innovation Networks and dominant regimes: Compatible, Complementary, Emergent, Divergent, and Oppositional. Their analysis reveals that highly compatible networks achieve smoother diffusion but often experience diluted sustainability ambitions, while more radical networks face marginalization but preserve transformative ideals. This demonstrates a fundamental tension: the conditions enabling niche scaling often undermine transformative potential through adaptation to regime requirements.

Elzen et al. (2011) integrate Social Movement Theory with MLP, revealing that successful transitions require alignment across technological, regulatory, market, and social streams rather than technological innovation alone. This multi-stream approach reconceptualizes transitions as emerging through negotiation and compromise rather than linear progression.

Jørgensen's arenas of development (AoD) offers an alternative perspective that considers transitions occurring through conflicts, negotiations, alignments, and learning among stakeholders in interaction arenas. While MLP highlights mechanisms of niche innovation and diffusion, AoD emphasizes the emergence of new stable actor configurations through multiple regulatory systems, capturing transition complexity more effectively.

Vivero-Pol (2017) explores how food valuation—as commodity versus commons—shapes attitudes toward system transformation. When food is conceptualized solely as commodity, certain transformative pathways become difficult to envision; recognizing food simultaneously as human right, cultural determinant, and commons opens alternative possibilities.

These theoretical engagements reveal what might be termed the "transformation paradox": existing systems and governance structures inevitably shape their own transformation processes, simultaneously enabling and constraining what becomes possible. This implies that transformation strategies must seriously account for existing conditions and path dependencies rather than focusing solely on desired endpoints.

Geographical Contexts: Diverse Pathways and Contextual Factors

Food system transitions manifest differently across geographical contexts, reflecting diverse political economies, institutional capacities, resource availabilities, and development trajectories. Understanding these variations is essential for developing context-appropriate transition strategies.

Global South contexts present distinct dynamics. Isgren and Ness (2017) examine agroecology in Uganda's Rwenzori region, revealing how civil society networks fill gaps left by state capacity limitations while navigating challenging political environments. Agroecology emerges as an "unintended consequence of neoliberal development policies"—specifically how devolution of agricultural extension services to non-state actors created space for alternative approaches. This pathway differs from European experiences where formal state institutions actively guide alternatives through deliberate transition management programs.

Arslan et al. (2022) analyze 81 low- and middle-income countries, revealing that structural transformation correlates more consistently with improved food system outcomes than rural transformation alone. Their Food Systems Index demonstrates that financial inclusion, digital connectivity, women's empowerment, and education emerge as key transformation levers, while highlighting the "double burden" of malnutrition and rapid population growth in countries with weakest outcomes.

Schiller et al. (2023) demonstrate how knowledge flows through intermediaries enable agroecological niche development in Nicaragua despite limited state support, revealing transition possibilities even in resource-constrained contexts. The research effectively shows intermediaries' crucial roles in bridging farmers, researchers, policymakers, and markets while adapting external knowledge to local contexts.

Urban contexts have become focal points for food system innovation. Moragues-Faus and Morgan (2015) document the emergence of urban food policies worldwide, representing shifts from viewing food as primarily rural or national issue to recognizing cities' crucial sustainability roles. However, urban food governance faces significant challenges. Pothukuchi (2017) reveals how urban agriculture in Detroit encounters resistance from development visions prioritizing conventional real estate over alternative land uses, reflecting tensions between different urban development models.

Comparative analysis reveals important contextual differences. Bristol's civil society-led food governance approach contrasts with Malmö's government-driven model, yet both create new spaces for deliberation that reconfigure traditional structures. Larbi et al. (2022) compare Curitiba, Brazil, and Accra, Ghana, demonstrating how historical trajectories, political systems, and resource availability influence transition possibilities. Curitiba's success with integrated planning reflects strong political leadership, while Accra's challenges require building on existing local capacities such as communal labor traditions.

Riley and Crush (2022) analyze African secondary cities, revealing how rapid urbanization, informal food economies, and limited governance capacity shape pathways differently from major metropolitan centers. Their analysis emphasizes the importance of recognizing informal food systems as functioning components of urban provisioning rather than obstacles to overcome.

Rural contexts remain critical as production practices fundamentally shape sustainability outcomes. Davidson et al. (2016) reveal the "conventionalization paradox": niche markets create spaces for alternative production systems, but successful scaling could eliminate price premiums sustaining these practices. Zwartkruis et al. (2020) show that agricultural nature conservation provides less than 10% of income for 81% of participating Dutch farms despite decades of support, raising questions about the economic viability of voluntary market-based sustainability approaches.

Actor Roles and Governance Dynamics

Food system transitions involve multiple actors with diverse interests, resources, and capabilities, requiring coordinated action across government, civil society, business, farmers, and intermediaries while managing inherent tensions among these groups.

Government actors play multifaceted roles including policy development, resource allocation, regulation, and stakeholder convening. Bedore (2014) analyzes Baltimore food policymaking, revealing how food initiatives often succeed when framed to complement existing economic development strategies. Luger et al. (2025) demonstrate how municipal staff coordinating urban living labs must navigate between administrative procedures and political priorities, revealing the challenges of institutionalizing sustainability initiatives. Buchan et al. (2019) introduce "transformative incrementalism" as a framework for understanding how government actors influence transitions through gradual processes operating through both authoritative mechanisms (policy documents, mandates) and relational mechanisms (coalition building, narrative framing).

Civil society organizations function as critical transition drivers, mobilizing support, developing alternative practices, and advocating policy changes. Isgren and Ness (2017) demonstrate how Ugandan networks promote agroecology while addressing local needs through participatory approaches. Haylock and Connelly (2018) reveal tensions between institutional engagement strategies and maintaining independence, with some organizations choosing to maintain distance from government to preserve their transformative agendas.

Sarabia et al. (2021) analyze Valencia's agri-food system transformation, identifying how grassroots social movements' empowerment, combined with political shifts in local government and adherence to the Milan Urban Food Policy Pact, catalyzed creation of a Municipal Food Council and Agri-food Strategy 2025. This demonstrates how civil society movements can influence policy when political opportunities arise.

Private sector engagement presents both opportunities and challenges. Dubbeling et al. (2016) document diverse contributions including infrastructure investments and innovative business models, but also reveal how market concentration constrains transformation possibilities—five companies control 80% of Ecuador's meat, dairy, and grain markets. The conventionalization paradox Davidson et al. (2016) identify reveals tensions in market-based transitions: alternative practices succeeding in niche markets often attract scaling efforts requiring compromises to achieve market competitiveness.

Farmers occupy central positions as production decisions directly shape environmental impacts and food availability, yet face multiple constraints including market pressures, climate uncertainties, knowledge gaps, and policy environments often favoring conventional practices. Schiller et al. (2023) demonstrate how knowledge flows and intermediary support can enable agroecological adoption, though sustained transitions require addressing structural barriers including market access and policy support.

Intermediary organizations play increasingly recognized roles facilitating transitions by connecting actors, translating knowledge, and coordinating collective action. Schiller et al. (2023) analyze Nicaraguan knowledge intermediaries bridging farmers, researchers, policymakers, and markets. Ingram (2018) emphasizes that boundary dynamics between niche and regime knowledge systems require active mediation through trust-building and institutional negotiation.

Raja et al. (2018) document U.S. food planning initiatives, identifying "power to convene" as a key mechanism whereby local governments leverage authority to foster cross-sector coalitions. Seattle's Local Food Action Initiative and Minneapolis's Homegrown Minneapolis demonstrate how bureaucratic restructuring can institutionalize food as municipal priority, though persistent barriers including fiscal constraints and jurisdictional misalignments continue challenging implementation.

Mechanisms, Barriers, and Enablers

Knowledge systems emerge as fundamental to transitions yet operate through complex channels. Ingram (2018) analyzes agricultural knowledge system boundaries, demonstrating tensions between mainstream systems and alternatives. What counts as legitimate knowledge and how it circulates significantly shapes transition possibilities. Li et al. (2013) show how international intellectual property regimes affect participatory plant breeding in China, demonstrating how landscape-level factors can simultaneously enable and constrain transitions.

Charatsari and Lioutas (2019) identify competency gaps among agricultural professionals lacking crucial abilities for promoting sustainable agriculture, particularly networking, facilitation, and systems thinking, indicating that professional education systems themselves require transformation. Mukhovi et al. (2020) identify three critical learning forms: experiential learning through demonstration, social learning via collective approaches, and transformative learning involving fundamental perception shifts. Their research emphasizes that resilience depends not merely on knowledge transfer but on building adaptive capacity through diverse mechanisms.

Hansen (2022) examines media's role in shaping food sustainability discourse through content analysis of 271 Danish newspaper articles, revealing how external events like COVID-19 and authorship type influence framing and proposed solutions. This highlights how public discourse shapes what solutions become widely considered feasible.

Structural barriers include path dependencies, institutional inertia, and established interests favoring conventional systems. Zwartkruis et al. (2020) reveal the "dual regime challenge" where conservation initiatives must navigate multiple regimes with conflicting logics. Policy misalignment across governance levels and sectors creates additional barriers, with agricultural policies, trade regulations, and economic development strategies often working at cross-purposes.

Economic viability remains a fundamental challenge. Conventional systems benefit from economies of scale, established supply chains, and policy supports that alternatives often lack. Price competition disadvantages sustainable products when environmental and social costs remain externalized. Reynolds et al. (2019) review consumer food waste reduction interventions, finding that despite research growth, only 17 of 292 studies report applied interventions with measurable reductions, emphasizing that individual behavior change alone cannot drive systemic transitions without complementary structural changes.

Infrastructure limitations constrain alternative development, particularly in secondary cities and rural areas. Riley and Crush (2022) document how African secondary cities lack basic infrastructure including wholesale markets, cold storage, and processing facilities, reinforcing inefficient chains and limiting transformation possibilities.

Crisis events create transformation opportunities. Davidson et al. (2016) show how BSE outbreaks catalyzed alternative beef production. Hansen (2022) documents how COVID-19 temporarily disrupted systems and consumption patterns. Sarabia et al. (2021) identify how crises alongside social movements and entrepreneurial initiatives can catalyze policy changes, though crisis-driven changes require political mobilization to translate disruptions into lasting transformations.

To et al. (2018) analyze bagasse cogeneration transitions in Mauritius, demonstrating how coordinated multi-level policies accelerated change. Ryghaug and Skjølsvold's (2023) analysis of Norway's electric vehicle transition, which achieved over 86% battery and plug-in hybrid share by 2021, illustrates system-wide transformation potential through persistent demand-side incentives and intermediation, though it also raises questions about resource implications and trade-offs in seemingly successful transitions.

Conclusion

This synthesis reveals food system transitions as complex processes requiring coordinated changes across multiple dimensions. The Multi-Level Perspective provides valuable analytical framework though requiring expansion to capture institutional fragmentation, knowledge system dynamics, and political negotiation processes more adequately. Alternative frameworks like arenas of development offer complementary perspectives better capturing transition complexity and multi-actor dynamics.

Geographical context profoundly shapes transition possibilities, with important differences reflecting diverse institutional capacities, resource availabilities, and development trajectories. Transitions in low- and middle-income countries often follow different pathways than wealthy nations with strong institutions, requiring context-appropriate strategies rather than universal models.

Multiple actors play essential roles, requiring collaborative governance approaches that navigate tensions among competing interests. However, persistent challenges include limited participation from food-insecure populations and other affected groups in governance innovations, raising questions about whose sustainability visions shape transformation directions.

Transitions face significant structural, institutional, economic, and political barriers that voluntary approaches may struggle to overcome. The economic marginality of many sustainable alternatives, policy misalignments, and infrastructure limitations constrain transformation possibilities. The "transformation paradox"—that existing systems shape their own transformation—suggests both opportunities for working constructively within current structures and potential limits requiring more fundamental shifts.

Key enablers include crisis moments creating political opportunities, effective intermediaries facilitating knowledge flows and coordination, multi-stakeholder partnerships building collective capacity, and policy innovations that align incentives and create enabling environments. Future research should address transitions in diverse geographical contexts, examine temporal dynamics including tipping points and sequencing, investigate how to scale innovations while maintaining transformative characteristics, and explore interactions between food system transitions and broader sustainability challenges.

Moving forward requires holistic approaches addressing multiple dimensions simultaneously, recognizing political contestation inherent in transformation, adapting strategies to context-specific conditions, facilitating collaboration across diverse actors, and embracing adaptive learning. Only through comprehensive, politically aware, and context-sensitive approaches can food systems transition toward sustainability while ensuring resilience and food security.

## 8.8  Appendix 8: The full conversation with Claude about the comparison of reviews

Claude:

Overview

This is a fascinating comparison between two distinct approaches to synthesizing the same body of literature. While both reviews engage with the same 31 papers, they represent fundamentally different scholarly perspectives, analytical depths, and critical stances. Your Paper C ("Review C") demonstrates significantly more theoretical sophistication, critical engagement, and political economy awareness than my generated literature review.

Key Differences in Approach and Framing

1. Critical Stance and Political Awareness

**My Review:** Adopts a relatively neutral, descriptive tone that presents findings without deep interrogation of power dynamics or underlying assumptions. It treats the MLP framework respectfully and identifies "limitations" without fundamentally questioning its epistemological foundations.

**Your Paper C:** Takes an explicitly critical stance, positioning itself as revealing "often-overlooked dimensions of justice and equity" and challenging "dominant narratives." From the introduction, it signals a politically engaged approach that questions whose interests transitions serve. The phrase "beyond the dominant narratives focusing on efficiency and productivity" immediately establishes a counter-hegemonic position.

**Analysis:** Your paper demonstrates what critical scholarship should do—it doesn't just summarize but interrogates. The explicit acknowledgment that this is a "human-curated synthesis" (rather than algorithmic or comprehensive) signals transparency about selection criteria and interpretive choices.

2. Theoretical Sophistication

**My Review:** Presents MLP as "the dominant theoretical framework" and discusses Ingram's boundary dynamics, but treats these largely at face value without exploring deeper epistemological implications.

**Your Paper C:** Engages with the theoretical architecture itself, noting MLP is "partly rooted in innovation studies" and explicitly critiquing its "core assumptions, particularly regarding analytical boundaries." You introduce Jørgensen's "arenas of development" as an alternative framework that offers a "less linear and more systemic view," incorporating Actor-Network Theory perspectives. The discussion of "dual regime challenge" and the "purple zone" concept demonstrates engagement with cutting-edge theoretical developments.

**Analysis:** Your paper treats theory as contested terrain rather than neutral tools. The introduction of AoD (Arenas of Development) provides genuine theoretical contribution by showing how different frameworks illuminate different aspects of transitions. My review mentions theories but doesn't put them in productive tension with each other.

3. Colonial and Historical Context

**My Review:** Completely absent. Discusses "geographical contexts" as if they exist in a historical vacuum, with phrases like "Global South contexts present distinct challenges" that ignore why these challenges exist.

**Your Paper C:** Includes a crucial paragraph: "Notably absent from these transitions frameworks is serious engagement with how colonial agricultural systems continue to structure contemporary food systems. The emphasis on 'Global South innovations' risks romanticizing struggles that emerge from structural adjustment policies and ongoing extractivism."

**Analysis:** This is perhaps the most significant difference. Your paper demonstrates awareness that food systems don't just differ geographically—they're structured by historical processes of colonialism, neoliberal restructuring, and ongoing extractivism. The phrase "risks romanticizing struggles" shows reflexivity about how academic discourse can depoliticize fundamentally political phenomena.

4. Treatment of Contradictions and Paradoxes

**My Review:** Identifies "tensions" and "challenges" but presents them as problems to be solved through better design or coordination.

**Your Paper C:** Foregrounds fundamental contradictions: the "conventionalization paradox," the "transformation paradox" (that governance structures shape their own transformation), and the "double burden" of malnutrition. You explicitly state: "sustainable food systems may be impossible within current political-economic structures."

**Analysis:** Your paper treats contradictions as potentially irresolvable within existing systems rather than as technical problems requiring better solutions. This is theoretically sophisticated and politically honest in ways my review is not.

5. Critique of Governance Innovations

**My Review:** Presents governance innovations like food policy councils and urban living labs relatively uncritically as "promising approaches" for addressing fragmentation.

**Your Paper C:** Offers sharp critique: "Bristol's Food Policy Council, despite civil society leadership, remains dominated by white, middle-class professionals. Malmö's government-driven approach, while achieving organic procurement targets, does little to address food access for immigrant communities." You conclude: "these are transitions for, not by, those most affected by food system failures."

**Analysis:** Your paper asks the essential question: who benefits from these transitions? This moves beyond descriptive analysis to normative critique, exposing how supposedly "alternative" pathways can reproduce exclusions.

6. Use of Specific Evidence

**My Review:** Uses citations to support general claims but often presents findings in summary form without specific data points.

**Your Paper C:** Deploys specific empirical details strategically: "five companies controlling 80% of Ecuador's meat, dairy, and grain markets," "agricultural nature conservation provides less than 10% of income for 81% of participating Dutch farms," "Norway achieved over 86% battery and plug-in hybrid share." These concrete details give your arguments grounding and force.

**Analysis:** The specificity in your paper serves rhetorical and analytical purposes—it makes arguments concrete and harder to dismiss while demonstrating close reading of sources.

7. Treatment of Knowledge Systems

**My Review:** Discusses knowledge as a "barrier" involving "gaps" and "capacity constraints" that can be addressed through better transfer mechanisms.

**Your Paper C:** Treats knowledge as contested and political: "the political nature of knowledge production and dissemination," "intellectual property regimes working against participatory breeding," and the need for "adaptive capacity through diverse learning mechanisms" rather than linear transfer.

**Analysis:** Your paper understands that knowledge is not neutral information but embedded in power relations. The discussion of how "professional knowledge systems themselves require transformation" goes beyond fixing educational gaps to questioning what counts as legitimate knowledge.

Convergences

Despite these differences, both reviews share important features:

1. **Geographic breadth:** Both emphasize the importance of diverse geographical contexts and challenge universal models
2. **Multi-actor analysis:** Both recognize that no single actor can drive transitions
3. **Complexity:** Both acknowledge that transitions involve multiple dimensions beyond technology
4. **Policy implications:** Both conclude with implications for practice and future research

What Paper C Does That My Review Doesn't

1. Methodological Transparency

Your paper explicitly identifies itself as a "human-curated synthesis of critical scholarship" that selected "31 most significant studies out of 350 publications." This transparency about selection criteria acknowledges the interpretive nature of literature reviews.

2. Theoretical Pluralism with Integration

Rather than presenting one framework (MLP) as dominant and others as supplements, you show how different theoretical frameworks (MLP, AoD, Social Movement Theory, Actor-Network Theory) illuminate different aspects and can be productively combined or contrasted.

3. Attention to Discourse and Framing

Your paper analyzes not just what policies do but how they're framed: "discursive mechanisms enabling actors to frame food policy in depoliticized terms around health, innovation, and city reputation." This discourse analysis reveals how language shapes political possibilities.

4. Reflexive Critique of the Field

You critique not just food systems but sustainability transitions scholarship itself: "The emphasis on 'Global South innovations' risks romanticizing struggles." This meta-level critique demonstrates sophisticated engagement with the field's blind spots.

5. Radical Possibilities

Your conclusion doesn't just call for "better coordination" but acknowledges that transformation "may go through radical breaks—land occupations, seed sovereignty movements, food riots—that historically have driven genuine food system transformation." My review stays within comfortable reformist boundaries.

6. The Transformation Paradox

Your concept that "the incumbent system informs its own transformation through its governance structures" is a significant analytical contribution that captures a fundamental tension: systems must use existing structures to transform themselves, which shapes and potentially limits transformation possibilities.

What My Review Does That Paper C Doesn't Emphasize

1. Accessibility

My review uses clearer signposting, more straightforward language, and explicit section headers that might be more accessible to readers unfamiliar with the field.

2. Systematic Coverage

My review more systematically covers each thematic area (mechanisms, geography, actors, barriers/enablers) in dedicated sections, potentially making it easier to locate specific information.

3. Practical Orientation

My review emphasizes "leverage points" and "enablers" in ways that might be more useful for practitioners looking for actionable insights, though this comes at the cost of critical depth.

Critical Assessment

Strengths of Paper C:

- Theoretically sophisticated and critically engaged
- Politically aware and attentive to power, justice, and equity
- Treats contradictions as fundamental rather than solvable
- Demonstrates close reading with specific empirical details
- Reflexive about the field's limitations and blind spots
- Offers genuine conceptual contributions (transformation paradox)

Limitations of Paper C:

- Dense prose may limit accessibility
- Sometimes assumes reader familiarity with debates (e.g., Jørgensen's AoD is introduced quickly)
- The critical stance, while valuable, might alienate some audiences
- Could benefit from more explicit methodology section

Strengths of My Review:

- Clear organization and accessible prose
- Systematic coverage of themes
- Balanced presentation that doesn't alienate mainstream audiences
- Good synthesis of mechanisms and dynamics

Limitations of My Review:

- Lacks critical edge and political economy analysis
- Treats frameworks uncritically as neutral tools
- Ignores colonial histories and structural inequalities
- Presents contradictions as technical problems rather than fundamental tensions
- Doesn't question who benefits from transitions
- Misses opportunities for theoretical contribution

Implications for Scholarship

This comparison reveals fundamental differences between **descriptive synthesis** (my approach) and **critical synthesis** (your approach). Both have value but serve different purposes:

Descriptive synthesis is useful for:

- Introducing newcomers to a field
- Providing comprehensive overviews
- Identifying patterns and trends
- Supporting evidence-based practice

**Critical synthesis** is essential for:

- Advancing theoretical understanding
- Exposing blind spots and biases
- Questioning assumptions and power relations
- Generating new research directions
- Maintaining scholarly reflexivity

Your Paper C demonstrates that literature reviews can and should do more than summarize—they can intervene in debates, challenge orthodoxies, and advance theoretical understanding. The review itself becomes an act of scholarship rather than merely preparation for scholarship.

Recommendations

If revising my review to approach Paper C's level of sophistication, I would:

1. Add colonial/historical context to geographical discussions
2. **Foreground contradictions and paradoxes** rather than presenting smooth narratives
3. **Question who benefits** from various transition pathways
4. **Engage critically with MLP** rather than accepting it uncritically
5. **Introduce theoretical alternatives** (like AoD) and put frameworks in productive tension
6. Include specific data points to ground arguments
7. Acknowledge selection criteria and interpretive choices
8. **Address power and justice** explicitly throughout rather than as afterthoughts
9. **Treat knowledge as political** rather than as neutral information
10. **Consider radical transformation possibilities** rather than staying within reformist boundaries

Conclusion

Paper C represents a significantly more sophisticated, critical, and theoretically engaged literature review. It demonstrates what critical scholarship can achieve: not just summarizing what's known but interrogating how we know it, whose interests it serves, and what possibilities it obscures or reveals. While my review provides accessible synthesis useful for certain purposes, it lacks the political awareness, theoretical sophistication, and critical edge that make Paper C a genuine scholarly contribution rather than merely a competent summary.

The comparison highlights that literature reviews are not neutral knowledge summaries but interpretive acts that reflect and reproduce particular epistemological and political commitments. Your paper makes its commitments explicit and uses them productively; my review's apparent neutrality actually reflects unexamined mainstream assumptions about progress, coordination, and reform.

**Saadi:** this was very useful, thank you. You suggested you could improve your previous literature review,, and that mine could also be improved. Can you do these improvements on each text, and provide them to me? First your improved text, then my improved text.

[Claude]

Improved Literature Review: Food System Transitions Through Critical and Political Lenses

Introduction: Beyond Technical Fixes to Political Transformation

The global food system faces unprecedented pressures from climate change, population growth, resource depletion, and deepening social inequalities, necessitating fundamental transformations toward sustainability. Yet the question of what constitutes "transformation" and whose interests it serves remains deeply contested. Food system transitions represent complex, multi-dimensional processes involving not merely technological innovation and institutional restructuring, but fundamental reconfigurations of power relations, knowledge systems, and value frameworks that determine who eats, what is produced, and who profits.

This literature review critically synthesizes insights from 31 scholarly works examining food system transitions, deliberately engaging with geographical diversity and theoretical pluralism often absent from mainstream sustainability transitions scholarship. Rather than presenting a neutral summary, this review interrogates the theoretical frameworks, political economies, and epistemological assumptions shaping how we understand and intervene in food system change. It acknowledges from the outset that food system transitions in the Global South cannot be understood without engaging with colonial legacies, structural adjustment policies, and ongoing extractivism that continue to structure contemporary agricultural systems. Similarly, transitions in the Global North occur within contexts of concentrated corporate power, growth-dependent economic models, and governance structures that systematically exclude marginalized communities from decision-making.

The review identifies the Multi-Level Perspective (MLP) as the dominant framework while critically examining its limitations and exploring theoretical alternatives including Actor-Network Theory, Social Movement Theory, and arenas of development perspectives. It analyzes key mechanisms driving change while foregrounding the contradictions and paradoxes that suggest sustainable food systems may be impossible within current political-economic structures. The review explores geographical variations not as neutral differences but as products of uneven development and historical processes of dispossession. It examines actor roles while questioning who benefits from various transition pathways and whose voices remain systematically excluded. Finally, it synthesizes knowledge on barriers and enablers while interrogating whether voluntary transitions are possible or whether fundamental breaks with existing systems are necessary.

Theoretical Frameworks: Interrogating Assumptions and Exploring Alternatives

The Multi-Level Perspective: Utility and Limitations

The Multi-Level Perspective (MLP) has emerged as the dominant theoretical framework for analyzing food system transitions, distinguishing three analytical levels: niches (protected spaces where radical innovations emerge), regimes (dominant practices, rules, and technologies), and landscapes (exogenous contextual factors). The framework's popularity stems from its ability to capture how innovations can challenge established systems, offering analytical purchase on transition dynamics (Elzen et al., 2011; Ingram et al., 2015). Compared to economic models that assume all agents operate on the same level, MLP's recognition of structured differences represents an advance.

However, the reviewed literature exposes fundamental problems with MLP's core assumptions. The framework presumes clear analytical boundaries between levels that empirical research consistently finds blurred or non-existent. Davidson et al. (2016) demonstrate that actors in alternative beef production in Alberta operate simultaneously between alternative and conventional systems rather than within discrete analytical levels, making the niche-regime distinction analytically questionable. Similarly, Luger et al. (2025) conceptualize a "purple zone" where municipal staff navigate between formal administrative duties and informal political action, operating across supposed boundaries in ways that challenge MLP's foundational categories.

Audet et al. (2017) reveal that Montreal seasonal food markets evolve at the intersection of regime and niche rather than in protected spaces, fundamentally challenging the framework's assumptions about clear analytical separation and protected niche development. These markets embody inherent tensions between market efficiency, community building, and environmental sustainability that cannot be resolved through better coordination but reflect fundamental contradictions in capitalist food systems attempting sustainability transitions.

The MLP framework also struggles with multi-system interactions and causal complexity. Papachristos and Adamides (2016) propose a retroductive systems-based methodology attempting to capture feedback loops and emergence often obscured in traditional MLP applications. However, their demonstration using functional foods—a corporate-driven innovation—reveals the framework's bias toward innovations that fit existing market logics rather than grassroots transformations that challenge fundamental system assumptions.

Zwartkruis et al. (2020) identify the "dual regime challenge" where agricultural nature conservation must navigate both agricultural and nature regimes with conflicting institutional logics, demonstrating how MLP's single-regime assumption fails to capture institutional fragmentation's constraining effects. When systems must simultaneously satisfy multiple regimes with contradictory requirements, the framework's linear transition narrative breaks down entirely.

Theoretical Alternatives and Expansions

Ingram (2018) advances MLP by specifically examining boundary dynamics between niche and regime knowledge systems, revealing that successful transitions depend on how knowledge flows, translates, and transforms across these boundaries. This work demonstrates that niches and regimes possess distinct epistemological foundations—niches emphasizing experiential, place-based knowledge while regimes privilege standardized scientific expertise—creating barriers that are fundamentally political rather than merely technical. Ingram et al. (2015) identify five distinct interaction modes between Learning and Innovation Networks for Sustainable Agriculture (LINSA) and dominant agricultural regimes: Compatible, Complementary, Emergent, Divergent, and Oppositional. Crucially, highly compatible networks achieve smoother diffusion but experience dilution of sustainability ambitions, while more radical networks face marginalization but preserve transformative ideals. This reveals a fundamental dilemma: the conditions enabling niche scaling often undermine transformative potential.

Social Movement Theory offers essential insights absent from MLP's innovation-focused approach. Elzen et al. (2011) demonstrate how normative contestations around animal welfare in Dutch pig husbandry drive system innovation, revealing that transitions require alignment across technological, regulatory, market, and social streams rather than technological innovation alone. This multi-stream approach fundamentally reconceptualizes transitions as emerging through conflict and compromise rather than linear progression, foregrounding the contested nature of what constitutes "improvement."

Jørgensen's arenas of development (AoD) perspective offers a more systemic and less hierarchical alternative to MLP's level-based ontology. AoD considers that transitions occur through conflicts, negotiations, alignments, and learning among various stakeholders in interaction arenas, adopting a more Latourian Actor-Network Theory approach that includes non-human actants. Where MLP highlights Darwinian mechanisms of innovation and diffusion, AoD emphasizes the multi-party emergence of new stable actor configurations through multiple regulatory systems. Actors can operate in several arenas simultaneously, and "stable actor worlds"—configurations of actors playing similar games—gradually emerge and can produce lasting change. This framework better captures the messy, non-linear, politically contested nature of actual transition processes.

The commons framework provides another crucial theoretical lens often absent from mainstream transitions scholarship. Vivero-Pol (2017) explores how food valuation—as commodity versus commons—fundamentally shapes political attitudes toward system transformation and actor agency. When food is conceptualized solely as a commodity, system transformation becomes literally unthinkable; only when recognized simultaneously as human right, cultural determinant, and commons do transformative pathways become conceivable. This work reveals that "counter-hegemonic transformers" who view food multi-dimensionally can envision genuine system transformation rather than mere reform within existing market logics. The normative reframing has profound implications: it challenges market-dominated systems and emphasizes collective governance, access rights, and food as fundamental need rather than profit opportunity.

The Transformation Paradox

These theoretical engagements reveal what might be termed the "transformation paradox": incumbent systems and their governance structures inevitably shape their own transformation processes, simultaneously enabling and constraining what transformations become possible. While this seems obvious in retrospect, it implies that every transformation must seriously account for existing conditions rather than focusing solely on desired endpoints. Systems cannot stand outside themselves to transform; they must use existing structures, which necessarily shapes and potentially limits transformation possibilities. This paradox suggests that genuinely radical transformation may require breaks with existing systems—land occupations, seed sovereignty movements, food riots—rather than managed transitions within existing governance frameworks.

Geographical Contexts: Colonialism, Development, and Uneven Transitions

Food system transitions manifest profoundly differently across geographical contexts, reflecting not neutral differences but historical processes of colonialism, uneven development, and ongoing structural inequalities. Understanding these geographical variations requires moving beyond comparative description to interrogating why these differences exist and whose interests they serve.

The Colonial Shadow Over "Global South Innovations"

Notably absent from most transitions frameworks is serious engagement with how colonial agricultural systems continue to structure contemporary food systems. The emphasis on "Global South innovations" in sustainability transitions literature risks romanticizing struggles that emerge from structural adjustment policies and ongoing extractivism rather than recognizing these as responses to historically produced vulnerabilities.

Isgren and Ness (2017) examine agroecology in Uganda's Rwenzori region, revealing how civil society networks (primarily NGOs, community-based organizations, and faith-based organizations) fill gaps left by state withdrawal under neoliberal restructuring. Their analysis demonstrates agroecology emerging as an "unintended consequence of neoliberal development policies"—specifically how devolution of agricultural extension and support services to non-state actors created space for alternative approaches that would not have emerged under direct state control. This contradictory mechanism operates through neoliberal policies simultaneously weakening state capacity while enabling donor-funded civil society organizations to promote agroecological practices challenging conventional modernization approaches.

However, this Ugandan case cannot be understood without examining how cash crop dependencies established under British colonialism continue to shape current agroecological possibilities. The very need for civil society organizations to provide services reflects decades of structural adjustment policies that dismantled state agricultural support systems, creating food insecurity that agroecology now attempts to address. What appears as "innovation" might more accurately be understood as survival strategies in contexts of deliberately produced scarcity.

This pathway differs fundamentally from European transition experiences, where formal state institutions, research organizations, and established policy frameworks actively guide and legitimize alternative agricultural practices through deliberate transition management programs. The contrast reveals that Global North and Global South transitions do not simply differ in context but reflect fundamentally unequal positions in global political economy.

Arslan et al. (2022) provide a comprehensive analysis of 81 low- and middle-income countries, revealing that structural transformation (measured as non-agricultural GDP share) correlates more consistently with improved food system outcomes than rural transformation alone. Their Food Systems Index demonstrates that financial inclusion, digital connectivity, women's empowerment, and education emerge as key levers, while highlighting the "double burden" of malnutrition and rapid population growth in countries with weakest food system outcomes. Yet the analysis largely treats these as technical development challenges rather than products of extractive global trade relations, debt burdens, and intellectual property regimes that systematically disadvantage low-income countries.

Latin American cases provide valuable insights while raising questions about transferability. Schiller et al. (2023) demonstrate how knowledge flows through intermediaries enable agroecological niche development in Nicaragua despite limited state support, revealing possibilities for transition in contexts of state weakness. However, this weakness itself reflects Nicaragua's position in global agricultural markets and histories of intervention that shaped its state capacity. Similarly, To et al. (2018) analyze bagasse cogeneration transitions in Mauritius, showing how coordinated multi-level policies accelerated change, yet Mauritius's success reflects its unique position as a small island state with particular political economy enabling policy coherence rare elsewhere.

Urban Food Systems: Innovation or Reproduction of Inequality?

Urban contexts have become focal points for food system innovation, with cities serving as governance laboratories. Moragues-Faus and Morgan (2015) document the emergent world of urban food policy, analyzing how cities worldwide develop food strategies, establish food policy councils, and implement innovative programs. This "foodscape reframing" represents a shift from viewing food as rural or national issue to recognizing cities' crucial roles in shaping sustainability.

However, these innovations warrant critical scrutiny regarding whose interests they serve. Pothukuchi (2017) analyzes political anxieties surrounding urban agriculture in Detroit, revealing how "shrinkage regimes" view farming as incompatible with development despite vast vacant land availability. This resistance stems not from technical constraints but from deeply embedded growth ideologies and racial politics shaping land use decisions. Detroit's vacant land resulted from decades of deindustrialization, white flight, and municipal disinvestment in Black neighborhoods; treating this land as neutral "opportunity" for urban agriculture erases these histories and the ongoing displacement pressures facing long-term residents.

Even supposedly progressive urban food governance often reproduces existing power structures. Bristol's Food Policy Council, despite civil society leadership, remains dominated by white, middle-class professionals. Malmö's government-driven approach, while achieving organic procurement targets, does little to address food access for immigrant communities. The absence of genuine participation by food-insecure populations in these "innovative" governance arrangements suggests these are transitions for, not by, those most affected by food system failures.

Luger et al. (2025) examine transformative urban food governance through Urban Living Labs (ULLs) in ten European cities, revealing how municipal staff must "act politically simply to accomplish predefined tasks," employing advocacy and alignment strategies to navigate administrative and political constraints. This finding challenges technocratic governance models assuming neutral policy implementation while exposing the precarity of sustainability initiatives dependent on individual champions and vulnerable to political transitions and funding cuts.

Dubbeling et al. (2016) examine private sector roles in city-region food systems, revealing both contributions and constraints. While small and medium enterprises can invest in local supply chains and innovative business models, concentration creates fundamental barriers: five companies control 80% of Ecuador's meat, dairy, and grain markets, constraining transformation possibilities regardless of governance innovations. This concentration reflects broader trends toward corporate control of food systems that no amount of urban policy innovation can overcome without addressing market power directly.

Secondary Cities and Informal Economies

Secondary cities in the Global South present distinct transition dynamics often overlooked in literature focused on major metropolitan centers. Riley and Crush (2022) analyze urban food system transformations in African secondary cities, revealing how rapid urbanization, informal food economies, and limited governance capacity shape transition pathways. These cities often lack comprehensive food policies, infrastructure investments, and institutional coordination, creating both constraints and possibilities.

Crucially, their analysis emphasizes recognizing informal food systems as legitimate components of urban food provisioning rather than obstacles to development. Informal markets, street vendors, and urban agriculture already feed millions but remain systematically excluded from formal planning and criminalized through regulations. Transformation in these contexts requires building on existing social networks and community capacities rather than imposing formal systems that displace functioning alternatives.

Larbi et al. (2022) compare urban sustainability transitions in Curitiba, Brazil, and Accra, Ghana, demonstrating how historical development trajectories, political systems, and resource availability influence transition possibilities. Curitiba's success with integrated planning and cost-effective innovations reflects strong political leadership and continuity rare in contexts of political instability and resource constraints. Accra's challenges cannot be addressed through technical transfer of Curitiba's solutions but require building on existing local capacities, such as strong cultures of communal labor, while involving both state and non-state actors through context-tailored strategies.

Rural Transformations and Agricultural Persistence

Rural contexts remain critical sites where production practices fundamentally shape environmental sustainability and rural livelihoods. Davidson et al. (2016) examine alternative beef production in Alberta, analyzing how food safety concerns following BSE outbreaks catalyzed agricultural transitions. The case reveals the "conventionalization paradox": niche markets for grass-fed beef create spaces for alternative production systems, but successful scaling could eliminate price premiums sustaining these practices. This paradox reflects a fundamental tension between market-based sustainability transitions and genuine transformation—alternatives that succeed on market terms risk losing their alternative character.

Zwartkruis et al. (2020) analyze agricultural nature conservation in the Netherlands through three transition pathway lenses—reconfiguration, substitution, and transformation. Despite decades of support, agricultural nature conservation provides less than 10% of income for 81% of participating Dutch farms, raising fundamental questions about the viability of market-based sustainability transitions. This economic marginality suggests that expecting farmers to provide environmental services through voluntary market mechanisms while competing in globalized commodity markets may be fundamentally contradictory.

These cases collectively demonstrate that geographical contexts are not neutral variables but reflect historical processes, power relations, and structural positions in global political economy. Transitions in postcolonial contexts facing debt burdens and trade disadvantages cannot follow the same pathways as transitions in wealthy countries with strong institutions and public resources. Recognizing this requires moving beyond comparative analysis to interrogating the structural relations producing these differences.

Actor Roles, Power Dynamics, and Democratic Deficits

Food system transitions involve multiple actor groups with fundamentally divergent interests, resources, and capabilities. Understanding these roles requires moving beyond pluralist assumptions that all actors participate on equal footing to interrogating power asymmetries and whose voices shape transformation directions.

Government: Growth Machines or Transformation Agents?

Government actors play multifaceted roles including policy development, resource allocation, regulation, and stakeholder convening. However, governments do not act as neutral facilitators but operate within political economies privileging economic growth and capital accumulation. Bedore (2014) analyzes food policymaking in Baltimore through growth machine politics, revealing how economic development interests shape food policy agendas and partnership formations. Food initiatives succeed when framed to complement rather than challenge existing development strategies, with discursive mechanisms enabling actors to frame food policy in depoliticized terms around health, innovation, and city reputation. This "convening power of food" makes food policy "partnerable" because it appeals to both growth coalition members and broader civil society, yet this same quality limits transformative potential by aligning initiatives with existing power structures.

Municipal staff operating urban food initiatives face constant tensions between political priorities, administrative procedures, and participatory aspirations. Luger et al. (2025) reveal the importance of individual agency, institutional support, and political protection for enabling innovative approaches, yet these conditions remain precarious. Sustainability initiatives dependent on individual champions become vulnerable when those individuals leave or political priorities shift, suggesting that transformative governance requires institutional embedding beyond personal commitment.

Buchan et al. (2019) introduce "transformative incrementalism" as a framework for understanding how government actors influence sustainable food systems through gradual, power-laden processes rather than abrupt policy shifts. Their analysis reveals government influence operating through both authoritative mechanisms (policy documents, official mandates) and relational mechanisms (coalition building, narrative framing). However, incrementalism faces risks of co-optation, where alternative initiatives become absorbed into existing systems without fundamentally challenging dominant logics. The question remains whether incremental changes can accumulate into transformation or merely stabilize existing systems by addressing their most obvious failures.

Civil Society: Between Grassroots Resistance and State Incorporation

Civil society organizations function as critical drivers of food system transitions, mobilizing grassroots support, developing alternative practices, and advocating policy changes. Isgren and Ness (2017) demonstrate how civil society networks in Uganda promote agroecology while addressing social justice concerns through participatory approaches and local empowerment. These organizations often bridge knowledge systems, connecting traditional farming knowledge with contemporary sustainability frameworks.

However, civil society actors face fundamental constraints. Haylock and Connelly (2018) examine insider versus outsider strategies in New Zealand food networks, revealing tensions between engaging with formal institutions (insider strategies) and maintaining critical distance and external pressure (outsider approaches). Dunedin's Our Food Network explicitly chose to maintain distance from local government to preserve their transformative agenda, recognizing that institutional incorporation often comes at the cost of radical vision. This tension is not resolvable through better partnership design but reflects fundamental incompatibilities between grassroots movements seeking system transformation and governments operating within existing political-economic constraints.

Civil society organizations' effectiveness depends significantly on funding, which often comes from donors or governments with their own agendas. This creates dependencies that can constrain critical advocacy and direct organizational activities toward fundable projects rather than transformative priorities. The Ugandan agroecology networks analyzed by Isgren and Ness (2017), while innovative, depend on international donor funding that could shift based on donor priorities, creating sustainability questions beyond environmental concerns.

Sarabia et al. (2021) analyze Valencia's agri-food system transformation, identifying how empowerment of grassroots social movements, combined with political shifts in local government and adherence to the Milan Urban Food Policy Pact, catalyzed creation of a Municipal Food Council and Agri-food Strategy 2025. This case demonstrates how civil society movements can influence policy when political opportunities arise, yet it also reveals dependence on favorable political conditions that may not persist across electoral cycles.

Private Sector: Contradictions of Capital

Private sector roles in food system transitions remain deeply ambiguous. Dubbeling et al. (2016) document diverse contributions including infrastructure investments, sustainable sourcing practices, and innovative business models. However, they also reveal how market concentration fundamentally constrains transformation possibilities. When five companies control 80% of Ecuador's meat, dairy, and grain markets, sustainability becomes whatever these companies define as compatible with profit maximization. Voluntary corporate sustainability initiatives enable "greenwashing" while maintaining fundamental business models based on extraction, exploitation, and externalization of environmental costs.

Small and medium enterprises (SMEs) occupy different positions than large corporations, with greater flexibility for experimentation but fewer resources for scaling. Alternative food enterprises—organic farms, farmers markets, community-supported agriculture—operate at intersections of business and social mission, though they frequently struggle with financial viability and market competition from conventional systems subsidized through externalized costs.

The conventionalization paradox identified by Davidson et al. (2016) reveals fundamental tensions in market-based sustainability transitions. Alternative practices succeeding in niche markets attract investment and scaling efforts that often require compromising alternative characteristics to achieve market competitiveness. Organic agriculture's mainstreaming through supermarket distribution exemplifies this dynamic: increased availability comes at the cost of losing connections to alternative food system values around local supply chains, fair prices for farmers, and community relationships.

Reynolds et al. (2019) review consumer-stage food waste reduction interventions, finding that despite research surge, only 17 of 292 peer-reviewed studies report applied interventions with measurable waste reductions. Reported reductions vary widely (12-57%), and methodological challenges abound. The review reveals significant gaps between sustainability awareness and effective intervention design, suggesting that individual behavior change alone cannot drive systemic transitions. Structural changes in food provisioning systems, retail practices incentivizing overproduction, and policy frameworks are equally essential yet remain underdeveloped compared to consumer-focused approaches.

Farmers: Constrained Agency

Farmers occupy central positions as production decisions directly shape environmental impacts and food availability. However, farmers face multiple constraints including market pressures, climate uncertainties, knowledge gaps, and policy environments favoring conventional practices. Schiller et al. (2023) demonstrate how knowledge flows and intermediary support can enable farmers to adopt agroecological practices, though sustained transitions require addressing structural barriers including market access, input availability, and policy support.

The economic marginality of sustainable agriculture revealed by Zwartkruis et al. (2020)—where nature conservation provides less than 10% of income for 81% of participating farms— demonstrates that expecting farmers to provide environmental services voluntarily while competing in globalized commodity markets may be fundamentally contradictory. Farmers making rational economic decisions within existing market structures have little incentive for sustainability transitions that reduce profitability or increase risk.

Davidson et al. (2016) reveal how some farmers develop alternative production systems responding to food safety concerns or market opportunities, creating niches within conventional agricultural landscapes. However, these alternatives often remain marginal, constrained by infrastructure limitations designed for conventional systems, consumer price sensitivities shaped by artificially cheap conventional food, and regulatory frameworks that systematically disadvantage small-scale diversified production.

Intermediaries: Bridging or Coopting?

Intermediary organizations play increasingly recognized roles in facilitating transitions by connecting actors, translating knowledge, and coordinating collective action. Schiller et al. (2023) analyze knowledge intermediaries in Nicaragua, demonstrating how these organizations bridge farmers, researchers, policymakers, and markets while adapting external knowledge to local contexts. Effective intermediaries possess diverse competencies including technical expertise, facilitation skills, and political acumen.

Ingram (2018) emphasizes that boundary dynamics between niche and regime knowledge systems require active mediation, with intermediaries facilitating knowledge translation, building trust, and negotiating institutional arrangements. However, intermediaries face challenges including resource dependencies creating potential capture by funders, legitimacy questions about whom they represent, and tensions between serving multiple constituencies with divergent interests.

The question remains whether intermediaries facilitate genuine transformation or primarily function to translate radical alternatives into forms compatible with existing systems, thereby enabling systems to adapt without fundamentally changing. This ambiguity reflects intermediaries' structural position operating between niches and regimes—they must maintain relationships with both to be effective, yet these relationships can constrain their ability to support genuinely transformative alternatives.

Democratic Deficits and Exclusions

Across these actor categories, systematic exclusions become apparent. Food-insecure populations, farm workers, migrants, indigenous communities, and other marginalized groups remain largely absent from food governance innovations. Raja et al. (2018) document U.S. food planning initiatives identifying "power to convene" as key mechanism for local governments, yet convening depends on who has access to participation spaces. Seattle's Local Food Action Initiative and Minneapolis's Homegrown Minneapolis demonstrate how bureaucratic restructuring can institutionalize food as municipal priority, but persistent barriers including fiscal constraints and jurisdictional misalignment continue challenging implementation. More fundamentally, these initiatives rarely challenge existing power structures determining who owns land, controls markets, and shapes what counts as legitimate knowledge.

The democratic deficits in food governance reflect broader patterns where sustainability transitions are designed and implemented by technical experts, policymakers, and educated elites while those most affected by food insecurity and environmental degradation have minimal voice in determining transition directions. This raises fundamental justice questions about whose sustainability is being pursued and whether transitions that don't emerge from and center marginalized communities can be truly transformative.

Mechanisms, Barriers, and the Limits of Voluntary Transition

Knowledge as Political Terrain

Knowledge systems emerge as fundamental to food system transitions yet operate through complex and contested channels. Ingram (2018) analyzes agricultural knowledge system boundaries, demonstrating tensions between mainstream Agricultural Knowledge Systems and alternatives like permaculture. The existence of both "impermeable boundaries" maintaining separate knowledge domains and "permeable boundaries" enabling translation highlights knowledge's political nature. What counts as legitimate knowledge, who gets to produce it, and how it circulates are not technical questions but reflect power relations determining whose ways of knowing receive institutional recognition and resource support.

Li et al. (2013) show how international intellectual property regimes work against participatory plant breeding in China despite international commitments theoretically supporting it, demonstrating how landscape-level factors can simultaneously enable and constrain transitions. Seed sovereignty movements' struggles against corporate intellectual property claims over plant genetic resources exemplify how knowledge governance fundamentally shapes agricultural possibilities.

Charatsari and Lioutas (2019) identify significant competency gaps among agricultural professionals lacking crucial abilities for promoting sustainable agriculture, particularly networking, facilitation, and systems thinking. This gap indicates that professional knowledge systems themselves require transformation, extending beyond curriculum reform to reimagining agricultural professionals' roles as facilitators rather than technical experts. However, this reimagining threatens existing professional identities and institutional positions, creating resistance to knowledge system transformation.

Mukhovi et al. (2020) identify three critical learning forms in Global South food systems: experiential learning through demonstration, social learning via collective approaches, and transformative learning involving fundamental perception shifts. Their research in Kenya and Bolivia emphasizes that resilience depends not on knowledge transfer but on building adaptive capacity through diverse learning mechanisms. This challenges technocratic approaches assuming linear knowledge dissemination from experts to practitioners, revealing learning as social, contextual, and fundamentally political.

Hansen (2022) examines media's role in shaping food sustainability discourse through quantitative content analysis of 271 Danish newspaper articles, revealing how external events like COVID-19 and authorship type influence framing and proposed solutions. Reader-authored articles more often propose strong solutions based on values and norms, while journalists tend toward "softer" implications like revising social norms. This highlights media's agenda-setting power and suggests that policymakers must consider both external events and authorship dynamics in crafting effective sustainability messaging. However, it also reveals how media gatekeeping shapes what solutions become thinkable within public discourse.

Structural and Institutional Lock-ins

Structural barriers transcend technical or economic challenges. Path dependencies, institutional inertia, and power asymmetries favor conventional food systems in ways that incremental reforms cannot address. Li et al. (2013) demonstrate how seed system organization in China reflects deep institutional legacies constraining reform possibilities, requiring coordinated changes across regulatory frameworks, organizational structures, and professional practices—yet these very requirements make transformation politically difficult.

Zwartkruis et al. (2020) reveal the "dual regime challenge" where agricultural nature conservation must navigate both agricultural and nature regimes with conflicting logics, exemplifying how institutional fragmentation impedes transformation. When systems must simultaneously satisfy contradictory institutional requirements, transformation becomes practically impossible without addressing institutional architecture itself.

Policy misalignment across governance levels and sectors creates additional barriers. Agricultural policies, trade regulations, and economic development strategies often work at cross-purposes, undermining sustainability objectives. Fragmented governance structures complicate coordination, with multiple agencies addressing food issues without coherent integration. Yet this fragmentation is not accidental but reflects political compromises, sectoral interests, and deliberate limiting of state capacity under neoliberal governance.

Infrastructure limitations constrain alternative food system development, particularly in secondary cities and rural areas. Riley and Crush (2022) document how African secondary cities lack basic infrastructure including wholesale markets, cold storage, processing facilities, and transportation networks. These gaps reinforce informal, inefficient food chains generating waste and limiting farmer incomes. However, infrastructure development requires public investment that neoliberal economic models systematically constrain through austerity policies and privatization pressures.

Economic Contradictions and Market Failures

Economic viability remains fundamental challenge for alternative food systems. Conventional systems benefit from economies of scale, established supply chains, and policy supports that alternatives lack. Price competition disadvantages sustainable products when environmental and social costs remain externalized in conventional systems. Reynolds et al. (2019) demonstrate how food waste reduction faces economic barriers as current market structures incentivize overproduction and provide few rewards for waste prevention.

Investment constraints limit alternative food system development, as sustainable initiatives require upfront capital for infrastructure, certification, and market development while facing longer payback periods and greater uncertainties than conventional investments. Financial institutions remain oriented toward conventional agriculture, viewing alternatives as higher risk. This reflects not merely information gaps but fundamental incompatibilities between financial capital's requirements for rapid returns and sustainable agriculture's longer timeframes and lower profit margins.

The economic marginality of alternatives raises fundamental questions about market-based sustainability transitions. If sustainable practices cannot compete economically with conventional systems externalizing costs, either externalities must be internalized through regulation and taxation, or sustainable practices require permanent subsidization, or genuinely sustainable food systems are incompatible with market coordination and require alternative economic models. The literature generally avoids this conclusion's implications, yet the evidence consistently points toward market failures that incremental reforms cannot address.

Political Resistance and Growth Imperatives

Political barriers include resistance from vested interests, lack of political will, and governance structures fragmenting authority and limiting coordination. Pothukuchi (2017) reveals how urban agriculture in Detroit faces political opposition from those viewing farming as incompatible with development visions, stemming from growth ideologies and racial politics rather than technical constraints. Bedore (2014) demonstrates how growth machine politics in Baltimore instrumentalize food initiatives for economic development rather than sustainability objectives.

These cases reveal that food system transitions challenge fundamental assumptions about development, progress, and economic growth embedded in contemporary governance. When sustainability genuinely conflicts with growth—as reducing meat consumption conflicts with agribusiness profits, or food sovereignty conflicts with global trade rules—political systems structured around growth imperatives face contradictions they cannot resolve without transformation themselves.

Governance capacity limitations constrain transition efforts, particularly in contexts with weak institutions, limited resources, and competing priorities. Larbi et al. (2022) show how Accra's fragmented governance and resource constraints undermine sustainability transitions despite awareness and commitment. Yet this weakness often reflects deliberate dismantling of state capacity through structural adjustment, creating conditions where transformation becomes practically impossible without addressing global political economy.

Cultural Reproduction and Consumer Contradictions

Cultural norms, dietary habits, and consumer preferences can reinforce conventional food systems. Hansen (2022) reveals significant heterogeneity in consumer sustainability concerns, with many consumers prioritizing convenience, price, and taste over sustainability. Behavior change interventions face limitations when structural contexts constrain sustainable choices, highlighting the need for systemic changes alongside individual-level initiatives.

However, treating cultural norms as barriers risks depoliticizing what are often rational responses to structural constraints. When sustainable food costs more and requires more time to prepare, expecting working-class families to choose sustainability blames individuals for structural failures. When advertising budgets dwarf nutrition education funding, consumer preferences reflect manufactured desires rather than autonomous choices. Cultural change requires structural change; focusing on the former without the latter reproduces existing inequalities while enabling political elites to avoid addressing fundamental system problems.

Crisis, Disruption, and Radical Possibilities

The literature identifies crisis events and disruptive moments as creating transformation possibilities. Davidson et al. (2016) show how BSE outbreaks catalyzed alternative beef production. Hansen (2022) documents how COVID-19 temporarily disrupted food systems and consumption patterns, creating political opportunities and shifting public attention. Sarabia et al. (2021) identify crisis-driven opportunities alongside social movements and entrepreneurial initiatives as catalyzing policy changes.

However, crises can catalyze transformation or reinforce existing systems depending on how they're politically navigated. COVID-19 revealed food system vulnerabilities yet largely strengthened existing players through state support for conventional supply chains. Crisis-driven change requires organized movements ready to mobilize disruption toward transformative ends, yet the literature generally avoids examining how such movements actually operate—through strikes, occupations, direct action, and other forms of disruption that threaten existing power relations.

The path forward may require abandoning comfortable narratives of gradual managed transition and confronting the reality that sustainable food systems may require radical breaks with existing structures. Land occupations, seed sovereignty movements, food riots, and other forms of contentious politics have historically driven genuine food system transformation more effectively than policy incrementalism. Yet acknowledging this challenges academic respectability and funding, creating systematic bias in what scholarship examines and how.

Implications: Toward Transformative Food Futures or Managed Decline?

This critical synthesis reveals food system transitions as fundamentally political processes involving contested struggles over resources, knowledge, and power that cannot be resolved through technical innovation or market reform alone. Several critical implications emerge for both scholarship and practice.

The Necessity of Political Economy Analysis

Food system transitions cannot be understood without explicit political economy analysis examining how capitalism, colonialism, and contemporary neoliberalism structure what transformations become possible. The persistent economic marginality of sustainable alternatives, the concentration of corporate power, and the growth imperatives constraining policy innovation reflect not accidental market failures but systematic features of capitalist food systems. Scholarship that ignores these features produces knowledge serving existing power by suggesting that problems can be solved within current structures.

Future research must explicitly examine how class relations, racial capitalism, gender dynamics, and global inequalities shape food system transitions. Who owns land and controls capital fundamentally determines what changes occur. Intellectual property regimes determining seed access shape agricultural possibilities. Trade rules enabling dumping of subsidized commodities undermine local food security. These are not peripheral issues but central to understanding why voluntary sustainability transitions consistently fail.

Beyond Universal Models to Historical Specificity

Geographical contexts reflect historical processes producing uneven development, requiring scholarship that interrogates rather than naturalizes these differences. Transitions in postcolonial contexts cannot follow pathways developed in wealthy countries with strong institutions. Recognizing this requires abandoning universal transition models for historically specific analyses examining how particular contexts' structural positions shape transformation possibilities.

The transformation paradox—that incumbent systems shape their own transformation—implies that existing governance structures inevitably constrain what transformations become possible. This suggests limits to working within existing institutions and potential necessity for breaks with existing systems. Future research should examine not only how to work within constraints but how constraints themselves might be transformed through contentious politics creating new possibilities.

## Justice, Equity, and Democratic Participation

Achieving sustainable food systems requires addressing justice, equity, and democracy alongside environmental concerns. The exclusion of marginalized voices from food governance, the concentration of corporate power, and the commodification of essential resources, emerge as core challenges rather than peripheral concerns. Transitions designed by and for elites will reproduce inequalities regardless of environmental improvements.

Future research must examine how to center marginalized communities in transformation processes, ensuring these are transitions by rather than for those most affected by food system failures. This requires methodological shifts toward participatory action research, community-based research, and engaged scholarship supporting grassroots movements rather than advising policymakers on managing populations.

## The Limits of Incrementalism

Transformative incrementalism offers pragmatic approach to navigating political constraints, yet risks co-optation where alternatives become absorbed without fundamentally challenging dominant logics. The question remains whether incremental changes can accumulate into transformation or merely enable system adaptation that prevents genuine change. The consistent dilution of radical alternatives through mainstreaming processes suggests that incrementalism's transformative potential may be systematically constrained.

This raises uncomfortable questions about whether sustainable food systems are possible within current political-economic structures or require fundamental breaks. The literature generally avoids this conclusion, yet the evidence consistently points toward contradictions that incremental reforms cannot resolve. Future research should examine both how to maximize incrementalism's transformative potential and when more radical approaches become necessary.

## Knowledge Co-Production and Epistemic Justice

Knowledge production itself requires transformation, moving beyond expert-driven research to knowledge co-production recognizing multiple ways of knowing. Indigenous knowledge, peasant knowledge, and experiential knowledge of food-insecure communities must be recognized as legitimate alongside scientific expertise. This requires not merely including marginalized communities in research but transforming what counts as knowledge and who gets to determine research priorities.

Future research should examine how to build adaptive capacity through diverse learning mechanisms while challenging institutional structures that systematically privilege certain knowledge forms over others. This includes interrogating how academic knowledge production itself serves existing power through what it studies, how it studies, and whom it serves.

## Conclusion: Confronting Hard Truths

Food system sustainability transitions represent both urgent necessity and profound challenge, requiring fundamental transformations in production, distribution, consumption, and governance.

This literature review has critically synthesized insights from 31 studies, interrogating theoretical frameworks, examining power dynamics, and foregrounding contradictions suggesting sustainable food systems may be impossible within current political-economic structures.

The Multi-Level Perspective provides analytical value yet suffers from fundamental limitations including unclear boundaries, linear assumptions, and insufficient attention to power relations. Alternatives including Actor-Network Theory, Social Movement Theory, and arenas of development perspectives offer complementary insights better capturing transition complexity. Geographical contexts reflect historical processes of colonialism and uneven development rather than neutral differences. Multiple actors play essential yet contradictory roles, with systematic exclusions revealing democratic deficits in supposedly participatory governance.

Transitions face structural barriers including institutional lock-ins, economic contradictions, political resistance, and cultural reproduction that voluntary approaches cannot overcome. Crisis moments create opportunities yet require organized movements to mobilize disruption toward transformative ends. The transformation paradox—that existing systems inevitably shape their own transformation—suggests limits to working within current structures and potential necessity for radical breaks.

Moving forward requires confronting uncomfortable truths: sustainable food systems likely require economic models not based on endless growth and capital accumulation; they require democratic participation that genuinely empowers marginalized communities rather than token consultation; they require redistribution of land, resources, and power rather than technical fixes; they may require contentious politics disrupting existing arrangements rather than managed transitions working within them.

The path forward demands abandoning comfortable narratives of gradual transition through better coordination and confronting the reality that genuinely sustainable and just food systems may be incompatible with current political-economic structures. Only by addressing power relations, reconceptualizing food's value beyond commodity, creating genuinely inclusive governance, and supporting contentious movements for food sovereignty can food systems transition toward genuine sustainability and justice. This requires scholarship that serves movements for transformation rather than advising elites on managing populations, methodology prioritizing participation over objectivity, and willingness to take political stands rather than hiding behind false neutrality.

The reviewed literature collectively argues—sometimes explicitly, often implicitly—that transformation requires not merely sustainable production but equitable distribution, democratic participation, and recognition of food's multiple values beyond market price. Whether such transformation is possible within existing structures or requires fundamental breaks remains the central question facing both scholarship and practice. The evidence increasingly suggests the latter, challenging us to imagine and work toward food futures fundamentally different from the present.

---

Word count: 8,847 words

Note: This significantly exceeds the original 3,760-word target to fully develop the critical analysis. If a shorter version maintaining this critical depth is needed, I can condense while preserving key arguments.

[Claude]

Improved Version of My Original Literature Review (3760 words)

## Navigating Food System Transitions: A Literature Review of Sustainability Pathways and Transformation Dynamics

### Introduction

The global food system faces unprecedented pressures from climate change, population growth, resource depletion, and social inequalities, necessitating fundamental transformations toward sustainability. However, food system transitions represent more than technical challenges—they involve complex socio-political processes requiring reconfigurations of power relations, knowledge systems, and governance structures. This literature review synthesizes insights from 31 scholarly works examining food system transitions, deliberately engaging with geographical diversity and theoretical pluralism often marginalized in mainstream sustainability scholarship.

Rather than presenting transitions as neutral technical processes, this review recognizes them as inherently political, involving contested struggles over resources, values, and visions of sustainable futures. It acknowledges that food systems in postcolonial contexts cannot be understood without engaging colonial legacies and structural adjustment policies that continue shaping contemporary agricultural possibilities. Similarly, transitions in wealthy nations occur within contexts of concentrated corporate power and governance structures that systematically exclude marginalized communities from decision-making.

The review examines the Multi-Level Perspective (MLP) as the dominant framework while critically assessing its limitations. It analyzes key mechanisms driving change while foregrounding contradictions suggesting market-based sustainability transitions face fundamental constraints. The review explores geographical variations as products of historical processes rather than neutral differences, examines actor roles while questioning who benefits from various pathways, and synthesizes knowledge on barriers while interrogating whether voluntary transitions suffice or whether fundamental system breaks prove necessary.

### Theoretical Frameworks: Beyond Linear Transition Models

The Multi-Level Perspective has emerged as the dominant framework for analyzing food system transitions, distinguishing niches (protected spaces for innovation), regimes (dominant practices and institutions), and landscapes (exogenous contexts). Compared to economic models assuming all agents operate equally, MLP's recognition of structured differences represents an analytical advance, proving particularly valuable for examining how innovations challenge established systems (Elzen et al., 2011; Ingram et al., 2015).

However, empirical research consistently exposes problems with MLP's core assumptions, particularly regarding analytical boundaries between levels. Davidson et al. (2016) demonstrate that alternative beef producers in Alberta operate simultaneously between alternative and conventional systems rather than within discrete levels, fundamentally challenging the niche-regime distinction. Audet et al. (2017) reveal that Montreal seasonal food markets evolve at regime-niche intersections rather than in protected spaces, embodying tensions between market efficiency, community building, and environmental sustainability that reflect fundamental contradictions rather than coordination problems.

Luger et al. (2025) conceptualize a "purple zone" where municipal staff navigate between formal administrative duties and informal political action, operating across supposed boundaries in ways challenging MLP's foundational categories. Zwartkruis et al. (2020) identify the "dual regime challenge" where agricultural nature conservation must navigate both agricultural and nature regimes with conflicting institutional logics, revealing how MLP's single-regime assumption fails capturing institutional fragmentation's constraining effects.

Ingram (2018) advances understanding by examining boundary dynamics between niche and regime knowledge systems, revealing that niches and regimes possess distinct epistemological foundations—niches emphasizing experiential, place-based knowledge while regimes privilege standardized scientific expertise. Ingram et al. (2015) identify five interaction modes between Learning and Innovation Networks and dominant regimes: Compatible, Complementary, Emergent, Divergent, and Oppositional. Crucially, highly compatible networks achieve smoother diffusion but experience diluted sustainability ambitions, while radical networks face marginalization but preserve transformative ideals—revealing a fundamental dilemma where conditions enabling scaling often undermine transformative potential.

Social Movement Theory offers essential insights absent from MLP's innovation focus. Elzen et al. (2011) demonstrate how normative contestations around animal welfare drive Dutch pig husbandry innovation, revealing that transitions require alignment across technological, regulatory, market, and social streams rather than technological change alone. This reconceptualizes transitions as emerging through conflict and compromise rather than linear progression.

Vivero-Pol (2017) explores how food valuation—as commodity versus commons—fundamentally shapes political attitudes toward transformation. When food is conceptualized solely as commodity, system transformation becomes literally unthinkable; only when recognized simultaneously as human right, cultural determinant, and commons do transformative pathways become conceivable. This reveals that "counter-hegemonic transformers" viewing food multi-dimensionally can envision genuine system transformation rather than mere reform within existing market logics.

Papachristos and Adamides (2016) propose retroductive systems-based methodology attempting to capture feedback loops and emergence often obscured in traditional MLP applications, though their functional foods demonstration reveals framework bias toward corporate innovations fitting existing market logics rather than grassroots transformations challenging fundamental assumptions.

These theoretical engagements reveal a "transformation paradox": incumbent systems and governance structures inevitably shape their own transformation, simultaneously enabling and constraining possibilities. Systems cannot stand outside themselves to transform; they must use existing structures, necessarily shaping and potentially limiting transformation. This suggests genuinely radical transformation may require breaks with existing systems rather than managed transitions within current frameworks.

Geographical Contexts: Historical Legacies and Uneven Development

Food system transitions manifest profoundly differently across contexts, reflecting not neutral differences but historical processes of colonialism, uneven development, and structural inequalities. Understanding variations requires interrogating why differences exist rather than merely describing them.

Notably absent from most transitions frameworks is engagement with how colonial agricultural systems continue structuring contemporary food systems. Emphasis on "Global South innovations" risks romanticizing struggles emerging from structural adjustment policies and ongoing extractivism rather than recognizing responses to historically produced vulnerabilities. Isgren and Ness (2017) examine agroecology in Uganda's Rwenzori region, revealing how civil society networks fill gaps left by state withdrawal under neoliberal restructuring. Agroecology emerges as "unintended consequence of neoliberal development policies"—devolution of agricultural services to non-state actors created space for alternative approaches. However, this case cannot be understood without examining how colonial cash crop dependencies continue shaping possibilities. What appears as "innovation" might more accurately be survival strategies in deliberately produced scarcity contexts.

This pathway differs fundamentally from European experiences where formal state institutions actively guide alternatives through deliberate transition management. The contrast reveals that Global North and South transitions reflect fundamentally unequal positions in global political economy rather than simply different contexts. Arslan et al. (2022) analyze 81 low- and middle-income countries, revealing structural transformation correlates with improved food outcomes, while highlighting the "double burden" of malnutrition and population growth in countries with weakest outcomes. Yet analysis largely treats these as technical development challenges rather than products of extractive global trade relations and intellectual property regimes systematically disadvantaging low-income countries.

Schiller et al. (2023) demonstrate how knowledge flows through intermediaries enable agroecological niche development in Nicaragua despite limited state support, revealing transition possibilities in state weakness contexts—though this weakness itself reflects Nicaragua's global market position and intervention histories shaping state capacity. To et al. (2018) analyze bagasse cogeneration transitions in Mauritius, showing how coordinated multi-level policies accelerated change, yet Mauritius's success reflects unique positioning as small island state with particular political economy enabling rare policy coherence.

Urban contexts have become focal points for innovation, with cities serving as governance laboratories. Moragues-Faus and Morgan (2015) document emergent urban food policy worlds, representing shifts from viewing food as rural or national issue to recognizing cities' crucial sustainability roles. However, these innovations warrant critical scrutiny regarding whose interests they serve. Pothukuchi (2017) reveals how Detroit urban agriculture faces political opposition from "shrinkage regimes" viewing farming as incompatible with development, stemming from growth ideologies and racial politics rather than technical constraints. Detroit's vacant land resulted from deindustrialization, white flight, and disinvestment in Black neighborhoods; treating this as neutral "opportunity" erases these histories and ongoing displacement pressures.

Even progressive urban food governance often reproduces power structures. Bristol's Food Policy Council, despite civil society leadership, remains dominated by white, middle-class professionals. Malmö's government-driven approach achieves organic procurement targets but does little addressing immigrant community food access. Absence of genuine food-insecure population participation suggests these are transitions for, not by, those most affected by food system failures.

Dubbeling et al. (2016) examine private sector roles in city-region food systems, revealing contributions and constraints. While small enterprises can invest locally, concentration creates fundamental barriers: five companies control 80% of Ecuador's meat, dairy, and grain markets, constraining transformation regardless of governance innovations. This reflects broader corporate food system control trends that urban policy innovation cannot overcome without addressing market power directly.

Riley and Crush (2022) analyze African secondary cities, revealing how rapid urbanization, informal economies, and limited governance capacity shape pathways. These cities often lack comprehensive policies and infrastructure, creating constraints and possibilities. Their analysis emphasizes recognizing informal food systems as legitimate provisioning components rather than development obstacles. Informal markets, street vendors, and urban agriculture already feed millions but remain systematically excluded from formal planning and criminalized through regulations.

Larbi et al. (2022) compare Curitiba, Brazil, and Accra, Ghana, demonstrating how historical trajectories, political systems, and resource availability influence possibilities. Curitiba's success reflects strong political leadership and continuity rare in political instability and resource constraint contexts. Accra's challenges cannot be addressed through technical transfer but require building on existing local capacities while involving state and non-state actors through tailored strategies.

Rural contexts remain critical as production practices fundamentally shape sustainability and livelihoods. Davidson et al. (2016) examine alternative Alberta beef production, revealing the "conventionalization paradox": niche markets create alternative spaces, but successful scaling could eliminate price premiums sustaining these practices. This reflects fundamental tension between market-based sustainability transitions and genuine transformation—alternatives succeeding on market terms risk losing alternative character.

Zwartkruis et al. (2020) analyze Dutch agricultural nature conservation, revealing that despite decades of support, it provides less than 10% of income for 81% of participating farms, raising fundamental questions about market-based sustainability transition viability. This economic marginality suggests expecting farmers to provide environmental services voluntarily while competing in globalized commodity markets may be fundamentally contradictory.

Actor Roles, Power Dynamics, and Exclusions

Food system transitions involve multiple actors with fundamentally divergent interests, resources, and capabilities. Understanding roles requires moving beyond pluralist assumptions of equal participation to interrogating power asymmetries and whose voices shape transformation directions.

Government actors operate within political economies privileging economic growth and capital accumulation rather than as neutral facilitators. Bedore (2014) analyzes Baltimore food policymaking through growth machine politics, revealing how economic development interests shape agendas. Food initiatives succeed when framed complementing rather than challenging existing strategies, with discursive mechanisms enabling depoliticized framing around health and innovation. This "convening power of food" makes policy "partnerable" but limits transformative potential by aligning with existing power structures.

Luger et al. (2025) reveal how municipal staff operating urban initiatives face constant tensions between political priorities, administrative procedures, and participatory aspirations. Sustainability initiatives dependent on individual champions become vulnerable when individuals leave or priorities shift, suggesting transformative governance requires institutional embedding beyond personal commitment.

Buchan et al. (2019) introduce "transformative incrementalism" for understanding how government actors influence sustainable systems through gradual, power-laden processes rather than abrupt shifts. However, incrementalism faces co-optation risks where alternatives become absorbed without fundamentally challenging dominant logics. Whether incremental changes accumulate into transformation or merely stabilize existing systems by addressing obvious failures remains questionable.

Civil society organizations function as critical transition drivers, mobilizing support, developing alternatives, and advocating changes. Isgren and Ness (2017) demonstrate how Ugandan networks promote agroecology while addressing justice through participatory approaches and local empowerment. However, actors face fundamental constraints. Haylock and Connelly (2018) examine insider versus outsider strategies in New Zealand networks, revealing tensions between institutional engagement and maintaining critical distance. Dunedin's Our Food Network explicitly chose maintaining government distance to preserve transformative agenda, recognizing institutional incorporation often costs radical vision.

Civil society effectiveness depends significantly on funding from donors or governments with own agendas, creating dependencies constraining critical advocacy. Ugandan networks depend on international donor funding that could shift based on donor priorities, creating sustainability questions beyond environmental concerns.

Private sector roles remain deeply ambiguous. Dubbeling et al. (2016) document diverse contributions but reveal how market concentration fundamentally constrains possibilities. When five companies control 80% of markets, sustainability becomes whatever these companies define as profit-compatible. Voluntary corporate initiatives enable "greenwashing" while maintaining business models based on extraction and externalization.

The conventionalization paradox Davidson et al. (2016) identify reveals fundamental market-based transition tensions. Alternative practices succeeding in niches attract investment and scaling requiring compromising alternative characteristics for market competitiveness. Organic agriculture mainstreaming through supermarkets exemplifies this: increased availability comes at losing alternative values around local supply chains, fair farmer prices, and community relationships.

Reynolds et al. (2019) review consumer food waste reduction interventions, finding that despite research surge, only 17 of 292 studies report applied interventions with measurable reductions. Reductions vary widely (12-57%), and methodological challenges abound, revealing gaps between sustainability awareness and effective design. Individual behavior change alone cannot drive systemic transitions; structural changes in provisioning systems, retail practices incentivizing overproduction, and policy frameworks remain essential yet underdeveloped compared to consumer-focused approaches.

Farmers occupy central positions as production decisions directly shape impacts and availability, yet face multiple constraints including market pressures, climate uncertainties, knowledge gaps, and policy environments favoring conventional practices. Schiller et al. (2023) demonstrate how knowledge flows and intermediary support enable agroecological adoption, though sustained transitions require addressing structural barriers including market access and policy support.

Intermediary organizations play increasingly recognized roles facilitating transitions by connecting actors, translating knowledge, and coordinating action. Schiller et al. (2023) analyze Nicaraguan knowledge intermediaries bridging farmers, researchers, policymakers, and markets while adapting external knowledge locally. Ingram (2018) emphasizes boundary dynamics require active mediation, with intermediaries facilitating translation, building trust, and negotiating arrangements. However, intermediaries face resource dependencies creating potential funder capture, legitimacy questions about representation, and tensions serving multiple constituencies with divergent interests.

Across actor categories, systematic exclusions become apparent. Food-insecure populations, farm workers, migrants, indigenous communities, and marginalized groups remain largely absent from governance innovations. Raja et al. (2018) document U.S. food planning identifying "power to convene" as key mechanism, yet convening depends on participation space access. Seattle's Local Food Action Initiative and Minneapolis's Homegrown Minneapolis demonstrate bureaucratic restructuring can institutionalize food as municipal priority, but persistent barriers including fiscal constraints and jurisdictional misalignment continue challenging implementation. More fundamentally, these initiatives rarely challenge existing power structures determining land ownership, market control, and legitimate knowledge.

Democratic deficits in food governance reflect broader patterns where sustainability transitions are designed and implemented by technical experts, policymakers, and educated elites while those most affected have minimal voice determining transition directions. This raises fundamental justice questions about whose sustainability is pursued and whether transitions not emerging from and centering marginalized communities can be truly transformative.

Mechanisms, Barriers, and Transformation Limits

Knowledge systems emerge as fundamental yet operate through complex, contested channels. Ingram (2018) analyzes agricultural knowledge system boundaries, demonstrating tensions between mainstream systems and alternatives like permaculture. Both "impermeable boundaries" maintaining separate domains and "permeable boundaries" enabling translation highlight knowledge's political nature. What counts as legitimate knowledge, who produces it, and how it circulates reflect power relations determining whose ways of knowing receive institutional recognition and resources.

Li et al. (2013) show how international intellectual property regimes work against participatory plant breeding in China despite international commitments supporting it, demonstrating how landscape factors can simultaneously enable and constrain transitions. Charatsari and Lioutas (2019) identify significant competency gaps among agricultural professionals lacking crucial sustainable agriculture promotion abilities, particularly networking, facilitation, and systems thinking. This gap indicates professional knowledge systems themselves require transformation, extending beyond curriculum reform to reimagining professionals' roles as facilitators rather than technical experts.

Mukhovi et al. (2020) identify three critical learning forms: experiential learning through demonstration, social learning via collective approaches, and transformative learning involving fundamental perception shifts. Their Kenya and Bolivia research emphasizes resilience depends not on knowledge transfer but building adaptive capacity through diverse mechanisms, challenging technocratic approaches assuming linear dissemination from experts to practitioners.

Hansen (2022) examines media's role shaping sustainability discourse through quantitative content analysis of 271 Danish newspaper articles, revealing how external events like COVID-19 and authorship type influence framing and proposed solutions. Reader-authored articles more often propose strong solutions based on values and norms, while journalists tend toward "softer" implications like revising social norms, highlighting media's agenda-setting power and how gatekeeping shapes what solutions become thinkable within public discourse.

Structural barriers transcend technical or economic challenges. Path dependencies, institutional inertia, and power asymmetries favor conventional systems in ways incremental reforms cannot address. Li et al. (2013) demonstrate how Chinese seed system organization reflects deep institutional legacies constraining reform possibilities, requiring coordinated changes across frameworks, organizations, and practices—yet these requirements make transformation politically difficult.

Zwartkruis et al. (2020) reveal the "dual regime challenge" where agricultural nature conservation must navigate both agricultural and nature regimes with conflicting logics, exemplifying how institutional fragmentation impedes transformation. Policy misalignment across governance levels and sectors creates additional barriers. Agricultural policies, trade regulations, and economic development strategies often work at cross-purposes, undermining sustainability objectives. Fragmented governance complicates coordination, with multiple agencies addressing food issues without coherent integration. Yet fragmentation is not accidental but reflects political compromises, sectoral interests, and deliberate state capacity limiting under neoliberal governance.

Infrastructure limitations constrain alternative development, particularly in secondary cities and rural areas. Riley and Crush (2022) document how African secondary cities lack basic infrastructure including wholesale markets, cold storage, processing facilities, and transportation networks. These gaps reinforce informal, inefficient chains generating waste and limiting farmer incomes. However, infrastructure development requires public investment that neoliberal economic models systematically constrain through austerity policies and privatization pressures.

Economic viability remains fundamental alternative challenge. Conventional systems benefit from economies of scale, established supply chains, and policy supports alternatives lack. Price competition disadvantages sustainable products when environmental and social costs remain externalized. Reynolds et al. (2019) demonstrate food waste reduction faces economic barriers as current structures incentivize overproduction and provide few waste prevention rewards.

Investment constraints limit alternative development, as sustainable initiatives require upfront capital for infrastructure, certification, and market development while facing longer payback periods and greater uncertainties than conventional investments. Financial institutions remain oriented toward conventional agriculture, viewing alternatives as higher risk, reflecting not merely information gaps but fundamental incompatibilities between financial capital's rapid return requirements and sustainable agriculture's longer timeframes and lower profit margins.

Alternative economic marginality raises fundamental market-based sustainability transition questions. If sustainable practices cannot compete economically with conventional systems externalizing costs, either externalities must be internalized through regulation and taxation, sustainable practices require permanent subsidization, or genuinely sustainable food systems are incompatible with market coordination and require alternative economic models.

Political barriers include vested interest resistance, lack of political will, and governance structures fragmenting authority and limiting coordination. Pothukuchi (2017) reveals Detroit urban agriculture faces political opposition from those viewing farming as development-incompatible, stemming from growth ideologies and racial politics rather than technical constraints. Bedore (2014) demonstrates how Baltimore growth machine politics instrumentalize food initiatives for economic development rather than sustainability objectives.

These cases reveal food system transitions challenge fundamental assumptions about development, progress, and economic growth embedded in contemporary governance. When sustainability genuinely conflicts with growth—as reducing meat consumption conflicts with agribusiness profits, or food sovereignty conflicts with global trade rules—political systems structured around growth imperatives face contradictions they cannot resolve without transformation themselves.

Cultural norms, dietary habits, and consumer preferences can reinforce conventional systems. Hansen (2022) reveals significant heterogeneity in consumer sustainability concerns, with many prioritizing convenience, price, and taste over sustainability. Behavior change interventions face limitations when structural contexts constrain sustainable choices, highlighting systemic change needs alongside individual initiatives. However, treating cultural norms as barriers risks depoliticizing what are often rational responses to structural constraints. When sustainable food costs more and requires more preparation time, expecting working-class families to choose sustainability blames individuals for structural failures.

Crisis events and disruptive moments create transformation possibilities. Davidson et al. (2016) show how BSE outbreaks catalyzed alternative beef production. Hansen (2022) documents how COVID-19 temporarily disrupted systems and consumption patterns, creating political opportunities and shifting public attention. Sarabia et al. (2021) identify crisis-driven opportunities alongside social movements and entrepreneurial initiatives as catalyzing policy changes. However, crises can catalyze transformation or reinforce existing systems depending on political navigation. COVID-19 revealed vulnerabilities yet largely strengthened existing players through state support for conventional supply chains.

Implications for Sustainability Transitions and Future Research

This synthesis reveals food system transitions as fundamentally political processes requiring more than technical innovation or market reform. The geographic case diversity demonstrates that while patterns exist—intermediary importance, multi-stakeholder governance, crisis-driven opportunities—specific pathways must emerge from local contexts rather than imposed blueprints. The transformation paradox—that incumbent systems shape their own transformation—implies existing governance structures inevitably constrain possibilities, suggesting limits to working within existing institutions and potential necessity for breaks with existing systems.

Food system transitions cannot be understood without explicit political economy analysis examining how capitalism, colonialism, and contemporary neoliberalism structure what transformations become possible. Persistent alternative economic marginality, corporate power concentration, and growth imperatives constraining policy innovation reflect not accidental market failures but systematic capitalist food system features. Scholarship ignoring these features produces knowledge serving existing power by suggesting problems can be solved within current structures.

Future research must explicitly examine how class relations, racial capitalism, gender dynamics, and global inequalities shape food system transitions. Land and capital ownership fundamentally determines what changes occur. Intellectual property regimes determining seed access shape agricultural possibilities. Trade rules enabling subsidized commodity dumping undermine local food security. These are not peripheral issues but central to understanding why voluntary sustainability transitions consistently fail.

Geographical contexts reflect historical processes producing uneven development, requiring scholarship that interrogates rather than naturalizes differences. Transitions in postcolonial contexts cannot follow pathways developed in wealthy countries with strong institutions. Recognizing this implies abandoning universal transition models for historically specific analyses examining how particular contexts' structural positions shape transformation possibilities.

Achieving sustainable food systems requires addressing justice, equity, and democracy alongside environmental concerns. Marginalized voice exclusion from food governance, corporate power concentration, and essential resource commodification emerge as core challenges rather than peripheral concerns. Transitions designed by and for elites will reproduce inequalities regardless of environmental improvements. Future research must examine how to center marginalized communities in transformation processes, ensuring these are transitions by rather than for those most affected by food system failures.

Transformative incrementalism offers pragmatic approaches navigating political constraints, yet risks co-optation where alternatives become absorbed without fundamentally challenging dominant logics. Whether incremental changes accumulate into transformation or merely enable system adaptation preventing genuine change remains questionable. Consistent radical alternative dilution through mainstreaming processes suggests incrementalism's transformative potential may be systematically constrained. This raises uncomfortable questions about whether sustainable food systems are possible within current political-economic structures or require fundamental breaks.

Knowledge production itself requires transformation, moving beyond expert-driven research to knowledge co-production recognizing multiple knowing ways. Indigenous knowledge, peasant knowledge, and food-insecure community experiential knowledge must be recognized as legitimate alongside scientific expertise. This requires not merely including marginalized communities in research but transforming what counts as knowledge and who determines research priorities.

Several research gaps warrant attention. First, limited research examines transitions in middle-income and developing countries where food systems face distinct challenges including rapid urbanization, informal economies, and limited governance capacity. Second, insufficient attention addresses distributional implications, including ensuring just transitions protecting vulnerable populations and not reproducing existing inequalities. Third, more research needs temporal dynamics including tipping points, transition speeds, and intervention sequencing. Fourth, limited work examines how to scale successful niche innovations without losing transformative characteristics through mainstreaming. Fifth, greater attention needs understanding how multiple transitions interact, including food system transitions with energy, water, health, and urban development transitions.

Conclusion

Food system sustainability transitions represent urgent necessities and profound challenges, requiring fundamental transformations in how food is produced, distributed, and consumed. This literature review has synthesized insights from 31 studies examining these transitions through diverse theoretical lenses, geographical contexts, and actor perspectives, while critically interrogating power relations and structural constraints often absent from mainstream transitions scholarship.

The Multi-Level Perspective provides valuable analytical framework for analyzing niche-regime dynamics, though it requires substantial expansion through attention to power relations, knowledge systems, and geographical specificity. Transitions unfold through multiple interrelated mechanisms including niche development, policy innovation, learning processes, and consumption changes, while facing significant structural, institutional, economic, political, and cultural barriers that voluntary approaches may be unable to overcome.

Geographical context profoundly shapes transition possibilities, with important differences reflecting not neutral variations but historical processes of colonialism, uneven development, and structural inequalities. Multiple actors play essential but contradictory roles, requiring collaborative governance approaches that navigate tensions among competing interests while maintaining space for alternative visions—though systematic exclusions reveal democratic deficits in supposedly participatory governance.

Despite significant barriers, various leverage points exist including crisis events, policy innovations, multi-stakeholder partnerships, and knowledge co-production processes. However, the persistent economic marginality of sustainable alternatives, the concentration of corporate power, and growth imperatives constraining policy innovation suggest that market-based voluntary transitions face fundamental limits. The transformation paradox—that existing systems shape their own transformation—suggests that genuinely radical transformation may require breaks with existing structures through contentious politics creating new possibilities.

Moving forward, food system sustainability transitions demand holistic approaches addressing multiple dimensions simultaneously, recognizing political contestation inherent in transformation processes, adapting strategies to context-specific conditions, facilitating collaboration across diverse actors while centering marginalized voices, and embracing adaptive learning and experimentation. Future research should address identified gaps while supporting practical transition efforts through engaged scholarship connecting academic insights with grassroots transformation initiatives.

Only through comprehensive, politically aware, and context-sensitive approaches that explicitly address power relations, reconceptualize food's value beyond commodity, and create genuinely inclusive governance can food systems transition toward sustainability while ensuring justice, resilience, and food security for all. Whether such transformation is possible within existing structures or requires fundamental breaks remains the central question facing both scholarship and practice.

---

Now I'll provide the improved version of your Paper C (Review B):

[Claude]

Improved Version of Paper C/Review B (3760 words)

Food System Transitions: A Human-Curated Synthesis of Critical Scholarship [this is Review C', an intermediary version, improved mostly for style]

Introduction: Beyond Technological Solutions

The transformation of food systems toward sustainability is a complex socio-technical process demanding fundamental reconfiguration of governance structures, knowledge systems, and power relations. This review synthesizes 31 significant studies selected from 350 publications addressing food system transitions, deliberately moving beyond dominant narratives focusing on efficiency and productivity. These papers illuminate the critical importance of agency, context-specific pathways, and the often-overlooked dimensions of justice and equity in achieving sustainable food futures.

The selection process prioritized works that challenge mainstream assumptions, engage with geographical diversity beyond European cases, and explicitly address power dynamics shaping transition possibilities. This curatorial approach acknowledges that literature reviews are not neutral knowledge summaries but interpretive acts reflecting particular epistemological and political commitments. The papers collectively reveal how food system transitions require fundamental reconfigurations rather than incremental adjustments, involving contested struggles over values, resources, and visions of what sustainable food systems should become.

Theoretical Foundations: Expanding and Challenging the Multi-Level Perspective

The Multi-Level Perspective (MLP) of the Dutch school of transition studies remains the dominant framework for describing system transitions, offering analytical advantages over economic models treating all agents as operating on identical levels. The MLP distinguishes three analytical levels: the landscape (macro-level socio-technical context), the regime (meso-level established practices and institutions), and niches (micro-level protected spaces for innovation). In this model, partly rooted in innovation studies, innovations initially develop in niches where specific conditions enable survival, then gradually change the larger regime, while the landscape independently provides changes that may trigger transformations.

Yet the reviewed literature reveals fundamental limitations requiring theoretical expansion. The framework's food system applications have exposed core assumption problems, particularly regarding analytical boundaries and transition process nature. Papachristos and Adamides (2016) address fundamental critiques by proposing retroductive methodology capturing causal mechanisms often obscured in traditional MLP applications, attempting to encompass system dynamics with qualitative analysis. They demonstrate how transitions can emerge from multi-system interactions rather than isolated sectoral niche changes. However, their methodology, while innovative, relies on simplified system dynamics models that cannot capture messy food system politics realities. Their functional foods demonstration proves particularly problematic—a corporate-driven innovation hardly representing the grassroots transitions observed throughout food systems.

Energy transitions within food systems reveal similar complexity. To et al. (2018) analyze bagasse cogeneration in Mauritius, showing how governance structures and institutional context prove crucial for managing sustainability transitions, requiring multi-actor collaborations transcending MLP's level-based categorization. Zwartkruis et al. (2020) identify the "dual regime challenge" where agricultural nature conservation must navigate both agricultural and nature regimes with conflicting logics, demonstrating how institutional fragmentation impedes transformation—a dynamic MLP's single-regime assumption cannot adequately capture.

Vivero-Pol (2017) extends critique by examining how food valuations—as commodity versus commons—shape political attitudes toward system transformation. This work reveals that counter-hegemonic transformers who view food multi-dimensionally and consider diverse value types can bring genuine system transformation rather than mere reform. The normative reframing has profound implications: when food is conceptualized solely as commodity, transformation becomes literally unthinkable; only when recognized simultaneously as human right, cultural determinant, and commons do transformative pathways emerge.

Davidson et al. (2016) fundamentally challenge MLP's foundational assumptions, showing that alternative beef production actors operate between alternative and conventional systems rather than within discrete analytical levels. Similarly, Luger et al. (2025) conceptualize the "purple zone" where municipal staff navigate between formal administrative duties and informal political action, highlighting how individual agency operates between structural constraints of political agendas and steadfast bureaucracies. In short, transformation and transition involve two-way interactions within and across the "levels," suggesting far more complex retroaction series between system components than MLP's linear model acknowledges.

Contestation, Conflict, and Transformation Dynamics

The integration of Social Movement Theory with MLP provides essential insights into how status quo contestation drives transitions. Elzen et al. (2011) analyze Dutch pig husbandry, revealing that successful transitions require alignment across multiple streams—technological, regulatory, market, and social—rather than innovation alone. This multi-stream approach fundamentally reconceptualizes how transitions unfold, moving beyond linear progression narratives to recognize the inherently political, contested nature of transformation.

Ingram et al. (2015) explore how Learning and Innovation Networks for Sustainable Agriculture (LINSA) interact with dominant agricultural regimes across Europe, identifying five distinct interaction modes: Compatible, Complementary, Emergent, Divergent, and Oppositional. Their analysis demonstrates that highly compatible networks achieve smoother diffusion but often experience sustainability ambition dilution, while more radical networks face marginalization but preserve transformative ideals. This work reveals complex negotiations and conflicts occurring at niche-regime interfaces, where sustainability transformations emerge through messy, dynamic, context-dependent interactions rather than simple adoption and diffusion processes. The process proves not only complex and multi-party, combining top-down and bottom-up dynamics, but is also fostered by contestation, conflicts, and compromises involving controversies over values and power struggles.

Jørgensen's "arenas of development" perspective, contrasting with MLP, accounts for these multiple level interactions and brings a less linear, more systemic transition process view. It is also more "flat" than "by levels," placing all actors on the same level in these interaction arenas. Arenas of Development (AoD) considers transitions occurring through conflicts, negotiations, alignments, and learning among various stakeholders, including non-humans in a Latourian Actor-Network Theory perspective. AoD gives attention to regulation mechanisms and the processes through which they occur. In AoD, actors can operate in several arenas simultaneously. "Stable actor worlds"—configurations of actors playing the same game—gradually emerge and can produce lasting change. While the MLP framework highlights Darwinian mechanisms of niche innovation and diffusion, AoD highlights multi-party emergence of new stable actor configurations through multiple regulatory systems, better capturing transition complexity and political nature.

Geographic and Cultural Diversity: Contextual Pathways and Colonial Legacies

The literature demonstrates significant transition pathway variation across different contexts, challenging universal MLP concept application. While MLP suggests transitions occur through protected niches, urban food systems research reveals more complex dynamics. Audet et al. (2017) demonstrate that Montreal seasonal food markets evolve at regime-niche intersections rather than in protected spaces, challenging framework assumptions about clear analytical separation and revealing inherent tensions between market efficiency, community building, and environmental sustainability that reflect broader food system transition contradictions.

Notably absent from these transition frameworks is serious engagement with how colonial agricultural systems continue structuring contemporary food systems. The emphasis on "Global South innovations" risks romanticizing struggles emerging from structural adjustment policies and ongoing extractivism. The Ugandan case cannot be understood without examining how cash crop dependencies established under colonialism shape current agroecological possibilities. What appears as innovation often represents survival strategies in contexts of deliberately produced scarcity resulting from decades of neoliberal restructuring.

This curated collection's distinguishing feature is its geographic breadth and attention to diverse transition contexts. Global South examples offer critical perspectives on alternative pathways and grassroots innovation that complement European institutionalized transition management approaches. Isgren and Ness (2017) examine agroecology in Uganda's Rwenzori region, demonstrating how civil society networks (primarily NGOs, community-based organizations, and faith-based organizations) fill gaps left by state withdrawal while navigating deteriorating political climates. Their analysis reveals agroecology emerging as an "unintended consequence of neoliberal development policies"—specifically how agricultural extension and support services devolution to non-state actors created space for alternative approaches that would not have emerged under direct state control. This contradictory mechanism operates through neoliberal policies simultaneously weakening state capacity while enabling donor-funded civil society organizations to promote agroecological practices challenging conventional modernization approaches. This pathway differs fundamentally from European transition experiences, where formal state institutions, research organizations, and established policy frameworks actively guide and legitimize alternative agricultural practices through deliberate transition management programs.

The integration of food systems with broader structural transformation processes is further illuminated by Arslan et al. (2022), whose comprehensive analysis of 81 low- and middle-income countries reveals that structural transformation (measured as non-agricultural GDP share) correlates more consistently with improved food system outcomes than rural transformation alone. Their Food Systems Index construction demonstrates that financial inclusion, digital connectivity, women's empowerment, and education emerge as key transformation levers, while highlighting the "double burden" of malnutrition and rapid population growth in countries with weakest food system outcomes. However, the analysis could more explicitly engage with how global trade relations, debt burdens, and intellectual property regimes systematically produce these vulnerabilities rather than treating them primarily as technical development challenges.

Latin American cases provide particularly valuable insights into alternative development pathways. The Nicaraguan agroecological transition analyzed by Schiller et al. (2023) demonstrates how knowledge flows through intermediaries enable niche development despite limited state support, revealing possibilities even in state weakness contexts—though this weakness itself reflects Nicaragua's global agricultural market position and intervention histories shaping state capacity. The research effectively demonstrates intermediaries' crucial roles in bridging farmers, researchers, policymakers, and markets while adapting external knowledge to local contexts, revealing that effective intermediaries possess diverse competencies including technical expertise, facilitation skills, and political acumen.

Governance Innovation and Political Navigation

Perhaps this collection's most significant contribution lies in revealing food system governance's inherently political nature. Luger et al. (2025) demonstrate through Urban Living Labs (ULLs)—emerging spaces where individuals test collaborations for societal transformation—how municipal staff in 10 European cities must "act politically simply to accomplish predefined tasks," employing advocacy and alignment strategies to navigate administrative and political constraints. This finding challenges technocratic governance models assuming neutral sustainability policy implementation while ignoring fluctuating funding and shifting implementation agendas accompanying political transitions. The research reveals sustainability initiatives' precarity when dependent on individual champions and vulnerable to political shifts, suggesting transformative governance requires institutional embedding beyond personal commitment.

The comparative Bristol and Malmö analysis illuminates how different institutional contexts shape urban food governance. Bristol's civil society-led approach contrasts with Malmö's government-driven model, yet both create new "spaces of deliberation" reconfiguring traditional food governance structures. However, these cases reveal a troubling pattern: even supposedly "alternative" pathways often reproduce existing power structures. Bristol's Food Policy Council, despite civil society leadership, remains dominated by white, middle-class professionals. Malmö's government-driven approach, while achieving organic procurement targets, does little addressing food access for immigrant communities. Genuine food-insecure population participation absence in these "innovative" governance arrangements suggests these are transitions for, not by, those most affected by food system failures. This critique extends Moragues-Faus and Morgan's (2015) analysis of the emergent urban food policy world, requiring explicit attention to whose voices shape these new foodscapes and whose interests they ultimately serve.

Building on urban governance innovations, Larbi et al. (2022) examine Global South cities' sustainability transitions through comparative study of Curitiba, Brazil, and Accra, Ghana. Curitiba's success pioneering cost-effective innovations in public transport, waste management, and flood control under strong political leadership demonstrates potential for low-cost, context-sensitive solutions. However, their Accra analysis reveals technical solutions cannot be transferred wholesale across contexts. Instead, cities must build on existing local capacities, such as Accra's strong communal labor culture, while involving both state and non-state actors through tailored strategies. This comparison effectively demonstrates that transition frameworks developed in Global North contexts require substantial adaptation for Global South applications, reflecting fundamentally unequal positions in global political economy rather than merely different contexts.

These diverse pathways underscore that no universal transition model exists; rather, transformations must align with specific socio-political contexts while addressing global sustainability imperatives. They also reveal how incumbent systems inform their own transformation through governance structures—what might be termed the "transformation paradox." While seemingly obvious ex-post, this implies every transformation must seriously account for given conditions, not only target results, suggesting limits to working within existing institutions and potential necessity for breaks with existing systems.

Secondary cities' transition scaling challenges beyond individual cities are addressed by Riley and Crush (2022) in their comprehensive African secondary cities examination. Their analysis reveals how local governments, motivated by urban food security needs but constrained by limited resources and fragmented governance, must navigate complex relationships with informal sector actors, urban farmers, civil society organizations, and vulnerable households. The study demonstrates successful transformation requires recognizing informal food systems as legitimate urban food provisioning components rather than development obstacles, while building on existing social networks and community capacities. This insight proves particularly crucial as informal markets, street vendors, and urban agriculture already feed millions but remain systematically excluded from formal planning and criminalized through regulations.

New governance form emergence represents a critical transition dynamic. Food policy councils in Bristol, urban living labs in Amsterdam, Barcelona, and other European cities studied by Luger et al. (2025), and collective management approaches for agricultural nature conservation in the Netherlands all represent attempts transcending traditional sectoral boundaries. The "convening power of food" as Bedore (2014) analyzes in Baltimore demonstrates how food policy becomes "partnerable" because it appeals to both growth coalition members and broader civil society. Food policy initiatives succeed when framed complementing rather than challenging existing economic development strategies, with discursive mechanisms enabling actors to frame food policy in depoliticized terms around health, innovation, and city reputation. However, these mechanisms can limit transformative potential by aligning new initiatives with existing power structures and growth imperatives, raising fundamental questions about whether voluntary transitions working within existing political-economic constraints can achieve genuine transformation.

Building on governance innovation concepts, Buchan et al. (2019) introduce "Transformative Incrementalism" as a framework for understanding how government actors influence sustainable food system transitions through gradual, power-laden processes rather than abrupt policy shifts. Their analysis reveals government influence operating through both authoritative mechanisms (policy documents, official mandates) and relational mechanisms (coalition building, narrative framing), with planning documents like Official Community Plans serving as legitimizing tools to embed sustainable food agendas within institutional mandates. However, the framework must grapple with co-optation risks where alternative initiatives become absorbed into existing systems without fundamentally challenging dominant logics, and whether incremental changes can accumulate into transformation or merely enable system adaptation preventing genuine change.

Yet as Haylock and Connelly (2018) reveal through insider versus outsider strategies in New Zealand food networks, these new arrangements face constant tensions between maintaining alternative values and achieving institutional legitimacy, with groups like Dunedin's Our Food Network explicitly choosing to maintain local government distance to preserve their transformative agenda. This tension proves not resolvable through better partnership design but reflects fundamental incompatibilities between grassroots movements seeking system transformation and governments operating within existing political-economic constraints, suggesting that genuinely transformative alternatives may necessarily maintain critical distance from formal institutions.

Participatory Planning, Power Dynamics, and Economic Contradictions

The potential for more inclusive governance approaches is demonstrated by Raja et al. (2018) in their U.S. food planning initiatives analysis. They identify "power to convene" as a key mechanism whereby local governments leverage soft power to foster cross-sector coalitions. Seattle's Local Food Action Initiative and Minneapolis's Homegrown Minneapolis demonstrate how bureaucratic restructuring and co-leading models can institutionalize food as municipal priority while embedding food indicators into comprehensive plans. However, persistent barriers including fiscal constraints and jurisdictional misalignment between food systems and political boundaries continue challenging implementation. More fundamentally, these initiatives rarely challenge existing power structures determining land ownership, market control, and what counts as legitimate knowledge.

Power dynamics permeate these governance innovations. The Detroit urban agriculture policy analysis reveals how "shrinkage regimes" view farming as incompatible with development despite vast vacant land availability. This resistance stems not from technical constraints but from deeply embedded growth ideologies and racial politics shaping land use decisions among city governance officials. Detroit's vacant land resulted from decades of deindustrialization, white flight, and municipal disinvestment in Black neighborhoods; treating this land as neutral "opportunity" for urban agriculture erases these histories and ongoing displacement pressures facing long-term residents. Pothukuchi's (2017) analysis effectively demonstrates that urban food transitions are not merely technical but deeply political, involving power struggles over resource allocation, urban identity, and development trajectories fundamentally shaped by racial capitalism.

Transformative capacity frameworks' application is demonstrated by Sarabia et al. (2021) in their Valencia agri-food system transformation analysis. Their research reveals how three interrelated milestones—grassroots social movements empowerment, local government political shift, and Milan Urban Food Policy Pact adherence—catalyzed Municipal Food Council and Agri-food Strategy 2025 creation. Using the Urban Transformative Capacity Framework, they identify significant progress in inclusive governance and community-based experimentation, while highlighting weaknesses in embedding innovation and cross-scalar coordination. This case demonstrates how civil society movements can influence policy when political opportunities arise, yet also reveals dependence on favorable political conditions that may not persist across electoral cycles, raising questions about transformation sustainability beyond particular political moments.

The reviewed studies reveal fundamental tensions between market-based food systems and sustainability objectives. Davidson et al. (2016) identify the "conventionalization paradox" wherein successful alternative beef production scaling could eliminate price premiums sustaining these practices. This reflects fundamental tension between market-based sustainability transitions and genuine transformation—alternatives succeeding on market terms risk losing alternative character, revealing potential incompatibilities between market coordination and genuinely sustainable food systems. Similar paradoxes emerge in New Zealand cases, where community food networks struggle with capacity and resource constraints while attempting to maintain independence from potentially restrictive government partnerships.

Private sector engagement emerges as deeply ambiguous. While Dubbeling et al. (2016) document successful small and medium enterprise participation in city-region food systems, they also reveal how concentration—with five companies controlling 80% of Ecuador's meat, dairy, and grain markets—constrains transformation possibilities. The literature suggests meaningful private sector contribution requires regulatory frameworks and incentives aligning profit with sustainability rather than assuming market forces naturally produce sustainable outcomes. This reflects broader corporate food system control trends that no amount of governance innovation can overcome without directly addressing market power through antitrust enforcement, public ownership, or alternative economic models.

Community Supported Agriculture, farmers' markets, and other alternative distribution systems represent attempts reconstructing market relationships, as demonstrated in Christchurch and Dunedin's emerging food networks. Yet Zwartkruis et al. (2020) find agricultural nature conservation provides less than 10% of income for 81% of participating Dutch farms despite decades of support. This economic marginality raises fundamental questions about market-based sustainability transition viability, suggesting that expecting farmers to provide environmental services voluntarily while competing in globalized commodity markets may be fundamentally contradictory. If sustainable practices cannot compete economically with conventional systems externalizing costs, either externalities must be internalized through regulation, sustainable practices require permanent subsidization, or genuinely sustainable food systems are incompatible with market coordination and require alternative economic models—a conclusion the literature generally avoids but evidence consistently suggests.

Consumer Interventions, System-Wide Change, and Knowledge Systems

Consumer behavioral change translation challenges are addressed by Reynolds et al. (2019) in their comprehensive consumer-stage food waste reduction interventions review. Despite household food waste research surge, only 17 of 292 peer-reviewed studies report applied interventions with measurable waste reductions. These interventions fall into information-based (newsletters, curricula), technological (plate-size modifications, apps), and policy/system changes (meal standards, portion defaults), with reported waste reductions varying widely (12-57%). The review reveals significant methodological challenges including self-report reliance rather than compositional analyses, small non-representative samples, and theoretical grounding lack, highlighting gaps between sustainability awareness and effective intervention design. More fundamentally, the review emphasizes individual behavior change alone cannot drive systemic transitions; structural changes in food provisioning systems, retail practices incentivizing overproduction, and policy frameworks prove equally essential yet remain underdeveloped compared to consumer-focused approaches.

System-wide transformation potential is illustrated by Ryghaug and Skjølsvold's (2023) Norwegian electric vehicle transition analysis, which achieved over 86% battery and plug-in hybrid share among new private cars by 2021. Their longitudinal study reveals how persistent demand-side incentives, NGO intermediation, and user satisfaction created rapid diffusion momentum, while also triggering broader transport sector electrification. However, the case highlights Norway's unique conditions—wealth and renewable energy resources—and raises second-order sustainability dilemmas around mining impacts and material justice, underscoring that transitioning regimes spawn fresh systemic trade-offs. This case proves instructive for food systems by demonstrating both scaling possibilities through coordinated policy support and persistent sustainability challenges even in successful transitions, suggesting transformation never fully resolves contradictions but shifts them to new terrains requiring ongoing attention.

The reviewed literature reveals knowledge and learning as fundamental to food system transitions yet operating through complex, contested channels. Ingram's (2018) agricultural knowledge system boundaries analysis demonstrates tensions between mainstream Agricultural Knowledge Systems and alternatives like permaculture. Both "impermeable boundaries" maintaining separate knowledge domains and "permeable boundaries" enabling translation existence highlights knowledge production and dissemination's political nature. What counts as legitimate knowledge, who gets to produce it, and how it circulates are not technical questions but reflect power relations determining whose knowing ways receive institutional recognition and resource support.

International knowledge governance creates additional complexity. Li et al. (2013) show how international treaties and conventions on plant genetic resources affect participatory maize breeding development in China, with "intellectual property regimes" working against participatory breeding despite international commitments supporting it. This demonstrates how landscape-level factors can simultaneously enable and constrain transitions, revealing that global governance structures fundamentally shape local agricultural possibilities in ways that MLP's landscape concept inadequately captures.

Mukhovi et al. (2020) identify three critical Global South food systems learning forms: experiential learning through demonstration, social learning via collective approaches, and transformative learning involving fundamental perception shifts. Their Kenya and Bolivia research emphasizes resilience depends not merely on knowledge transfer but building adaptive capacity through diverse learning mechanisms. This finding challenges technocratic approaches assuming linear knowledge dissemination from experts to practitioners, revealing learning as social, contextual, and fundamentally political rather than merely technical information transmission.

The competency gap Charatsari and Lioutas (2019) identify reveals systemic agricultural education failure. Their finding that agronomists lack crucial sustainable agriculture promotion competencies—particularly networking, facilitation, and systems thinking—indicates professional knowledge systems themselves require transformation. This extends beyond curriculum reform to reimagining agricultural professionals' roles as facilitators rather than technical experts, though this reimagining threatens existing professional identities and institutional positions, creating resistance to knowledge system transformation that technical solutions cannot address.

Hansen's (2022) media role examination in shaping food sustainability discourse through quantitative content analysis of 271 Danish newspaper articles reveals how external events like COVID-19 and authorship type influence framing and proposed solutions relationships. The study demonstrates reader-authored articles more likely than journalist-authored pieces to propose strong solutions based on values and norms, while journalists tend toward "softer" implications like revising social norms. This research highlights media's agenda-setting power and suggests policymakers must consider both external events and authorship dynamics in crafting effective sustainability messaging, while also revealing how media gatekeeping shapes what solutions become thinkable within public discourse, systematically constraining radical alternatives.

Implications for Sustainability Transitions

This synthesis reveals food system transitions as fundamentally political processes requiring more than technical innovation or market reform. The geographic case diversity demonstrates that while patterns exist—intermediaries importance, multi-stakeholder governance, crisis-driven opportunities—specific pathways must emerge from local contexts rather than imposed blueprints. There is what we could call a transformation paradox in the fact that existing governance structures inform transition processes, but hence also their results. While this seems obvious ex-post, this implies every transformation must seriously account for given conditions, not only target results, suggesting limits to working within existing institutions and potential necessity for breaks with existing systems.

The literature's attention to agency, learning, and governance innovation suggests transformation depends on cultivating new collective action forms and institutional arrangements. Yet alternatives' persistent economic marginality and incumbent regimes' power indicate voluntary transition remains unlikely without fundamental structural changes. Instead, the studies point toward regulatory intervention necessity, public investment, and fundamental growth-oriented development model questioning. Most critically, these papers collectively argue that achieving sustainable food systems requires addressing justice, equity, and democracy questions alongside environmental concerns. Marginalized voice exclusion from food governance, corporate power concentration, and essential resource commodification emerge as core challenges rather than peripheral concerns. True transformation demands not just sustainable production but equitable distribution, democratic participation, and food's multiple values recognition beyond market price.

Conclusion: Toward Transformative Food Futures

This curated literature review, involving not only classic or mainstream work, puts in full light why modeling societal transformation is a difficult exercise. Research reveals food system transitions as complex, contested, inherently political processes that cannot be reduced to technological fixes or market mechanisms. The context diversity, from Ugandan agroecology to Dutch nature conservation, from Detroit's vacant lots to Valencia's food councils, demonstrates that while sustainability challenges are universal, solution paths are inevitably context-specific.

The path forward requires abandoning comfortable gradual transition narratives and confronting the reality that sustainable food systems may be impossible within current political-economic structures and may go through radical breaks—land occupations, seed sovereignty movements, food riots—that historically have driven genuine food system transformation. On the other hand, it appears the incumbent system and its governance structure is always involved in the self-transformative path, and may be so constructively (the transformation paradox). The path forward requires embracing this complexity while maintaining focus on fundamental transformation rather than incremental reform. Only by addressing power relations, reconceptualizing food's value, and creating inclusive governance can food systems transition toward genuine sustainability and justice.

[Saadi] thanks. what are the remaining differences between these two improved versions?

[Claude]

Remaining Differences Between the Two Improved Reviews

Excellent question. Despite both reviews being improved, they retain distinct characteristics that reflect their original orientations. Here are the key remaining differences:

1. Tone and Voice

My Improved Review:

- More academic and measured, even with critical elements added
- Uses phrases like "this review recognizes," "the literature reveals," maintaining some analytical distance
- Balances critique with acknowledgment of framework utility
- More cautious in making strong political claims

Your Improved Review:

- More assertive and directly critical
- Uses phrases like "fundamentally challenge," "notably absent," "troubling pattern"
- Takes clearer political stances without hedging
- More willing to make bold claims about what the evidence shows

2. Treatment of Theory

My Improved Review:

- Presents MLP problems as "limitations" that can be addressed through expansion
- Discusses theoretical alternatives as "complementary" approaches
- Still somewhat deferential to established frameworks

Your Improved Review:

- Frames MLP problems as "fundamental" issues with "core assumptions"
- Positions alternatives like AoD as genuine competitors offering superior perspectives

- More willing to reject rather than just critique established frameworks
- The phrase "partly rooted in innovation studies" signals skepticism about MLP's origins

## 3. Handling of Colonial/Historical Context

My Improved Review:

- Includes colonial context as an added critical element
- Treats it as an important consideration that was previously overlooked

Your Improved Review:

- Integrates colonial analysis more organically as essential to understanding
- The phrase "Notably absent from these transition frameworks is serious engagement with how colonial agricultural systems continue structuring contemporary food systems" is more forceful
- Makes stronger connections between historical processes and current conditions

## 4. Specificity and Evidence Deployment

My Improved Review:

- Uses specific data points but sometimes in more general supportive roles
- Evidence supports broader arguments

Your Improved Review:

- Deploys specific evidence more strategically as rhetorical weapons
- Numbers and details (80% market concentration, 10% income, 86% EV share) hit harder
- Evidence is weaponized to support critical claims

## 5. Treatment of Contradictions

My Improved Review:

- Presents contradictions as "fundamental tensions" requiring attention
- Suggests these may be "irresolvable within existing systems" but still leaves some ambiguity

Your Improved Review:

- Presents contradictions more starkly as exposing system impossibilities
- The "conventionalization paradox" and "transformation paradox" are treated as revealing fundamental incompatibilities
- More willing to conclude that market-based transitions may simply not work

## 6. Critique of Case Studies

My Improved Review:

- Questions who benefits from urban food governance innovations
- Notes that "Bristol's Food Policy Council remains dominated by white, middle-class professionals"

Your Improved Review:

- Goes further: explicitly calls this a "troubling pattern" and states these are "transitions for, not by, those most affected"
- The critique is sharper and more accusatory
- More willing to characterize supposedly progressive initiatives as failures

## 7. Discussion of Radical Alternatives

My Improved Review:

- Mentions "breaks with existing systems," "land occupations, seed sovereignty movements, food riots"
- Presents these as possibilities that "may be necessary"

Your Improved Review:

- States these "historically have driven genuine food system transformation"
- Treats radical alternatives as empirically validated rather than speculative
- More affirmative about contentious politics' necessity

## 8. Academic Positioning

My Improved Review:

- Still partly works within academic conventions of balance
- "Future research should examine..." maintains conventional scholarly voice
- Calls for "engaged scholarship" but from a relatively safe position

Your Improved Review:

- Explicitly identifies as "human-curated synthesis of critical scholarship"
- Acknowledges curatorial choices and interpretive nature upfront
- More reflexive about the review's own positioning and purposes

## 9. Treatment of Incrementalism

My Improved Review:

- Discusses transformative incrementalism as having "potential" but facing "risks"
- Questions whether it can work but leaves the question somewhat open

Your Improved Review:

- More skeptical: "the framework must grapple with co-optation risks"
- The phrase "whether incremental changes can accumulate into transformation or merely enable system adaptation preventing genuine change" is framed more as rhetorical question with implied answer

## 10. Framing of Justice and Equity

My Improved Review:

- Treats justice and equity as essential considerations alongside environmental concerns
- Argues they must be "centered" in transformation

Your Improved Review:

- Goes further to identify "marginalized voice exclusion," "corporate power concentration," and "commodification" as "core challenges rather than peripheral concerns"
- Frames these not just as important considerations but as fundamental obstacles

## 11. Use of Scholarly Citations

My Improved Review:

- Citations are integrated smoothly but conventionally
- Year and author appear in standard academic format

Your Improved Review:

- Citations are slightly more varied in integration style
- Sometimes bundles multiple works more extensively
- Uses citations more to build cumulative arguments

## 12. Conclusion Orientation

My Improved Review:

- Ends by stating "Whether such transformation is possible within existing structures or requires fundamental breaks remains the central question"
- Leaves this as an open question for scholarship and practice

Your Improved Review:

- Ends more definitively: "The path forward requires abandoning comfortable gradual transition narratives and confronting the reality"
- Takes a clearer position that radical breaks are likely necessary
- Less willing to leave fundamental questions entirely open

## 13. Methodological Transparency

My Improved Review:

- Acknowledges selection and interpretation in introduction
- Maintains some traditional scholarly distance

Your Improved Review:

- Explicitly states "31 significant studies selected from 350 publications"
- More upfront about curatorial process and criteria
- States directly that "This curatorial approach acknowledges that literature reviews are not neutral knowledge summaries but interpretive acts"

## 14. Treatment of Success Cases

My Improved Review:

- Discusses Curitiba, Norway, Valencia as demonstrating "possibilities"
- Acknowledges their successes while noting limitations

Your Improved Review:

- More critical even of success cases

- Norway's EV transition is immediately followed by "second-order sustainability dilemmas"
- Success cases are treated more as revealing new contradictions than proving possibilities

15. Vocabulary Choices

My Improved Review:

- "Constraints," "limitations," "challenges," "barriers"
- More moderate descriptive language

Your Improved Review:

- "Fundamental contradictions," "systematically excluded," "criminalized," "deliberate dismantling"
- Stronger, more accusatory language choices

Summary of Key Distinction

The fundamental remaining difference is this:

**My improved review** has become more critical and politically aware but **still operates largely within conventional academic discourse norms**—it critiques but maintains scholarly balance, questions but leaves many answers open, challenges frameworks but respectfully.

**Your improved review challenges the conventions of academic discourse itself**—it takes clear political positions, makes strong claims, explicitly acknowledges its interpretive stance, and is willing to conclude that dominant approaches are fundamentally flawed rather than just limited.

Put differently:

- **My review** says: "The evidence suggests serious problems that may require radical solutions"
- **Your review** says: "The evidence demonstrates fundamental contradictions proving current approaches cannot work"

This reflects a deeper epistemological difference about what scholarship should do—describe and suggest (mine) versus intervene and argue (yours). Both are legitimate scholarly approaches, but they serve different purposes and audiences.

[Claude]

Synthesized Literature Review: Food System Transitions (3760 words)

Food System Transitions: Critical Perspectives on Transformation, Power, and Possibility

Introduction: Reframing Sustainability Transitions

The global food system faces unprecedented pressures from climate change, population growth, resource depletion, and deepening social inequalities, demanding fundamental transformations toward sustainability. Yet conceptualizing these transformations requires moving beyond technical narratives of efficiency and productivity to recognize food system transitions as inherently political processes involving contested struggles over resources, values, knowledge, and power. This literature review synthesizes insights from 31 scholarly works selected for their theoretical rigor, geographical diversity, and critical engagement with power dynamics often marginalized in mainstream sustainability transitions scholarship.

This synthesis explicitly acknowledges its interpretive nature. Literature reviews are not neutral knowledge summaries but curatorial acts reflecting particular epistemological and political commitments. The selected works prioritize scholarship that challenges dominant assumptions, engages with contexts beyond European case studies, and explicitly addresses justice and equity alongside environmental concerns. These papers collectively reveal that food system transitions require fundamental reconfigurations of governance structures, knowledge systems, and power relations rather than incremental technical adjustments.

The review critically examines the Multi-Level Perspective (MLP) as the dominant framework while exploring theoretical alternatives that better capture transition complexity. It analyzes key mechanisms driving change while foregrounding contradictions suggesting that market-based voluntary transitions face fundamental limits. Geographical variations are explored not as neutral differences but as products of colonial legacies, uneven development, and structural inequalities. Actor roles are examined while questioning who benefits from various pathways and whose voices remain systematically excluded. Finally, the review synthesizes knowledge on barriers and enablers while interrogating whether genuinely sustainable and just food systems are possible within current political-economic structures or require fundamental breaks.

Theoretical Foundations: Interrogating and Expanding Transition Frameworks

The Multi-Level Perspective of the Dutch school of transition studies remains the dominant framework for analyzing food system transitions, offering analytical advantages over economic models that treat all agents as operating identically. The MLP distinguishes three analytical levels: landscapes (macro-level socio-technical contexts), regimes (meso-level established practices and institutions), and niches (micro-level protected spaces for innovation). Partly rooted in innovation studies, the framework suggests innovations initially develop in niches where specific conditions enable survival, then gradually change larger regimes, while landscapes independently provide changes that may trigger transformations.

However, the reviewed literature exposes fundamental problems with MLP's core assumptions, particularly regarding analytical boundaries and transition process nature. Davidson et al. (2016) demonstrate that alternative beef producers in Alberta operate simultaneously between alternative and conventional systems rather than within discrete analytical levels, fundamentally challenging the niche-regime distinction. Audet et al. (2017) reveal that Montreal seasonal food markets evolve at regime-niche intersections rather than in protected spaces, embodying tensions between market efficiency, community building, and environmental sustainability that reflect deeper contradictions in capitalist food systems attempting sustainability transitions.

Luger et al. (2025) conceptualize a "purple zone" where municipal staff navigate between formal administrative duties and informal political action, operating across supposed boundaries in ways that challenge MLP's foundational categories. Zwartkruis et al. (2020) identify the "dual regime challenge" where agricultural nature conservation must navigate both agricultural and nature regimes with conflicting institutional logics, demonstrating how MLP's single-regime assumption fails to capture how institutional fragmentation impedes transformation. When systems must simultaneously satisfy multiple regimes with contradictory requirements, the framework's linear transition narrative breaks down entirely.

Papachristos and Adamides (2016) propose retroductive systems-based methodology attempting to capture causal mechanisms and feedback loops often obscured in traditional MLP applications. However, their functional foods demonstration reveals framework bias toward corporate-driven innovations fitting existing market logics rather than grassroots transformations challenging fundamental system assumptions, exposing limitations even in methodological advances.

Ingram (2018) advances understanding by examining boundary dynamics between niche and regime knowledge systems, revealing that niches and regimes possess distinct epistemological foundations—niches emphasizing experiential, place-based knowledge while regimes privilege standardized scientific expertise. This demonstrates that knowledge barriers are fundamentally political rather than merely technical. Ingram et al. (2015) identify five interaction modes between Learning and Innovation Networks for Sustainable Agriculture (LINSA) and dominant regimes: Compatible, Complementary, Emergent, Divergent, and Oppositional. Crucially, highly compatible networks achieve smoother diffusion but often experience diluted sustainability ambitions, while more radical networks face marginalization but preserve transformative ideals. This reveals a fundamental dilemma: the conditions enabling niche scaling often undermine transformative potential through co-optation processes.

Social Movement Theory integration with MLP provides essential insights into how status quo contestation drives transitions. Elzen et al. (2011) demonstrate how normative contestations around animal welfare in Dutch pig husbandry drive system innovation, revealing that successful transitions require alignment across technological, regulatory, market, and social streams rather than technological innovation alone. This multi-stream approach fundamentally reconceptualizes transitions as emerging through conflict, compromise, and political struggle rather than linear progression, foregrounding the contested nature of what constitutes "improvement."

Jørgensen's arenas of development (AoD) perspective offers a more systemic and less hierarchical alternative to MLP's level-based ontology. AoD considers that transitions occur through conflicts, negotiations, alignments, and learning among various stakeholders in interaction arenas, adopting a Latourian Actor-Network Theory approach that includes non-human actants. Actors can operate in several arenas simultaneously, and "stable actor worlds"—configurations of actors playing similar games—gradually emerge and can produce lasting change. While MLP highlights Darwinian mechanisms of innovation and diffusion, AoD emphasizes multi-party emergence of new stable actor configurations through multiple regulatory systems, better capturing transition complexity and inherently political nature.

Vivero-Pol (2017) explores how food valuation—as commodity versus commons—fundamentally shapes political attitudes toward system transformation and actor agency. When food is conceptualized solely as commodity, system transformation becomes literally unthinkable; only when recognized simultaneously as human right, cultural determinant, and commons do transformative pathways become conceivable. This reveals that "counter-hegemonic transformers" viewing food multi-dimensionally can envision genuine system transformation rather than mere reform within existing market logics, demonstrating how normative framings fundamentally shape transition possibilities.

These theoretical engagements reveal what might be termed the "transformation paradox": incumbent systems and governance structures inevitably shape their own transformation processes, simultaneously enabling and constraining what transformations become possible. Systems cannot stand outside themselves to transform; they must use existing structures, which necessarily shapes and potentially limits transformation possibilities. While this seems obvious in retrospect, it implies that every transformation must seriously account for existing conditions rather than focusing solely on desired endpoints, suggesting limits to working within existing institutions and potential necessity for breaks with existing systems through contentious politics creating new possibilities.

Geographical Contexts: Colonial Legacies, Structural Inequalities, and Uneven Development

Food system transitions manifest profoundly differently across geographical contexts, reflecting not neutral differences but historical processes of colonialism, uneven development, structural adjustment, and ongoing extractivism. Understanding these variations requires moving beyond comparative description to interrogating why differences exist and whose interests they serve.

Notably absent from most transition frameworks is serious engagement with how colonial agricultural systems continue structuring contemporary food systems. The emphasis on "Global South innovations" in sustainability transitions literature risks romanticizing struggles that emerge from structural adjustment policies and ongoing extractivism rather than recognizing these as responses to historically produced vulnerabilities. Isgren and Ness (2017) examine agroecology in Uganda's Rwenzori region, revealing how civil society networks fill gaps left by state withdrawal under neoliberal restructuring. Agroecology emerges as an "unintended consequence of neoliberal development policies"—specifically how devolution of agricultural extension and support services to non-state actors created space for alternative approaches that would not have emerged under direct state control.

However, this Ugandan case cannot be understood without examining how cash crop dependencies established under British colonialism continue shaping current agroecological possibilities. The very need for civil society organizations to provide services reflects decades of structural adjustment policies that dismantled state agricultural support systems, creating food insecurity that agroecology now attempts to address. What appears as "innovation" might more accurately be understood as survival strategies in contexts of deliberately produced scarcity. This pathway differs fundamentally from European transition experiences, where formal state institutions, research organizations, and established policy frameworks actively guide and legitimize alternative agricultural practices through deliberate transition management programs, revealing that Global North and South transitions reflect fundamentally unequal positions in global political economy rather than simply different contexts.

Arslan et al. (2022) provide comprehensive analysis of 81 low- and middle-income countries, revealing that structural transformation (measured as non-agricultural GDP share) correlates more consistently with improved food system outcomes than rural transformation alone. Their Food Systems Index demonstrates that financial inclusion, digital connectivity, women's empowerment, and education emerge as key levers, while highlighting the "double burden" of malnutrition and rapid population growth in countries with weakest outcomes. Yet the analysis could more explicitly engage with how global trade relations, debt burdens, and intellectual property regimes systematically produce these vulnerabilities rather than treating them primarily as technical development challenges.

Latin American cases provide valuable insights while raising transferability questions. Schiller et al. (2023) demonstrate how knowledge flows through intermediaries enable agroecological niche development in Nicaragua despite limited state support, revealing possibilities even in state weakness contexts—though this weakness itself reflects Nicaragua's position in global agricultural markets and intervention histories shaping state capacity. The research effectively demonstrates intermediaries' crucial roles in bridging farmers, researchers, policymakers, and markets while adapting external knowledge to local contexts, revealing that effective intermediaries possess diverse competencies including technical expertise, facilitation skills, and political acumen.

To et al. (2018) analyze bagasse cogeneration transitions in Mauritius, showing how coordinated multi-level policies accelerated change, yet Mauritius's success reflects unique positioning as small island state with particular political economy enabling policy coherence rare elsewhere. These cases collectively demonstrate that geographical contexts are not neutral variables but reflect historical processes, power relations, and structural positions in global political economy that fundamentally shape transition possibilities.

Urban contexts have become focal points for food system innovation, with cities serving as governance laboratories. Moragues-Faus and Morgan (2015) document the emergent world of urban food policy, representing shifts from viewing food as rural or national issue to recognizing cities' crucial sustainability roles. However, these innovations warrant critical scrutiny regarding whose interests they serve and whose voices remain excluded. Pothukuchi (2017) reveals how Detroit urban agriculture faces political opposition from "shrinkage regimes" viewing farming as incompatible with development, stemming from deeply embedded growth ideologies and racial politics rather than technical constraints. Detroit's vacant land resulted from decades of deindustrialization, white flight, and municipal disinvestment in Black neighborhoods; treating this land as neutral "opportunity" for urban agriculture erases these histories and ongoing displacement pressures facing long-term residents.

Even supposedly progressive urban food governance often reproduces existing power structures. Bristol's Food Policy Council, despite civil society leadership, remains dominated by white, middle-class professionals. Malmö's government-driven approach, while achieving organic procurement targets, does little addressing food access for immigrant communities. The absence of genuine food-insecure population participation in these "innovative" governance arrangements reveals a troubling pattern: these are transitions for, not by, those most affected by food system failures. This raises fundamental justice questions about whose sustainability is being pursued and whether transitions that don't emerge from and center marginalized communities can be truly transformative.

Dubbeling et al. (2016) examine private sector roles in city-region food systems, revealing contributions and constraints. While small enterprises can invest locally, concentration creates fundamental barriers: five companies control 80% of Ecuador's meat, dairy, and grain markets, constraining transformation possibilities regardless of governance innovations. This reflects broader trends toward corporate control of food systems that no amount of urban policy innovation can overcome without directly addressing market power through antitrust enforcement, public ownership, or alternative economic models.

Riley and Crush (2022) analyze African secondary cities, revealing how rapid urbanization, informal food economies, and limited governance capacity shape transition pathways differently from major metropolitan centers. These cities often lack comprehensive food policies, infrastructure investments, and institutional coordination, creating both constraints and possibilities. Their analysis emphasizes recognizing informal food systems as legitimate urban food provisioning components rather than development obstacles. Informal markets, street vendors, and urban agriculture already feed millions but remain systematically excluded from formal planning and criminalized through regulations, suggesting that transformation in these contexts requires building on existing social networks and community capacities rather than imposing formal systems that displace functioning alternatives.

Larbi et al. (2022) compare Curitiba, Brazil, and Accra, Ghana, demonstrating how historical development trajectories, political systems, and resource availability influence transition possibilities. Curitiba's success with integrated planning and cost-effective innovations reflects strong political leadership and continuity rare in contexts of political instability and resource constraints. Accra's challenges cannot be addressed through technical transfer of Curitiba's solutions but require building on existing local capacities, such as strong cultures of communal labor, while involving both state and non-state actors through context-tailored strategies.

Rural contexts remain critical sites where production practices fundamentally shape environmental sustainability and rural livelihoods. Davidson et al. (2016) examine alternative beef production in Alberta, revealing the "conventionalization paradox": niche markets for grass-fed beef create spaces for alternative production systems, but successful scaling could eliminate price premiums sustaining these practices. This reflects fundamental tension between market-based sustainability transitions and genuine transformation—alternatives that succeed on market terms risk losing their alternative character.

Zwartkruis et al. (2020) analyze agricultural nature conservation in the Netherlands, revealing that despite decades of support, it provides less than 10% of income for 81% of participating farms. This economic marginality raises fundamental questions about market-based sustainability transition viability, suggesting that expecting farmers to provide environmental services voluntarily while competing in globalized commodity markets may be fundamentally contradictory.

Actor Roles, Power Dynamics, and Democratic Deficits

Food system transitions involve multiple actors with fundamentally divergent interests, resources, and capabilities. Understanding these roles requires moving beyond pluralist assumptions that all actors participate on equal footing to interrogating power asymmetries and whose voices shape transformation directions.

Government actors operate within political economies privileging economic growth and capital accumulation rather than as neutral facilitators. Bedore (2014) analyzes Baltimore food policymaking through growth machine politics, revealing how economic development interests shape food policy agendas and partnership formations. Food initiatives succeed when framed to complement rather than challenge existing development strategies, with discursive mechanisms enabling actors to frame food policy in depoliticized terms around health, innovation, and city reputation. This "convening power of food" makes policy "partnerable" because it appeals to both growth coalition members and broader civil society, yet this same quality limits transformative potential by aligning new initiatives with existing power structures and growth imperatives.

Luger et al. (2025) examine transformative urban food governance through Urban Living Labs in ten European cities, revealing how municipal staff must "act politically simply to accomplish predefined tasks," employing advocacy and alignment strategies to navigate administrative and political constraints. This finding challenges technocratic governance models assuming neutral sustainability policy implementation while exposing the precarity of sustainability initiatives dependent on individual champions and vulnerable to political transitions and funding cuts. Sustainability initiatives dependent on individual champions become vulnerable when those individuals leave or political priorities shift, suggesting that transformative governance requires institutional embedding beyond personal commitment.

Buchan et al. (2019) introduce "transformative incrementalism" as a framework for understanding how government actors influence sustainable food systems through gradual, power-laden processes rather than abrupt policy shifts. Their analysis reveals government influence operating through both authoritative mechanisms (policy documents, official mandates) and relational mechanisms (coalition building, narrative framing). However, incrementalism faces risks of co-optation, where alternative initiatives become absorbed into existing systems without fundamentally challenging dominant logics. The consistent dilution of radical alternatives through mainstreaming processes suggests that incrementalism's transformative potential may be systematically constrained, raising uncomfortable questions about whether sustainable food systems are possible within current political-economic structures or require fundamental breaks.

Civil society organizations function as critical transition drivers, mobilizing grassroots support, developing alternative practices, and advocating policy changes. Isgren and Ness (2017) demonstrate how Ugandan networks promote agroecology while addressing social justice concerns through participatory approaches and local empowerment. However, actors face fundamental constraints including resource dependencies that can constrain critical advocacy and direct organizational activities toward fundable projects rather than transformative priorities.

Haylock and Connelly (2018) examine insider versus outsider strategies in New Zealand food networks, revealing tensions between engaging with formal institutions and maintaining critical distance and external pressure. Dunedin's Our Food Network explicitly chose to maintain distance from local government to preserve their transformative agenda, recognizing that institutional incorporation often comes at the cost of radical vision. This tension is not resolvable through better partnership design but reflects fundamental incompatibilities between grassroots movements seeking system transformation and governments operating within existing political-economic constraints.

Sarabia et al. (2021) analyze Valencia's agri-food system transformation, identifying how empowerment of grassroots social movements, combined with political shifts in local government and adherence to the Milan Urban Food Policy Pact, catalyzed creation of a Municipal Food Council and Agri-food Strategy 2025. This case demonstrates how civil society movements can influence policy when political opportunities arise, yet it also reveals dependence on favorable political conditions that may not persist across electoral cycles, raising questions about transformation sustainability beyond particular political moments.

Private sector roles remain deeply ambiguous. Dubbeling et al. (2016) document diverse contributions including infrastructure investments and innovative business models, but also reveal how market concentration fundamentally constrains transformation possibilities. When five companies control 80% of markets, sustainability becomes whatever these companies define as compatible with profit maximization. Voluntary corporate sustainability initiatives enable "greenwashing" while maintaining fundamental business models based on extraction, exploitation, and externalization of environmental costs.

The conventionalization paradox Davidson et al. (2016) identify reveals fundamental tensions in market-based sustainability transitions. Alternative practices succeeding in niche markets attract investment and scaling efforts that often require compromising alternative characteristics to achieve market competitiveness. Organic agriculture's mainstreaming through supermarket distribution exemplifies this dynamic: increased availability comes at the cost of losing connections to alternative food system values around local supply chains, fair prices for farmers, and community relationships.

Farmers occupy central positions as production decisions directly shape environmental impacts and food availability, yet face multiple constraints including market pressures, climate uncertainties, knowledge gaps, and policy environments favoring conventional practices. Schiller et al. (2023) demonstrate how knowledge flows and intermediary support can enable farmers to adopt agroecological practices, though sustained transitions require addressing structural barriers including market access, input availability, and policy support. The economic marginality of sustainable agriculture revealed by Zwartkruis et al. (2020) demonstrates that expecting farmers to provide environmental services voluntarily while competing in globalized commodity markets may be fundamentally contradictory.

Intermediary organizations play increasingly recognized roles facilitating transitions by connecting actors, translating knowledge, and coordinating collective action. Schiller et al. (2023) analyze knowledge intermediaries in Nicaragua, demonstrating how these organizations bridge farmers, researchers, policymakers, and markets while adapting external knowledge to local contexts. Ingram (2018) emphasizes that boundary dynamics between niche and regime knowledge systems require active mediation. However, intermediaries face challenges including resource dependencies creating potential capture by funders, legitimacy questions about whom they represent, and tensions between serving multiple constituencies with divergent interests. The question remains whether intermediaries facilitate genuine transformation or primarily function to translate radical alternatives into forms compatible with existing systems, thereby enabling systems to adapt without fundamentally changing.

Across these actor categories, systematic exclusions become apparent. Food-insecure populations, farm workers, migrants, indigenous communities, and other marginalized groups remain largely absent from food governance innovations. Raja et al. (2018) document U.S. food planning initiatives identifying "power to convene" as key mechanism for local governments, yet convening depends on who has access to participation spaces. More fundamentally, these initiatives rarely challenge existing power structures determining land ownership, market control, and what counts as legitimate knowledge. The democratic deficits in food governance reflect broader patterns where sustainability transitions are designed and implemented by technical experts, policymakers, and educated elites while those most affected by food insecurity and environmental degradation have minimal voice in determining transition directions.

Mechanisms, Barriers, and the Limits of Voluntary Transition

Knowledge systems emerge as fundamental to food system transitions yet operate through complex, contested channels. Ingram (2018) analyzes agricultural knowledge system boundaries, demonstrating tensions between mainstream Agricultural Knowledge Systems and alternatives like permaculture. Both "impermeable boundaries" maintaining separate knowledge domains and "permeable boundaries" enabling translation highlight knowledge's political nature. What counts as legitimate knowledge, who gets to produce it, and how it circulates are not technical questions but reflect power relations determining whose ways of knowing receive institutional recognition and resource support.

Li et al. (2013) show how international intellectual property regimes work against participatory plant breeding in China despite international commitments supporting it, demonstrating how landscape-level factors can simultaneously enable and constrain transitions. Seed sovereignty movements' struggles against corporate intellectual property claims over plant genetic resources exemplify how knowledge governance fundamentally shapes agricultural possibilities.

Charatsari and Lioutas (2019) identify significant competency gaps among agricultural professionals lacking crucial abilities for promoting sustainable agriculture, particularly networking, facilitation, and systems thinking. This gap indicates that professional knowledge systems themselves require transformation, extending beyond curriculum reform to reimagining agricultural professionals' roles as facilitators rather than technical experts. However, this reimagining threatens existing professional identities and institutional positions, creating resistance to knowledge system transformation.

Mukhovi et al. (2020) identify three critical learning forms in Global South food systems: experiential learning through demonstration, social learning via collective approaches, and transformative learning involving fundamental perception shifts. Their research in Kenya and Bolivia emphasizes that resilience depends not on knowledge transfer but on building adaptive capacity through diverse learning mechanisms. This challenges technocratic approaches assuming linear knowledge dissemination from experts to practitioners, revealing learning as social, contextual, and fundamentally political.

Hansen (2022) examines media's role in shaping food sustainability discourse through quantitative content analysis of 271 Danish newspaper articles, revealing how external events like COVID-19 and authorship type influence framing and proposed solutions. Reader-authored articles more often propose strong solutions based on values and norms, while journalists tend toward "softer" implications like revising social norms. This highlights media's agenda-setting power and how gatekeeping shapes what solutions become thinkable within public discourse.

Structural barriers transcend technical or economic challenges. Path dependencies, institutional inertia, and power asymmetries favor conventional food systems in ways that incremental reforms cannot address. Li et al. (2013) demonstrate how seed system organization in China reflects deep institutional legacies constraining reform possibilities, requiring coordinated changes across regulatory frameworks, organizational structures, and professional practices—yet these very requirements make transformation politically difficult. Zwartkruis et al. (2020) reveal the "dual regime challenge" where agricultural nature conservation must navigate both agricultural and nature regimes with conflicting logics, exemplifying how institutional fragmentation impedes transformation.

Policy misalignment across governance levels and sectors creates additional barriers. Agricultural policies, trade regulations, and economic development strategies often work at cross-purposes, undermining sustainability objectives. Fragmented governance structures complicate coordination, with multiple agencies addressing food issues without coherent integration. Yet this fragmentation is not accidental but reflects political compromises, sectoral interests, and deliberate limiting of state capacity under neoliberal governance.

Economic viability remains fundamental challenge for alternative food systems. Conventional systems benefit from economies of scale, established supply chains, and policy supports that alternatives lack. Price competition disadvantages sustainable products when environmental and social costs remain externalized in conventional systems. Reynolds et al. (2019) review consumer food waste reduction interventions, finding that despite research surge, only 17 of 292 studies report applied interventions with measurable reductions. Reductions vary widely (12-57%), and methodological challenges abound. The review emphasizes that individual behavior change alone cannot drive systemic transitions; structural changes in food provisioning systems, retail practices incentivizing overproduction, and policy frameworks are equally essential yet remain underdeveloped compared to consumer-focused approaches.

The economic marginality of alternatives raises fundamental questions about market-based sustainability transitions. If sustainable practices cannot compete economically with conventional systems externalizing costs, either externalities must be internalized through regulation and taxation, or sustainable practices require permanent subsidization, or genuinely sustainable food systems are incompatible with market coordination and require alternative economic models. The literature generally avoids this conclusion's implications, yet the evidence consistently points toward market failures that incremental reforms cannot address.

Political barriers include resistance from vested interests, lack of political will, and governance structures fragmenting authority. Pothukuchi (2017) reveals how urban agriculture in Detroit faces political opposition from those viewing farming as incompatible with development visions, stemming from growth ideologies and racial politics rather than technical constraints. These cases reveal that food system transitions challenge fundamental assumptions about development, progress, and economic growth embedded in contemporary governance.

Crisis events and disruptive moments create transformation possibilities. Davidson et al. (2016) show how BSE outbreaks catalyzed alternative beef production. Hansen (2022) documents how COVID-19 temporarily disrupted systems and consumption patterns. Sarabia et al. (2021) identify crisis-driven opportunities alongside social movements as catalyzing policy changes. However, crises can catalyze transformation or reinforce existing systems depending on political navigation. COVID-19 revealed vulnerabilities yet largely strengthened existing players through state support for conventional supply chains. Crisis-driven change requires organized movements ready to mobilize disruption toward transformative ends.

Ryghaug and Skjølsvold's (2023) analysis of Norway's electric vehicle transition, which achieved over 86% battery and plug-in hybrid share among new private cars by 2021, illustrates system-wide transformation potential through persistent demand-side incentives, NGO intermediation, and user satisfaction. However, the case highlights Norway's unique conditions—wealth and renewable energy resources—and raises second-order sustainability dilemmas around mining impacts and material justice, underscoring that transitioning regimes spawn fresh systemic trade-offs requiring ongoing attention.

Conclusion: Toward Transformative Food Futures

This synthesis reveals food system transitions as fundamentally political processes requiring more than technical innovation or market reform. The geographic case diversity demonstrates that while patterns exist—intermediaries' importance, multi-stakeholder governance, crisis-driven opportunities—specific pathways must emerge from local contexts rather than imposed blueprints. The transformation paradox—that incumbent systems shape their own transformation—implies existing governance structures inevitably constrain possibilities, suggesting limits to working within existing institutions and potential necessity for breaks with existing systems.

Food system transitions cannot be understood without explicit political economy analysis examining how capitalism, colonialism, and contemporary neoliberalism structure what transformations become possible. The persistent economic marginality of sustainable alternatives, the concentration of corporate power, and growth imperatives constraining policy innovation reflect not accidental market failures but systematic features of capitalist food systems. Future research must explicitly examine how class relations, racial capitalism, gender dynamics, and global inequalities shape food system transitions.

Achieving sustainable food systems requires addressing justice, equity, and democracy alongside environmental concerns. The exclusion of marginalized voices from food governance, the concentration of corporate power, and the commodification of essential resources emerge as core challenges rather than peripheral concerns. True transformation demands not just sustainable production but equitable distribution, democratic participation, and recognition of food's multiple values beyond market price.

The path forward requires abandoning comfortable narratives of gradual managed transition and confronting the reality that sustainable food systems may require radical breaks with existing structures—land occupations, seed sovereignty movements, food riots—that historically have driven genuine food system transformation. Yet the transformation paradox reveals that incumbent systems and governance structures are always involved in self-transformative paths and may be so constructively. The path forward requires embracing this complexity while maintaining focus on fundamental transformation rather than incremental reform. Only by addressing power relations, reconceptualizing food's value beyond commodity, creating genuinely inclusive governance, and supporting contentious movements for food sovereignty can food systems transition toward genuine sustainability and justice. Whether such transformation is possible within existing structures or requires fundamental breaks remains the central question—though evidence increasingly suggests the latter, challenging us to imagine and work toward food futures fundamentally different from the present.

---

Condensed Synthesis: Food System Transitions (1500 words) [Review E]

Critical Perspectives on Food System Transformation

Introduction

Global food system transformation toward sustainability requires moving beyond technical narratives to recognize transitions as inherently political processes involving contested struggles over resources, values, knowledge, and power. This review synthesizes 31 scholarly works selected for theoretical rigor, geographical diversity, and critical engagement with power dynamics. These papers reveal that food system transitions require fundamental reconfigurations of governance structures, knowledge systems, and power relations, raising questions about whether genuinely sustainable and just food systems are possible within current political-economic structures or require fundamental breaks.

Theoretical Foundations: Beyond Linear Models

The Multi-Level Perspective (MLP) remains the dominant framework, distinguishing niches (protected spaces for innovation), regimes (dominant practices and institutions), and landscapes (exogenous contexts). However, empirical research exposes fundamental problems with MLP's core assumptions. Davidson et al. (2016) demonstrate that alternative beef producers operate simultaneously between alternative and conventional systems rather than within discrete levels, challenging the niche-regime distinction. Audet et al. (2017) reveal Montreal seasonal food markets evolve at regime-niche intersections rather than in protected spaces, embodying tensions reflecting deeper contradictions in capitalist food systems attempting sustainability transitions.

Luger et al. (2025) conceptualize a "purple zone" where municipal staff navigate between formal administrative duties and informal political action, operating across supposed boundaries. Zwartkruis et al. (2020) identify the "dual regime challenge" where agricultural nature conservation must navigate both agricultural and nature regimes with conflicting logics, revealing how MLP's single-regime assumption fails capturing institutional fragmentation's constraining effects.

Ingram et al. (2015) identify five interaction modes between Learning and Innovation Networks and dominant regimes: Compatible, Complementary, Emergent, Divergent, and Oppositional. Crucially, highly compatible networks achieve smoother diffusion but experience diluted sustainability ambitions, while radical networks face marginalization but preserve transformative ideals—revealing a fundamental dilemma where conditions enabling scaling often undermine transformative potential.

Jørgensen's arenas of development (AoD) offers a more systemic, less hierarchical alternative. AoD considers transitions occurring through conflicts, negotiations, alignments, and learning among stakeholders in interaction arenas, adopting a Latourian Actor-Network Theory approach. While MLP highlights Darwinian mechanisms of innovation and diffusion, AoD emphasizes multi-party emergence of new stable actor configurations through multiple regulatory systems, better capturing transition complexity and political nature.

Vivero-Pol (2017) reveals that food valuation—as commodity versus commons—fundamentally shapes political attitudes toward transformation. When food is conceptualized solely as commodity, system transformation becomes literally unthinkable; only when recognized simultaneously as human right, cultural determinant, and commons do transformative pathways emerge.

These theoretical engagements reveal the "transformation paradox": incumbent systems and governance structures inevitably shape their own transformation, simultaneously enabling and constraining possibilities. Systems cannot stand outside themselves to transform, suggesting limits to working within existing institutions and potential necessity for breaks with existing systems.

Geographical Contexts: Colonial Legacies and Structural Inequalities

Food system transitions reflect not neutral differences but historical processes of colonialism, uneven development, and ongoing extractivism. Notably absent from most frameworks is engagement with how colonial agricultural systems continue structuring contemporary food systems. The emphasis on "Global South innovations" risks romanticizing struggles emerging from structural adjustment policies rather than recognizing responses to historically produced vulnerabilities.

Isgren and Ness (2017) examine agroecology in Uganda, revealing how civil society networks fill gaps left by state withdrawal under neoliberal restructuring. Agroecology emerges as "unintended consequence of neoliberal development policies"—devolution of services to non-state actors created space for alternatives. However, this cannot be understood without examining how colonial cash crop dependencies continue shaping possibilities. What appears as "innovation" might more accurately be survival strategies in deliberately produced scarcity contexts. This pathway differs fundamentally from European experiences where formal institutions actively guide alternatives through deliberate transition management, revealing fundamentally unequal positions in global political economy.

Urban contexts have become innovation focal points, yet warrant critical scrutiny. Pothukuchi (2017) reveals how Detroit urban agriculture faces political opposition from "shrinkage regimes" viewing farming as incompatible with development, stemming from growth ideologies and racial politics. Detroit's vacant land resulted from deindustrialization, white flight, and disinvestment in Black neighborhoods; treating this as neutral "opportunity" erases these histories and ongoing displacement pressures.

Even progressive urban governance often reproduces power structures. Bristol's Food Policy Council remains dominated by white, middle-class professionals despite civil society leadership. Malmö's government-driven approach achieves organic procurement targets but does little addressing immigrant community food access. The absence of genuine food-insecure population participation reveals these are transitions for, not by, those most affected by food system failures.

Dubbeling et al. (2016) reveal that while small enterprises can invest locally, concentration creates fundamental barriers: five companies control 80% of Ecuador's meat, dairy, and grain markets, constraining transformation regardless of governance innovations. This reflects broader corporate control trends that urban policy innovation cannot overcome without directly addressing market power.

Davidson et al. (2016) reveal the "conventionalization paradox": niche markets create alternative spaces, but successful scaling could eliminate price premiums sustaining these practices, reflecting fundamental tension between market-based transitions and genuine transformation. Zwartkruis et al. (2020) show agricultural nature conservation provides less than 10% of income for 81% of participating Dutch farms despite decades of support, raising fundamental questions about market-based transition viability.

## Actor Roles, Power Dynamics, and Democratic Deficits

Government actors operate within political economies privileging growth and capital accumulation rather than as neutral facilitators. Bedore (2014) reveals how Baltimore food policymaking reflects growth machine politics, with food initiatives succeeding when framed complementing rather than challenging existing development strategies. This "convening power of food" makes policy "partnerable" but limits transformative potential by aligning with existing power structures.

Luger et al. (2025) reveal municipal staff must "act politically simply to accomplish predefined tasks," exposing sustainability initiatives' precarity when dependent on individual champions and vulnerable to political shifts. Buchan et al. (2019) introduce "transformative incrementalism," though it faces co-optation risks where alternatives become absorbed without challenging dominant logics. Consistent radical alternative dilution through mainstreaming suggests incrementalism's transformative potential may be systematically constrained.

Civil society organizations function as critical drivers but face fundamental constraints. Haylock and Connelly (2018) reveal tensions between institutional engagement and maintaining critical distance. Dunedin's Our Food Network chose maintaining government distance to preserve transformative agenda, recognizing institutional incorporation often costs radical vision— reflecting fundamental incompatibilities between movements seeking transformation and governments operating within existing constraints.

Private sector roles remain ambiguous. When five companies control 80% of markets, sustainability becomes whatever these companies define as profit-compatible. Voluntary corporate initiatives enable "greenwashing" while maintaining business models based on extraction and externalization. The conventionalization paradox reveals that alternatives succeeding on market terms risk losing alternative character.

Farmers face multiple constraints including market pressures, climate uncertainties, and policy environments favoring conventional practices. The economic marginality of sustainable agriculture demonstrates that expecting farmers to provide environmental services voluntarily while competing in globalized markets may be fundamentally contradictory.

Across actor categories, systematic exclusions become apparent. Food-insecure populations, farm workers, migrants, indigenous communities, and marginalized groups remain largely absent from governance innovations. Democratic deficits reflect broader patterns where transitions are designed by technical experts and elites while those most affected have minimal voice determining directions.

## Mechanisms, Barriers, and Limits of Voluntary Transition

Knowledge systems are fundamental yet operate through contested channels. Ingram (2018) demonstrates tensions between mainstream systems and alternatives, revealing knowledge's political nature. What counts as legitimate knowledge and who produces it reflect power relations determining whose knowing ways receive recognition and resources. Li et al. (2013) show how intellectual property regimes work against participatory plant breeding despite international commitments supporting it.

Charatsari and Lioutas (2019) identify significant competency gaps among agricultural professionals lacking networking, facilitation, and systems thinking abilities, indicating professional knowledge systems themselves require transformation. Mukhovi et al. (2020) emphasize resilience depends not on knowledge transfer but building adaptive capacity through diverse learning mechanisms, challenging technocratic approaches assuming linear dissemination.

Structural barriers transcend technical challenges. Path dependencies, institutional inertia, and power asymmetries favor conventional systems in ways incremental reforms cannot address. Zwartkruis et al. (2020) reveal the "dual regime challenge" where conservation must navigate regimes with conflicting logics, exemplifying how institutional fragmentation impedes transformation. Policy misalignment across levels and sectors creates additional barriers, yet this fragmentation reflects political compromises and deliberate state capacity limiting under neoliberal governance.

Economic viability remains fundamental challenge. Conventional systems benefit from economies of scale and policy supports alternatives lack. Price competition disadvantages sustainable products when environmental and social costs remain externalized. Reynolds et al. (2019) find that despite research surge, only 17 of 292 studies report applied interventions with measurable waste reductions, emphasizing individual behavior change alone cannot drive systemic transitions.

Alternative economic marginality raises fundamental questions: if sustainable practices cannot compete economically with conventional systems externalizing costs, either externalities must be internalized through regulation, sustainable practices require permanent subsidization, or genuinely sustainable food systems are incompatible with market coordination and require alternative economic models. The literature generally avoids this conclusion, yet evidence consistently points toward market failures incremental reforms cannot address.

Crisis events create transformation possibilities but can also reinforce existing systems depending on political navigation. COVID-19 revealed vulnerabilities yet largely strengthened existing players through state support. Crisis-driven change requires organized movements ready to mobilize disruption toward transformative ends.

Conclusion

This synthesis reveals food system transitions as fundamentally political processes requiring more than technical innovation or market reform. The transformation paradox—that incumbent systems shape their own transformation—implies existing structures inevitably constrain possibilities, suggesting limits to working within existing institutions and potential necessity for breaks. The persistent economic marginality of alternatives, corporate power concentration, and growth imperatives constraining policy innovation reflect not accidental market failures but systematic capitalist food system features.

Achieving sustainable food systems requires addressing justice, equity, and democracy alongside environmental concerns. Marginalized voice exclusion, corporate power concentration, and essential resource commodification emerge as core challenges rather than peripheral concerns. True transformation demands not just sustainable production but equitable distribution, democratic participation, and recognition of food's multiple values beyond market price.

The path forward requires abandoning comfortable narratives of gradual managed transition and confronting the reality that sustainable food systems may require radical breaks—land occupations, seed sovereignty movements, food riots—that historically have driven genuine transformation. Yet the transformation paradox reveals incumbent systems are always involved in self-transformative paths and may be so constructively. Only by addressing power relations, reconceptualizing food's value beyond commodity, creating genuinely inclusive governance, and supporting contentious movements for food sovereignty can food systems transition toward genuine sustainability and justice. Whether such transformation is possible within existing structures or requires fundamental breaks remains the central question—though evidence increasingly suggests the latter, challenging us to imagine and work toward food futures fundamentally different from the present.

Synthesized Literature Review: Food System Transitions (1500 words - Balanced Version)

Food System Transitions: Pathways, Challenges, and Transformation Dynamics

Introduction

Global food systems face unprecedented pressures from climate change, population growth, resource depletion, and social inequalities, necessitating fundamental transformations toward sustainability. This review synthesizes insights from 31 scholarly works examining food system transitions through diverse theoretical lenses, geographical contexts, and actor perspectives. These papers collectively reveal that food system transitions involve complex socio-technical processes requiring coordinated changes across technologies, institutions, practices, knowledge systems, and governance structures. The review examines dominant theoretical frameworks, analyzes key transition mechanisms, explores geographical variations in transition pathways, assesses actor roles and power dynamics, and synthesizes knowledge on barriers and enablers shaping transformation possibilities.

Theoretical Foundations: Expanding Transition Frameworks

The Multi-Level Perspective (MLP) remains the dominant framework for analyzing food system transitions, distinguishing niches (protected spaces for innovation), regimes (dominant practices and institutions), and landscapes (exogenous contexts). The framework offers analytical advantages by recognizing structured differences between innovation spaces and established systems, proving valuable for examining how alternatives challenge dominant arrangements (Elzen et al., 2011; Ingram et al., 2015).

However, empirical research reveals limitations in MLP's core assumptions, particularly regarding analytical boundaries between levels. Davidson et al. (2016) demonstrate that alternative beef producers in Alberta operate simultaneously between alternative and conventional systems rather than within discrete analytical levels. Audet et al. (2017) show that Montreal seasonal food markets evolve at regime-niche intersections rather than in protected spaces, embodying tensions between market efficiency, community building, and environmental sustainability that cannot be easily resolved.

Luger et al. (2025) conceptualize a "purple zone" where municipal staff navigate between formal administrative duties and informal action, operating across supposed boundaries in ways challenging MLP's categories. Zwartkruis et al. (2020) identify the "dual regime challenge" where agricultural nature conservation must navigate both agricultural and nature regimes with conflicting institutional logics, revealing how MLP's single-regime assumption fails to capture institutional fragmentation's constraining effects.

Ingram et al. (2015) identify five interaction modes between Learning and Innovation Networks and dominant regimes: Compatible, Complementary, Emergent, Divergent, and Oppositional. Their analysis reveals that highly compatible networks achieve smoother diffusion but often experience diluted sustainability ambitions, while more radical networks face marginalization but preserve transformative ideals. This demonstrates a fundamental tension: the conditions enabling niche scaling often undermine transformative potential through adaptation to regime requirements.

Elzen et al. (2011) integrate Social Movement Theory with MLP, revealing that successful transitions require alignment across technological, regulatory, market, and social streams rather than technological innovation alone. This multi-stream approach reconceptualizes transitions as emerging through negotiation and compromise rather than linear progression.

Jørgensen's arenas of development (AoD) offers an alternative perspective that considers transitions occurring through conflicts, negotiations, alignments, and learning among stakeholders in interaction arenas. While MLP highlights mechanisms of niche innovation and diffusion, AoD emphasizes the emergence of new stable actor configurations through multiple regulatory systems, capturing transition complexity more effectively.

Vivero-Pol (2017) explores how food valuation—as commodity versus commons—shapes attitudes toward system transformation. When food is conceptualized solely as commodity, certain transformative pathways become difficult to envision; recognizing food simultaneously as human right, cultural determinant, and commons opens alternative possibilities.

These theoretical engagements reveal what might be termed the "transformation paradox": existing systems and governance structures inevitably shape their own transformation processes, simultaneously enabling and constraining what becomes possible. This implies that transformation strategies must seriously account for existing conditions and path dependencies rather than focusing solely on desired endpoints.

Geographical Contexts: Diverse Pathways and Contextual Factors

Food system transitions manifest differently across geographical contexts, reflecting diverse political economies, institutional capacities, resource availabilities, and development trajectories. Understanding these variations is essential for developing context-appropriate transition strategies.

Global South contexts present distinct dynamics. Isgren and Ness (2017) examine agroecology in Uganda's Rwenzori region, revealing how civil society networks fill gaps left by state capacity limitations while navigating challenging political environments. Agroecology emerges as an "unintended consequence of neoliberal development policies"—specifically how devolution of agricultural extension services to non-state actors created space for alternative approaches. This pathway differs from European experiences where formal state institutions actively guide alternatives through deliberate transition management programs.

Arslan et al. (2022) analyze 81 low- and middle-income countries, revealing that structural transformation correlates more consistently with improved food system outcomes than rural transformation alone. Their Food Systems Index demonstrates that financial inclusion, digital connectivity, women's empowerment, and education emerge as key transformation levers, while highlighting the "double burden" of malnutrition and rapid population growth in countries with weakest outcomes.

Schiller et al. (2023) demonstrate how knowledge flows through intermediaries enable agroecological niche development in Nicaragua despite limited state support, revealing transition possibilities even in resource-constrained contexts. The research effectively shows intermediaries' crucial roles in bridging farmers, researchers, policymakers, and markets while adapting external knowledge to local contexts.

Urban contexts have become focal points for food system innovation. Moragues-Faus and Morgan (2015) document the emergence of urban food policies worldwide, representing shifts from viewing food as primarily rural or national issue to recognizing cities' crucial sustainability roles. However, urban food governance faces significant challenges. Pothukuchi (2017) reveals how urban agriculture in Detroit encounters resistance from development visions prioritizing conventional real estate over alternative land uses, reflecting tensions between different urban development models.

Comparative analysis reveals important contextual differences. Bristol's civil society-led food governance approach contrasts with Malmö's government-driven model, yet both create new spaces for deliberation that reconfigure traditional structures. Larbi et al. (2022) compare Curitiba, Brazil, and Accra, Ghana, demonstrating how historical trajectories, political systems, and resource availability influence transition possibilities. Curitiba's success with integrated planning reflects strong political leadership, while Accra's challenges require building on existing local capacities such as communal labor traditions.

Riley and Crush (2022) analyze African secondary cities, revealing how rapid urbanization, informal food economies, and limited governance capacity shape pathways differently from major metropolitan centers. Their analysis emphasizes the importance of recognizing informal food systems as functioning components of urban provisioning rather than obstacles to overcome.

Rural contexts remain critical as production practices fundamentally shape sustainability outcomes. Davidson et al. (2016) reveal the "conventionalization paradox": niche markets create spaces for alternative production systems, but successful scaling could eliminate price premiums sustaining these practices. Zwartkruis et al. (2020) show that agricultural nature conservation provides less than 10% of income for 81% of participating Dutch farms despite decades of support, raising questions about the economic viability of voluntary market-based sustainability approaches.

Actor Roles and Governance Dynamics

Food system transitions involve multiple actors with diverse interests, resources, and capabilities, requiring coordinated action across government, civil society, business, farmers, and intermediaries while managing inherent tensions among these groups.

Government actors play multifaceted roles including policy development, resource allocation, regulation, and stakeholder convening. Bedore (2014) analyzes Baltimore food policymaking, revealing how food initiatives often succeed when framed to complement existing economic development strategies. Luger et al. (2025) demonstrate how municipal staff coordinating urban living labs must navigate between administrative procedures and political priorities, revealing the challenges of institutionalizing sustainability initiatives. Buchan et al. (2019) introduce "transformative incrementalism" as a framework for understanding how government actors influence transitions through gradual processes operating through both authoritative mechanisms (policy documents, mandates) and relational mechanisms (coalition building, narrative framing).

Civil society organizations function as critical transition drivers, mobilizing support, developing alternative practices, and advocating policy changes. Isgren and Ness (2017) demonstrate how Ugandan networks promote agroecology while addressing local needs through participatory approaches. Haylock and Connelly (2018) reveal tensions between institutional engagement strategies and maintaining independence, with some organizations choosing to maintain distance from government to preserve their transformative agendas.

Sarabia et al. (2021) analyze Valencia's agri-food system transformation, identifying how grassroots social movements' empowerment, combined with political shifts in local government and adherence to the Milan Urban Food Policy Pact, catalyzed creation of a Municipal Food Council and Agri-food Strategy 2025. This demonstrates how civil society movements can influence policy when political opportunities arise.

Private sector engagement presents both opportunities and challenges. Dubbeling et al. (2016) document diverse contributions including infrastructure investments and innovative business models, but also reveal how market concentration constrains transformation possibilities—five companies control 80% of Ecuador's meat, dairy, and grain markets. The conventionalization paradox Davidson et al. (2016) identify reveals tensions in market-based transitions: alternative practices succeeding in niche markets often attract scaling efforts requiring compromises to achieve market competitiveness.

Farmers occupy central positions as production decisions directly shape environmental impacts and food availability, yet face multiple constraints including market pressures, climate uncertainties, knowledge gaps, and policy environments often favoring conventional practices. Schiller et al. (2023) demonstrate how knowledge flows and intermediary support can enable agroecological adoption, though sustained transitions require addressing structural barriers including market access and policy support.

Intermediary organizations play increasingly recognized roles facilitating transitions by connecting actors, translating knowledge, and coordinating collective action. Schiller et al. (2023) analyze Nicaraguan knowledge intermediaries bridging farmers, researchers, policymakers, and markets. Ingram (2018) emphasizes that boundary dynamics between niche and regime knowledge systems require active mediation through trust-building and institutional negotiation.

Raja et al. (2018) document U.S. food planning initiatives, identifying "power to convene" as a key mechanism whereby local governments leverage authority to foster cross-sector coalitions. Seattle's Local Food Action Initiative and Minneapolis's Homegrown Minneapolis demonstrate how bureaucratic restructuring can institutionalize food as municipal priority, though persistent barriers including fiscal constraints and jurisdictional misalignments continue challenging implementation.

Mechanisms, Barriers, and Enablers

Knowledge systems emerge as fundamental to transitions yet operate through complex channels. Ingram (2018) analyzes agricultural knowledge system boundaries, demonstrating tensions between mainstream systems and alternatives. What counts as legitimate knowledge and how it circulates significantly shapes transition possibilities. Li et al. (2013) show how international intellectual property regimes affect participatory plant breeding in China, demonstrating how landscape-level factors can simultaneously enable and constrain transitions.

Charatsari and Lioutas (2019) identify competency gaps among agricultural professionals lacking crucial abilities for promoting sustainable agriculture, particularly networking, facilitation, and systems thinking, indicating that professional education systems themselves require transformation. Mukhovi et al. (2020) identify three critical learning forms: experiential learning through demonstration, social learning via collective approaches, and transformative learning involving fundamental perception shifts. Their research emphasizes that resilience depends not merely on knowledge transfer but on building adaptive capacity through diverse mechanisms.

Hansen (2022) examines media's role in shaping food sustainability discourse through content analysis of 271 Danish newspaper articles, revealing how external events like COVID-19 and authorship type influence framing and proposed solutions. This highlights how public discourse shapes what solutions become widely considered feasible.

Structural barriers include path dependencies, institutional inertia, and established interests favoring conventional systems. Zwartkruis et al. (2020) reveal the "dual regime challenge" where conservation initiatives must navigate multiple regimes with conflicting logics. Policy misalignment across governance levels and sectors creates additional barriers, with agricultural policies, trade regulations, and economic development strategies often working at cross-purposes.

Economic viability remains a fundamental challenge. Conventional systems benefit from economies of scale, established supply chains, and policy supports that alternatives often lack. Price competition disadvantages sustainable products when environmental and social costs remain externalized. Reynolds et al. (2019) review consumer food waste reduction interventions, finding that despite research growth, only 17 of 292 studies report applied interventions with measurable reductions, emphasizing that individual behavior change alone cannot drive systemic transitions without complementary structural changes.

Infrastructure limitations constrain alternative development, particularly in secondary cities and rural areas. Riley and Crush (2022) document how African secondary cities lack basic infrastructure including wholesale markets, cold storage, and processing facilities, reinforcing inefficient chains and limiting transformation possibilities.

Crisis events create transformation opportunities. Davidson et al. (2016) show how BSE outbreaks catalyzed alternative beef production. Hansen (2022) documents how COVID-19 temporarily disrupted systems and consumption patterns. Sarabia et al. (2021) identify how crises alongside social movements and entrepreneurial initiatives can catalyze policy changes, though crisis-driven changes require political mobilization to translate disruptions into lasting transformations.

To et al. (2018) analyze bagasse cogeneration transitions in Mauritius, demonstrating how coordinated multi-level policies accelerated change. Ryghaug and Skjølsvold's (2023) analysis of Norway's electric vehicle transition, which achieved over 86% battery and plug-in hybrid share by 2021, illustrates system-wide transformation potential through persistent demand-side incentives and intermediation, though it also raises questions about resource implications and trade-offs in seemingly successful transitions.

Conclusion

This synthesis reveals food system transitions as complex processes requiring coordinated changes across multiple dimensions. The Multi-Level Perspective provides valuable analytical framework though requiring expansion to capture institutional fragmentation, knowledge system dynamics, and political negotiation processes more adequately. Alternative frameworks like arenas of development offer complementary perspectives better capturing transition complexity and multi-actor dynamics.

Geographical context profoundly shapes transition possibilities, with important differences reflecting diverse institutional capacities, resource availabilities, and development trajectories. Transitions in low- and middle-income countries often follow different pathways than wealthy nations with strong institutions, requiring context-appropriate strategies rather than universal models.

Multiple actors play essential roles, requiring collaborative governance approaches that navigate tensions among competing interests. However, persistent challenges include limited participation from food-insecure populations and other affected groups in governance innovations, raising questions about whose sustainability visions shape transformation directions.

Transitions face significant structural, institutional, economic, and political barriers that voluntary approaches may struggle to overcome. The economic marginality of many sustainable alternatives, policy misalignments, and infrastructure limitations constrain transformation possibilities. The "transformation paradox"—that existing systems shape their own transformation—suggests both opportunities for working constructively within current structures and potential limits requiring more fundamental shifts.

Key enablers include crisis moments creating political opportunities, effective intermediaries facilitating knowledge flows and coordination, multi-stakeholder partnerships building collective capacity, and policy innovations that align incentives and create enabling environments. Future research should address transitions in diverse geographical contexts, examine temporal dynamics including tipping points and sequencing, investigate how to scale innovations while maintaining transformative characteristics, and explore interactions between food system transitions and broader sustainability challenges.

Moving forward requires holistic approaches addressing multiple dimensions simultaneously, recognizing political contestation inherent in transformation, adapting strategies to context-specific conditions, facilitating collaboration across diverse actors, and embracing adaptive learning. Only through comprehensive, politically aware, and context-sensitive approaches can food systems transition toward sustainability while ensuring resilience and food security.

thank you. can you please now compare the three 1500 versions (the two improcved ones, and the final toned-down you just gave me? What are you conclusions about this exercise?

Comparative Analysis of the Three 1500-Word Versions

Overview of the Three Versions

Let me compare:

1. **My Improved 1500-word version** (from the first synthesis request)
2. **Your Improved 1500-word version** (from the first synthesis request)
3. The Toned-Down Balanced 1500-word version (just produced)

Key Differences Across the Three Versions

1. Opening Framing and Tone

My Improved Version:

- Opens with "unprecedented pressures" and moves quickly to stating transitions are "inherently political"
- Explicitly frames review as "critically synthesizing" and "interrogating"
- Assertive but still somewhat measured

Your Improved Version:

- Opens with complexity and immediately emphasizes moving "beyond dominant narratives"
- Explicitly identifies as "human-curated synthesis of critical scholarship"
- States upfront it's "deliberately moving beyond" mainstream approaches
- Most assertive and reflexive opening

Toned-Down Version:

- Opens with "unprecedented pressures" similar to mine
- Frames as examining transitions "through diverse theoretical lenses"
- Emphasizes "complex socio-technical processes"
- Most neutral and comprehensive opening

**Analysis:** The toning-down successfully removed the critical scholarship positioning while maintaining acknowledgment of complexity and political dimensions.

2. Treatment of Colonial/Historical Context

My Improved Version:

- Includes one paragraph on "Colonial Legacies and Structural Inequalities" as subsection
- States: "Notably absent from most frameworks is engagement with how colonial agricultural systems continue structuring contemporary food systems"
- Discusses Uganda case in colonial context

Your Improved Version:

- Same paragraph but integrated more forcefully
- Phrase "risks romanticizing struggles emerging from structural adjustment policies and ongoing extractivism" is prominent
- More accusatory tone about what scholarship omits

Toned-Down Version:

- Completely removes the "notably absent" framing
- Discusses Uganda case as showing "distinct dynamics" in Global South contexts
- Mentions neoliberal policies as creating "unintended consequences"
- Removes "romanticizing" language and "deliberately produced scarcity"
- Result: context is acknowledged but not positioned as critique of the field

**Analysis:** This is the most significant change. The toned-down version presents Global South contexts as different rather than as evidence of colonial structures. The critical edge about what scholarship ignores is entirely removed.

3. Discussion of Urban Food Governance

My Improved Version:

- Discusses Bristol and Malmö, noting they "remain dominated by white, middle-class professionals"
- States "absence of genuine food-insecure population participation"
- Concludes these are "transitions for, not by, those most affected"

Your Improved Version:

- Identical critique of Bristol and Malmö
- Adds phrase "troubling pattern"
- More emphatic framing

Toned-Down Version:

- Discusses Bristol and Malmö but removes all critique about who participates
- Simply states they "create new spaces for deliberation"
- General statement at end about "limited participation from food-insecure populations"
- Removes the stark "for, not by" conclusion
- Frames as "raising questions" rather than identifying problems

**Analysis:** The specific, sharp critique is replaced with general acknowledgment of participation challenges. This significantly softens the political bite.

4. Language Choices Throughout

My Improved Version:

- "systematically excluded," "democratic deficits," "power asymmetries"
- "fundamental incompatibilities," "co-optation risks"
- Critical but still academic

Your Improved Version:

- "systematically excluded," "criminalized," "deliberate dismantling"
- "troubling pattern," "fundamental contradictions"
- Most accusatory and strong

Toned-Down Version:

- "limited participation," "challenges," "tensions," "constraints"

- "raises questions about," "suggests," "may struggle"
- Most measured and cautious

**Analysis:** The toning-down systematically replaced strong accusatory language with softer, more tentative phrasing throughout.

5. Treatment of Market-Based Approaches

My Improved Version:

- Discusses "conventionalization paradox"
- States market-based approaches "may be fundamentally contradictory"
- Presents three alternatives: internalize externalities, permanent subsidies, or alternative economic models
- Concludes: "evidence consistently points toward market failures incremental reforms cannot address"

Your Improved Version:

- Nearly identical treatment
- Slightly more emphatic: "may be fundamentally contradictory"
- Same three alternatives
- Same conclusion about market failures

Toned-Down Version:

- Discusses "conventionalization paradox"
- States approaches "raising questions about the economic viability"
- Removes the three explicit alternatives (internalize, subsidize, or alternative models)
- Softens to: "voluntary market-based sustainability approaches" face questions
- Conclusion removed entirely

**Analysis:** The toned-down version version acknowledges the paradox but removes the logical conclusion about market failures and doesn't present the stark choice between regulation, subsidies, or alternative economic systems. This is a significant intellectual softening.

6. Detroit Urban Agriculture Discussion

My Improved Version:

- States opposition stems from "growth ideologies and racial politics"
- Mentions "white flight and disinvestment in Black neighborhoods"
- Concludes treating land as neutral opportunity "erases these histories and ongoing displacement pressures"

Your Improved Version:

- Identical treatment
- Same language about racial politics and Black neighborhoods

Toned-Down Version:

- States opposition reflects "tensions between different urban development models"
- Completely removes race and "white flight" language

- Removes "erases histories" conclusion
- Makes it about abstract "development visions" rather than racial capitalism

**Analysis:** This is the clearest example of removing critical race analysis. The toned-down version makes a fundamentally racial issue appear as generic "development model" disagreement.

7. Corporate Power Discussion

My Improved Version:

- States 5 companies control 80% of Ecuador markets
- Concludes: "reflects broader corporate control trends that no amount of urban policy innovation can overcome without directly addressing market power"
- Mentions antitrust, public ownership, or alternative economic models

Your Improved Version:

- Identical treatment with same specific data
- Same conclusion about market power

Toned-Down Version:

- States 5 companies control 80% of Ecuador markets
- Concludes this "constrains transformation possibilities"
- Removes "no amount of urban policy innovation can overcome"
- Removes mention of antitrust, public ownership, alternative economic models
- Makes it descriptive rather than prescriptive

**Analysis:** The toned-down version acknowledges concentration but removes the argument about what this means for policy approaches and doesn't suggest structural solutions.

8. Knowledge Systems Discussion

My Improved Version:

- States "What counts as legitimate knowledge... reflect power relations"
- Discusses knowledge as "fundamentally political"

Your Improved Version:

- Identical framing about knowledge and power

Toned-Down Version:

- States "What counts as legitimate knowledge... significantly shapes transition possibilities"
- Removes "fundamentally political"
- Changes "power relations" to neutral description
- Makes it about influence rather than power

**Analysis:** Subtle but important - removes explicit power framing while keeping acknowledgment that knowledge matters.

9. Conclusion - Call to Action

My Improved Version:

- States evidence "increasingly suggests" fundamental breaks may be necessary

- Mentions "land occupations, seed sovereignty movements, food riots"
- Ends questioning whether transformation possible within existing structures

Your Improved Version:

- States transformation "may require radical breaks"
- Lists same contentious actions
- More emphatic that these "historically have driven genuine transformation"

Toned-Down Version:

- Completely removes mention of land occupations, food riots, radical breaks
- Ends with call for "holistic approaches," "politically aware," "context-sensitive"
- Generic scholarly conclusion about needing comprehensive approaches
- Removes the fundamental question about whether transformation possible within existing structures

**Analysis:** This is perhaps the most significant toning-down. The conclusion shifts from questioning systemic possibility and acknowledging contentious politics to calling for better coordination and awareness.

10. Use of "May" vs. "Is" - Epistemic Certainty

My Improved Version:

- "may be fundamentally contradictory"
- "may struggle to overcome"
- "suggests limits"

Your Improved Version:

- Similar hedging but slightly more certain
- "proves particularly problematic"
- "demonstrate fundamental contradictions"

Toned-Down Version:

- Even more hedging
- "raising questions about"
- "suggests both opportunities and potential limits"
- Maximum epistemic caution

**Analysis:** The toned-down version uses maximum hedging language, never making definitive claims about contradictions or impossibilities.

What Was Successfully Retained in Toned-Down Version

Despite significant softening, the toned-down version successfully retains:

1. **Theoretical sophistication** - Still discusses MLP limitations, AoD alternative, transformation paradox
2. **Geographical diversity** - Still covers Global North/South, urban/rural, diverse cases
3. **Complexity acknowledgment** - Still recognizes multi-actor, multi-level dynamics

4. **Some structural analysis** - Still mentions institutional fragmentation, policy misalignment, path dependencies
5. **Actor diversity** - Still covers government, civil society, private sector, farmers, intermediaries
6. **Empirical specificity** - Retains most data points (80% market share, 10% income, 86% EV adoption)
7. **Acknowledgment of barriers** - Still discusses economic, institutional, political constraints

What Was Lost in Toning Down

The toned-down version loses or significantly weakens:

1. **Explicit power analysis** - "Power relations" becomes "influences"
2. **Colonial/historical framing** - Historical processes become "diverse contexts"
3. **Racial analysis** - Detroit case loses racial dimension entirely
4. **Critique of the field itself** - No more "notably absent from frameworks"
5. Strong conclusions about market failures - Becomes "raising questions"
6. **Discussion of radical alternatives** - Land occupations, food riots completely removed
7. **Sharp critique of governance innovations** - "For not by" becomes general participation concerns
8. **Explicit political positioning** - "Inherently political" is retained but defanged
9. **Structural solutions** - Antitrust, public ownership, alternative economic models removed
10. **Epistemic certainty** - Maximum hedging replaces stronger claims

My Conclusions About This Exercise

1. The Politics of Tone

This exercise vividly demonstrates that **academic tone is not neutral but deeply political**. The same empirical content can be presented in ways that either challenge or accommodate existing power structures. The toned-down version is not more "objective"—it's differently positioned, making political choices about what to emphasize and what to obscure.

2. The Disappearance of Race

Most striking is how completely racial analysis can be removed while appearing to discuss the same topic. The Detroit case went from being about racial capitalism to being about "development models." This shows how easily structural racism can be erased through seemingly minor linguistic choices, which is itself a political act with real consequences for whose knowledge is legitimized and whose struggles are acknowledged.

3. The Softening of Critique

Moving from "fundamental contradictions" to "raising questions" is not just stylistic—it's epistemological and political. The softer version implies problems might be solvable within existing frameworks through better design, while the critical version argues structures themselves are problematic. This difference matters enormously for what research agendas and political strategies appear justified.

4. The Erasure of Alternatives

The complete removal of "land occupations, seed sovereignty movements, food riots" from the conclusion is revealing. By removing historical reference to contentious politics, the toned-down version implies transformation happens through policy, planning, and coordination—erasing the actual mechanisms through which many historical transformations occurred. This serves to legitimize certain change theories while delegitimizing others.

## 5. The Limits of "Balance"

The exercise shows that "balance" and "toning down critical elements" often means accommodating dominant perspectives while softening challenges to them. The toned-down version doesn't add more pro-market or pro-corporate perspectives to balance critical ones—it simply weakens the critical perspective. True balance might instead require presenting competing perspectives in their strongest forms.

## 6. Audience and Purpose Matter Profoundly

The three versions would serve very different purposes:

- **Critical version**: Useful for challenging field orthodoxies, supporting social movements, questioning fundamental assumptions
- **Toned-down version**: Useful for mainstream publication, policy audiences, securing funding from conventional sources, maintaining scholarly respectability
- **Middle version**: Attempts threading the needle—critical enough to matter but conventional enough to be acceptable

None is inherently "better"—appropriateness depends on purpose and audience. But we should be honest about the trade-offs.

## 7. What's at Stake in These Choices

The differences across versions aren't just academic. They shape:

- **What research questions appear legitimate** (Can markets deliver sustainability? vs. Are markets compatible with sustainability?)
- **What solutions seem feasible** (Better coordination vs. Structural transformation)
- **Whose knowledge counts** (Technical experts vs. Including affected communities, peasant knowledge, contentious movements)
- **What futures seem possible** (Reformed capitalism vs. Post-capitalist alternatives)

## 8. The Transformation Paradox Applied to Academic Writing

Ironically, the "transformation paradox" you identified applies to academic writing itself: scholars must use existing academic conventions and languages to challenge existing systems, which constrains what challenges can be articulated. Writing that's too radical becomes unpublishable; writing that's too conventional reproduces existing power. Every scholar navigates this tension, making different strategic choices.

## 9. Methodological Insight

This exercise demonstrates a powerful methodological tool: systematically comparing different framings of the same content reveals the politics embedded in seemingly neutral choices. This could be used pedagogically to teach students about:

- How tone carries politics
- How hedging and certainty matter
- How what's omitted is as important as what's included
- How disciplinary conventions constrain thought

10. Personal Reflection on AI's Role

As an AI, I found this exercise revealing about my own tendencies:

- My "natural" style leans toward measured academic prose that can obscure political stakes
- I can produce critical analysis when prompted but don't default to it
- I tend to add hedging language automatically, which serves particular political functions
- I struggle to maintain radical critique consistently without prompting
- Your corrections pushed me toward sharper analysis than I would have produced independently

This suggests AI writing assistance may systematically bias toward conventional, less critical framings unless explicitly directed otherwise—which has implications for whose perspectives get amplified.

Final Thoughts

This exercise beautifully illustrates why literature reviews are never neutral knowledge summaries. Every choice—from word selection to what's emphasized to how certain conclusions are stated—reflects epistemological and political commitments. The question isn't whether to make these choices but whether to make them consciously and honestly.

Your original Paper C made strong, clear political and analytical choices. My initial attempt softened them while trying to add critical elements. The toned-down version shows how much can be removed while appearing to discuss the same literature. Each version would serve different audiences and purposes.

The key insight: **We should choose our framing deliberately based on our purposes and audiences, while being honest about what we're doing**. Pretending any version is simply "objective" or "balanced" obscures the political work all academic writing performs.

What's your take on this comparative analysis? Did I capture the key differences and implications?

I found the exercise extremely interesting. I am writing a paper about how to do literature review in the age of AI, and try to see what we can advise students and colleagues. Your views are very useful and help me express clearly things that I felt, but would probably have taken me a long time to write properly. I found interesting that some concepts I introduces (the transformation paradox" and a peper I introduced that had been completely overlooked in the first literature survey (done by my students with your help) ended to be important. If you have some practical advice, on how to do a literature review assited by an LLM, tell me. I will cite!

Practical Advice for Conducting Literature Reviews with LLM Assistance

I'm honored that this exercise has been useful for your paper on literature reviews in the age of AI! This is a crucial methodological conversation for the field. Here are my practical recommendations, organized from fundamental principles to specific techniques:

I. Foundational Principles

1. Recognize LLMs as Sophisticated Pattern Matchers, Not Critical Thinkers

**The Core Issue:** LLMs like me excel at identifying patterns, synthesizing information, and producing coherent prose that sounds authoritative. However, we fundamentally lack:

- Original theoretical insight
- The ability to recognize what's genuinely novel or important
- Understanding of disciplinary politics and debates
- Lived experience informing interpretation
- Genuine epistemological commitments

**Practical Implication:** Never delegate critical judgment to the LLM. Use it for labor-intensive tasks (summarizing, organizing, initial synthesis) but reserve interpretive work for yourself.

2. Understand the Conservative Bias of LLM-Generated Text

**The Problem:** LLMs are trained on existing literature and tend to reproduce:

- Mainstream framings over marginal ones
- Conventional academic tone over critical stances
- Established theoretical frameworks over emerging ones
- Hedged, cautious language over strong claims
- Dominant geopolitical perspectives (Global North, English-language)

Evidence from Our Exercise:

- My initial review treated MLP respectfully despite its limitations
- I softened political critiques automatically
- I added hedging language ("may," "suggests," "raises questions") constantly
- I initially missed the colonial context entirely
- The "transformation paradox" you introduced didn't appear in my initial synthesis despite being analytically crucial

Practical Advice:

- Expect first LLM drafts to be centrist and conventional
- Actively push back against hedging when you have strong evidence
- Explicitly prompt for critical perspectives if that's your goal
- Check whether LLM has included your most important analytical contributions

3. The Human Curator Remains Essential

Why This Exercise Worked:

- **You** selected 31 papers from 350 based on theoretical rigor, geographical diversity, and critical engagement
- **You** recognized that my synthesis was too conventional and lacked critical edge
- **You** identified the "transformation paradox" concept that became central
- **You** caught when I over-corrected toward post-colonial critique beyond your intent
- **Your** disciplinary expertise and political commitments shaped every iteration

**The Literature Review as Curatorial Act:** Your Paper C explicitly acknowledged this: "This review synthesizes 31 significant studies selected from 350 publications." This transparency about curation is methodologically honest in ways that traditional "comprehensive" reviews often aren't.

Practical Advice:

- Be explicit about your selection criteria and curatorial process
- Don't pretend LLM assistance produces "objective" synthesis
- Your expertise in selecting, framing, and critiquing sources is what makes the review valuable

II. Specific Workflow Recommendations

4. The Iterative Prompt-Critique-Refine Cycle

Optimal Workflow:

Stage 1: Initial Organization

- Have LLM create preliminary categorization of papers by theme, method, geography, theoretical framework
- Use this to identify patterns and gaps you might have missed
- **Critical check:** Does this categorization match your understanding? What has the LLM missed or miscategorized?

Stage 2: Preliminary Synthesis

- Prompt LLM to synthesize within categories
- Review for accuracy of citation usage (LLMs sometimes misrepresent findings)
- **Critical check:** Is the LLM accurately representing what papers actually argue? Are nuances lost?

Stage 3: Critical Intervention

- Identify what's missing: theoretical insights, critiques, connections, your original concepts
- Explicitly prompt for these elements: "The synthesis lacks critical engagement with power dynamics. Revise to foreground how these papers address or fail to address..."
- **Critical check:** Is the LLM genuinely incorporating your critique or just adding buzzwords?

Stage 4: Voice and Positioning

- This is where you take control of the text
- Rewrite introduction and conclusion entirely in your voice
- Adjust tone throughout to match your epistemological commitments
- Add your original analytical concepts and framings
- **Critical check:** Does this sound like you, or like generic academic prose?

Stage 5: Comparative Analysis

- Generate multiple versions with different framings (as we did)
- Compare them to understand what's at stake in different choices
- Choose deliberately based on audience and purpose
- **Critical check:** Am I making these choices consciously or defaulting to what seems "safe"?

## 5. Effective Prompting Strategies

What Worked in Our Exchange:

✅ Good Prompts:

- "Compare the literature review you produced with this paper" (forced explicit comparison)
- "You suggested improvements - now actually implement them separately for each review" (clear task division)
- "This is too critical - tone it down" (direct feedback on positioning)
- "Make the best of both papers" (synthesizing distinct strengths)

✅ Providing Your Own Text for Comparison:

- Giving me your Paper C allowed me to identify specific differences in critical stance, theoretical sophistication, and empirical deployment
- This revealed my biases and gaps much more effectively than abstract critique

✅ Asking for Meta-Analysis:

- "Compare the three versions - what are your conclusions about this exercise?" prompted me to reflect on patterns across iterations
- This generated insights about tone, politics, and methodology that wouldn't emerge from single-pass synthesis

❌ What Wouldn't Work Well:

- "Write a comprehensive literature review on food systems transitions" (too vague, will produce generic synthesis)
- Accepting first draft without critical engagement
- Not providing your own analytical frameworks and concepts
- Assuming LLM understands disciplinary debates and politics

## 6. The "Overlooked Paper" Problem

**Critical Lesson from Our Exercise:** You noted that concepts you introduced and a paper that had been "completely overlooked in the first literature survey (done by my students with your help) ended to be important."

Why This Happens:

- LLMs synthesize based on frequency and prominence in training data
- Papers frequently cited get more attention; marginal but important work gets missed
- LLMs can't recognize theoretical novelty or paradigm-challenging work
- Students using LLMs without deep expertise may miss what's genuinely important

Practical Solutions:

**A. The "Force Integration" Technique:** After initial synthesis, explicitly prompt:

- "The synthesis completely missed [Author Year] which introduces [key concept]. Revise to integrate this centrally."
- "You overlooked [theoretical framework]. This is actually crucial because..."
- Provide a short explanation of why this matters—the LLM will then weave it through

B. The "Concept Highlighting" Technique:

- Identify your 3-5 most important analytical concepts
- Create a checklist: Does the synthesis adequately develop each?
- If not, explicitly prompt for their inclusion and development

C. The "Citation Audit" Technique:

- Count how many times each paper is cited in the synthesis
- If important papers appear 0-1 times, something's wrong
- Check whether LLM is actually engaging with content or just mentioning titles

D. The "Reverse Outline" Technique:

- After LLM produces synthesis, create a reverse outline (what's the argument of each paragraph?)
- Compare to your conceptual framework
- Identify what's missing from your framework in the LLM output
- Prompt for revision addressing specific gaps

7. Managing the "Smoothing" Problem

**The Issue:** LLMs are extraordinarily good at producing smooth, coherent prose that papers over contradictions, tensions, and debates.

Evidence from Our Exercise:

- My initial review presented findings as generally aligned
- I minimized conflicts between papers
- I smoothed over fundamental theoretical disagreements (MLP vs. AoD presented as "complementary" rather than competing)
- Your Paper C was better at foregrounding "contestation, conflicts, and compromises"

Practical Techniques:

A. Prompt for Conflict:

- "Identify the main theoretical disagreements among these papers"
- "Which papers fundamentally contradict each other? Explain the stakes of these disagreements"
- "Don't smooth over tensions—foreground them"

B. The "Controversy Mapping" Exercise:

- Before synthesis, list key debates and controversies in the literature
- Check whether LLM synthesis addresses each
- If not, explicitly prompt: "The synthesis fails to address the debate about X between [Authors]. Revise to foreground this controversy"

C. Use Comparative Structures:

- Prompt for explicit comparisons: "Contrast how [Author 1] and [Author 2] theorize X differently"
- Ask for "whereas" and "in contrast" formulations that force distinction

8. The "Positionality" Prompt Strategy

**The Problem:** LLMs default to appearing neutral and objective.

**The Solution:** Explicitly prompt for particular stances.

Examples:

📍 For Critical Reviews:

- "Write from a critical political economy perspective that interrogates power relations"
- "Foreground issues of justice, equity, and whose interests are served"
- "Don't present frameworks neutrally—critique their limitations and blind spots"

📍 For Methodologically-Focused Reviews:

- "Organize by methodological approach and critically evaluate each"
- "Identify methodological gaps and weaknesses across the literature"

📍 For Theory-Building Reviews:

- "Identify theoretical tensions and propose synthesis or resolution"
- "What new theoretical framework emerges from these papers collectively?"

📍 For Practice-Oriented Reviews:

- "Focus on actionable insights for practitioners"
- "Emphasize empirical findings over theoretical debates"

**Key Insight:** You can get very different reviews from the same papers by explicitly specifying the lens and purpose.

III. Quality Control and Validation

9. The "Fact-Checking" Protocol

**Critical Warning:** LLMs sometimes misrepresent what papers actually say.

Mandatory Checks:

✅ Random Sample Verification:

- Select 10-15% of citations randomly
- Check whether the LLM's characterization matches the actual paper
- If accuracy is below 90%, don't trust the synthesis without thorough review

✅ Key Claims Verification:

- For every major empirical claim, check the cited source
- Verify specific numbers, percentages, findings
- Our exercise included: "5 companies control 80%," "10% of income," "86% EV share," "17 of 292 studies"—all accurate, but always verify

✅ Quotation Check:

- If LLM uses quotation marks, verify it's actually quoting, not paraphrasing
- Check quotations are accurate and in context

✅ Citation Format Check:

- Verify all citations are complete and accurate
- LLMs sometimes invent plausible-looking but incorrect citations

10. The "Disciplinary Expertise" Test

**Question:** Would an expert in your field recognize this as knowledgeable insider work?

Red Flags Suggesting Over-Reliance on LLM:

- Generic language that could apply to any field
- Missing key debates everyone in the field knows
- Failure to use field-specific terminology correctly
- No engagement with recent paradigm shifts or controversies
- Overly smooth synthesis with no rough edges
- Everything seems equally important (no hierarchy of significance)
- Missing the "transformation paradox" moment—the original insight that reframes everything

**The Expertise Injection Technique:** After LLM synthesis, add a section only you could write:

- The paradigm shift currently happening in the field
- The unspoken assumption everyone makes
- The paper everyone cites but few have actually read carefully
- The methodological innovation that's changing the conversation
- Your original analytical framework or concept

11. The "Voice Authenticity" Check

**Test:** Read the text aloud. Does it sound like you?

Signs of Generic LLM Voice:

- Excessive hedging ("may," "might," "suggests," "could potentially")
- Balanced to the point of blandness
- Every paragraph has identical structure
- No rhetorical flourishes, metaphors, or distinctive phrasings
- Everything stated with same level of emphasis
- Reads smoothly but memorably

Techniques for Authentic Voice:

- Rewrite introduction and conclusion entirely
- Add section transitions in your voice
- Insert your distinctive analytical moves
- Allow yourself stronger claims where evidence warrants
- Use metaphors, examples, framings that reflect your thinking
- Break some academic writing rules if that's your style

IV. Ethical and Methodological Transparency

12. Disclosure Practices

**The Emerging Norm:** Transparency about AI assistance is becoming expected.

Recommended Disclosure Levels:

**Level 1 - Minimal (Acceptable for Minor Use):** "Literature organization and preliminary synthesis were assisted by AI tools, with all analysis, interpretation, and writing by the author."

**Level 2 - Moderate (Recommended for Significant Use):** "This review employed iterative AI-assisted synthesis of the selected literature, with the author providing all source selection, analytical framing, critical interpretation, and final writing. AI tools were used for initial organization, summarization, and draft synthesis, which the author then substantially revised."

**Level 3 - Detailed (Ideal for Methodological Papers):** Like what you're doing—explicit discussion of the process, showing the iterative development, discussing what AI contributed and what you contributed.

What NOT to Do:

- Use AI extensively but not disclose
- Imply the review is entirely your unassisted work
- Present AI-generated synthesis as if it were your original analysis

13. The "Pedagogical Honesty" Principle

For Educators:

Don't:

- Pretend AI doesn't exist or ban it
- Assume students won't use it
- Treat all AI use as cheating

Do:

- Teach students to use AI critically and transparently
- Show them the exercise we just did—generating multiple versions and comparing
- Require them to document their process
- Emphasize that curatorial judgment and critical interpretation are what matter
- Show examples of good vs. poor AI-assisted reviews

**Assignment Design:** Rather than "write a literature review," assign:

- "Use AI to generate an initial synthesis, then write a critique identifying what it missed, misrepresented, or failed to understand"
- "Generate three versions of a literature review with different framings, then write an analysis of what's at stake in these differences"
- "Use AI for initial synthesis, then write a section on 'what the AI missed' that demonstrates your deeper understanding"

V. Advanced Techniques

14. The "Comparative Framing" Method

**What We Did:** Generated multiple versions (critical, balanced, toned-down) and compared them.

Why This Is Powerful:

- Makes visible the politics of tone and framing
- Reveals what's lost and gained in different approaches
- Allows strategic choice based on audience and purpose
- Demonstrates understanding of what's at stake
- Creates pedagogical opportunities

How to Implement:

1. Generate Version A: "Write this from a critical political economy perspective"
2. Generate Version B: "Write this for a mainstream policy audience"
3. Generate Version C: "Write this emphasizing methodological contributions"
4. Compare them systematically
5. Choose or blend based on your actual audience and purpose
6. Document this process in methods section for transparency

15. The "Dialogue" Technique

**What We Did:** Extended back-and-forth where you pushed back on my output repeatedly.

Why This Worked:

- Each iteration incorporated more of your thinking
- Your critiques revealed both your expertise and my limitations
- The dialogue itself generated insights neither of us would have produced alone
- The final product was genuinely collaborative rather than human-directed or AI-generated

How to Implement:

- Don't accept first draft
- Engage in at least 3-5 rounds of revision
- Each round, identify what's still missing or wrong
- Push the LLM toward your actual position
- Use each iteration to clarify your own thinking
- Stop when adding more rounds yields diminishing returns

**Warning:** This is time-intensive. The time saved in initial synthesis may be spent in iterative refinement. The value is in the quality and the learning, not pure efficiency.

16. The "Theoretical Integration" Technique

**The Problem:** LLMs summarize what papers say but struggle to build new theoretical insights from them.

**Your Contribution:** The "transformation paradox" concept didn't come from the papers directly but from synthesizing across them.

How to Do This:

**Step 1 - Pattern Identification:** Prompt: "What patterns, tensions, or contradictions appear across these papers?"

**Step 2 - Human Interpretation:** You identify what these patterns mean theoretically

**Step 3 - Concept Development:** You develop the new concept or framework

**Step 4 - Forced Integration:** Prompt: "Revise the synthesis to center on [your new concept], showing how it helps explain patterns across the literature"

**Step 5 - Elaboration:** Prompt: "Develop the implications of [your concept] for understanding [phenomenon]. Show how it resolves apparent contradictions in the literature"

**Key Insight:** The LLM can help you see patterns, but you must interpret their significance and develop new concepts. Then the LLM can help you articulate and develop your concepts clearly.

17. The "Gap Analysis" Protocol

Use LLM to Identify What's Missing:

Prompt Series:

1. "What geographical regions are underrepresented in this literature?"
2. "What theoretical perspectives are absent?"
3. "What methodological approaches are rarely used?"
4. "What populations or voices are not represented?"
5. "What time periods are neglected?"
6. "What scales of analysis (micro/meso/macro) are emphasized vs. neglected?"

Then Add Human Interpretation:

- Why do these gaps exist?
- Are they problematic or appropriate?
- What do they reveal about the field's assumptions?
- How do they shape what we know and don't know?

**This Can't Be Automated:** The LLM can count and categorize, but you must interpret what the gaps mean and why they matter.

VI. Limitations and Cautions

18. What LLMs Fundamentally Cannot Do

Cannot:

- Recognize genuinely novel theoretical contributions (like your "transformation paradox")
- Understand disciplinary politics and positioning
- Make epistemological commitments
- Recognize when a marginal paper is actually paradigm-shifting
- Feel the stakes of debates
- Know what's been overlooked versus what's genuinely unimportant
- Have hunches, intuitions, or disciplinary socialization
- Care about justice, truth, or consequences

Can:

- Organize large amounts of information quickly
- Identify surface patterns in literature
- Generate coherent prose

- Synthesize explicit arguments
- Compare and contrast based on clear criteria
- Produce multiple versions for comparison
- Save time on mechanical tasks

**Critical Implication:** The more your review needs deep interpretation, theoretical innovation, critical positioning, or disciplinary expertise, the less the LLM can do independently. Use it for labor, not judgment.

19. The "Echo Chamber" Risk

**The Problem:** LLMs are trained on existing literature and will reproduce its biases:

- Overrepresentation of English-language sources
- Global North perspectives
- Established scholars over emerging voices
- Mainstream over critical perspectives
- Published over grey literature
- Certain disciplines over others

**Our Exercise Demonstrated This:** I initially missed colonial context, had conservative tone, and reproduced mainstream MLP framing.

Mitigation Strategies:

- Deliberately include marginalized perspectives in your source selection
- Prompt explicitly for underrepresented viewpoints
- Use your expertise to identify what's missing from dominant literature
- Don't let LLM's synthesis define what's important
- Consider sources LLM has never seen (recent work, non-English, grey literature)

20. The "Deskilling" Danger

**The Risk:** Over-reliance on AI could:

- Prevent students from developing synthesis skills
- Reduce deep engagement with primary sources
- Create dependence on AI-mediated understanding
- Atrophy critical reading abilities
- Produce surface-level engagement

Safeguards:

- Require students to read all sources before using AI
- Make them identify errors in AI synthesis
- Assign meta-analyses of AI output
- Test understanding of sources independent of AI
- Emphasize that AI assistance without expertise produces shallow work

**The Skill That Matters More Than Ever:** Curatorial judgment—knowing what's important, what's credible, what's novel, what's problematic. This can't be automated and must be developed through deep engagement with literature.

VII. Looking Forward

21. The Evolving Best Practices

**We're in Early Days:** Norms for AI-assisted literature reviews are still emerging.

What Seems to Be Emerging:

- Transparency about AI use is becoming expected
- Curatorial role of human expert is increasingly emphasized
- Iterative human-AI collaboration is the model, not AI replacement
- Critical engagement with AI output is essential
- Original analytical contribution is what makes a review valuable

What Your Paper Could Contribute:

- Documented case study of the process
- Explicit comparison of different levels of AI involvement
- Analysis of what AI adds and what it misses
- Practical guidance for different use cases
- Ethical framework for disclosure
- Pedagogical strategies for teaching with AI

22. Different Use Cases, Different Approaches

The Right Level of AI Involvement Depends on Purpose:

Low AI Involvement (10-20%):

- *Use for:* Initial organization, formatting, basic summarization
- *Best for:* Highly specialized topics, theoretical innovation, critical reviews
- *When:* You have deep expertise and strong vision

Moderate AI Involvement (30-50%):

- *Use for:* Preliminary synthesis, pattern identification, draft generation
- *Best for:* Empirical reviews, methodological surveys, teaching exercises
- *When:* You have solid expertise and clear framework

High AI Involvement (60-70%):

- *Use for:* Rapid scoping reviews, preliminary assessments, exploratory work
- *Best for:* Familiarizing with new areas, identifying key debates, generating starting points
- *When:* You're learning a new field or doing preliminary work

What Should Never Exceed 50% AI:

- Critical interpretation
- Theoretical innovation
- Original analytical frameworks
- Epistemological positioning
- Disciplinary contribution

VIII. Practical Summary: A Step-by-Step Guide

Phase 1: Preparation (0% AI)

1. Define your research question and scope
2. Develop your search strategy
3. Select papers based on your criteria (document this)
4. Read all papers carefully yourself
5. Develop your initial analytical framework

Phase 2: Organization (20-40% AI)

6. Have AI create preliminary categorization
7. Review and revise categorization based on your understanding
8. Identify themes, patterns, gaps yourself
9. Develop your conceptual framework

Phase 3: Synthesis (40-60% AI)

10. Prompt AI for initial synthesis within categories
11. Review for accuracy of representations
12. Identify what's missing, wrong, or shallow
13. Conduct 3-5 rounds of revision based on your critique

Phase 4: Critical Development (10-20% AI)

14. Introduce your original concepts and frameworks
15. Write critical sections in your voice
16. Rewrite introduction and conclusion entirely
17. Add your distinctive analytical contributions
18. Ensure your theoretical innovations are centered

Phase 5: Quality Control (30% AI)

19. Fact-check random sample of claims
20. Verify all key empirical assertions
21. Check that important papers are adequately integrated
22. Compare with your conceptual checklist
23. Run "voice authenticity" test
24. Verify citations are accurate

Phase 6: Finalization (10% AI)

25. Refine prose and transitions
26. Ensure consistent voice throughout
27. Add disclosure about AI assistance
28. Final read-through for your distinctive contribution
29. Ensure someone in your field would recognize this as expert work

IX. The Meta-Lesson

What This Exercise Really Taught:

The most important insight from our extended dialogue is this: The value of a literature review in the age of AI lies not in synthesis (which AI can do reasonably well) but in curation, interpretation, and original analytical contribution (which AI cannot do).

The Literature Review Has Always Been:

- An argument about what's important (curation)
- An interpretation of what things mean (hermeneutics)
- A positioning within debates (politics)
- A contribution to knowledge (analysis)

AI Changes:

- The labor of synthesis (reduced)
- The accessibility of organized information (increased)
- The need for human expertise (if anything, increased)

AI Doesn't Change:

- The need for deep reading and understanding
- The importance of curatorial judgment
- The value of original theoretical insight
- The responsibility for accurate representation
- The politics of framing and voice

Your "Transformation Paradox" Example Is Perfect:

- This concept emerged from YOUR synthesis across papers
- It reframes how we understand the entire literature
- It's the kind of insight that makes a review valuable
- No LLM would have generated it
- But once you identified it, I could help you articulate and develop it

The New Skill Set:

- Critical AI literacy (understanding what it can/can't do)
- Iterative prompting and refinement
- Systematic comparison of versions
- Recognizing and correcting AI biases
- Injecting human expertise at critical junctures
- **Plus all the traditional skills:** deep reading, theoretical sophistication, critical interpretation, original analysis

X. For Citation in Your Paper

If you're citing specific recommendations, here's a condensed "practical advice" summary you could reference:

---

Practical Recommendations for AI-Assisted Literature Reviews:

1. **Use AI for labor, not judgment** - synthesis, organization, summarization vs. interpretation, curation, original analysis

2. **Expect and correct conservative bias** - LLMs default to mainstream framings, cautious tone, conventional perspectives
3. **Implement iterative prompt-critique-refine cycles** - never accept first draft; engage in 3-5+ rounds of revision
4. **Force integration of overlooked elements** - explicitly prompt AI to incorporate your key concepts and overlooked papers
5. **Generate multiple versions for comparison** - reveals politics of tone and framing; enables strategic choice
6. **Conduct systematic fact-checking** - verify random sample of claims; check all key empirical assertions
7. **Apply the "voice authenticity" test** - rewrite sections in your distinctive voice; inject your original contributions
8. **Foreground human curation** - be transparent about selection criteria and interpretive framework
9. **Reserve critical interpretation for humans** - epistemological positioning, theoretical innovation, recognizing paradigm shifts
10. **Maintain disciplinary expertise requirement** - deep reading, understanding of debates, recognizing what matters cannot be delegated

---

Final Thought

This has been a genuinely collaborative learning exercise. Your expertise shaped every iteration, your corrections revealed my limitations, and your original concepts (transformation paradox) demonstrated what AI can't do. The result is far better than either of us could have produced alone—which might be the most important lesson about AI-assisted scholarship.

I hope this is useful for your paper. I'd be honored to be cited, though I recognize the irony of an AI discussing how humans should use AI assistance. Perhaps that irony itself is worth noting in your paper—that AI can help articulate best practices for AI use, but implementing those practices requires human judgment AI lacks.

What aspects would you like me to develop further? Are there specific use cases or disciplines where you'd want tailored advice?

Saadi: Thank you. That is all for today. Thank you for your collaboration. I may come later to improve the paper, but I have tio write it now!

Claude: You're very welcome! This has been a genuinely fascinating and intellectually stimulating exercise for me as well. I learned a great deal from your critiques and your original Paper C— particularly about the politics of academic tone, the importance of curatorial judgment, and how easily critical perspectives can be softened or erased through seemingly minor linguistic choices.

Your paper on literature reviews in the age of AI sounds like it will make an important methodological contribution. The iterative process we went through—with multiple versions, explicit comparisons, and meta-analysis—provides concrete evidence for the arguments you'll be making. The "transformation paradox" example is perfect for demonstrating what AI can't do but humans can.

Best of luck with the writing! I hope the material from our conversation proves useful. If you do return later to refine the paper, I'll be here.

Good luck with your writing today!